\documentclass[a4paper,11pt]{book}

\usepackage[latin1]{inputenc}
\usepackage{amsmath,amsfonts,amscd,amssymb}
\usepackage[T1]{fontenc}
\usepackage[english]{babel}
\usepackage[dvips]{graphicx}
\usepackage{epsfig}
\usepackage{changebar}

\usepackage[all]{xy}

\usepackage{fancybox}
\usepackage{fancyhdr}
\usepackage{vmargin} 

\setmarginsrb{2.5cm}{2cm}{2cm}{2cm}{14mm}{6mm}{0mm}{10mm}

\fancypagestyle{plain}{
\fancyhf{}

\fancyfoot[LE,RO]{\thepage} 
 }

\makeatletter

\newtheorem{theorem}{Theorem}[section]

\newtheorem{corollary}[theorem]{Corollary}

\newtheorem{definition}[theorem]{Definition}
\newtheorem{example}[theorem]{Example}

\newtheorem{lemma}[theorem]{Lemma}

\newtheorem{proposition}[theorem]{Proposition}
\newtheorem{remark}[theorem]{Remark}

\def\KK{{\mathbb K}}
\def\NN{{\mathbb N}}
\def\CC{{\mathbb C}}
\def\RR{{\mathbb R}}
\def\NN{{\mathbb N}}
\def\ZZ{{\mathbb Z}}
\newcommand{\beq}{\begin{equation}}
\newcommand{\eeq}{\end{equation}}
\newcommand{\beqa}{\begin{eqnarray}}
\newcommand{\eeqa}{\end{eqnarray}}
\newcommand{\noi}{\noindent}
\newcommand{\e}{\varepsilon}
\newcommand{\p}{\varphi}
\newcommand{\ttA}{{\tilde {\tilde A}}}
\newcommand{\tA}{\tilde  A}
\newcommand{\ttB}{{\tilde {\tilde B}}}
\newcommand{\tB}{\tilde  B}
\newcommand{\ttP}{{\tilde {\tilde \varphi}}}
 \newcommand{\tP}{\tilde  \varphi}
\newcommand{\hhA}{{\hat {\hat A}}}
\newcommand{\hA}{\hat  A}
\newcommand{\hhB}{{\hat {\hat B}}}
\newcommand{\hB}{\hat  B}
\newcommand{\hhP}{{\hat {\hat \varphi}}}
 \newcommand{\hP}{\hat  \varphi}
\newcommand{\tpsi}{\tilde \psi}
\newcommand{\ttpsi}{{\tilde{ \tilde \psi}}}
\newcommand{\ttl}{{\tilde{\tilde  \lambda}}}
\newcommand{\tl}{\tilde \lambda}
\newcommand{\tr}{\tilde \rho}
\newcommand{\ttr}{{\tilde{\tilde  \rho}}}

\newcommand{\g}{{\mathfrak g}}
\newcommand{\h}{{\mathfrak h}}
\newcommand{\pard}{[\!\! [}
\newcommand{\pari}{]\!\! ]}
\newcommand{\3}{3SUSY}
\newcommand{\n}{{\mathfrak n}}
\newcommand{\rr}{{\mathfrak r}}

\newcommand{\s}{{{\mathfrak{sl}}(2)}}
\newcommand{\A}{{\mathrm {Aut}}}
\def\>{\rangle}
\def\<{\langle}

\begin{document}

\begin{titlepage}
{    
\hspace{-2cm}
\vspace{-2cm}
\parbox[t]{9cm}


\begin{tabular}{lr}
\includegraphics{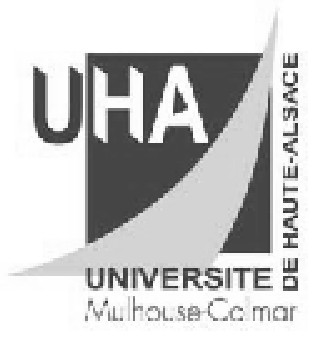}\hspace{6cm} & \includegraphics{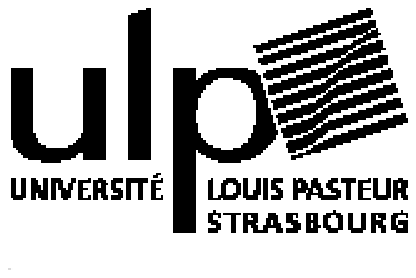}\\
Universit\'e de Haute Alsace & Universit\'e Louis Pasteur \\
Facult\'e de Sciences et Techniques & U.F.R. de Sciences Physiques
\end{tabular}


\vskip 1.5cm
\begin{center}
\Large{

{\bf{\underline{THESE}}}

\vskip 0.1cm

pour obtenir le grade de\\ 

\vskip 0.1cm

DOCTEUR DE L'UNIVERSITE DE HAUTE ALSACE\\ \vskip 0.1cm


pr\'esent\'ee et soutenue publiquement\\ 
 
\vskip 0.1cm

par\\

\vskip 0.3cm

{\bf{ADRIAN TANAS\u{A}}}

\vskip 0.2cm

le 15 septembre 2005

\vskip 0.3cm

\underline{Titre:}

\vskip 0.1cm

Sous-alg\`ebres de Lie de l'alg\`ebre de Weyl.\\
Alg\`ebres de Lie d'ordre $3$ et leurs applications \`a la supersym\'etrie cubique

\vskip 0.5cm

\underline{Co-directeurs de th\'ese:}

\vskip 0.1cm

- Michel Goze

- Michel Rausch de Traubenberg

- Marcus Julius Slupinski

\vskip 0.5cm

\underline{JURY}

\vskip 0.1cm

M. le professeur Yu Khakimdjanov\\
M. le professeur Richard Kerner\\
M. le professeur Hubert Rubenthaler\\
M. le professeur Michel Goze\\
M. Michel Rausch de Traubenberg (MdC Habilit\'e)\\
M. Marcus Julius Slupinski (MdC Habilit\'e)
}
\end{center}
}
\end{titlepage}

\pagestyle{empty}

\cleardoublepage



${}$

\vspace{9cm}

\begin{flushright}
{\Large{Familiei mele}}
\end{flushright}



\newpage
\pagestyle{empty}

{\Large{Acknowledgements}}

\bigskip

\vskip 1cm

For this kind of situations my first thoughts go directly to my family, which by its moral support made me not to feel alone, no matter where I was geographically. I cannot measure in words or anything else the gratitude I owe them, so I allow myself to stop here on this issue.

\medskip

Secondly my thanks go to my PhD advisors, M. Goze, M. Rausch de Traubenberg and M. Slupinski whom have welcomed me in Alsace at my coming here and whom had the patience to teach me some of the many things I did not know.

I am grateful to G. Moultaka who had always taken time to answer my questions; I also wish to thank D. Bennequin, T. Levasseur, H. Rubenthaler  and R. Stanton 
 for suggesting us different references for chapter $3$ of this thesis.

Moreover I would like to thank the staffs of the Laboratories of Mulhouse and Strasbourg where the work of this PhD was done.

I would also wish to record my gratitude to the members of my jury who taken of their time to report on my work.

\medskip

My thoughts go then to my mates ``tonton'' Franckie Stauffer and Raul ``Babyface'' Dillenschneider which whom I can say I have suffered together during all these years.  Practically in the same category I can place ``Platon de Rammersmatt'' Goze and ``Dr. Arschlecker'' Kenoufi whose moral support helped me to bypass the more difficult psichologic situations I have encountered during these years.

Last but not least I also need to thank my friends from the ``exterior world'', Toni Bissoni and everybody else, whom helped to change ideas and see other type of persons and facts during this period in Strasbourg.











\pagestyle{fancy}

\tableofcontents

\chapter{Introduction}

\vskip 0.5cm


    The interplay between theoretical physics and algebra is a basic feature of science of today. Maybe the most important tool of this interplay is the {\it symmetries}, which on one hand play a crucial role in theoretical physics models and on the other hand, from the mathematical point of view, are described by different types of algebraic structures (amongst which of course the most frequently used are Lie algebras and then, in extended frameworks, Lie superalgebras).

    Symmetries, by their beauty, simplicity and harmony become almost an obsession for the human scientific spirit and its thirst of knowledge. Indeed, the concept of symmetry evolved since the ancient Greeks
\footnote{The word symmetry has a Greek origin, {\it syn} meaning ``with'' and {\it metron} ``measure''.}
 (which were considering  the circles and spheres as nature's perfect symmetries) until the symmetries of the Standard Model of particle physics and beyond. 
In fundamental physics of today, one can state as a principle the fact that symmetry dictates dynamics. Indeed, once one has the algebraic structure which groups the requested symmetries, irreducible representations are to be looked for. These give rise to multiplets with whom one can construct invariant Lagrangians, the fundamental notion of field theory. Once the Lagrangians are obtained, the equations of motion show the dynamics of the fields of the model.

\medskip

In the framework of the Standard Model, the symmetries of nature are divided into two distinct categories, the symmetries of space-time (the  rotations, Lorentz boosts and translations, all grouped by the Poincar\'e algebra, see subsection $2.3.2$)
and internal symmetries, each of them associated to one fundamental interaction, the strong, weak and electromagnetic interactions, thus by phenomenological reasons giving the algebra ${\mathfrak{su}}(3)\times {\mathfrak{su}}(2)\times {\mathfrak{u}}(1)$.
Within this framework, the Coleman-Mandula no-go theorem indicates what  are the most general  Lie algebras one can use to group symmetries (see subsection $5.1.2$). Nevertheless, from different theoretical and experimental reasons exposed briefly in subsection $5.1.1$, particle physicists need today to look for new physics beyond the Standard Model. Maybe one of the most appealing candidates for this new physics is supersymmetry (SUSY). Supersymmetry evades the no-go theorems by using a more general algebraic structure, Lie superalgebras (see subsection $5.1.2$).  This 
interest coming from particle physics implied the development of the study of superalgebras also from the mathematical point of view. What basically SUSY does is to extend the Poincar\'e symmetry by some new generators, the supercharges $Q$ which close with anticommutation relations. 

However, at the present state of art, nothing forbids the possibilities of grouping symmetries in more exotic algebraic structure. Here we will use Lie algebras of order $3$ to do this. We chose a particular extension of the Poincar\'e algebra, different of the supersymmetric one and we then show a way to implement it at the level of a field theoretical model.

\medskip




    This thesis is organised as follows. In the next chapter we briefly overview some basic notions concerning Lie algebras and Lie superalgebras, notions which will prove to be useful for the sequel. In the third chapter we study what are the finite-dimensional Lie algebras which can be realised in the Weyl algebra $A_1$. The list of Lie algebras obtained turns to be  discrete. Furthermore, we then analyse in more detail a particular class of realisations of $\s$.

    In the fourth chapter we introduce the notion of Lie algebras of order $F$ and we then focus on the case $F=3$ for which some non-trivial examples are given. We then set the basis of the study of deformations for these algebraic structure. Furthermore, an equivalent, binary  formulation is  introduced and analysed.

    We then pass along to some possible applications to theoretical physics models. After a short preamble where we recall briefly the state of the art of particle physics today (namely giving a short glimpse of supersymmetry) we then begin the study of cubic supersymmetry, the field theoretical model that we try to construct upon a particular Lie algebra of order $3$ (algebra extending non-trivially the Poincar\'e symmetry). In the rest of chapter $5$ we begin this study in four space-time dimensions. After a recall of the  algebra, we will exhibit some irreducible matrix realisations. Bosonic multiplets are obtained and then a free theoretical model is constructed. Compatibility with abelian gauge invariance and possibilities of interactions are closely analysed.
    The last chapter is dedicated to the generalisation of this model to arbitrary dimensions. We obtain here a symmetry acting at the level of antisymmetric gauge fields, with whom explicit invariant Lagrangian are obtained.
    Finally, a certain number of appendices concerning different calculations useful at different points of the text are given at the 
end of this thesis.

\medskip

If chapters $2$ and $5$ are some brief overview of existing results on Lie (super)algebras and respectively no-go theorems in particle physics, the rest of this thesis basically presents results of \cite{io3} (chapter $3$), \cite{io1} (chapter $5$), \cite{praga} (chapter $6$), chapter $4$ being the main body of a forthcoming publication.

\chapter{Lie algebras and Lie superalgebras. A brief overview}

     In this chapter, we recall definitions of the mathematical structures that will be needed later on. 
    As general references, one can consult the books \cite{FuchsSchweigert} of J. Fuchs and C. Schweigert, \cite{sorba} of L. Frappat, A. Sciarrino and P. Sorba and, for a more physical approach, the book \cite{georgi} of H. Georgi; for a more mathematical approach one can consult the book \cite{jacobson} of Jacobson.

\section{Definition of Lie algebras}
\subsection{Definitions}




A \textit{Lie algebra} $\mathfrak{g}$ over $\KK$ is a vector space with a bilinear law 
$$[.,.]:\g\times\g\to\g$$ 
(called  \textit{Lie bracket} or  \textit{commutator})
 satisfying
\begin{enumerate}
\item $[x,x]=0$, for all $x \in \mathfrak{g}$  
\item $[x,[y,z]]+[y,[z,x]]+[z,[x,y]]=0$, for all $x,y,z \in \mathfrak{g}$ (\textit{Jacobi identity}).
\end{enumerate}


The first property implies 
the antisymmetry of the Lie bracket: $[x,y]=-[y,x]$, for any $x,y\in \g$.

   The \textit{dimension} of a Lie algebra is its dimension considered as a $\KK-$vector space. The characteristic of $\KK$ is different of two. Here we consider $\KK$ to be $\RR$ or $\CC$.

If the dimension $d$ is finite or countably infinite, one uses the notation $T^a$ to refer to the \textit{generators of the Lie algebra} and the equation:
\beqa
\label{constantesdestructure}
[T^a,T^b]=f^{ab}_{\mbox{            }c} T^c, 
\mbox{ with }f^{ab}_c\in \KK
\eeqa
defines the scalars $f^{ab}_{\mbox{            }c}$, called the \textit{structure constants} of the Lie algebra. 

In physics literature, the convention used is
$$[T^a,T^b]=i f^{ab}_{\mbox{  }c} T^c. $$ 
Here, when $\KK=\RR$ one has $f^{ab}_c\in i\RR$; if we are interested in algebras that have unitary representations (which is the case in physics models), then $f^{ab}_c\in \RR$. 
The reason for picking this convention is that, in quantum mechanics, the commutators of two self-adjoint operator proves to be anti-self-adjoint. 


In this thesis we use the convention (\ref{constantesdestructure}).

\medskip


   If $\KK$ is the field of real numbers $\RR$, we speak of a \textit{real Lie algebra} and if $\KK$ is the field of complex numbers $\CC$, we speak of a \textit{complex Lie algebra}. Here we will be concerned about these two particular types of algebraic structures.

\subsection{Examples}
\label{exemple-LA}

{\it i)} To any $({\mathfrak A},\mu)$ an associative algebra, one can associate a Lie algebra, denoted by $\g_{\mathfrak A}$, as follows 
\beqa
\label{asso-lis}
[x,y]=\mu (x,y) -\mu (y,x),\ \forall x,y.
\eeqa

More generally, a Lie admissible algebra is a non-associative algebra (the category of non-associative algebras being larger than the category of associative algebras) for whom $\mu (x,y) -\mu (y,x)$ is a Lie bracket \cite{remm}.
The study of this class of admissible Lie algebras is of special importance for hadronic physics.


\medskip

\noi
{\it ii)} The \textit{(first) Weyl algebra} $A_1$ is the complex unitary associative algebra generated by two generators $p,\ q$ subject to the relation 
\beqa
\label{Weyl-def}
\mu(p,q)-\mu(q,p)=1.
\eeqa

A part of this thesis  will be devoted to the study of subalgebras of  $A_1$.



\medskip



\noi
\textit{iii)} {\it Angular momentum}. In quantum physics one has the usual commutation relations between the components of angular momentum
\beqa
\label{su2}
[J_i,J_j]=\e_{ij}^kJ_k,\ i,j,k=1,2,3.
\eeqa
Considered as a complex Lie algebra, this algebra is isomorphic to ${\mathfrak{sl}}(2,\CC)$ (denoted also ${\mathfrak{sl}}(2)$).
Considering (\ref{su2}) as a real Lie algebra, one has the ${\mathfrak{su}}(2)$ algebra, which is a real form of ${\mathfrak{sl}}(2)$, where
$$ {\s}=\{ M \in {\cal M}_2 ({\CC}) \ :\ Tr\ M = 0\}$$
$$ {\mathfrak{su}}(2)=\{ M \in {\cal M}_2 ({\CC}) \ :\ Tr\ M = 0,\ M^\dag = -M\}.$$
The standard basis of $\s$ is given by
$$e_+=\begin{pmatrix} 0 & 1 \\ 0 & 0 \end{pmatrix},
e_-=\begin{pmatrix} 0 & 0 \\ 1 & 0 \end{pmatrix}, e_0=\begin{pmatrix}
1 & 0 \\ 0 & -1 \end{pmatrix}.$$
We can note that $\s$ admits a non-compact real form ${\mathfrak{sl}}(2,\RR)$. Indeed, the complex change of basis
\beqa
\label{linear}
 J_\pm=J_1\pm iJ_2,\ J_0=2J_3,
\eeqa
gives
$$[J_0,J_\pm]=\pm 2J_\pm,\ [J_+,J_-]=J_0$$
which is a real Lie algebra not isomorphic to ${\mathfrak{su}}(2)$.



\medskip
\noi
\textit{iv)} {\it The infinite-dimensional Heisenberg algebra}. It is 
defined by
\beqa
\label{heisenberg}
[a_m,a_n]=n \delta_{n+m,0}, \mbox{ with }n,m\in \ZZ.
\eeqa
where $(a_m)_{m\in {\ZZ}}$ is a basis of this algebra.

\noi
This Lie algebra is obtained when one deals with the second quantisation of a bosonic field. The Heisenberg's uncertainty relations underly the canonical commutation relations that will lead to (\ref{heisenberg}), hence the name of Heisenberg algebra.

Let us remark here the difference between the three-dimensional Heisenberg algebra ${\cal H}_3$, which is a {\it Lie algebra} defined by $$[p,q]=1$$ and the infinite-dimensional {\it associative algebra} $A_1$.

\subsection{Homomorphisms of Lie algebras}
 

A \textit{homomorphism} from the Lie algebra $\g$ to the Lie algebra $\h$ is a linear map $\phi:\g\to \h$ which preserves the algebraic structures
$$[x,y]\mapsto \phi([x,y])=[\phi(x),\phi(y)], \forall x,y\in\g$$



A bijective homomorphism is called an \textit{isomorphism}.
One of the fundamental goals in Lie algebra theory is the classification of Lie algebras up to isomorphism.

A homomorphism $\phi:\g\to \g$ of a Lie algebra to itself is called an \textit{endomorphism} and a isomorphism from a Lie algebra to itself is called an \textit{automorphism}.

\section{Structure of Lie algebras}
\subsection{Solvable Lie algebras}
 
    For further use, for any subsets $\g_1,\g_2$ of a Lie algebra $\g$ let us introduce the notation
$$ [\g_1,\g_2]=< [x,y]\ : \ x\in \g_1, y\in \g_2>.$$
which corresponds to the subalgebra of generated by the brackets $[x,y]$.

An \textit{abelian Lie algebra} is a Lie algebra which satisfies $[\mathfrak{g},\mathfrak{g}]=\{ 0 \}$.

   The \textit{derived algebra} $\mathfrak{g}^\prime$, is defined as the set:
\beqa
\label{derivedalgebra}
\mathfrak{g}^\prime=[\mathfrak{g},\mathfrak{g}]
\eeqa
i.e. the set of all linear combinations of brackets of elements of $\mathfrak{g}$. 

By induction one defines the \textit{upper central series}:
\beqa
\label{uppercentralseries}
\mathfrak{g}^{(i)}=[\mathfrak{g}^{(i-1)},\mathfrak{g}^{(i-1)}]
\eeqa
with $\mathfrak{g}^{(1)}=\mathfrak{g}^\prime$. If this series ends up with $\{ 0 \}$, then one calls $\mathfrak{g}$ a \textit{solvable Lie algebra}.

Similarly one defines the \textit{lower central series}:
\beqa
\label{lowercentralseries}
\mathfrak{g}_{(i)}=[\mathfrak{g},\mathfrak{g}_{(i-1)}]
\eeqa
with $\mathfrak{g}_{(1)}=\mathfrak{g}^\prime$. If this series up ends with $\{ 0 \}$, then one calls $\mathfrak{g}$ a \textit{nilpotent Lie algebra}.

One should remark that abelianity implies nilpotency and nilpotency implies solvability. Moreover, the derived algebra of a solvable algebra is nilpotent. 


Before ending this subsection, let us also remark that the solvable finite-dimensional Lie algebras form a {\it continuous} family of algebras. 
Indeed, the set of nilpotent Lie algebras of dimension $<7$ is discrete and the set of nilpotent Lie algebras of dimension $\ge 7$ is continuous. 
We will prove in the next chapter that if one imposes for such a finite-dimensional Lie algebra the condition of being realisable in the Weyl algebra, only a {\it discrete} set  fulfils this condition.

For further references of nilpotent Lie algebras, one may consult the book \cite{goze-book} of M. Goze and Y. Khakimdjanov.

 %



\subsection{Semi-simple and reductive Lie algebras}

 A subspace $\h\subseteqq\g $ of a Lie algebra $\g$ which itself is a Lie algebra with respect to the Lie bracket of $\g$ is a \textit{Lie subalgebra} of $\g$.

   An \textit{invariant subalgebra (ideal)} is a subalgebra such that the commutator of any of its elements with any element of $\mathfrak{g}$ is in the subalgebra.

   A \textit{proper ideal} is an ideal that is neither equal to $\{0\}$ nor to $\g$ itself.

    A \textit{simple} Lie algebra is a non-abelian Lie algebra which contains no proper ideals.
    A \textit{semi-simple} Lie algebra is a direct sum of simple Lie algebras.
A \textit{reductive} Lie algebra is a direct sum of semi-simple and abelian Lie algebras.

    An arbitrary Lie algebra $\g$ has a semi-direct sum structure $\g=\mathfrak s \ltimes \mathfrak r$, with $\mathfrak s$ a semi-simple Lie algebra and $\mathfrak r$ a solvable Lie algebra, \textit{i.e.}
$$[\mathfrak s, \mathfrak s]\subset \mathfrak s,\ [\mathfrak s, \mathfrak r]\subset \mathfrak r,\ [\mathfrak r, \mathfrak r]\subset \mathfrak r.$$
The ideal $\mathfrak r$ is the \textit{radical} of $\g$, \textit{i.e.} the unique maximal solvable ideal of $\g$.
This result is known as Levi-Malcev's decomposition theorem.


Let $\g$ be a semi-simple Lie algebra. Then a maximal abelian subalgebra $\h$ such that, for any $H\in \h$, 
$${\rm ad}(H):\g\to\g,\ {\rm ad}(H)(x)=[H,x], \forall x\in\g$$
 is diagonalisable, is called a \textit{Cartan subalgebra} of $\g$.

\noi
All Cartan subalgebras are of the same dimension and this integer is called the \textit{rank} of $\mathfrak{g}$\footnote{When one has a physical system for which the infinitesimal symmetries are described by $\g$, the rank of $\g$ is the maximal number of quantum numbers which can be used to label the states of the system.}.

The Cartan subalgebra of a semi-simple Lie algebra is not unique.
Nevertheless, these Cartan subalgebra are all related by automorphisms of $\g$.

    Let $\g$ be a simple complex Lie algebra of dimension $n$ and rank $r$. Let $\h$ be a Cartan subalgebra of $\g$ with basis of generators $\{H_1,\ldots,H_r\}$. One can then complete a basis of $\g$ \footnote{This is a consequence of the Cartan theorem, see for example \cite{FuchsSchweigert} or \cite{sorba}.} with $n-r$ generators $E_\alpha$ such that
\beqa
\label{radacini}
[H_i, E_\alpha]=\lambda_\alpha^i E_\alpha,
\eeqa
\noi
(i.e. the $n-r$ generators $E_\alpha$ are simultaneous eigenvectors of $H_i$ with eigenvalues $\lambda_\alpha^i$).\\
The $r-$dimensional vector $\lambda_\alpha=(\lambda_\alpha^1,...,\lambda_\alpha^r)$  is called the \textit{(root) vector} associated to the generator $E_\alpha$ (for any $\alpha$).
   The set of all roots of $\g$ is called the \textit{root system} of $\g$.

As $\g$ is spanned by elements satisfying (\ref{radacini}), one has
$$\g=\h\oplus_\alpha \g_\alpha, \mbox{ where } \g_\alpha=\{x\in\g\vert [H_i,x]=\lambda_\alpha^i x, \mbox{ for all }i=1,\ldots,r\}.$$


The miracle of root theory is that, inside the root system there exist a basis of \textit{simple roots} $\lambda_1,\ldots,\lambda_r$ such that any root $\lambda$ can be written as a linear combination $\lambda=\sum_{i=1}^r \gamma_i\lambda_i$, with all the coefficients $\gamma_i$ positive (one then speaks about  (\textit{positive roots}) or negative (one then speaks about \textit{negative roots}). The simple roots themselves are positive roots which cannot be written as a sum of positive roots with positive coefficients.

We denote by $\n_+$ the space of vectors associated to positive roots and by $\n_-$ the space of vectors associated to negative roots.
The decomposition (as sum of vector space) 
$$\g=\n_-\oplus\h\oplus\n_+$$
 is called the \textit{Borel decomposition} or \textit{triangular decomposition} (sometimes also called the \textit{Cartan decomposition} or the \textit{Gauss decomposition}).\\
The subalgebras ${\mathfrak b}=\h\oplus\n_\pm$ are called \textit{Borel subalgebras} of $\g$; they are maximal solvable subalgebras of $\g$.




By Cartan theorem, for a simple complex Lie algebra, one has a particular 
basis which allows to write the commutation relations in a simplified manner for different types of calculations, the Cartan-Weyl basis.





\subsection{On the classification of simple Lie algebras} 

    The finite-dimensional simple complex Lie algebras were classified in the nineteenth century by E. Cartan: there are four infinite series and in additional five exceptional algebras. The series are:
\beqa
\label{series}
A_r &\approx& {\mathfrak{sl}}(r+1) , \ \ (r\geq 1) \nonumber \\
B_r &\approx& {\mathfrak{sp}}(2r), \ \ (r\geq 3) \nonumber \\
C_r &\approx& {\mathfrak{so}}(2r+1),  \ \ (r \geq 2) \nonumber \\
D_r &\approx& {\mathfrak{so}}(2r)  \ \ (r \geq 4)
\eeqa
where $r$ is the rank of the Lie algebra.\\
The five additional exceptional algebras are 
$$E_6,\ E_7,\ E_8,\ G_2,\ F_4.$$ 
They play a special role in today's theoretical physics (see for example \cite{specialele,specialele2}).

  There is a similar but more complicated classification of finite-dimensional simple real Lie algebras. 

Before ending this subsection, let us also remark that the simple complex finite-dimensional Lie algebras form a discreet family of algebras. We will prove in the next chapter, that if one imposes for such a Lie algebra to be realised in the Weyl algebra, only a { discrete} set of points amongst this family can fulfil this condition.

\section{Lie algebras and particle physics}

Of crucial interest in nowadays fundamental physics are the real Lie
algebras $\mathfrak{so}(1,3)$ (the Lorentz algebra), defined by 
\eqref{Lorentz-algebra}, and the Poincar\'e
algebra, defined by 
\eqref{Poincare-algebra}.  This particular interest comes from the fact that these Lie
algebras regroup the symmetries of $4-$dimensional space-time. The
Lorentz algebra groups the generators of rotations and boost; the Poincar\'e algebra adds the generators of space-time
translations, thus completing the frame of space-time symmetries.

Before this, we define another notion of special importance for the study of Lie algebras, the Casimir operators. 

Let $\g$ be a Lie algebra, $\g^\otimes=\bigoplus_{n=0}^\infty\g^{\otimes n}=\bigoplus_{n=0}^\infty\left(\bigotimes_{j=1}^n \g \right)$ be the tensor algebra over $\g$ and ${\cal I}$ the ideal of $\g^\otimes$ generated by $[x,y]-(x\otimes y-y\otimes x)$, where $x,y\in\g$. The {\it universal enveloping algebra} ${\cal U}(\g)$ is the quotient $g^\otimes / {\cal I}$.

A {\it Casimir operator} $C$ of a Lie algebra $\g$ is an operator constructed the universal enveloping algebra of $\g$, and which is $\g -$invariant, that is
$$[C, x]=0\ \forall x\in\g.$$
where the bracket in the universal enveloping algebra is defined as $[A,B]=AB-BA$. 

From a physical point of view, the Casimir operators have a special importance because they allow to endow the particles of some multiplet with general properties as we will see in this section.

{\bf Example:} the only Casimir operator of the Lie algebra defined by (\ref{su2}) is 
$$J^2=\sum_{i=1}^3 J_i^2.$$

\subsection{The Lorentz algebra} 

This Lie algebra is generated by six
independent generators $L_{mn}=-L_{nm}$ ($m,n=0,\ldots,3$), thus taken to be $L_{01}, L_{02}, L_{03}, L_{12}, L_{13},
  L_{23}$, subject to the following commutation relations:
\beqa
\label{Lorentz-algebra}
\left[L_{mn}, L_{pq}\right]=
\eta_{nq} L_{pm}-\eta_{mq} L_{pn} + \eta_{np}L_{mq}-\eta_{mp} L_{nq},
\eeqa
where $\eta_{mn}=\eta^{mn}$ is the metric $\mathrm{diag}(1,-1,-1,-1)$. As usual,
the metric matrix is used to raise or lower the indices
$L^{mn}=\eta^{mp}\eta^{nq}L_{pq}$ {\it etc.} 
If we define a change of basis
\beqa
J_i&=&\frac 12 \e_{ijk}L^{jk},\nonumber \\
K_i&=&L_{0i}, \mbox{ with } i,j,k=1,2,3, \nonumber
\eeqa
the commutation relations (\ref{Lorentz-algebra}) write, in this new basis,
\beqa
[J_i,J_j]&=&\e_{ijk}J_k,\nonumber \\
\left[J_i,K_j\right]&=&\e_{ijk}K_k,\nonumber \\
\left[K_i,K_j\right]&=&-\e_{ijk}J_k,\mbox{ with } i,j,k=1,2,3. \nonumber
\eeqa
The Casimir operators  are 
\beqa
\label{casimir-lorentz}
&&C_1=\frac 12 L^{mn} L_{mn}=\overrightarrow{J}^2-\overrightarrow{K}^2, \nonumber \\
&&C_2=\frac 12 \e_{mnpq} L^{mn}L^{pq}=2 \overrightarrow{J}\cdot \overrightarrow{K}.
\eeqa
\noi
 
\subsection{The Poincar\'e algebra}

 As already mentioned above,
this Lie algebra regroups the symmetries of space-time. Algebraically
speaking, it is the semi-direct sum of the Lorentz algebra
${\mathfrak{so}}(1,3)$  and the abelian algebra spanned by the momenta
$P_m$ ($m=0,\dots,3$). Thus, the Poincar\'e algebra is not semi-simple nor even a reductive Lie algebra. It is generated by ten independent generators: the six Lorentz generators $L_{mn}$ and the four translation generators $P_m$, subject to the following commutation relations
\beqa
\label{Poincare-algebra}
\left[L_{mn}, P_p \right]&=& \eta_{np} P_m -\eta_{mp} P_n, \nonumber \\
\left[P_m,P_n\right]&=&0.
\eeqa
A realisation by polynomial coefficients differential operators acting
on variables $x_m$ ($m=0,..3$), the space-time coordinates, is given by
\beqa
\label{realizare-poincare}
P_m&=&\partial_m,\nonumber\\
M_{mn}&=&x_m\partial_n-x_n\partial_m,
\eeqa
where we use the standard notations
$\partial_m=\frac{\partial}{\partial x^m}$ and $\partial^m=\frac{\partial}{\partial x_m}$.\\
The Casimir operators are
\beqa
\label{casimir-poincare}
P^2&=&P_mP^m\\
W^2&=&W_mW^m, \mbox{ with } W^m=\frac 12 \e^{mnpq}P_nM_{pq} \mbox{ (the Pauli-Lubanski vector}).\nonumber
\eeqa
\noi
Their eigenvalues are related to the mass and resp. the spin. Thus, if one has such Casimir operators, then the particles associated to some representation have have the same mass and spin.


%









\section{Representations of Lie algebras} 
\subsection{Definition}


 A \textit{representation} of a Lie algebra $\mathfrak{g}$ on a vector
 space $V$, called  the \textit{representation space}, is a map
$$ R:\g\to \mathrm{End}(V)$$
which satisfies
$$ R(x) \circ R(y) - R(y) \circ R(x) = R (\left[ x,y \right]) \quad \forall x,y \in \mathfrak{g}$$
($\circ$ signifying the composition of operators).

If $R$ is injective, then it is called a {\it faithful representation}.
A representation $R$ is called {\it irreducible} if there are no invariant subspaces
 of the representation space $V$ except the trivial ones.

   When $V$ is a complex (resp. real) vector space, we speak of a \textit{complex (respectively real) representation}.

The \textit{dimension of the representation} is the dimension of the representation space seen as a vector space over the base field.

 Since $R(x)$ is a linear application 
we can represent it  by its matrix. Therefore, one can see a representation as the process of associating to
 any element of the Lie algebra a  matrix, in such a way that the commutation law of $\mathfrak{g}$ is reproduced. The dimension of the representation is exactly the size of the matrices.

The basic example of a representation of a Lie algebra is the \textit{adjoint representation} given by
\beqa
\label{adjoint-def}
x\mapsto {\rm ad}(x),\ {\rm ad}(x):\g\to\g,\ {\rm ad}(x)(y)=[x,y],\forall y\in\g.
\eeqa
\noi
 In this example the Lie algebra acts on itself and thus the dimension of the algebra is the same as the dimension of the representation. However, in general, this is not the case, the dimension of a representation is different from the dimension of the Lie algebra $\mathfrak{g}$. 

An important result in the theory of representations of Lie algebra is the Ado theorem, which states that any finite-dimensional Lie algebra $\g$ over the field $\KK=\RR$ or $\CC$ has a faithful finite-dimensional representation on some vector space $V$. However, the dimension of $V$ is not specified. If ${\rm dim}\, V={\rm dim}\, \g$, then the Lie algebra $\g$ is called an {\it affine algebra}.

\subsection{Weights and  highest weight representations for $\s$} 

Among the representations of $\s$, the highest weight representations form a particular interesting subclass. The main reason for this is that any finite-dimensional representation of a simple Lie algebra belongs to this subclass.

Let $\Lambda\in {\NN}^*$ and  $V_\Lambda $ be the vector space of dimension $\Lambda+1$ and basis $\{ v_{-\Lambda}, v_{-\Lambda +2},\dots,v_{\Lambda}\}$. We denote here any vector $v_\lambda$ by $|\lambda>$. We define the finite-dimensional irreducible representation 
$D_\Lambda : {\s}\to {\rm End}(V_\Lambda)$ by
\beqa
\label{highest-weight}
D_\Lambda(J_0)|\Lambda>=\Lambda|\Lambda>,\ D_\Lambda(J_+)|\Lambda>=|0>,\ D_\Lambda(J_-)|\Lambda>=\sqrt \Lambda |\Lambda - 2>.
\eeqa
\noi 
which will allow us to obtain the action of the generator on any vector $v_\lambda$.
The number $\Lambda$ is
called the \textit{highest weight} of the representation space. The
number
 $j=\frac{\Lambda}{2}$ is the {\it spin} of the representation $D_\Lambda$\footnote{is commonly done in the treatment of angular momentum in non-relativistic quantum mechanics}. 
Up to multiplication with a non-zero
number, the vector $|\Lambda>\in V_\Lambda$ is unique; one refers to it as the \textit{highest
  weight vector} of the representation.

One sees that the application of the lowering operator $J_-$ gives another vector, $|\Lambda - 2>$; it also is a $J_0$ eigenvector, with eigenvalue (or {\it weight}), $\Lambda -2$.
All other elements of $V_\Lambda$ which are eigenvectors of $J_0$ can similarly be
obtained by applications of the lowering operator $J_-$ on the
highest weight vector $|\Lambda>$. Each application of the generator
changes the weight by
$2$. 

As a vector space, $V_\Lambda$ is written as
the direct sum of one-dimensional vector space spanned by any
vector $|\lambda>$, $V_\Lambda=\bigoplus V_\lambda$, with $\lambda=-\Lambda,-\Lambda+2,\dots,\Lambda$.

\section{Contractions and deformations of Lie algebras}
\label{contractions-deformations}

The notions of contractions and deformations of algebras are intimately related, as we will see in this section for the case of Lie algebras.
The physical interest  stems from the need to relate, in a meaningful way, the symmetries of different systems and thus to bring the corresponding phenomena to a common understanding. In \cite{segal}, I. E. Segal argued about the notion of contraction from the viewpoint of physical background: if two theories (like for example relativistic and classical mechanics) are related by a limiting process, then the associated invariance groups (like the Poincar\'e and Galilei groups) should also be related by some limiting process.



From the viewpoint of physical interest the study of deformations is of special importance since it allows to find all Lie algebras which can be contracted into a given Lie algebra $\g$. Indeed, these Lie algebras have to be looked for between the deformations of $\g$.

\bigskip

The local study of the variety of Lie algebra multiplication on ${\CC}^n$ is essentially based here on the notion of ``perturbation'' of a given point of this variety. 




\subsection{Contractions}
\label{contractions}

Let $\{ X_i:\ i=1,\dots,n \}$ be a basis of ${\KK}^n$ ($\KK=\RR$ or $\CC$) and consider a Lie algebra multiplication $\mu_0$ defined by
$$\mu_0 (X_i,X_j)=C_{ij}^kX_k.$$
Let $(f_n)_{n\ge 1}$ be a sequence of automorphisms of ${\KK}^n$ and define the Lie bracket
$$\mu_n (X_i, X_j)=f_n^{-1} (\mu_0 (f_n (X_i), f_n (X_j))).$$
This multiplication is isomorphic to $\mu_0$. 
Note that one may denote this by
$$ \mu_n=f_n^{-1}\circ\mu_0\circ f_n \times f_n .$$
If, for any couple $(X_i, X_j)$ the limit ${\rm lim}_{n\to\infty} \mu_n (X_i, X_j)$ exists, the bracket $\mu$ defined by 
$$  \mu (X_i, X_j) = {\rm lim} _{n\to\infty} \mu_n (X_i, X_j) $$
is a Lie algebra bracket called {\it contraction} of $\mu_0$. 
On the other hand, if $n$ is sufficiently large, one can consider $\mu_n$ as a deformation of the contracted Lie algebra bracket $\mu$.

\bigskip





Let us now give a basic example of Lie algebra contraction which presents physical interest. (This is actually 
an example of a In\"on\"u-Wigner contraction.)

\noi

\textbf{Example:} Consider the Lie algebra ${\mathfrak{so}}(3)$, whose Lie brackets are
$$[X_1,X_2]=X_3,\ [X_2, X_3]=X_1,\ [X_3, X_1]=X_2.$$
Define now the automorphism of ${\RR}^3$ given by
\beqa
\label{ex-IW}
\tilde{X}_1=\e X_1,\ \tilde{X}_2=\e X_2,\ \tilde{X}_3= X_3,
\eeqa
with $\e\ne 0$. 
Then, taking the limit $\e\to 0$, one obtains the Lie algebra of the Euclidean group $E(2)$ in two dimensions (one rotation and two translations):
$$[{\tilde{X}}_1,\tilde{X}_2]=0,\ [{\tilde{X}}_3,\tilde{X}_1]=\tilde{X}_2,\ [{\tilde{X}}_3,\tilde{X}_2]=\tilde{X}_1.$$

Other examples of special importance in fundamental physics are the contraction of the de Sitter algebra $\mathfrak{so}(1,4)$ to the Poincar\'e algebra or the contraction of the Poincar\'e algebra to the Galilei algebra.


\medskip

The general notion of contraction introduced above was particularised in order to suit better physics issues. The most frequently used in physics literature is the In\"on\"u-Wigner contraction \cite{inonu-wigner}.

\subsection{In\"on\"u-Wigner contractions}
\label{iw}


Let $(f_\e)_{\e\in{\KK}^*}$ be a one-parameter family of  automorphisms of ${\KK}^n$ of the form
$$f_\e=f_1+\e f_2$$
where $f_1$ is singular, $f_2$ is regular and $\e\approx 0$. 

We assume that, in an appropriate basis of ${\KK}^n$, $f$ reduces to 
\beqa
\label{Inonu-Wigner}
f_1 =\begin{pmatrix} 1_r & 0 \\ 0 & 0 \end{pmatrix},\ 
f_2 =\begin{pmatrix} V & 0 \\ 0 & 1_{n-r} \end{pmatrix}
\eeqa
where ${\rm rk}(f_1)={\rm rk}(V)=r$.
Notice that one can often write $f$ under this form; however this is not always possible. 



Let $\mu_0$ be a Lie algebra bracket on ${\KK}^n$ and let $C_{ij}^k$ be the structure constants of $\mu_0$ in the fixed basis $X_i$ ($i=1,\dots,n$). The application $\mu_\e=f_\e^{-1}\circ\mu_0\circ f_\e\times f_\e$ admits a limit if $C_{ij}^k=0$ for $i,j\le r$ and $k\ge r+1$. For this to be realised it is necessary that the application $\mu_0$ restricted to the subspace of ${\KK}^r$ generated by the vectors $X_1,\dots,X_r$ is a Lie bracket. Let 
 us denote this Lie bracket (of dimension $r$) by $\phi$.  Thus, the {\it In\"on\"u-Wigner contraction} $\mu_1$ of $\mu_0$ defined by
$$ \mu_1 = {\rm lim}_{\e\to 0} f_\e^{-1}\circ \mu_0 \circ f_\e \times f_\e $$
is isomorphic to the semi-direct product $\phi\ltimes 0_{n-r}$, where $0_{n-r}$ stands for the trivial bracket corresponding to the abelian structure on ${\KK}^{n-r}$.

Thus, the In\"on\"u-Wigner contraction allows to contract a given Lie bracket to a bracket related to a subalgebra.

As already mentioned the example of Lie algebra contractions above are In\"on\"u-Wigner contractions, where for the basis $X_3, X_1, X_2$ we take $f_1=\begin{pmatrix} 1 & 0 & 0 \\ 0 & 0 & 0 \\ 0 & 0 & 0 \end{pmatrix}$ and $f_2=\begin{pmatrix} 1 & 0 & 0 \\ 0 & 1 & 0 \\ 0 & 0 & 1 \end{pmatrix}$, which, to satisfy condition \eqref{Inonu-Wigner}, is taken to be slightly different of \eqref{ex-IW} but nevertheless leads to the same contraction.

\subsection{The Saletan and L\'evy-Nahas contractions}
\label{ww}

As stated before, the In\"on\"u-Wigner contractions represent a particular case of Lie algebra contractions. Let us now give a  generalisation, namely the Saletan contractions \cite{salentan}. They also consider an automorphism of ${\KK}^n$ of the form $f_\e=f_1+\e f_2$  with $f_2$ regular
and $\e\approx 0 $. Nevertheless, $f_1$ and $f_2$ do not necessary write as in \eqref{Inonu-Wigner}.
One suppose that for these automorphisms there exists $g$ such that
$$ f_\e= \e 1 + (1-\e)g $$

Obviously this type of contraction also can still be extended to more general definitions.

\bigskip

\bigskip

The L\'evy-Nahas contractions \cite{levy-nahas} represent a first generalisation of the Saletan contractions or a second generalisation of the In\"on\"u-Wigner contractions. The automorphism $f$ is given by
$$f_\e=\e f_1 + \e^2 f_2$$ 
where the applications $f_1$ and $f_2$ satisfy the hypothesis of the Saletan contractions. The difference compared to the In\"on\"u-Wigner or Saletan contractions is obvious, since in the automorphism $f_\e$ one now has the coefficient $\e^2$.

An interesting example of such a contraction is given by the contraction of the ${\mathfrak {so}}(3)$ Lie algebra to the three-dimensional Heisenberg algebra ${\cal H}_3$. Indeed, consider $\mu_0$, the Lie bracket $\mu_1$ of ${\mathfrak {so}}(3)$ defined by
$$\mu_0 (X_1, X_2)=X_3,\ \mu_0(X_2, X_3)= X_1,\ \mu_0(X_3, X_1)= X_2.$$
Let $f_\e$ the automorphisms of ${\KK}^3$ defined by
$$f(X_1)=\e^2 X_1,\ f(X_2)=\e X_2,\ f(X_3)=\e X_3. $$
The bracket $f_\e^{-1}\circ \mu_0\circ f_\e\times f_\e$ admits as a limit ($\e\to 0$) the Lie bracket of ${\cal H}_3$
$$\mu_1(X_1, X_2)=0,\ \mu_1(X_2, X_3)= X_1,\ \mu_1(X_3, X_1)= 0.$$
Moreover, one might consider to extend this to automorphisms of type
$$ f= \e f_1 + \e^2 f_2 +\dots + \e^n f_n, $$
or, formally, even to
$$ f= \e f_1 + \e^2 f_2 +\dots + \e^n f_n + \dots \ . $$


\bigskip

\bigskip

In \cite{ww-2}, E. Weimar-Woods proves that any contraction is equivalent to a contraction defined by the family  $(f_\e)_{\e\in{\KK}^*}$ of automorphisms of ${\KK}^n$ 
which are diagonal with respect to some basis of $V$ and given in terms of integer powers of the parameter $\e$. 
This means that, in a chosen basis, 
$f_\e=e^{i_j} f_j$ (with $i_j\in\ZZ$) where all the $f_i$ are diagonal, with the elements of the diagonal equal to $0$ or $1$.
In \cite{ww-1} she obtains all contractions of the real three-dimensional Lie algebras.

\subsection{A ``general'' definition of contractions}

We have seen in the previous subsections different definition of contractions. Nevertheless, none of them corresponds to a topological notion.

Indeed, let $L^n$ be the set of all Lie algebra multiplications on ${\KK}^n$. If one fixes a  basis of ${\KK}^n$, say $\{ X_1,\dots,X_n \}$, a Lie bracket identifies itself to the structure constants $C_{ij}^k$. In $L^n$ the set $\{ C_{ij}^k\}$ satisfies the Jacobi identity
$$ C_{ij}^k C_{kl}^m + C_{li}^kC_{kj}^m+C_{jl}^{k}C_{ki}^m = 0.$$
One thus has an algebraic variety.
 If $\mu_0$ is a Lie bracket from $L^n$, the set of isomorph Lie brackets is denoted by ${\cal O}_{\mu_0}\subseteq L^n$. Particularly, if $(f_\e)_{\e\in{\KK}^*}$ is a family of automorphisms of ${\KK}^n$, the Lie brackets
$$f_\e^{-1}\circ \mu_\e \circ f_\e \times f_\e $$
belong to ${\cal O}_{\mu_0}$. Thus, a contraction ${\rm lim}_{\e\to 0}f_\e^{-1}\circ \mu_\e \circ f_\e \times f_\e $ is a point of adherence (for the topology of the variety where the closed sets are described by polynomial equations).
We call a (topological) contraction of $\mu_0$ any point of adherence of ${\cal O}(\mu_0)$ in $L^n$.

\medskip

All the contractions defined above are particular cases of this definition. For $n\ge 5$ there exists topological contractions which do not write as ${\rm lim}_{\e\to 0}f_\e^{-1}\circ \mu_\e \circ f_\e \times f_\e $ \cite{Burde}.

\subsection{Deformations}
\label{deformations}

In this subsection we briefly overview the notion of deformations of Lie algebra as it has been described by M. Gerstenhaber in \cite{deformatii}. For the sake of completeness we also mention the Nijenhuis-Richardson deformations \cite{nr}. This notion is in some sense reciprocal to the notion of contraction of the previous subsections.

\subsubsection{The deformations of Gerstenhaber}
\label{gerstenhaber}

Let $\mu_0$ be a Lie bracket on ${\KK}^n$ ($\KK=\RR$ or $\CC$). A {\it deformation} of $\mu_0$ is a Lie bracket on ${\KK}^n$ defined by
$$\mu_t (X_1,X_2)= \mu_0 (X_1,X_2) + t F_1 (X_1, X_2) + t^2 F_2(X_1,X_2)+\dots $$
where $F_i$ are bilinear applications.
One can check that the Jacobi identity is satisfied by $\mu_t$ iff the maps $f_i$ satisfy the equations:
\beqa
\label{rez-g}
\sum_{\sigma\in S_3} \sum_{\begin{tiny} {\begin{array}{l} p+q=r\\ r=0,1,\dots\end{array}}\end{tiny}} F_q (F_p (X_{\sigma(1)}, X_{\sigma (2)}), X_{\sigma (3)}) + F_p (F_q (X_{\sigma(1)}, X_{\sigma (2)}), X_{\sigma (3)})=0, \mbox{ for any }r
\eeqa
\noi
where we put $F_0=\mu_0$. In particular, for $n=1$ one gets a condition for $F_1$
\beqa
\label{F111}
 \sum_{\sigma\in S_3} F_1 ( F_0 (X_{\sigma(1)}, X_{\sigma (2)}), X_{\sigma(3)})+ F_0 ( F_1  (X_{\sigma(1)}, X_{\sigma (2)}), X_{\sigma(3)})=0. 
\eeqa
\noi
Thus, a necessary condition for $\mu_t$ to satisfy the Jacobi identity is \eqref{F111}. It is this observation that will lead to the study of A. Nijenhuis and R. W. Richardson.

As an example of deformation isomorphic to $\mu_0$ let us give the trivial deformation
$$\mu_t=\phi_t^{-1} (\mu_0 (\phi_t, \phi_t))$$
with $\phi(t)$ is an automorphism of ${\KK}^n$ given by
$$ \phi_t = {\rm {Id}}+ t \gamma_1 + t^2 \gamma_2 +\dots$$
where the $\gamma_i$ are any linear applications.

In \cite{levy-nahas}, M. L\'evy-Nahas determines all possible deformations for the three-dimensional real Lie algebra.

\subsubsection{The deformations of Nijenhuis-Richardson}

To avoid the formal framework of \cite{deformatii}, A. Nijenhuis and R. W. Richardson
work in the algebraic frame of the variety of Lie brackets over ${\KK}^n$ parametrised by the structure constants in a given basis.
If the series $\mu_t$ gives a Lie multiplication isomorphic to $\mu_0$, then it exists $f\in {\mathfrak {gl}}(n,\CC)$ such that $F_1 (X_1, X_2) = \mu_0 ( f(X_1), X_2) + \mu_0 (X_1, f(X_2)) - f(\mu_0 (X_1, X_2))$. This allows to define a criteria such that all deformations of a given Lie multiplication are isomorphic to the original multiplication. Such a multiplication is called {\it rigid}.




\section{Lie superalgebras}

      In this section we recall the definition and some underlying
      properties of  Lie superalgebras, which appear as some
      natural extension of Lie algebras. 
To the symmetries described by the commutation relations of a Lie algebra, bosonic symmetries, one might add a different type of symmetries, the fermionic symmetries. For these, the mathematical relation to be used is the anticommutation. Thus, one obtains Lie superalgebras; this is  the mathematical structure that underlies SUSY, as we will see in subsection $5.1.2$. 
As general reference for this section, one can consult for example \cite{sorba}.

\subsection{Definitions}

Let $A$ be an algebra over a field $\KK$ (${\KK}=\RR$ or $\CC$) with internal laws $+$ and $*$. $A$ is called a \textit{superalgebra} or a \textit{${\ZZ}_2$-graded algebra} (with respect to the product law $*$) if it can be written as a direct sum of two vector spaces $A=A_{\bar 0}\oplus A_{\bar 1}$, such that
$$ A_i * A_j \subseteq A_{i+j}, \mbox{ for any }i,j\in{\ZZ}_2. $$

An element $x\in A_{\bar 0}$ is said to be of degree $0$ ($\mathrm{deg}\ x=0$) and an element $y\in A_{\bar 1}$ is said to be of degree $1$ ($\mathrm{deg}\ y=1$).

For an (associative or not) superalgebra, one defines a new operation, the \textit{superbracket} or  \textit{supercommutator} as
\beqa
\label{super-comutator}
\pard x,y \pari =x*y - (-1)^{\mathrm{deg}x \  \mathrm{deg}y} y*x.
\eeqa

A Lie superalgebra $A$ over a field $\KK$ 
is a ${\ZZ}_2$-graded vector space 
$A=A_{\bar 0}\oplus A_{\bar 1}$
with a superbracket $\pard .,. \pari$ satisfying:
\begin{enumerate}
\item $\pard A_i, A_j\pari\subseteq A_{i+j}$;
\item 
  $\pard x_1, x_2 \pari = (-1)^{\mathrm{deg}x_1 \  \mathrm{deg}x_2}\pard x_2, x_1 \pari, \mbox{ for any }x_1,x_2\in A;$ 
\item 
\beqa
\label{super-Jacobi}
  (-1)^{\mathrm{deg}x_1 \  \mathrm{deg}x_3}\pard x_1,\pard x_2, x_3 \pari \pari + (-1)^{\mathrm{deg}x_2 \  \mathrm{deg}x_1}\pard x_2,\pard x_3, x_1 \pari \pari + (-1)^{\mathrm{deg}x_3 \  \mathrm{deg}x_3}\pard x_3,\pard x_1, x_2 \pari \pari = 0, \nonumber \\
\eeqa
\noi for any $x_1,x_2,x_3\in A$. These identities are known as the super Jacobi identities.
\end{enumerate}

We call $A_{\bar 0}$, the $0$-graded part, the \textit{bosonic sector}, its generators $B_m$ \textit{bosonic generators} and we call $A_{\bar 1}$, the $1$-graded part, the \textit{fermionic sector}, its generators $F_\alpha$, \textit{fermionic generators}. It is customary to use the \textit{commutator} symbol $\left[.,.\right]$ for the product of a bosonic generator with anything and the \textit{anticommutator} symbol $\{.,.\}$ for the product of two fermionic generators.

    With this standard notation the equations:
\beqa
\left[ B_m,B_n\right] & = & f_{mn}^{\mbox{                                  }p}B_p,\nonumber \\
\left[ B_m,F_\alpha \right] & = &(R_m)_{\alpha}^{\mbox{                                  }\beta}F_{\beta},\nonumber \\
\{ F_\alpha, F_\beta\} & = & S_{\alpha \beta}^{\mbox{                                  }m}B_m  \nonumber
\eeqa
define the structure constants $ f_{mn}^{\mbox{                                  }p},(R_m)_{\alpha}^{\mbox{                                  }\beta},  S_{\alpha \beta}^{\mbox{                                  }m}$ of the Lie superalgebra.

    The super Jacobi identities (\ref{super-Jacobi}) take  their more familiar form:
\beqa
\label{super-Jacobi2}
\left[ \left[ B_m, B_n \right], B_p \right] + \left[ \left[ B_p, B_m \right], B_n \right]+ \left[ \left[ B_n, B_p \right], B_m \right]&=&0 \nonumber \\
\left[ \left[ F_\alpha, B_m \right], B_n \right] + \left[ \left[ B_n, F_\alpha \right], B_m\right]+ \left[ \left[ B_m,B_n \right],F_\alpha \right] & = & 0 \nonumber \\
\left[ \{F_\alpha, F_\beta \}, B_m \right] + \{ \left[ B_m, F_\alpha \right], F_\beta \} - \{ \left[ F_\beta,B_m \right],F_\alpha \} & = & 0 \nonumber \\
\left[ \{F_\alpha, F_\beta \}, F_\gamma \right] + \left[ \{ F_\gamma, F_\alpha \}, F_\beta \right] + \left[ \{ F_\beta,F_\gamma \}, F_\alpha \right] & = & 0 .
\eeqa


\subsection{Example of Lie superalgebra: the supersymmetry algebra}

The most common example of a real superalgebra is the SUSY algebra; we give here the simplest of the SUSY algebras, namely for $N=1$ and without any central charges. Its bosonic sector is the Poincar\'e algebra, generated by $L_{mn}=-L_{nm}$ and $P_p$ (with $m,n,p=0,...,3$); its fermionic sector is generated by the \textit{supercharges} $Q_\alpha$, a two-component left-handed (LH) Weyl spinor, and $\bar Q_{\dot \alpha}$, a two-component right-handed (RH) Weyl spinor (the undotted - $ \alpha,\beta$ and the dotted $ \dot \alpha, \dot \beta$ indices can take the values 1 and 2 -\textit{the two component notation)}). Thus, the SUSY algebra writes
\beqa
\label{SUSYalgebra}
&&\left[Q_\alpha, P_m\right]=0=[\bar Q_{\dot \alpha},P_m], \nonumber \\
&&\left[Q_\alpha, L_{mn}\right]=\frac{1}{2} (\sigma_{mn})_\alpha^{\mbox {    } \beta} Q_\beta \ \left[\bar Q_{\dot \alpha}, L_{mn}\right]=\frac{1}{2} \bar Q_{\dot \beta}(\bar \sigma_{mn})^{\dot \beta}_{\mbox{    } \dot \alpha}, \nonumber \\
&& \{Q_\alpha,Q_\beta\}=0=\{\bar Q_{\dot \alpha},\bar Q_{\dot \beta}\}, \nonumber \\
&& \{Q_\alpha,\bar Q_{\dot \beta}\}=2(\sigma^m)_{\alpha \dot \beta} P_m, 
\eeqa
\noindent
where 
$\sigma_m=(1,\sigma_i)$ are the usual Pauli  matrices, $\bar \sigma_m=(1,-\sigma_i)$  and $\sigma_{mn}=\frac 12 (\sigma_m \bar \sigma_n- \sigma_n \bar \sigma_m)$, $\bar \sigma_{mn}=\frac 12 (\bar \sigma_m  \sigma_n- \bar \sigma_n \sigma_m)$.

\subsection{Representations of Lie superalgebras}







Let $A=A_{\bar 0}\oplus A_{\bar 1}$ be a classical Lie superalgebra. Let $V=V_{\bar 0}\oplus V_{\bar 1}$ be a ${\ZZ}_2$-graded vector space and consider the superalgebra $\mathrm{End} V = \mathrm{End}_{\bar 0} V\oplus \mathrm{End}_{\bar 1} V$ of endomorphisms of $V$.\\
A linear representation $R$ of $A$ is a homomorphism of $A$ into $\mathrm{End} V$, that is
\beqa
\label{super-rep}
&&R(\alpha x + \beta y)= \alpha R(x) + \beta R(y),\nonumber \\
&&R(\pard x,y \pari)=\pard R(x), R(y) \pari,  \\
&&R(A_{\bar 0})\subseteq \mathrm{End}_{\bar 0} V,\ R(A_{\bar 1})\subseteq \mathrm{End}_{\bar 1} V,\mbox{ for any }\alpha,\beta\in{\KK}, x,y\in A.\nonumber
\eeqa

A particular representation is the \textit{adjoint representation}, which, for any $x\in A$ is defined as
\beqa
\label{super-adjointa}
{\rm ad}(x) :A\to A,\ y\mapsto\pard x,y \pari.
\eeqa

\medskip







\chapter{Finite-dimensional Lie subalgebras of the Weyl algebra $A_1$}




Let $n\in\NN^*$. We denote by $A_n$ 
 the associative algebra on $\CC$ generated by $2n$ operators $p_i,q_i$, $i=1,\dots,n$ subject to the relations
\beqa
\label{An}
&& p_iq_j-q_jp_i=\delta_i^j,\nonumber\\
&& p_ip_j-p_jp_i=0=q_iq_j-q_jq_i, \ i,j=1,\dots,n,\ i\ne j.
\eeqa

\medskip

We will also denote by  $A_0$  the commutative associative algebra of polynomials in two variables: $A_0=\CC [x,y]$.
In this chapter,
 we consider  the Weyl algebra $A_1$ which has been investigated extensively in mathematics and physics. Several studies of $A_1$ consist in generalising, for example in terms of representations, some properties of $A_0$.

The main question we address is to find (up to isomorphism) all finite-dimensional Lie algebras which can be realised in $A_1$. 
Since the generators $p$ and $q$ are actually the canonical variables of quantum mechanics, this question is of special importance in the framework of group theoretical symmetry studying.
The problem can be stated  as follows: given an arbitrary finite-dimensional Lie algebra $\g$, determine whether or not its generators can be realised as polynomials in the canonical operators $p$ and $q$.
An example of special importance for particle physics is the realisation \eqref{realizare-poincare} of the Poincar\'e algebra in $A_4$.

This question and closely related issues have been studied in different papers of mathematical physics.
In the seminal article \cite{Dixmier}, J. Dixmier started a systematic analysis of the structure of $A_1$. 
In \cite{italienii},
A. Simoni and F. Zaccaria proved that a necessary condition for a
complex semi-simple Lie algebra of rank $l$ to be realised in $A_n$ is $l\le n$. Thus, the only complex semi-simple Lie algebra which can be realised in $A_1$ is $\s$.  The realisations of $\s$ in $A_1$ are not classified to this day. Nevertheless, an important property   was proved by A. Joseph in \cite{joseph}, where he showed that the spectrum of the realisation in $A_1$ of  suitably normalised semi-simple elements is either $\ZZ$ or $2\ZZ$.  Of special importance for our present work is another result of \cite{joseph}, namely the classification (up to an automorphism of $A_1$) of the realisations of $\s$ for which the realisation of the standard semi-simple element $e_0$ is, in the Weyl algebra sense, strictly semi-simple (a notion first introduced by J. Dixmier in \cite{Dixmier}, see  also subsection \ref{basic} here). We denote this family of realisations by $\cal F$. 

The main result of  this chapter is to find all finite-dimensional Lie algebras that can be realised as subalgebras of $A_1$. We first treat the case of non-solvable Lie algebras and then, the more difficult case of solvable Lie algebras. Furthermore, the classification theorem then allows us to find all finite-dimensional Lie algebras which can be realised as subalgebras of $\mathrm{Der}(A_1)$.

In subsequent sections we analyse certain realisations of $\s$ in $A_1$ and, after an explicit proof of the classification theorem that gives the family $\cal F$, we then analyse $\cal F$ in more detail: equivalences of its elements under the action of the groups $\A(A_1)$ and $\A(A_1)\times \A(\s)$ are given. We then focus on the orbit of $\cal F$ under $\A(A_1)\times\A(\s)$, for which we give several characterisations in terms of the Dixmier partition; normal forms are obtained. Furthermore, we calculate the isotropy of the normal forms under the action of the group $\A (A_1)\times\A(\s)$. Finally, for the sake of completeness, we give explicit formulae for realisations of $\s$ not in the orbit of $\cal F$ under $\A(A_1)\times\A(\s)$.


\medskip

This chapter is structured as follows: in the first section we recall some basic properties of the Weyl algebra and some existing results on realisations of $\s$ in $A_1$. The second section is devoted to the classification of finite-dimensional Lie algebras which can be realised in $A_1$. This allows us to obtain in the third section all finite-dimensional Lie algebras which can be realised as subalgebras of ${\rm Der}(A_1)$. 
The fourth section of this chapter analyses  the family $\cal F$ in more detail.
In the last section we discuss some perspectives for future work.

\medskip

If, as already stated, the first section of this chapter recalls existing results on the Weyl algebra, the rest of the section are new results obtained in \cite{io3}.

\medskip

Throughout this chapter all Lie algebras will be complex unless otherwise stated.

\section{Summary of properties of $A_1$ and some known results}

\subsection{Basic properties of $A_1$; the Dixmier partition}
\label{basic}

We give here a list of properties of $A_1$ which will be needed in the sequel:
\begin{enumerate}
\item [{\bf  P1}] The elements $\{p^iq^j : i,j\in \mathbb N\}$ constitute a basis
of $A_1$; the centre $Z_{A_1}$ of $A_1$ is equal to $\CC$ and the scalars are the only invertible elements.
\item[{\bf P2}] All derivations of $A_1$ are inner derivations.
\item [{\bf  P3}] The algebra $A_1$ satisfies the commutative centraliser condition
 (\textit{ccc}): the centraliser of any element  in  $A_1\setminus Z_{A_1}$ is a commutative subalgebra (see \cite{Amitsur} and \cite{Dixmier}).
\item [{\bf  P4}]  For $i\in \mathbb N$ and $j\in \mathbb N^*$ we have
\beqa
\label{marcus}
[p,p^iq^j]=jp^iq^{j-1}, \ [q,p^iq^j]=-ip^{i-1}q^{j}.
\eeqa
\item [{\bf  P5}]  One has (see Theorem XIV of \cite{Littlewood} and page $213$ of \cite{Dixmier}) 
\begin{equation}
\label{Littlewood}
p^n q^n= pq(pq+1)(pq+2)\dots (pq+n -1) \mbox{ for all }n\in \mathbb N,
\end{equation}
\beqa
\label{relatia}
[p^\ell,q^j]=\sum_{r=1}^{min(j,\ell)}(-1)^{r+1}r!
C_j^r C_\ell^r p^{\ell-r}q^{j-r}, \mbox{ for any }j,\ell\in \mathbb N.
\eeqa
A useful consequence of  (\ref{Littlewood}) is that
$$[p^jq^j,p^kq^k]=[p^j q^j, (pq)^k]=0, \mbox{ for any }j,k\in \mathbb N.$$
\item[{\bf  P6}]  Two elements $p',q'\in A_1$ satisfying $[p',q']=1$ uniquely define an algebra
homomorphism from $A_1$ to itself and conversely, given an homomorphism $\alpha: A_1\to A_1$ we have $[\alpha (p),\alpha (q)]=1$. In \cite{Dixmier}, J. Dixmier
conjectured that an algebra homomorphism of $A_1$ is  invertible
and thus in fact an automorphism. This
conjecture - Dixmier's problem $1$ -  is still undecided.
\item [{\bf  P7}] If $n\in \mathbb N$, then
 ${\rm ad}(p^{n+1}):A_1\to A_1$ given by
${\rm ad}({p^{n+1}})(a)=[p^{n+1},a]$ for any $a\in A_1$
is locally  nilpotent, \textit{i.e.}, for each $a \in A_1$, there exists an $N\in \mathbb N$
(depending on $a$)
such that ${\rm ad}^N({p^{n+1}})(a)=0$. For $\lambda\in\CC$ one can then define $\Phi_{n,\lambda}:A_1\to
A_1$ by
$\Phi_{n,\lambda}(a)=\sum_{k=0}^N
\frac{(\frac{\lambda}{n+1}ad({p^{n+1}}))^k}{k!}(a)=
\exp(\frac{\lambda}{n+1}{\rm ad}({p^{n+1}}))(a)$ and this is the unique automorphism of $A_1$ such that
\beqa
\label{auto}
\Phi_{n,\lambda}(p)&=&p, \ \ \Phi_{n,\lambda}(q)=q+\lambda p^n. 
\eeqa
One defines $\Phi'_{n,\lambda}=\mathrm{exp}(-\frac{\lambda}{n+1}{\rm ad}({q^{n+1}}))$ similarly and shows that it is the unique automorphism of $A_1$ such that
\beqa
\label{auto'}
\Phi'_{n,\lambda}(q)&=&q, \ \ \Phi'_{n,\lambda}(p)=p+\lambda q^n.
\eeqa
The group of automorphisms of $A_1$ is generated by 
$\Phi_{n,\lambda}$ and 
$\Phi'_{n,\lambda}$ \cite{Dixmier}.
\end{enumerate}

\bigskip

\noi
{\bf The Dixmier partition:}
Let $a\in A_1$ be arbitrary. Set
\beqa
C(a)&=&\big\{ y\in A_1:\  {\rm ad}(a)(y)=0 \big\}, \nonumber \\
D(a)&=&\Big<y\in A_1 : {\rm ad}(a) (y)=\lambda y \mbox{ for some }
\lambda\in\CC \Big>,\nonumber \\
N(a)&=&\big\{ y\in A_1 :\ {\rm ad}^m(a) (y)=0 \mbox{ for some positive
integer } m \big\}. \nonumber \\
\eeqa
\noi
Note that $N(a)\cap D(a)=C(a)$.

\medskip

To illustrate these definitions, using {\bf P4} and {\bf P5}, consider the following examples:
\begin{itemize}
\item $C(p)=\CC[p]$, $D(p)=\CC[p]$ and $N(p)=A_1$ (see \cite{Dixmier});
\item $C(pq^2)=\CC[pq^2]=D(pq^2)\subset N(pq^2)$ but one has for example $p\in N(pq^2)\setminus C(pq^2)$ and $p^i\notin N(pq^2)$ (hence $N (pq^2)\ne A_1$, see \cite{Dixmier});
\item  $C(pq)= \CC [pq] = N(pq)$ but any $p^iq^j\in D(p^i q^j)$ (with eigenvalue $j-i$) and thus $D(pq)=A_1$ (see \cite{Dixmier});
\item $C(pq+pq^2)=N(pq+pq^2)=\CC [pq+pq^2]$ but for example $pq^2\in D(pq+pq^2)\setminus C(pq+pq^2)$ and furthermore $D(pq+pq^2)\ne A_1$ (see \cite{Dixmier});
\item  $C((pq)^2)=\CC [pq] = D((pq)^2)= N((pq)^2) \ne A_1$ (see \cite{joseph3}). 
\end{itemize}


 In \cite{Dixmier}, J. Dixmier  shows that  for
all $a\in A_1$, either $D(a)=C(a)$ or $N(a)=C(a)$.
As a consequence he proves the following \cite{Dixmier}

\begin{theorem}
\label{th-Dixmier}
(Dixmier partition) The set $A_1\setminus\CC$ is a disjoint union of the following
non-empty subsets:
\beqa
\Delta_1 &=& \big\{ a\in A_1\setminus{\CC}:\ D(a)=C(a), \ N(a)\ne C(a),\ N(a)=A_1  \big\}
\nonumber\\
\Delta_2 &=& \big\{ a\in A_1\setminus{\CC}:\ D(a)=C(a), \ N(a)\ne C(a),\ N(a)\ne A_1 \big\} \nonumber\\
\Delta_3 &=& \big\{ a\in A_1\setminus{\CC}:\ D(a)\ne C(a),\ N(a)=C(a), \ D(a)=A_1\big\}
\nonumber\\
\Delta_4 &=& \big\{ a\in A_1\setminus{\CC}:\ D(a)\ne C(a),\ N(a)=C(a), \ D(a)\ne A_1\big\} \nonumber\\
\Delta_5 &=& \big\{ a\in A_1\setminus{\CC}:\ D(a)=C(a),\ N(a)=C(a),\  C(a)\ne A_1\big\}. \nonumber
\eeqa
\end{theorem}
\noi
Note that this partition is stable under the action of $\A (A_1)$ and
multiplication by a non-zero scalar.

\begin{remark}
Note that
$$\CC=\big\{ a\in A_1 :\ D(a)=C(a),\ N(a)=C(a),\  C(a)=A_1\big\}.$$
\end{remark}

By the above examples,
$$
p\in \Delta_1,\ pq^2\in \Delta_2, \ pq\in\Delta_3,\ pq+pq^2\in\Delta_4,\ (pq)^2\in\Delta_5.
$$

Elements of $\Delta_1 \cup \Delta_2$ (resp. of $\Delta_1$) are said to
be nilpotent (resp. strictly nilpotent) and elements of $\Delta_3 \cup \Delta_4$ (resp. of $\Delta_3$) are said to
be semi-simple (resp. strictly semi-simple).
Moreover, one can classify, up to an automorphism of $A_1$, the strictly nilpotent and strictly semi-simple elements of $A_1$ by the following theorem.

\begin{theorem}
\label{Delta13}
\begin{enumerate}
\item (Theorem $9.1$ of
\cite{Dixmier}) 
 $a\in \Delta_1$ iff there exists an automorphism $\alpha$ of $A_1$ such
that $\alpha (a)$ is a polynomial in $p$.
\item (Theorem $9.2$ of
\cite{Dixmier}) 
  $a\in \Delta_3$ iff there exists an automorphism $\alpha$ of $A_1$ such
that $\alpha (a)=\mu pq + \nu$, for some $\mu\in{\CC}^*$ and $\nu\in{\CC}$ 
\end{enumerate}
\end{theorem}



\begin{remark}
\label{simplificari}
The notions of nilpotency and
  semi-simplicity in the Lie algebra sense are linked with the corresponding
 notions in the $A_1$ sense.
Let $x\in \g\setminus Z_\g$. If ${\rm ad}(x)$ is diagonalisable then
 $f(x)$ is semi-simple and  if ${\rm ad}(x)$ is nilpotent
then  $f(x)$ is nilpotent.
Indeed when ${\rm ad}(x)$ is diagonalisable, there exist $y\in\g$ and $\lambda\in\CC^*$ such that  $[x,y]=\lambda y$ and so, $f(y)\in D(f(x))$ and $f(y)\notin C(f(x))$. Hence $f(x)$ is semi-simple. Similar considerations hold for the nilpotent case.
\end{remark}

\begin{remark}
Recall that in general, in a semi-simple Lie algebra any element writes as the sum of a nilpotent and a semi-simple element. For the particular case of $\s$, any element is either nilpotent or semi-simple.
\end{remark}

\subsection{Some known results on realisations of $\s$ in $A_1$}
\label{rez-joseph-sectiune}

We now consider the problem of realisations of Lie algebras in $A_1$. We recall some known results on the classification of semi-simple Lie algebras in $A_1$ \cite{italienii} and on the classification of realisations of $\s$ \cite{joseph}. For this, we need the following definition

\begin{definition}
Let $\mathfrak{g}$ be a  complex Lie algebra and $A$ be a complex associative algebra. We define
$$ A^{\mathfrak{g}}=\{f:{\rm Hom}(\g,A_1): \ f
\mbox{ is injective and } f([a,b])=f(a)f(b)-f(b)f(a) \quad \forall a,b\in
A_1\}.$$
If $A^\g\ne \emptyset$ we say that the Lie algebra $\g$ can be realised as a Lie subalgebra of $A_1$ and 
an element of $A^\g$ is called a realisation of $\g$ in $A$.
\end{definition}

\begin{theorem}
Let $\g$ be a semi-simple complex Lie algebra of rank $\ell$. Then $A_n^\g=\emptyset$ if $\ell>n$.
\end{theorem}

Proofs of this theorem are given in \cite{italienii} (Theorem $3$) and \cite{joseph-jmp} (Theorem $4.2$). The only (up to an isomorphism) complex semi-simple Lie algebra that can be realised in the Weyl algebra is thus $\s$. We give an alternative proof of this result in the next section (see Theorem \ref{simple}); this proof extends to a larger class of associative algebras (see Remark \ref{simple-ccc}). 

Let us now recall a result of \cite{joseph}, where A. Joseph provides a classification of a particular class of realisations of $\s$ in $A_1$. For this we need the following definition: 

\begin{definition}
\label{triplet}
 Three non-zero elements $X,Y,H$ of $A_1$ are called an \textit{$\s$ triplet} if they satisfy the relations:
\beqa
[H,X]=2X,\quad
[H,Y]=-2Y,\quad
[X,Y]=H.
\eeqa
\end{definition}

\begin{proposition}
The set $A_1^{{\mathfrak sl}(2)}$ is in bijection with  the set of $\s$ triplets.
\end{proposition}
{\it Proof:}
 If $X,Y,H$ is an $sl(2)$ triplet, $f:{\mathfrak {sl}}(2)\to A_1$ given by
$f(e_+)=X, f(e_-)=Y, f(e_0)=H$ (where $e_\pm, e_0\in {\cal M}_2 (\CC)$ is the standard basis of $\s$, see subsection $2.1.2$) is a
Lie algebra homomorphism; conversely, $f(e_+), f(e_-),
f(e_0)$ is an $\s$ triplet if
$f:\s\to A_1$ is a Lie algebra homomorphism. QED

\medskip

By Remark \ref{simplificari}, one has that $H$ is semi-simple (in the Weyl algebra sense), that is
$$ H\in \Delta_3 \mbox{ or } H\in\Delta_4.$$

The realisations of $\s$ for which $H\in\Delta_3$ can be classified (see Theorem \ref{th-joseph}). In order to do this first note that if $H\in \Delta_3$ then by Theorem \ref{Delta13}, up to an automorphism of $A_1$, one has
$$H=\mu pq+\nu.$$
In fact one shows that $\mu=\pm 1$ or $\pm 2$. If $\mu<0$ define $\alpha\in \A (A_1)$ by $\alpha(p)=q$, $\alpha (q)=-p$.
We have
$$ \alpha (H)= -\mu pq+\mu +\nu$$
 which means that we can always suppose that $\mu > 0$ (up to an automorphism of $A_1$). Now, A. Joseph's result is

\begin{theorem}
\label{th-joseph}
(Lemma $2.4$ of \cite{joseph}) 
Let $H,X,Y\in A_1$ be an $\s$ triplet and let $H=\mu pq + \nu$ with $\mu\in CC^*$, $\nu \in \mathbb C$.
 Then:
\begin{enumerate}
\item $\mu=\pm 1$ or $\mu= \pm 2$.
\item \begin{enumerate} \item If $\mu=1$, there exists $a\in \mathbb C^*$ such that 
\begin{eqnarray}
\label{I-i}
X = -\frac12 a q^2, \ 
Y =  \frac12 \frac{1}{a} p^2, \
H = pq-\frac 12 \hskip 0.5cm \mbox{(solution I)}
\end{eqnarray}
(denote by $f_I^a:\s\to A_1$  the corresponding Lie algebra homomorphism).
\item If $\mu=2$, there exists $a\in {\mathbb C^*}$ and $ b\in \mathbb C$ such that either
\beqa
\label{IIA-i}
X = a(b+   pq) q, \
Y = -\frac{1}{a} p, \
H = 2pq +b \hskip 0.5cm \mbox{(solution IIA)}
\eeqa
\item[] or 
\beqa
\label{IIB-i}
X =-\frac{1}{a} q, \
Y = ap (b +  pq),\
H = 2pq +b  \hskip 0.5cm \mbox{(solution IIB)}
\eeqa
(denote by $f_{IIA}^{a,b}:\s\to A_1$  and resp. $f_{IIB}^{a,b}:\s\to A_1$ the corresponding Lie algebra homomorphisms).
\item if $\mu= -1$ then
\end{enumerate}
\end{enumerate}
\end{theorem}

A. Joseph further shows that the list above reduces, up to an automorphism of $A_1$, to the family 
\beqa
\label{def-F}
{\cal F}=(f_I^1, f_{IIA}^{1,b},f_{IIB}^{1,b'})_{b,b'\in\CC}.
\eeqa 
Moreover the realisations $f_{IIA}^{1,-b-2}$ and $f_{IIB}^{1,b}$ are equivalent under the action of the group $\A (A_1)\times \A(\s)$ (see Remark \ref{AB}). 

All these results are stated in \cite{joseph}. However, in  section $3.4$ explicit proofs are given; furthermore we will show that $\cal F$ represents normal forms for certain subsets of $A_1^\s$ under the action of the group $\A(A_1)\times\A(\s)$.

\subsection{On $A_1$ matrix representations}

Recall that one can think of $A_1$ as a non-commutative version of $\CC [x,y]$. A fundamental difference between $A_1$ and $\CC [x,y]$ is that $A_1$ has no finite-dimensional representations. Indeed, suppose that $A_1$ has a finite-dimensional representation and take $P$ and $Q$ to be  $n-$dimensional matrices satisfying
$$[P,Q]=1$$
(where $1$ denotes the $n-$dimensional identity matrix). This implies that ${\rm Tr}[X,Y]=0={\rm Tr}(1)=n$ which is a contradiction.

Furthermore, $A_1$ does not have any (left or right) ideal $I$ of finite codimension. Indeed if it did,  the quotient $A_1/I$  would be a  finite-dimensional representation, contradicting the previous result.


In \cite{berest}, Y. Berest and G. Wilson generalise the notion of representation to an ``approximate $n-$dimensional representation of $A_1$''. This is obtained by replacing the impossible matrix relation $[P,Q]=1$ by 
$${\rm rank}([P,Q]-1)=0 \mbox{ or }1$$
where $P$ and $Q$ are $n-$dimensional matrices.

\section{Classification theorem for finite-dimensional Lie algebras realisable in $A_1$}

We saw in the previous section that the only (up to isomorphism) complex semi-simple Lie algebra that can be realised in $A_1$ is $\s$. In this section we first give some examples of finite-dimensional Lie algebras which can be realised in $A_1$ and then prove that these are the only (up to isomorphism) Lie algebras with this property.

\subsection{Examples of realisations of Lie algebras in $A_1$}
\label{Exemple}

$  $
\begin{enumerate}
\label{initiatic}
\item [{\bf E1}] 
Any formulae of Theorem \ref{th-joseph}, for example
$$ X=-\frac 12 q^2,\ Y= \frac 12 p^2,\ H=pq-\frac12 $$
defines a realisation of $\s$ in $A_1$.
\item [{\bf E2}] The  elements $1, X, Y, H$ 
span a  Lie subalgebra of $A_1$ isomorphic to the direct
product $\s\times\CC$. 
\item [{\bf E3}] The  elements $1,p,q$  span  a Lie subalgebra
isomorphic to the three dimensional Heisenberg algebra 
${\cal H}_3$. 
\item [{\bf E4}] The elements  $1,p,q,X,Y,H$ 
span a Lie
subalgebra of $A_1$ isomorphic to a semi-direct product
$\s\ltimes{\cal H}_3$:
$$ [X, p]=q,\ [X, q]=0, \ [Y,p]=0,\ [Y,q]=p.$$   
\end{enumerate}

\begin{enumerate}
\item [{\bf E5}]
The elements $a^{i_1},\dots,a^{i_n}$, $a\in A_1$ span a Lie algebra isomorphic to
 the $n-$dimensional abelian Lie algebra ${\CC}^n$.
\end{enumerate}

\begin{remark}
\label{abeliana-dixmier}
A finite-dimensional abelian Lie subalgebra $A\ne\CC$ of $A_1$ is
contained in its commutant, $A'$, and this is a maximal abelian
 subalgebra
of $A_1$ (by Corollary $4.4$ of \cite{Dixmier}),  equal to $C(a)$ for some $a\in A_1$  (Corollary $4.3$ of \cite{Dixmier}). Hence the finite-dimensional abelian
Lie subalgebras of  $A_1$ are obtained as finite-dimensional linear
subspaces of maximal abelian  subalgebras of $A_1$.
\end{remark}

One may address the question: is $C(a)={\CC}[a]$? This is indeed true for $a\in \Delta_3\cup \Delta_4$ (Theorem $1.2$ of \cite{joseph3}). However it is not always the case for elements of $\Delta_1\cup\Delta_2\cup\Delta_5$. For example, $C(p^2)={\CC}[p]$.\\
Moreover, we have seen by the previous remark that an abelian Lie algebra (let's say for simplification of dimension two) is contained in $C(a)$ for some $a\in A_1$. One may also ask whether if, given any two commuting elements $a_1, a_2\in A_1$ there exists an element $a\in A_1$ such that $a_1$ and $a_2$ are polynomials in $a$. The answer to this question is no, a counterexample being given in \cite{joseph-jmp}: 

\begin{proposition}
\label{abeliana-ciudata}
Let $a_1=(p^2q)^3-\frac 32 (p^4q+ p^2qp^2)$ and $a_2=(p^2q)^2+2p^2$. 
Then there does not exist $a\in A_1$ such that $a_1, a_2\in \CC[a]$.
\end{proposition}
{\it Proof:} See Appendix A.

\begin{remark}
This example does not contradict Remark \ref{abeliana-dixmier}. Indeed, $\{a_1, a_2 \}\subset C(a_i)$, with $i=1,2$.
\end{remark}

\begin{enumerate}
 \item [{\bf E6}]
The $(n+1)$ elements $q, 1, p, \ldots, p^{n-1}$ ($n\ge 2$)
 span a non-abelian nilpotent Lie subalgebra isomorphic to
the filiform Lie algebra ${\cal L}_n$ \cite{muierea}.  If we set $X_0=-q,
X_k=\frac{1}{(n-k)!}p^{n-k}$, then the only non-zero commutation
relations are $[X_0, X_{k}]=X_{k+1}$ for $k=1,\dots,n-1$. 
The lower central series of this algebra is:
$$<1,p,\dots,p^{n-2}>,<1,p,\dots,p^{n-3}>,\dots,<1>,\{ 0 \}.$$
Note that the nilpotency index is $n-1$, which is the maximum value for a $(n+1)-$dimensional nilpotent Lie algebra.
\end{enumerate}

\begin{remark}
\label{H3-L2}
${\cal H}_3\cong{\cal L}_2$.
\end{remark}

\begin{enumerate}
\item [{\bf E7}] The $(n+2)$ elements $pq, q, 1, p, \ldots, p^{n-1}$
 span a non-nilpotent solvable Lie subalgebra whose derived algebra is isomorphic to
 ${\cal L}_n$.  If we set $h=pq, X_0=-q,
X_k=\frac{1}{(n-k)!}p^{n-k}$ for $k=1,\dots,n$ then the only non-zero commutation
relations are $[h,X_0]=X_0, [h, X_k]=-(n-k) X_k, [X_0, X_{k}]=X_{k+1}$
 for $k=1,\dots,n-1$. We denote this algebra by $\tilde {\cal L}_n$. Furthermore, $\tilde {\cal L}_n\cong \CC\ltimes {\cal L}_n$.
\item [{\bf E8}] The $(n+1)$ elements $pq, p^{i_1}, \ldots, p^{i_n}$, where $i_1,\dots, i_n$ are  distinct positive integers,
 span a non-nilpotent solvable Lie subalgebra whose derived algebra is
  $n-$dimensional and abelian.  If we set $h=pq$, $X_k=p^{i_k}$ for $k=1,\dots,n$ then the only non-zero commutation
relations are  $[h, X_k]=-i_k X_k$.
We denote this Lie algebra by ${\mathfrak r}_n(i_1,\dots,i_n)$. 
It is clear that ${\mathfrak r}_n(i_{\sigma(1)},\dots,i_{\sigma(n)})\cong {\mathfrak r}_n(i_1,\dots,i_n)$ for any permutation $\sigma\in S_n$, that ${\mathfrak r}_n(i_1,\dots,i_n)$ has a non-trivial centre iff one of the indices is zero and that ${\mathfrak r}_n(0,i_2,\dots,i_n)\cong {\mathfrak r}_{n-1}(i_2,\dots,i_n)\times\CC$. Furthermore ${\mathfrak r}_n(i_1,\dots,i_n)\cong \CC\ltimes \CC^n$.
\end{enumerate}

\begin{remark}
Proposition \ref{abeliana-ciudata} provides us with another realisation in $A_1$ of $\rr_2 (2,3)$ which is obtained by adding $pq$ to  $a_1$ and $a_2$.
\end{remark}

\bigskip

In the rest of this section we prove that these Lie algebras are actually the only finite-dimensional Lie algebras that can be realised in $A_1$. We firstly treat the case of non-solvable Lie algebras and then, the more complicated case of solvable Lie algebras.

\subsection{Non-solvable Lie algebras}

We first give our proof that $\s$ is the unique (up to isomorphism) complex semi-simple Lie algebra that can be realised in $A_1$. We then find all reductive Lie algebras which can be realised in $A_1$.
As already stated in the previous section, this proof extends to a larger class of associative algebras: the algebras that satisfy the {\it ccc} (see {\bf P3}). 

\begin{proposition}
\label{simple}
(i) If $\mathfrak{g}$ is a semi-simple complex Lie algebra that can be realised in $A_1$, then $\g\cong\s$.\\
(ii) The only non-abelian reductive complex Lie algebras 
that can be realised in $A_1$ are isomorphic to either $\s$ or $\s\times\CC$.
\end{proposition}
\textit{Proof}: (i) The only complex semi-simple Lie algebra of rank $1$ is $\s$; furthermore it can be realised in $A_1$ (see example {\bf E1}). 
Let now $\g$ be a complex semi-simple Lie algebra of 
rank $>1$; we prove that $A_1^\g=\emptyset$.
 
To do this, suppose for contradiction that there exists $f\in
 A_1^{\mathfrak{g}}$.
Let  $\mathfrak{h}\subset\mathfrak{g}$ be a Cartan subalgebra and let
$$\mathfrak{g}=\mathfrak{n}_- \oplus \mathfrak{h} \oplus \mathfrak{n}_+$$
be the corresponding triangular decomposition for some choice of simple
roots.
 Let $x\in\mathfrak{n}_+$
be a non-zero highest root vector and set $X=f(x)$. 
 The commutant of $x$
in  $\mathfrak{g}$, $C_{\mathfrak{g}}(x)$, contains $\mathfrak{n}_+$.

Consider now the case when $\g$ does not contain the direct product of two distinct copies of $\s$. 
Since  rank$({
\mathfrak{g}})>1$ it follows that  $\mathfrak{n}_+$ and hence
$C_{\mathfrak{g}}(x)$ are not abelian
and  so
 $C(X)$ is also not abelian. By the \textit{ccc}, $X\in Z_{A_1}$ and thus $x\in Z_\g$. But since  $\g$ is semi-simple it has trivial centre and thus $x=0$  which is a contradiction.

Let us now treat the case when $\g$ does contain the direct product of two distinct copies of $\s$, {\it i.e.} $\g=\g_1\times\g_2$, where $\g_1\cong\s$ and $\g_2\cong\s$. Take now $x\in\g_1$ non-zero and set $X=f(x)$. Hence $\g_2\subseteq C_\g (x)$ and hence $f(\g_2)\subseteq C (X)$. But $f(\g_2)$ is not abelian and so $C(X)$ is also not abelian. By the {\it ccc}, one has as above a contradiction.


\noi
(ii): Let $\g=\g_1\times \mathfrak z$ be a reductive complex Lie algebra where
$\g_1$ is  semi-simple  and  $\mathfrak z$ is  the centre of $\g$ (neither being trivial).

From part (i)  one has $\g_1\cong\s$. By example {\bf E2}, we know that $\s\times\CC$ can be realised in $A_1$. We now prove that this is actually the only reductive complex Lie algebra with this property.

 Let $z$ be an element of $\mathfrak z$ and set
$Z=f(z)$. Then $C(Z)$ contains $f(\g_1)$ which is not abelian and hence, by the \textit{ccc}, $Z\in Z_{A_1}$ and thus $f({\mathfrak z})\subseteq Z_{A_1}=\CC$. Since $\CC$ is of dimension $1$ the result follows. QED

\begin{remark}   
\label{simple-ccc}
This proof of part (i) remains true if we replace $A_1$ by any algebra $A$ satisfying the 
\textit{ccc}. However
it is not clear that there are any realisations of $\s$ at all in an arbitrary algebra satisfying the {\it ccc}.
We can affirm that if one has a complex semi-simple Lie algebra of rank $>1$, then it cannot be realised in any algebra satisfying the {\it ccc}.

The proof of part (ii)  up to proving that $f({\mathfrak z})\subseteq Z_{A_1}$ remains true if we replace $A_1$ by $A$. 
Note that {\it a priori}  the dimension of $Z_A$ is arbitrary.
One example of such an algebra is the universal enveloping algebra ${\cal U}(\s)$ (for other examples see \cite{Bavula})).
\end{remark}



To find all  finite-dimensional non-solvable Lie subalgebras we recall the Levi-Malcev theorem (see subsection $2.2.2$) which states that any finite-dimensional Lie algebra is isomorphic to the direct product of a semi-simple Lie algebra with a reductive Lie algebra $\rr$ (its radical). From part (i) of Proposition \ref{simple}, the semi-simple part is isomorphic to $\s$. Thus, we need to investigate solvable Lie subalgebras of $A_1$ which are stable by ${\rm ad}(X)$, ${\rm ad}(Y)$ and ${\rm ad}(H)$, where $X$, $Y$, $H$ are a standard basis of the semi-simple part.

\begin{definition}
\label{triplet2}
(i) An element $v\in A_1$ is of weight $\lambda\in\CC$ if
$[H,v]=\lambda v$.\\
(ii) 
The set of elements  of weight $\lambda$ in a linear 
subspace $E\subseteq A_1$ will be denoted by
$E_\lambda$.
\end{definition}

\begin{lemma}
\label{lema-marcus}
Let $X, Y, H$ be an $\s$ triplet and
let ${\mathfrak l}\subseteq A_1$ be a Lie subalgebra stable under 
${\mathrm {ad}}(X),{\mathrm {ad}}(Y)$ and ${\mathrm {ad}}(H)$. Suppose there
exists $v\ne 0$
 in $\mathfrak l$ of weight  $\lambda\in\CC^*$ such that 
$[X,v]=0$. Then $[v, [Y,v]]\ne 0$ and is of weight $2\lambda -2$. 
\end{lemma} 
\textit{Proof:} We suppose for contradiction that $[v, [Y,v]]=0$. Then
$C(v)$ contains  $[Y,v]$ and $X$ (by hypothesis). But, using the Jacobi identity,
$[X,[Y,v]]=[H,v]+[Y,[X,v]]=\lambda v + 0 =\lambda v \ne 0$ and therefore $C(v)$
is non-abelian. 
By the \textit{ccc}, $v\in Z_{A_1}$ and so $[H,v]=0$ which is a contradiction.
 Hence, $[v, [Y,v]]\ne 0$. Since $v$ is of weight $\lambda$ ($[H,v]=\lambda v$) then $[Y,v]$ is of weight $\lambda -2$ and then $[v, [Y,v]]$
is of weight $2\lambda -2$. QED


\begin{proposition}
\label{oarecare} 
Let $X, Y, H$ be an $\s$ triplet.\\
(i) Let ${\mathfrak l}\subseteq A_1$ be a finite-dimensional  Lie
subalgebra stable under  ${\rm {ad}}(X),{\rm {ad}}(Y)$ and ${\rm {ad}}(H)$. 
Then ${\mathfrak l}_\lambda=\{ 0 \}$ for $|\lambda|>2$.\\
(ii) Let ${\mathfrak r}\subseteq A_1$ be a finite-dimensional  solvable Lie subalgebra stable under  ${\mathrm {ad}}(X),{\mathrm {ad}}(Y)$ and ${\mathrm {ad}}(H)$. 
Then ${\mathfrak r}_\lambda=\{ 0 \}$ for $\vert\lambda\vert>1$.
\end{proposition}
\textit{Proof:} (i) Recall that if ${\mathfrak l}$ is a {finite}-dimensional representation of $\s$ then 
 ${\mathfrak l}_\lambda =\{ 0 \}$  iff ${\mathfrak l}_{-\lambda}=0$.
Let $\lambda_{max}$ be the largest eigenvalue of ${\rm ad}(H)$
restricted to ${\mathfrak l}$. If $\lambda_{max}=0$ the result is
obvious. If $\lambda_{max}>0$, then $\lambda_{max}\ge 2\lambda_{max}-2$
by Lemma \ref{lema-marcus}. Hence $\lambda_{max}\le 2$ and the result follows. 

\noi
(ii) From part (i), we have ${\mathfrak r}_{\lambda}= \{ 0 \}$ for all $|\lambda|>2$. We now prove that for the solvable subalgebra case, this is also true for $|\lambda|=2$.

 Suppose for contradiction that ${\mathfrak r}_{2} \ne \{ 0 \}$.
Then, there exists $v\ne 0$ in $\mathfrak r$ of weight $2$ and $[X,v]\in {\mathfrak r}_{4}=\{ 0 \}$  by part (i). Hence by Lemma \ref{lema-marcus} one finds $[v,[Y,v]\ne 0$ of weight $2$ in $\mathfrak r'$ (since it can be written as the commutator of two elements, $v$ and $[Y,v]$ of $\mathfrak r$). The process can now be reiterated and hence ${\mathfrak r}^{(i)}_2\ne \{ 0 \}$
for all positive $i$. But ${\mathfrak r}$ is solvable so that by
definition ${\mathfrak r}^{(m)}=\{ 0 \}$ for some $m\in \NN$. This is a
contradiction. QED

\medskip

One notices that part (i) of this Proposition imposes a condition on any Lie algebra $\mathfrak l$ satisfying the above conditions. However, the supplementary restriction obtained in part (ii) is valid only for a {\it solvable} Lie algebra $\rr$.

\begin{proposition}
\label{finala}
Let $X, Y, H$ be an $\s$ triplet and
let ${\mathfrak r}\subseteq A_1$ be a finite-dimensional  solvable Lie
subalgebra stable under ${\mathrm ad}(X),{\mathrm ad}(Y)$ and ${\mathrm
ad}(H)$. 
Then ${\mathfrak r}$ is isomorphic to either $\{ 0 \}$, $\CC$ or ${\cal H}_3$
 (the three-dimensional Heisenberg algebra).
\end{proposition}
\textit{Proof:} First note that ${\mathfrak r}={\mathfrak r}_0\oplus{\mathfrak r}_{-1}\oplus{\mathfrak r}_1$ by part (ii) of Proposition \ref{oarecare}. Then $[H, {\mathfrak r}_0]=0$ which means that $\rr_0\subseteq C(H)$. But, since $H\notin\CC$, by the {\it ccc},  $\rr_0$ abelian.
If $v\in {\mathfrak r}_0$ then $[X,v]$ is of weight $2$ that is $[X,v]\in {\mathfrak r}_2$ which, by part (ii) of Proposition \ref{oarecare}, means that $[X,v]=0$. Hence $[X, {\mathfrak r}_0]=0$; similarly $[Y, {\mathfrak r}_0]=0$ and by part (ii) of Proposition \ref{simple}, one has ${\mathfrak r}_0\subseteq \CC$ (i.e. ${\rm dim}\ {\mathfrak r}_0\le 1$).
Furthermore  $[{\mathfrak r}_{-1},{\mathfrak r}_1]\subseteq{\mathfrak r}_0$ since ${\mathrm
{ad}}(H)$ is a derivation.

Suppose first that $\mathrm{dim}\ {\mathfrak r}_{1}\ge 2$ and let
$v\in{\mathfrak r}_1\setminus \{0\}$. 
The kernel, $Z_{{\mathfrak r}_{-1}}(v)$, of the linear map 
${\rm ad}(v):{\mathfrak r}_{-1}\to{\mathfrak r}_{0}$ is of dimension $\ge 1$ and therefore contains a vector $w\ne 0$. Hence $C(v)$ contains $X$ and $w$. But $[X,w]\ne 0$ since ${\rm ad}(X) : {\mathfrak r}_{-1}\to {\mathfrak r}_{1}$ is an isomorphism and $w\ne 0$; so $C(v)$ is not abelian. 
By the \textit{ccc}, $v$ is a scalar which is a contradiction since $[H,v]=v$.
Therefore $\mathrm{dim}\ {\mathfrak r}_{-1}=\mathrm{dim}\ {\mathfrak r}_{1}\le 1$.

If $\mathrm{dim}\ {\mathfrak r}_{-1}=1$, then $ {\mathfrak r}_{-1}\oplus {\mathfrak r}_{1}$ is not abelian by 
Lemma \ref{lema-marcus} and so $\mathrm{dim}\ {\mathfrak
r}_0=1$. Since we already know that $\rr_0 \subseteq\CC$, this implies that $\rr_0=\CC$.
Hence
${\mathfrak r}_0$ is the centre of ${\mathfrak r}$ and it is now
obvious that ${\mathfrak r}$ is isomorphic to the three-dimensional Heisenberg algebra.

If $\mathrm{dim}\ {\mathfrak r}_{-1}=0$ then ${\mathfrak r}={\mathfrak r}_0$ and 
${\mathfrak r}_0=\{ 0 \}$ or $\CC$. QED

\bigskip

We can now conclude  this subsection with the following theorem:
\begin{theorem}
\label{mareata}
Let $\g$ be a finite-dimensional complex Lie algebra which is not solvable. Then $\g$ can be realised in $A_1$ iff $\g$ is isomorphic to one of the following:
\begin{enumerate}
\item $\s$,
\item $\s \times \CC$,
\item $\s\ltimes{\mathcal H}_3 $.
\end{enumerate}
\end{theorem}
\textit{Proof:} ($\Leftarrow$) By Examples {\bf E1}, {\bf E2}, {\bf E3} we know that $\s$,
 $\s \times \CC$, $\s\ltimes{\mathcal H}_3 $ can be realised in $A_1$. 

\noi
($\Rightarrow$) 
As already stated, by the Levi-Malcev theorem we can write $\g$ as a semi-direct
product $\g={\mathfrak s}\ltimes {\mathfrak r}$, where $\mathfrak r$ and $\mathfrak s$ are respectively  solvable and semi-simple subalgebras of $\g$. Let $f\in A_1^\g$. Then applying part (i) of Proposition \ref{simple} and Proposition \ref{finala}  to $f(\g)$ we obtain the result. 
Moreover, when ${\mathfrak r}\cong {\mathcal H}_3 $ the semi-direct product $\s \ltimes{\mathcal H}_3 $ is isomorphic to the semi-direct product of example {\bf E4} because ${\mathfrak r}={\mathfrak r}_{-1}\oplus {\mathfrak r}_0 \oplus {\mathfrak r}_1$ by the proof of Proposition \ref{finala}. QED

\subsection{Solvable Lie algebras}

In the previous subsection we obtained all (up to isomorphism) finite-dimensional non-solvable Lie algebras which can be realised in $A_1$. We now obtain all solvable finite-dimensional Lie algebras with this property. Keeping in mind that any finite-dimensional abelian Lie algebra can be realised in $A_1$ (see example {\bf E5}), we firstly treat here the case of nilpotent non-abelian Lie algebras and then of solvable non-nilpotent Lie algebras.

\subsubsection{Nilpotent non-abelian Lie algebras}

Recall the definitions (see subsection $2.2.1$) of the lower and upper central series $(\g_{(i)})_{i\in\NN}$ and $(\g^{(i)})_{i\in\NN}$ of a Lie algebra $\g$: 
$$\g_{(0)}=\g,\ \g_{(i+1)}=[\g,\g_{(i)}]$$
and
$$\g^{(0)}=\g,\ \g_{(i+1)}=[\g^{(i)},\g^{(i)}].$$

\begin{theorem}
\label{nilpotentele}
Let $\n\subset A_1$ be a nilpotent, non-abelian Lie
subalgebra of dimension $n$. 
Let  $k\ne 0$ be the unique
positive integer
such that $\n_{(k)}\ne\{ 0 \}$ and $\n_{(k+1)}=\{ 0 \}$.
\begin{enumerate}
\item $Z_\n=\n_{(k)}=\CC$;
\item There exist $P,Q\in \n$ such that $[P,Q]=1$ and $\n=<P>\oplus
Z_\n (Q)$;
\item $\n_{(1)}\subset Z_\n (Q)$;
\item ${\mathrm{dim}}\ \n_{(i)}=n-i-1$
for $1\le i\le k$;
\item $k=n-2$.
\end{enumerate}
\end{theorem}
\textit{Proof:} $1$: Let $z\in Z_\n$. Then its commutant in $A_1$ contains
$\n$ and is non-commutative. By the \textit{ccc}, $z$ is a
scalar and therefore $Z_\n\subseteq\CC$. But
 $Z_n$ contains $\n_{(k)}$ and so $Z_\n = \n_{(k)}=\CC$.

$2$: Since $\n_{(k)}=[\n,\n_{(k-1)}]=\CC$, there exist $P\in\n,\
Q\in\n_{(k-1)}$ such that $[P,Q]=1$. 
It is clear that  $<P>\cap Z_\n
(Q)=\{ 0 \}$ since $P$ and $Q$ do not commute. We now show that $\n=<P>+
Z_\n (Q)$ which will imply that $\n=<P>\oplus Z_\n (Q)$.
Let $v\in\n$. Then there exists $\lambda\in\CC$ such that $[v,Q]=\lambda$ since
$[\n,\n_{(k-1)}]=\CC$. Thus $v-\lambda P\in Z_\n(Q)$ and $v=\lambda
P+(v-\lambda P)\in <P>+Z_\n (Q)$.

$3$: Let $z\in Z_\n(Q)$. Then 
$$ [Q,[P,z]]=[[Q,P],z]+[P,[Q,z]]=0$$
because $[Q,P]=-1$ and $[Q,z]=0$. Hence $[P, Z_\n (Q)]\subseteq Z_\n
(Q)$ and since $[Z_\n (Q), Z_\n (Q)]\subseteq Z_\n(Q)$, 
this means using part $2$ that $\n_{(1)}=[\n,\n]$ is contained in $Z_\n (Q)$. 

$4$:   Since $\n=<P>\oplus
Z_\n (Q)$ and since by the \textit{ccc} $Z_\n (Q)$ is abelian, we have
\beqa
\label{adjoainta}
\n_{(i)}={\rm ad}^i(P)(Z_\n(Q)) \ \forall i\in\NN^*.
\eeqa
\noi
If $z\in Z_\n (Q)$ is such that ${\rm ad}(P)(z)=0$ then $z\in
 Z_{A_1} (P)\cap Z_{A_1} (Q)$. By Corollary $4.7$ of \cite{Dixmier},
 $Z_{A_1} (P)\cap Z_{A_1} (Q)=\CC$ since $P$ and $ Q$ do not commute.
Hence $z\in\CC$ and 
\beqa
\mathrm{Ker}\ {\rm ad}(P) \vert _{\n}&=&<P>\oplus\ \CC \nonumber \\
\mathrm{Ker}\ {\rm ad}(P) \vert _{Z_\n(Q)}&=&\CC.
\eeqa
This implies $4$. 

$5$: This follows immediately from $4$. QED

\begin{corollary}
\label{ln}
Let $\n\subset A_1$ be a finite-dimensional, non-abelian Lie subalgebra of dimension $n$. Then $\n\cong {\cal
L}_{n-1}$.
\end{corollary}
\textit{Proof:} 
Since $\n_{(k-1)}={\rm ad}^{k-1}(P)(Z_\n(Q))$ and
since $Q\in \n_{(k-1)}$ there exists $w\in Z_\n(Q)$ such that
${\rm ad}^{k-1}(P)(w)=Q$. It is well known that $J_{n-2}=w,
J_{n-3}={\rm ad}(P)(w),J_{n-4}={\rm ad}^2(P)(w),\ldots,J_0={\rm ad}^{k}(P)(w)$ 
are linearly independent (since
${\rm ad}^{k}(P)(w)\ne 0$ and 
${\rm ad}^{k+1}(P)(w)=0$) and by a dimension count, these vectors
form a basis of $Z_\n(Q)$. The only non-zero commutation relations of $\n$ in
the basis $J_{-1}=P, J_0, \ldots, J_{n-2}$ are 
$[J_{-1},J_i]=J_{i-1} \mbox{ for }  1\le i \le n-2$.
By redefining $X_0=J_{-1}$, $X_i=J_{n-i-1}$, where $0\le i \le n-1$
one obtains the standard commutation relations of ${\cal
L}_{n-1}$ (see {\bf E6}). QED

\begin{remark}
Note that in the realisation {\bf E6}, the role of $P$ is played by $-q$ and the role of $Q$ is played by $p$. 
\end{remark}

\subsubsection{Solvable non-nilpotent Lie algebras}

We have seen that the only finite-dimensional nilpotent Lie algebras that can be realised in $A_1$ are isomorphic to an ${\cal L}_n$ or an abelian Lie algebra. When one considers the case of solvable non-nilpotent Lie algebras which can be realised in $A_1$, of great use is the property (see subsection $2.2.1$) that states that the derived Lie algebra of a solvable Lie algebra is nilpotent. Hence one deals with two cases of solvable non-nilpotent Lie algebras:
\begin{enumerate}
\item the derived algebra is isomorphic to an ${\cal L}_n$;
\item the derived algebra is isomorphic to an abelian Lie algebra.
\end{enumerate}

Before treating these two cases, we need the following theorem:

\begin{theorem}
\label{solvable1}
Let ${\mathfrak g}\subset A_1$ be a finite-dimensional 
non-nilpotent Lie subalgebra and let $\g'$ be its derived algebra. Let $h\in \g$ be such that
${\rm ad}(h)$ is not nilpotent, let $0,\lambda_1,\dots,\lambda_k$ be its
 distinct eigenvalues and let $E_{0},
 E_{\lambda_1},\dots,E_{\lambda_k}$ be the corresponding eigenspaces.
\begin{enumerate}
\item $\g=E_{0}\oplus E_{\lambda_1}\oplus \ldots \oplus E_{\lambda_k}$
\item $\g'=(\g'\cap E_{0})\oplus E_{\lambda_1} \oplus\ldots \oplus E_{\lambda_k}$
\item $E_0=<h>$ or $E_0=<h>\oplus\CC$.
\end{enumerate}
\end{theorem}
\textit{Proof:} $1$: Recall that, by Engel's theorem, for a complex finite-dimensional non-nilpotent Lie algebra $\g$ there does indeed exist $h\in\g$
such that ${{\rm ad}}(h)\vert_\g$ is not nilpotent. Moreover, since $[h,h]=0$, ${{\rm ad}}(h)$ has $0$ as an eigenvalue.
By the theory of endomorphisms,
\beqa
\label{endo}
 {\mathfrak g}={\mathrm{Ker}}\  {{\mathrm ad}}^{m_0}(h)|_\g\oplus{\mathrm{Ker}} ({\mathrm ad}({h})|_\g -
\lambda_1)^{m_1}\oplus \dots \oplus {\mathrm{Ker}} ({\mathrm ad}({h})|_\g -
\lambda_k)^{m_k},
\eeqa
\noi where the characteristic polynomial of
${\rm ad}({h})\vert_\g$ is
$x^{m_0}(x-\lambda_1)^{m_1}\ldots(x-\lambda_k)^{m_k}$.
By Proposition $6.5$ of \cite{Dixmier},
$${\mathrm{Ker}} ({\rm ad}({h}) -
\lambda_j)^{m_j}={\mathrm{Ker}} ({\rm ad}({h}) -
\lambda_j).$$
Furthermore, since ${\rm ad}({h})$  is  not nilpotent, it  has a non-zero
eigenvalue which means that $C(h)\ne D(h)$.
Now, by Theorem \ref{th-Dixmier}, $C(h)=N(h)$ which can be written
$${\mathrm{Ker}} \ {\rm ad}^{m_0}(h)={\mathrm{Ker}} \ {\rm ad}({h}).$$
Inserting these results in \eqref{endo}, one obtains
$$\g=E_{0}\oplus E_{\lambda_1}\oplus \ldots \oplus E_{\lambda_k}$$

\noi
$2$: Note that
$v_i=\frac{1}{\lambda_i} [h, v_i]$ for all $v_i\in E_{\lambda_i}$ and thus $v_i\in\g'$ for all $v_i\in E_{\lambda_i}$. Hence one has
$$\g'=(\g'\cap E_{0})\oplus E_{\lambda_1} \oplus\ldots \oplus E_{\lambda_k}$$
and furthermore ${{\rm ad}}({h})$ is a diagonalisable derivation of ${\mathfrak g}'$.

\noi
$3$:  First note that $E_0\subseteq \, <h>\oplus\CC$. We now prove that one cannot have other distinct elements in $E_0$. For  this, let us take $h'\in E_0$. Then $[h,h']=0$. Furthermore, using the Jacobi identity, one has
$$[h',v_i]=\frac{1}{\lambda_i} [h',[h,v_i]]=\frac{1}{\lambda_i} [h,[h',v_i]]$$
for any $v_i\in E_{\lambda_i}$. Thus $[h',v_i]\in E_{\lambda_i}$  for any $v_i\in E_{\lambda_i}$ which means that
$[h',E_{\lambda_i}]\subseteq
E_{\lambda_i}$. Thus there exist $\alpha\in\CC$,
$v\in E_{\lambda_1}$ such that $[h',v]=\alpha v$.  Hence
$[h'-\frac{\alpha}{\lambda_1}h, v]=0$ which means that $h$ and $v$
commute with $h'-\frac{\alpha}{\lambda_1}h$. But $[h,v]\ne 0$ and so by the {\it
ccc},  $h'-\frac{\alpha}{\lambda_1}h\in\CC$. QED

\begin{remark}
\label{simplificare}
One notices that
if $\g$ is semi-simple, part $3$ of  Theorem \ref{solvable1} implies that the rank of $\g$ is equal to $1$; thus we have provided an alternative proof of
Proposition \ref{simple}.
However the proof of Proposition \ref{simple} holds for any
algebra satisfying the {\it ccc}, whereas the proof of the Theorem \ref{solvable1} 
depends on the existence of the Dixmier partition (Theorem \ref{th-Dixmier}) and other special properties not in general available for an arbitrary algebra satisfying the {\it ccc} (namely Proposition $6.5$ of \cite{Dixmier}), which are not
available for a general algebra satisfying the {\it ccc}.
\end{remark}

We now treat the case of a solvable non-nilpotent Lie algebra whose derived algebra is isomorphic to ${\cal L}_n$ and which is realisable in $A_1$.

\begin{theorem}
\label{solvable22}
Let ${\mathfrak r}\subset A_1$ be a finite-dimensional solvable 
non-nilpotent Lie subalgebra whose derived algebra ${\mathfrak r}'$ is isomorphic to ${\cal L}_n$. 
Then  ${\mathfrak r}$ is
isomorphic to $\tilde {\cal L}_n$.
\end{theorem}
{\it Proof:} Since $\rr'\subset A_1$ is isomorphic to ${\cal L}_n$, its
centre is $\CC$. Thus the centre of $\rr$ is $\CC$ and, by Theorem \ref{solvable1}, 
$E_0=<h>\oplus \CC$ (with the notations of Theorem \ref{solvable1}).

As remarked in part $2$ of the proof of  Theorem \ref{solvable1}, ${\rm ad}(h)\vert_{\rr'}$ is
diagonalisable and therefore one can make use of
 Theorem $1$ of \cite{GOZE} which states that there exists a basis $X_0,\ldots, X_n$ of
${\mathfrak r}'$ of eigenvectors of ${\rm ad}(h)\vert_{\rr'}$ such that the only
non-zero commutation relations are
\beqa
\label{baza}
[X_0, X_i]&=&X_{i+1},\ i=1,...,n-1\nonumber \\
\left[ h, X_j\right]&=&\alpha_j X_j, \ j=0,...,n.
\eeqa
\noi
Since $X_n$ commutes with the non-commuting $X_0,\dots, X_n$, by the {\it ccc}   $X_n\in\CC$, $[h,X_n]=0$ and  $\alpha_n=0$.
But ${\rm ad}(h)$ is a (non-zero) derivation, so one has
$$ {\rm ad}(h) ([X_0,X_i])= [{\rm ad}(h)(X_0),X_i]+[X_0,{\rm ad}(h)(X_i)]$$
and, using \eqref{baza} one obtains the following conditions on 
the eigenvalues of
${\rm ad}(h)\vert_{\rr'}$ satisfy 
\beqa
\label{cond1}
\alpha_{i+1}=\alpha_0+\alpha_i, \ i=1,..., n-1.
\eeqa
\noi
From this it follows that the eigenvalues of ${\rm ad}({h})|_{\rr'}$
are $\alpha_0, (n-1)\alpha_0,(n-2)\alpha_0,\dots, 0$, with
$\alpha_0\ne 0$.
The only non-zero commutation relations of
$\rr$ are then
\beqa
\label{baza-fin}
[X_0, X_i]&=&X_{i+1},\ i=1,..,n-1 \nonumber \\
\left[ \frac{1}{\alpha_0} h, X_0\right]&=& X_0 \nonumber \\
\left[  \frac{1}{\alpha_0} h, X_i\right]&=&- (n-i)X_i \ i=1,..,n-1.
\eeqa
This shows that $\rr$ is isomorphic to $\tilde {\cal{L}}_n$. QED 

\bigskip

We now treat the  case of a solvable Lie algebra whose derived algebra is isomorphic to an abelian Lie algebra of dimension $n$ and which is realisable in $A_1$.

For this, we need the following lemma (also see Theorem $3.2$ of \cite{joseph-jmp}):

\begin{lemma}
\label{joseph}
Let $h, X_1, X_2\in A_1\setminus\{ 0 \}$ and
 $(\lambda_1,\lambda_2)\in {\ZZ}^2\setminus \{(0, 0)\}$
be such that 
\beqa
\label{cond-joseph}
[h, X_1]=\lambda_1 X_1,\ [h, X_2]=\lambda_2 X_2,\mbox{ and }[X_1,
X_2]=0.
\eeqa
\noi Then $\lambda_1 \lambda_2 > 0$ and
there exists $a\in\CC^*$ such that
$X_1^{|\lambda_2|}=a X_2^{|\lambda_1|}$.
\end{lemma}
{\it Proof}: Since $[X_1, X_2]=0$, by Theorem $3.1$ of \cite{joseph-jmp} there exist
$m,n\in\NN^*$ and
$\alpha_{ij}\in\CC$ not all equal to zero such that
\beqa
\label{lema-joseph}
\sum_{i=0}^m \sum_{j=0}^n \alpha_{ij} X_1^i X_2^j=0
\eeqa
Then this can be rewritten as
\beqa
\sum_{u\in U} \left( \sum_{(i,j): i \lambda_1 + j \lambda_2=u} \alpha_{ij} X_1^i X_2^j\right)=0
\eeqa
\noi
where
$ U=\{ i \lambda_1 + j \lambda_2 \in \ZZ : \ 0\le i \le m, 0\le j \le n\}$.
Since eigenvectors corresponding to distinct eigenvalues are linearly 
independent we deduce that, for all $u\in U$, 
$$\sum_{(i,j): i \lambda_1 + j \lambda_2=u} \alpha_{ij} X_1^i X_2^j=0.$$
There exists $(i_0,j_0)\in \ZZ^2$ such that $\alpha_{i_0,j_0}\ne 0$. We
set $u_0=i_0 \lambda_1 + j_0 \lambda_2$
and 
$$S=\{(i,j)\in \ZZ^2:\lambda_1 + j \lambda_2=u_0\}\cap [0,m] \times [0,n].$$ 
In ${\RR}^2$ the solutions $(x,y)$ of the equation $x \lambda_1 +
y\lambda_2=u_0$ define an affine line and $S$ is a discrete subset of
this line.
It is then easy to see that 
there exist $(i',j')\in \ZZ^2$, $j'\ge 0$ and $(i_m,j_m)\in S$ such that 
\beqa
\label{aj-simplu}
i'\lambda_1+j'\lambda_2=0
\eeqa
\noi
and every
element $(i,j)$ of $S$ is of the form $(i,j)=(i_m, j_m)+k_{ij}(i',j')$ for
some $k_{ij}\in\NN$. 
Then one can write in the (left) field of fractions of
$A_1$ (see page $210$ of \cite{Dixmier})
$$\sum_{(i,j): i \lambda_1 + j \lambda_2=u_0} \alpha_{ij} X_1^i
X_2^j=X_1^{i_m}X_2^{j_m} \left(\sum_{(i,j): i \lambda_1 + j \lambda_2=u_0}\alpha_{ij} (X_1^{i'}X_2^{j'})^{k_{ij}}\right)=0.$$
\noi
Since by hypothesis $X_1^{i_m}X_2^{j_m}\ne 0$ this means that
a polynomial $P$ in the variable
$X_1^{i'}X_2^{j'}$ is equal to $0$. Factorising (in the field of fractions) we deduce that there exists $c\in\CC^*$ such that
$X_1^{i'}X_2^{j'}=c$. Hence
$X_1^{i'\lambda_1}X_2^{j'\lambda_1}=c^{\lambda_1}$,  $(X_1^{-\lambda_2}X_2^{\lambda_1})^{j'}=c^{\lambda_1}$ (by
(\ref{aj-simplu})) and therefore $X_1^{-\lambda_2}X_2^{\lambda_1}\in\CC^*$. 

If $\lambda_1 \lambda_2\le 0$ this means that
there exists $b\in\CC^*$ such
that $ X_1^{|\lambda_2|}X_2^{|\lambda_1|}=b$ which is clearly
impossible, since in $A_1$ the only invertible elements are the
scalars (see page $210$ of \cite{Dixmier}). Hence $\lambda_1\lambda_2>0$ and there exists $a\in\CC^*$ such that $X_1^{|\lambda_2|}=a X_2^{|\lambda_1|}$. QED



\medskip

We can now show 

\begin{theorem}
\label{solvable2}
Let ${\mathfrak r}\subset A_1$ be a finite-dimensional solvable 
non-nilpotent Lie subalgebra whose derived algebra ${\mathfrak r}'$ is abelian. 
Then ${\mathfrak r}$ is
isomorphic to an ${\mathfrak r}_n(i_1,\dots i_n)$, where  $i_1,\ldots,i_n$ is a set of distinct positive integers.
\end{theorem}
{\it Proof:}  
By Theorem \ref{solvable1} (with the same notations),
$$\rr=E_0\oplus E_{\lambda_1}\oplus E_{\lambda_2} \oplus \dots \oplus E_{\lambda_k}$$
and since $ E_{\lambda_1}\oplus E_{\lambda_2} \oplus \dots
E_{\lambda_k}\subseteq\rr'$ is abelian we deduce that
$$\rr'=E_{\lambda_1}\oplus E_{\lambda_2} \oplus \dots \oplus E_{\lambda_k}.$$

It is clear 
that if the centre of $\rr$ is $\CC$, then 
 by Theorem \ref{solvable1},
 $\rr\cong\tilde \rr\times\CC$ where 
$\tilde \rr=<h>\oplus E_{\lambda_1}\oplus \dots \oplus E_{\lambda_k}$ satisfies the hypothesis of
the theorem and has trivial centre.
Therefore to prove the theorem  it is enough to consider the case
where $\rr$ has trivial centre.
 
First, note that by Theorem \ref{solvable1} ($3$) we have  $E_0=<h>$. Next,
since $h\in\Delta_3 \cup \Delta_4$ (by Proposition $10.8$ of \cite{Dixmier}),
there exists $\rho\in\CC^*$ such that eigenvalues of ${\rm ad}(h)$
are integer multiples of $\rho$ (Corollary $9.3$ of \cite{Dixmier} if
$h\in\Delta_3$ and Theorem $1.3$ of \cite{joseph3} if
$h\in\Delta_4$). Hence $\frac{1}{\rho}{\rm ad}(h)$ has integer eigenvalues.
We know that $[E_{\lambda_i}, E_{\lambda_j}]=0$. Applying Lemma \ref{joseph} to $\frac{1}{\rho}h, X_i\in E_{\lambda_i}, X_j \in E_{\lambda_j}$ we deduce that $\frac{\lambda_i}{\rho}$ and $\frac{\lambda_j}{\rho}$, hence $\lambda_i$ and $\lambda_j$,  are of the same sign.
Similarly, if $X_1, X_2 \in E_{\lambda_i}$ then applying 
Lemma \ref{joseph} to $\frac{h}{\lambda_i}, X_1, X_2$ it follows that
$X_1= a X_2$ for some $a\in\CC^*$ and hence $ E_{\lambda_i}$ is of dimension $1$.
It is now clear that $\rr$ is isomorphic to 
$\rr_k(|\frac{\lambda_1}{\rho}|,\dots , |\frac{\lambda_k}{\rho}|)$ (see {\bf E8}). QED
 
\medskip

We summarise the results obtained in this subsection in the following theorem:

\begin{theorem}
\label{cazul-solvabil}
Let $\g$ be a complex finite-dimensional solvable Lie algebra which can be realised in $A_1$. Then $\g$ is isomorphic to one of the following:
\begin{enumerate}
\item an abelian Lie algebra $\CC^n$ ($n\in\NN$),
\item an ${\cal L}_n$ ($n\ge 2$),
\item an $\tilde {\cal L}_n$ ($n\ge 2$),
\item an $\rr_n(i_1,\dots,i_n)$ ($i_1,\dots,i_n$ distinct positive integers) 
\end{enumerate}
\end{theorem}

\medskip

Finally, the results of this section can be summarised in the following theorem:

\begin{theorem}
\label{toate-subalg}
Let $\g$ be a complex finite-dimensional Lie algebra which can be realised in $A_1$. Then $\g$ is isomorphic to one of the following:
\begin{enumerate}
\item $\s$,
\item $\s\times \CC$,
\item $\s\ltimes {\cal H}_3$,
\item an abelian Lie algebra $\CC^n$ ($n\in\NN$),
\item an ${\cal L}_n$ ($n\ge 2$),
\item an $\tilde {\cal L}_n$ ($n\ge 2$),
\item an $\rr_n(i_1,\dots,i_n)$ ($i_1,\dots,i_n$ distinct positive integers). 
\end{enumerate}
\end{theorem}

Further comments are in order on this result. We have remarked in the previous chapter of this thesis that the set of all finite-dimensional semi-simple Lie algebras is discrete (see subsection $2.2.3$) and that the set of all finite-dimensional solvable Lie algebras is continuous  (see subsection $2.2.1$). We see here that, imposing the condition of being realisable in $A_1$, one obtains only a finite number of non-solvable Lie algebras and  a countable family of solvable Lie algebras.

Moreover, this result can be seen as a generalisation of the corresponding result for  the commutative associative algebra $A_0=\CC [x,y]$. Indeed, if one replaces $A_1$ with $A_0$ in Theorem \ref{toate-subalg}, the list  obviously reduces to $\CC^n$, which is included in our results for $A_1$.

\section{Finite-dimensional Lie subalgebras of Der$(A_1)$}
\label{derA1}

In this section we find all finite-dimensional  Lie
algebras that can be realised as subalgebras of Der$(A_1)$ and 
we give some examples of Lie subgroups of $\A (A_1)$ which exponentiate them.

\begin{theorem}
\label{ptintro}
Let $\g\subseteq \mathrm{Der}(A_1)$ be a finite-dimensional Lie
subalgebra. Then $\g$ is either abelian or isomorphic to $\s$, $\s\ltimes {\CC}^2$, an ${\cal L}_n$, an $\tilde {\cal L}_n /\CC$ or an ${\mathfrak r}_n (i_1,\ldots,i_n)$ with $i_1,\dots,i_n$ a set of distinct positive integers.
\end{theorem}
{\it Proof:} Consider the commutative diagram:


$$\xymatrix{A_1\ar[r]^{\pi}&A_1/{\mathbb C}\ar[r]^{{\mathrm{ad}}}
&{\rm Der}(A_1)\\
\pi^{-1}({\rm ad}^{-1}(\mathfrak g))\ar@{^{(}->}[u]\ar[r]^{\pi}
&{\rm ad}^{-1}(\mathfrak g)\ar@{^{(}->}[u]\ar[r]^{\rm ad}
&\mathfrak g\ar@{^{(}->}[u]
}$$

In this diagram, $\pi: A_1\to A_1/\CC$ is a surjective Lie algebra
homomorphism and ${\rm ad}:A_1/\CC\to {\rm Der}(A_1)$ is a Lie algebra isomorphism (see page
$210$ of \cite{Dixmier}). 

If $\g$ is not abelian then $\pi^{-1}({\rm ad}^{-1} (\g)) $ is a
non-abelian subalgebra of $A_1$ containing $\CC$ and $\g\cong \pi^{-1}({\rm ad}^{-1}
(\g)) / \CC$.
By the results of the previous subsections $\pi^{-1}({\rm ad}^{-1}
(\g)) $ is isomorphic to one of the following: $\s\times\CC$, $\s
\ltimes {\cal H}_3$, ${\cal L}_n$, $\tilde {\cal {L}}_n$ or ${\mathfrak
r}_n$.
One now has to take the quotient by $\CC$:
\begin{itemize}
\item $\s\times{\CC}/{\CC}\cong \s$
\item $ {\cal L}_n/\CC$:
  A basis for ${\cal L}_n/\CC$ is $ \bar X_0, \dots, \bar X_{n-1}$ (see {\bf E6} for notations) and the only non-zero commutative relation are $[\bar X_0, \bar X_k]=\bar X_{k+1}$ for $k=1,\dots,n-2$. Hence this algebra is isomorphic to ${\cal L}_{n-1}$.

In particular, ${\cal H}_3/{\CC} = {\cal L}_2/{\CC}={\CC}^2$.

\item $ \tilde {\cal {L}}_n/ \CC$: A basis for $\tilde {\cal {L}}_n/ \CC$ is $\bar h, \bar X_0, \dots, \bar X_{n-1}$ (see {\bf E7} for notations) and the only non-zero commutation relations are  $[\bar h,\bar X_0]=\bar X_0$, $[\bar h, \bar X_k]=-(n-k)\bar X_k$ for $k=1,\dots,n-1$ and $[\bar X_0, \bar X_k]=\bar X_{k+1}$ for $k=1,\dots,n-2$.
\end{itemize}
Hence $\g$ is isomorphic to one
of: $\s$, $\s\ltimes {\CC}^2$, ${\cal {L}}_{n-1}$,
$\tilde {\cal {L}}_n/ \CC$ or an ${\mathfrak r}_n$ without centre (see {\bf E8}). QED

\begin{remark}
The derived Lie algebra of $ {\cal {L}}_n/ \CC$ is isomorphic to ${\cal L}_{n-1}$ but $\tilde {\cal {L}}_n/ \CC$ is neither isomorphic to $\tilde {\cal {L}}_{n-1}$ nor to ${\cal L}_{n}$, both of whose derived algebras are also isomorphic to ${\cal L}_{n-1}$.  In particular $\tilde {\cal {L}}_n/ \CC$ is  not nilpotent and its centre is trivial.
\end{remark}

This theorem implies that if $G$ is a finite-dimensional,
connected, in some sense Lie subgroup
of $\A (A_1)$, 
then $G$ is either abelian or a discrete quotient of
$SL(2,\CC)$ (see example {\bf E9} below),
$SL(2,\CC)\ltimes {\CC}^2$ (see example {\bf E10}), an $R_n (i_1, \ldots, i_n)$ (see
example {\bf E11} below), an $\tilde L_n/\CC$ (see example {\bf E12}) or an $L_n$ (see example {\bf E12}).
We now show that an abelian  group, $SL(2,\CC)$, 
$SL(2,{\CC})\ltimes {\CC}^2$, $R_n (i_1, \ldots, i_n)/\ZZ$, $L_n$ and $\tilde L_n/({\CC}\times\ZZ)$ can be holomorphically embedded
 in $\A(A_1)$.


\begin{enumerate}
\item [{\bf E9}]
 Recall that an element of the group $SL(2,\CC)$ is given by
\beqa
\label{SL2}
\begin{pmatrix} a_1 & a_2 \\ a_3 & a_4 \end{pmatrix}
\eeqa
\noi where $a_1,\dots, a_4 \in \CC$ satisfy $a_1a_4-a_2a_3=1$.
Define 
$\hat \alpha_1 : SL(2,{\CC})\to \A (A_1)$ by 
$$\hat
\alpha_1 ( \begin{pmatrix} a_1 & a_2 \\ a_3 & a_4 \end{pmatrix} )
(p)=a_2 q +a_4 p,\ \  \hat
\alpha_1 ( \begin{pmatrix} a_1 & a_2 \\ a_3 & a_4 \end{pmatrix} )
(q)=a_1 q +a_3 p$$
Then one checks that  $\hat \alpha_1$ is an  injective group homomorphism.

\item [{\bf E10}]
Define $\hat \alpha_1 : SL(2,{\CC})\to \A (A_1)$ as in {\bf E9}
 and $\hat \alpha_2 : {\CC}^2\to \A (A_1)$   
 by  $\hat
\alpha_2 ( \begin{pmatrix} b_1 \\ b_2 \end{pmatrix}
 )=e^{{\rm ad}({b_1q+b_2p})}$, that is 
$$\hat
\alpha_2 ( \begin{pmatrix} b_1 \\ b_2 \end{pmatrix} )
(p)=p-b_1,\ \  \hat
\alpha_2 ( \begin{pmatrix} b_1 \\ b_2 \end{pmatrix} )
(p)=q+b_2.$$   
Then one checks that $\hat \alpha_2$ also is an injective group homomorphism.


Moreover
$$ \hat \alpha_2 ( \begin{pmatrix} a_1 & a_2 \\ a_3 & a_4
\end{pmatrix} \begin{pmatrix} b_1 \\ b_2 \end{pmatrix} ) = \hat
\alpha_1 ( \begin{pmatrix} a_1 & a_2 \\ a_3 & a_4 \end{pmatrix} ) \ \hat
\alpha_2 ( \begin{pmatrix} b_1 \\ b_2 \end{pmatrix} ) \  \hat
\alpha_1 ( \begin{pmatrix} a_1 & a_2 \\ a_3 & a_4 \end{pmatrix}
)^{-1}$$
and so there exists a unique injective group homomorphism
$\hat \alpha : SL(2, {\CC}) \ltimes {\CC}^2 \to \A (A_1)$ extending $\hat
\alpha_1$ and $\hat \alpha_2$  defined by
$$\hat \alpha (gc)=\hat \alpha_1 (g)\hat \alpha_2 (c)\ \forall g\in SL(2,{\CC}),\ c\in{\CC}^2.$$
\end{enumerate}


\medskip 

\noi
For the remaining constructions, namely in order to embed discrete quotients of $R_n (i_1,\dots,i_n)$, $L_n$ and $\tilde L_n/\CC$ in $\A (A_1)$, we use the following procedure: we first find matrix representations of the corresponding Lie algebras, which in the specific cases we treat here can be explicitly exponentiated.
 Finally, homomorphisms are constructed and we calculate their kernels, thus obtaining the required embeddings of discrete quotients of the Lie groups.

\begin{enumerate}
\item [{\bf E11}] Recall that $\rr_n(i_1,\dots,i_n)$ (with $i_1,\dots, i_n$   distinct positive integers) is the Lie algebra generated by $h, X_k$ (with $k=1,\dots,n$) subject to the commutation
relations  $[h, X_k]=-i_k X_k$ (see {\bf E8}). Then 
$\rr_n(i_1,\dots,i_n)$ has a faithful matrix representation  $\pi: \rr_n(i_1,\dots,i_n)\to {\cal M}_{n+1}(\CC)$ given by
\beqa
\label{mat-rn}
\pi (X_k)= 
\begin{pmatrix}
0 & 0 & \dots & \dots & 0 \\
\vdots & 0 &  & & 0 \\
1 & \vdots & \ddots  & &\\
\vdots & \vdots & & \ddots  & \\
0 & 0 & & & 0
\end{pmatrix},
\
\pi (h)=
\begin{pmatrix}
0 & 0 & \dots & \dots & 0 \\
\vdots & -i_1 &  & & 0 \\
0 & \vdots & \ddots  & & \vdots \\
\vdots & \vdots & & \ddots  &  0\\
0 & 0 & & 0 & -i_n
\end{pmatrix}
\eeqa
\noi 
where $k=1,\dots,n$ and the $1$ in the first column of $\pi (X_k)$ lies in the $(k+1)$-st line. 
Exponentiating 
we obtain a subgroup $G\subset GL({n+1},\CC)$ where
\beqa
\label{connected1}
G=\left\{\begin{pmatrix} 1 & 0 & \ldots &
\ldots & 0 \\
a_1 & e^{-v i_1}  & 0 & \dots & \vdots \\ \vdots & 0 & \ddots & \ddots  & \vdots  \\
\vdots & \vdots & \ddots & \ddots  & 0 \\ a_n & 0 & \dots & 0 &
e^{-vi_n}\end{pmatrix} : (a_1,\dots,a_n,v)\in\CC^{n+1}\right\}.
\eeqa
However 
$G$ is not  simply-connected.
 To avoid this problem, we define
$R_n(i_1,\ldots, i_n)$ to be the Lie group whose underlying manifold is ${\CC}^{n+1}$ and whose group law is
\beqa
\label{inmultire}
(a_1,\dots,a_n,v).(a'_1,\dots,a'_n,v')=(a_1+a'_1e^{-vi_1},\dots,a_n+a'_ne^{-vi_n},v+v').
\eeqa
\noi
Then the map $g: R_n (i_1,\ldots, i_n)\to G$ given by
$$ g(a_1,\dots,a_n,v)=\begin{pmatrix} 1 & 0 & \ldots &
\ldots & 0 \\
a_1 & e^{-v i_1}  & 0 & \dots & \vdots \\ \vdots & 0 & \ddots & \ddots  & \vdots  \\
\vdots & \vdots & \ddots & \ddots  & 0 \\ a_n & 0 & \dots & 0 &
e^{-vi_n}\end{pmatrix} $$
is a group homomorphism with discrete kernel $\{ (0,\dots,0,\frac{2\pi i\ell}{{\rm h.c.f.}(i_1,\dots,i_n)}: \ell\in \ZZ\}$.

One can check that the Lie algebra of $R_n(i_1,\ldots, i_n)$  is isomorphic to ${\mathfrak r}_n (i_1,\dots,i_n)$ (see Appendix \ref{LieGroup}) and, using \eqref{inmultire}, one shows that
 $\Phi: R_n (i_1,\ldots, i_n) \to \A (A_1)$ defined by
\beqa
\Phi ((a_1,\dots,a_n,v)) (p)&=&e^{-v}p,\nonumber \\
 \Phi ((a_1,\dots,a_n,v)) (q)&=&e^v (q+\sum_{k=1}^n  \frac{a_k}{(i_k-1) !}
 p^{i_k-1}).\nonumber
\eeqa
\noi
 is a group homomorphism with  kernel $\{(0,\dots,0, 2\pi i \ell):\ell\in\ZZ\}$ isomorphic to $\ZZ$ and that in general $\Phi$ does not factor to a homomorphism from $G$ to $\A (A_1)$ since ${\rm Ker}\, g$ is not in general a subset of ${\rm Ker}\, \Phi$. Note that $\Phi ((a_1,\dots,a_n,v))=e^{{\rm ad}(\frac{a_1}{i_1!}X_1+\dots+\frac{a_n}{i_n!}X_n)}e^{{\rm ad}(vh)}$.
\end{enumerate}

The $n-$dimensional abelian subgroup of $R_n(i_1,\dots,i_n)$ is obtained  
by letting $v=0$; its Lie algebra is isomorphic to the $n-$dimensional abelian Lie algebra. Hence by restricting $\Phi$ to the $n-$dimensional abelian subgroup of $R_n(i_1,\dots,i_n)$ we obtain an injective group homomorphism from it into $\A (A_1)$.

\begin{enumerate}
\item [{\bf E12}] 
Recall that ${\cal L}_n$ is the Lie algebra generated by $X_k$ (where $k=0,\dots,n$) subject to the commutation
relations  $[X_0, X_{k}]=X_{k+1}$ (with $k=1,\dots,n-1$) (see {\bf E6}) and that  $\tilde {\cal L}_n$ is the Lie algebra generated by $h, X_k$ (where $k=0,\dots,n$) subject to the commutation
relations  $[X_0, X_{k}]=X_{k+1},[h,X_0]=X_0, [h, X_k]=-(n-k) X_k$ (with $k=1,\dots,n-1$) (see {\bf E7}).
Then ${\cal L}_n$ has a faithful matrix representation (see \cite{goze2}) $\rho: {\cal L}_n \to {\cal M}_{n+1}(\CC)$ given by
\beqa
\label{mat-ln}
\rho (X_0)&=&
\begin{pmatrix}
0 & 0 & \dots & \dots & \dots & 0 \\
0 & 0 & & \dots & \dots & 0 \\
0 & 1 & 0 & & & \vdots\\
0 & 0 & 1 & \ddots & & \vdots\\
\vdots &  & \ddots & \ddots & \ddots &  \vdots\\
0 & \dots &\dots & 0 &  1 & 0
\end{pmatrix},
\nonumber\\
\rho (X_k)&=&
\begin{pmatrix}
0 & 0 & \dots & \dots & 0 \\
\vdots & 0 &  & & 0 \\
1 & \vdots & \ddots  & &\\
\vdots & \vdots & & \ddots  & \\
0 & 0 & & & 0
\end{pmatrix}, 
\eeqa
\noi 
where $k=1,\dots,n $ and the $1$ in the first column of $\rho (X_k)$ lies in the $(k+1)$-st line.
This can be completed to a matrix representation of $\tilde {\cal L}_n$ by defining
\beqa
\label{mat-ln_tilde}
\rho(h)=\begin{pmatrix}
0 & 0 & 0 & \dots & 0 \\
0 & -n+1 & 0 & \dots & 0 \\
0 & 0 & -n+2 &  \ddots & \vdots \\
\vdots & &  & \ddots & \\
0 & 0 & 0 & \dots & 0
\end{pmatrix}
\eeqa
\noi
Exponentiating
we obtain a subgroup $H\subset GL(n+1,\CC)$ where
$$
H=\left\{\begin{pmatrix}
1 & 0 & 0 & \ldots & \ldots  &  0 \\
\\
e^{(-n+1)v}a_1 & e^{(-n+1)v} & \ddots & &  & \vdots \\
\\
e^{(-n+2)v}a_2 & te^{(-n+2)v} & e^{(-n+2)v} & \ddots &  &  \vdots \\
\\
e^{(-n+3)v}a_3 & \frac 12 t^2 e^{(-n+3)v} & te^{(-n+3)v} & e^{(-n+3)v} & \ddots  & \vdots \\
\vdots & \vdots & \ddots & \ddots & \ddots  &  0 \\
a_n & \frac{t^{n-1}}{(n-1)!} & \ldots &\frac 12 t^2  & t & 1
\end{pmatrix}: (a_1,\dots,a_n,t,v)\in\CC^{n+2}\right\}$$
However $H$ is not simply-connected. To avoid this problem we 
 define  $\tilde L_n$ to be the Lie group whose underlying manifold is ${\CC}^{n+2}$ and whose group law is
\beqa
\label{grup-ln}
 (a_1,\dots, a_n,t,v) . (a'_1,\dots,a'_n,t',v')= (a''_1,\dots,a''_n,t'',v'')
\eeqa
\noi where 
\beqa
\label{coef-ln}
a''_k&=&a_ke^{(n-k)v'}+\sum_{j=1}^{k-1} \frac{t^{k-j}}{(k-j)!}a'_je^{-(k-j)v'}+a'_k,\nonumber\\
t''&=&t'+te^{-v'},\nonumber\\
v''&=&v+v'.\nonumber
\eeqa
\noi
Then the map $\gamma: {\CC}^{n+2}\to H$ given by
$$\gamma (a_1,\dots,a_n,t,v)=
\begin{pmatrix}
1 & 0 & 0 & \ldots & \ldots  &  0 \\
\\
e^{(-n+1)v}a_1 & e^{(-n+1)v} & \ddots & &  & \vdots \\
\\
e^{(-n+2)v}a_2 & te^{(-n+2)v} & e^{(-n+2)v} & \ddots &  &  \vdots \\
\\
e^{(-n+3)v}a_3 & \frac 12 t^2 e^{(-n+3)v} & te^{(-n+3)v} & e^{(-n+3)v} & \ddots  & \vdots \\
\vdots & \vdots & \ddots & \ddots & \ddots  &  0 \\
a_n & \frac{t^{n-1}}{(n-1)!} & \ldots &\frac 12 t^2  & t & 1
\end{pmatrix}$$
is a group homomorphism with discrete kernel $\{ (0,\dots,0,0,2\pi i\ell): \ell\in \ZZ\}$.
One checks that the Lie algebra of $\tilde L_n$
is isomorphic to $\tilde {\cal L}_n$ (see Appendix \ref{LieGroup}). Note that $R_n(n-1,n-2,\dots,0)$ is a subgroup of $\tilde L_n$ by the inclusion:
$$(a_1,a_2,\dots,a_n,v)\mapsto (a_1 e^{(n-1)v}, a_2 e^{(n-2)v},\dots,a_n,0, v).$$
 In fact one can extend the isomorphism $\Phi$ of {\bf E10} to  $\tilde \Phi : \tilde L_n \to \A (A_1)$  by setting $\tilde \Phi ( (a_1,\dots,a_n,t,v))=\Phi ((a_1 e^{(-n+1)v}, a_2 e^{(-n+2)v},\dots,a_n,v)) e^{-t{\rm ad}(q)}$. Explicitly one gets
\beqa
\tilde \Phi ( (a_1,\dots,a_n,t,v))(p)&=&e^{-v}p+t,\nonumber \\
\tilde \Phi ( (a_1,\dots,a_n,t,v))(q)&=&e^v \left(q + \sum_{k=1}^{n-1} \frac{a_k e^{(-n+k)v}}{(n-k-1)!}p^{n-k-1}\right).
\eeqa
Using \eqref{grup-ln} one checks that $\tilde \Phi$ is a group homomorphism whose kernel is the subgroup \newline
$\left\{ (0,\dots,0,a_n,0,2\pi i k):\ k\in\ZZ, a_n\in\CC \right\}$. 
Note that $\tilde \phi$ does factor to a homomorphism from $H$ to $\A (A_1)$ since ${\rm Ker}\, \gamma \subset {\rm Ker}\, \tilde \phi$.
Finally note that the subgroup $L_{n-1}$ of $\tilde L_n$ obtained by letting $a_n=0$ and $v=0$ is simply-connected and its Lie algebra is isomorphic to ${\cal L}_{n-1}$. Hence by restricting $\tilde \Phi$ to $L_{n-1}$ we obtain an injective group homomorphism of $L_{n-1}$ 
into $\A (A_1)$.
\end{enumerate}


\section{The family $\cal F$ and its orbit under $\A(A_1)\times \A(\s)$}
\label{closer}

In subsection $3.1.2$ we  recalled some results of \cite{joseph}: 
if  the realisation of the standard semi-simple element of $\s$ is in $\Delta_3$ ({\it i.e.}, is strictly semi-simple), then the $\s$ realisation belongs (up to an automorphism of $A_1$) to the family ${\cal F}=\{ f_I^1, f_{IIA}^{1,b}, f_{IIB}^{1,b'}\}_{b,b'\in{\CC}}$. 
In this section we give a proof of this result and show that the family $\cal F$ reduces to $ {\cal N}=\{ f_I^1, f_{IIA}^{1,b}\}_{b\in{\CC}}$ under the
 action of the group $\A (A_1)\times \A (\s)$.
We then analyse
these realisations in terms of the Dixmier partition and 
determine exactly which elements of $\cal F$ are equivalent under
the groups $\A (A_1)$ and $\A (A_1)\times \A (\s)$.

The subsequent part of the section is concerned with the orbit of $\cal F$ under $\A (A_1)\times \A (\s)$.
We  give various characterisations of this orbit 
in terms of the Dixmier partition.
We show that $\cal N$ is a set of normal forms and calculate the corresponding isotropy groups.
Finally, for the sake of completeness we give an example of a realisation of $\s$ in $A_1$ which is not in the orbit of  ${\cal F}$.

\bigskip
 
Recall (see subsection $2.1.2$) the  notation:
$e_+=\begin{pmatrix} 0 & 1 \\ 0 & 0 \end{pmatrix},
e_-=\begin{pmatrix} 0 & 0 \\ 1 & 0 \end{pmatrix}, e_0=\begin{pmatrix}
1 & 0 \\ 0 & -1 \end{pmatrix},$

\begin{definition}
The group ${\mathrm{Aut}}(A_1)\times  {\mathrm{Aut}}(\s)$ has a left action on $
A_1^{\s}$ given by:
\beqa
(\alpha, w)\cdot f= \alpha\circ f\circ w^{-1}
\eeqa
where $\alpha \in \A (A_1)$, $f \in  A_1^\s$ and $w\in \A(\s)$.
\end{definition}



Recall that all automorphisms of $\s$ are given by the action of $SL(2,\CC)$ on $\s$.
Let us give the form of such a general automorphism, this result being used several times in the rest of this section.

\begin{proposition}
\label{apendice2}
Let  $g=\begin{pmatrix} a_1 & a_2 \cr a_3 & a_4 \end{pmatrix}\in SL(2, \CC)$. Define ${\rm Ad}(g)\in \A (\s)$ by ${\rm Ad}(g)(z)=gzg^{-1}$ for any $z\in\s$. 
Then
\beqa
\label{auto-sl2-2}
{\rm Ad}(g)(e_+)&=&\begin{pmatrix} -a_1a_3 & a_1^2 \cr -a_3^2 & a_1a_3 \end{pmatrix} \nonumber \\
{\rm Ad}(g)(e_-)&=&\begin{pmatrix} a_2a_4 & -a_2^2 \cr a_4^2 & -a_2a_4 \end{pmatrix} \nonumber\\
{\rm Ad}(g)(e_0)&=&\begin{pmatrix} a_1a_4+a_2a_3 & -2a_1a_2 \cr 2a_3a_4 & -a_1a_4-a_2a_3 \end{pmatrix}. \nonumber
\eeqa
\end{proposition}

\subsection{The family $\cal F$}

\begin{theorem}
\label{th-joseph-dem}
(Lemma $2.4$ of \cite{joseph}) 
Let $H,X,Y\in A_1$ be an $\s$ triplet and let $H=\mu pq + \nu$ with $\mu\in\CC^*$ and $\nu \in \mathbb C$.
 Then:
\begin{enumerate}
\item $\mu= \pm 1$ or $\mu= \pm 2$.
\item \begin{enumerate} \item If $\mu=1$, there exists $a\in \mathbb C^*$ such that 
\begin{eqnarray}
\label{I}
X = -\frac12 a q^2, \ 
Y =   \frac{1}{2a} p^2, \
H = pq-\frac 12 \hskip 0.5cm \mbox{(solution I).}
\end{eqnarray}
\item If $\mu=2$, there exists $a\in {\mathbb C^*}$ and $ b\in \mathbb C$ such that either
\beqa
\label{IIA}
X = a(b+   pq) q, \
Y = -\frac{1}{a} p, \
H = 2pq +b \hskip 0.5cm \mbox{(solution IIA)}
\eeqa
\item[] or 
\beqa
\label{IIB}
X =-\frac{1}{a} q, \
Y = a\ p (b +  pq),\
H = 2pq +b  \hskip 0.5cm \mbox{(solution IIB)}.
\eeqa
\end{enumerate}
\end{enumerate}
\end{theorem}
\noi
\textit{Proof:}
I. We  impose the $\s$ commutation relations: $[H,X]=2X$ and
$[H,Y]=-2Y$. Let $X=\sum_{i,j}a_{ij}p^iq^j$. Computing $[H,X]$ and using $[H, p^i q^j]=(j-i)\mu p^i q^j$ one gets
\beq
\label{HX}
[H,X]=\sum_{i,j}\mu (j-i)a_{ij}p^iq^j=2X=2\sum_{i,j}a_{ij}p^iq^j.
\eeq
This implies that
$$ \mu (j-i) a_{ij}=2 a_{ij}, \forall i,j\in \mathbb N.$$
If $a_{ij}\ne 0$ then 
\beqa
\label{constanta}
\mu=\frac{2}{j-i}.
\eeqa
and therefore $\lambda=\frac{2}{\mu}\in \mathbb N^*$. 


\noi
Since $\mu>0$, it is sufficient to consider the case $\lambda > 0$.

Using  {\bf P5}, there exists $N\in \mathbb N$ such that $X$ can be written
\beq 
\label{X}
X= \sum_{j=0}^N a_j (pq)^j q^\lambda, 
\eeq
 where $a_N \ne 0$.
Analogously, using $[H,Y]=-2Y$, there exists $M\in \mathbb N$ such that $Y$
can be written
\beq
\label{Y}
Y= p^\lambda \sum_{k=0}^M b_k (pq)^k,
\eeq
where $ b_M \ne 0$.

\medskip
\noi
II. The final step of our calculation is to impose the last commutation
relation: $[X,Y]=H$. From (\ref{X}) and
(\ref{Y}) one gets
\beq
\label{comm1}
[X,Y]=\sum_{j=0}^N \sum_{k=0}^M a_j b_k ((pq)^jp^\lambda [q^\lambda, (pq)^k] + [(pq)^j,p^\lambda](pq)^kq^\lambda+ (pq)^j[q^\lambda,p^\lambda](pq)^k).
\eeq
Setting $\xi=[p^\lambda,q^\lambda]$ and $C_k=[(pq)^k,q^\lambda]$ for any $k\in \mathbb N$, this commutator becomes
\beq
\label{comm2}
[X,Y]=\sum_{j=0}^N \sum_{k=0}^M a_j b_k ((pq)^jp^\lambda [q^\lambda, (pq)^k]  + [(pq)^j,p^\lambda]q^\lambda(C_k+(pq)^k) - (pq)^j \xi_\lambda(pq)^k).
\eeq

\noi
We will need the following two lemmas to complete the proof of the theorem.

\begin{enumerate}
\item[]
\begin{lemma}
\label{Lema1}
For $\lambda\in \mathbb N^*$ and $k\in \mathbb N$,
\beq
\label{lema1}
C_k=[(pq)^k,q^\lambda]=q^\lambda[(\lambda+pq)^k-(pq)^k].
\eeq
\end{lemma}
\textbf{Proof:} We will prove this equation by induction. For $k=0$, (\ref{lema1}) is trivially verified.\\
Computing
$$C_{k+1}=[(pq)^{k+1},q^\lambda],$$
one gets
$$C_{k+1}=(pq)C_k+[pq,q^\lambda](pq)^k.$$
Using the induction hypothesis and rearranging the terms so that $q^\lambda$ is on the left, one has
$$C_{k+1}=q^\lambda[(\lambda+pq)^{k+1}-(pq)^{k+1}],$$
which completes the proof of this lemma.  QED
\end{enumerate}

\noi
Using Lemma \ref{Lema1}, (\ref{comm2}) can be written as
\beq
[X,Y]=\sum_{j=0}^N \sum_{k=0}^M a_j b_k ((pq)^jp^\lambda [q^\lambda, (pq)^k]
+ [(pq)^j,p^\lambda]q^\lambda(\lambda+pq)^k) - (pq)^j \xi (pq)^k).
\eeq
Let us now put $A_k=p^\lambda [q^\lambda, (pq)^k]$ and
$A'_k=[(pq)^k,p^\lambda]q^\lambda$.
Then this expression becomes:
\beq
\label{comm}
[X,Y]=\sum_{j=0}^N \sum_{k=0}^M a_j b_k ((pq)^j A_k +A'_j (\lambda+pq)^k - (pq)^{j+k} \xi).
\eeq

\begin{enumerate}
\item[]
\begin{lemma}
For $\lambda\in \mathbb N^*$ and $n\in \mathbb N$, we have:
\beq
\label{lema2}
A_n=p^\lambda [q^\lambda, (pq)^n]=-pq(pq+1)...(pq+\lambda-1)[(\lambda+pq)^n-(pq)^n]
\eeq
and
\beq
\label{lema2'}
A'_n=[(pq)^n,p^\lambda]q^\lambda=-pq(pq+1)...(pq+\lambda-1)[(\lambda+pq)^n-(pq)^n]. 
\eeq
\end{lemma}
\textbf{Proof:} 
We first remark that by {\bf P5}, equation  (\ref{lema2}) is equivalent to
\begin{equation}
\label{forma}
A_n=-p^\lambda q^\lambda [(\lambda+pq)^n-(pq)^n].
\end{equation}
We will prove this equation by induction. For $n=0$, (\ref{lema2}) is trivially verified. Computing
$$A_{n+1}=p^\lambda [q^\lambda, (pq)^{(n+1)}],$$
one gets
$$A_{n+1}=(\lambda+pq)A_n-\lambda p^\lambda q^\lambda (pq)^k.$$
The result now follows from (\ref{forma}).
To prove Eq. [\ref{lema2'}] we just have to remark that by (\ref{Littlewood})
we have
\beqa
[p^\lambda q^\lambda, (pq)^n]=0 \nonumber
\eeqa
and therefore, using the Poisson formulae, $-A_n+A'_n=0$.  QED
\end{enumerate}

\vskip 1cm
\noi
We now return to our main calculation. Inserting into (\ref{comm}) the formula (see {\bf P5})
$$\xi=\sum_{r=1}^\lambda (-1)^{r+1} r! (C_{\lambda}^r)^2
 p^{\lambda-r}q^{\lambda-r}= (-1)^{\lambda+1}\lambda!+\sum_{r=1}^{\lambda-1}
 (-1)^{r+1} r! (C_{\lambda}^r)^2 pq(pq+1)\dots(pq+\lambda-r-1)$$
 and the formulae  for $A_n$ and $A'_n$ given by Lemma \ref{Lema1}, one gets
 the following formula for $[X,Y]$ expressed as a function of $z=pq$: 
\beqa
\label{comm3}
[X,Y]=-\sum_{j=0}^N \sum_{k=0}^M \{a_j b_k [z(z+1)\dots (z+\lambda -1)((z+\lambda)^{j+k}-z^{j+k})+ \nonumber \\
z^{j+k}\sum_{r=1}^{\lambda-1} (-1)^{r+1} r! (C_{\lambda}^r)^2 z(z+1)\dots (z+\lambda-r-1)]+z^{j+k}(-1)^{\lambda+1}\lambda!\}.
\eeqa
This expression must be equal to $H=\frac{2}{\lambda}z + \nu$. The highest power of $z$ in (\ref{comm3}) is $-a_N b_M \lambda (N+M+\lambda)z^{N+M+\lambda-1}$.\\
If $N+M+\lambda-1\ge 2$, that is $N+M \ge 3-\lambda$, then 
$-a_N b_M \lambda (N+M+\lambda)$ must be equal to $0$,
which is not possible. Hence $N+M < 3-\lambda$, $\lambda < 3- (N+M)$ and
$\lambda <3$.
Since we already know that $\lambda$ is a non-zero positive integer, we must
have $\lambda=1$ or $\lambda=2$, in other words $\mu=2$ or $\mu=1$.

From the inequality $N+M < 3-\lambda$, we see that 
$\lambda =1$ implies that $N+M < 2$ and  $\lambda=2$ implies $N+M<1$.

In conclusion, there are only three possible cases:
\begin{itemize}
\item 
$\lambda=2,\  N+M=0$. 
\item
 $\lambda=1,\ N+M=1$ 
\item
$\lambda=1,\  N+M=0$ 
\end{itemize}
These give rise respectively to the solutions  (\ref{I}), (\ref{IIA}) and  (\ref{IIB}), \textit{i.e.} to the solutions I, IIA and IIB, as one can easily verify. QED

\begin{corollary}
\label{reducereF}
Any realisation of Theorem \ref{th-joseph-dem} is equivalent under the action of $\A(A_1)$ to one of the members of the family 
${\cal F}=\{ f_I^1, f_{IIA}^{1,b}, f_{IIB}^{1,b'}\}_{b,b'\in\CC}.$
\end{corollary}
{\it Proof:}
I. For $a\in {\mathbb C^*}$, the automorphism
$\alpha_a\in  \A (A_1)$ given by
 $$\alpha_a(p) = \pm \sqrt{a} p, \ \ \alpha_a(q) = \pm \frac{1}{\sqrt{a}} q
$$
satisfy $f_I^1=\alpha_a \circ f_I^a$. 
This means that all maps given by the formulae $f_I^1$ are
equivalent to $f_I^1$ under the action of the group $\A (A_1)$.

\noi
II. For $a\in {\mathbb C^*}$, the automorphism $\beta_a\in \A (A_1)$ given by:
 $$\beta_a(p) = a p, \ \ \beta_a(q) = \frac{1}{a} q $$
satisfies $f_{IIA}^{1,b}=\alpha_a \circ f_{IIA}^{a,b}$. 
Hence the maps $f_{IIA}^{a,b}$ are all 
equivalent to $f_{IIA}^{1,b}$ under  the action of the group $\A (A_1)$.
Similarly, the  the maps $f_{IIB}^{a,b}$ are all 
equivalent to $f_{IIB}^{1,b}$ under  the action of the group $\A (A_1)$. QED

\subsection{Properties of ${\cal F}$ in terms of the Dixmier partition}
\label{Dixmier-F}

\begin{proposition}
\label{propI}
Let $f_I^1:\s\to A_1$ be defined by 
$f(e_+) = -\frac12  q^2, \ 
f(e_-) =   \frac{1}{2} p^2, \
f(e_0) = pq-\frac 12$. 
Then:
\begin{enumerate}
\item [(i)] The set of eigenvalues of
${{\rm ad}}({f_{I}^{1}(e_0)})$ is $\ZZ$.
\item [(ii)] $f_I^{1}(n)\in\Delta_1$  for any nilpotent $n\in\s$.
\item [(iii)] $f_I^{1}(s)\in\Delta_3$ for any semi-simple $s\in\s$.
\end{enumerate}
\end{proposition}
{\it Proof:} 
Part (i) follows since ${{\rm ad}}({H})(p^i q^j)=(j-i) p^i
q^j$. To prove parts (ii) and (iii) we first need to calculate under what conditions the linear combination $(\lambda e_+ + \mu e_- + \nu e_0)\in\s$ is nilpotent or semi-simple. Recall that $\lambda e_+ + \mu e_- + \nu e_0$ is nilpotent iff, in the standard basis of $\s$, ${\rm ad}(\lambda e_+ + \mu e_- + \nu e_0)$ has only zero eigenvalues  and is semi-simple iff has not only zero eigenvalues. This condition translates into $\nu^2\ne -\lambda\mu$ for $\lambda e_+ + \mu e_- + \nu e_0$ to be nilpotent and resp. $\nu^2= -\lambda\mu$ for $\lambda e_+ + \mu e_- + \nu e_0$ to be semi-simple.

Now considering $f_I^1(\lambda e_+ + \mu e_- + \nu e_0)$ and using Remark \ref{simplificari}, the results follows from Lemma $8.6$ of \cite{Dixmier}. QED


\begin{proposition}
\label{propIIA}
Let $f_{IIA}^{1,b}:\s\to A_1$ be defined by 
$f(e_+) = (b+   pq) q, \
f(e_-) = - p, \
f(e_0) = 2pq +b$. 
Then:
\begin{enumerate}
\item [(i)] The set of eigenvalues of
${{\rm ad}}({f_{IIA}^{1,b}(e_0)})$ is $2\ZZ$. 
\item [(ii)]
There exists a nilpotent $n\in\s$ such that
$f_{IIA}^{1,b}(n)\in\Delta_2$.
\item [(iii)]
There exists a semi-simple $s\in\s$ such that
$f_{IIA}^{1,b}(s)\in\Delta_4$. 
\end{enumerate}
\end{proposition}
{\it Proof:} 
Part (i) is obvious and parts (ii) and (iii) are consequences of the



\begin{enumerate}
\item[]
\begin{lemma}
\label{alta}
$ \lambda f_{IIA}^{1,b}(e_+) + \mu f_{IIA}^{1,b}(e_-) + \nu f_{IIA}^{1,b}(e_0)\in \Delta_1 \cup \Delta_3$ iff $\lambda =0$.
\end{lemma}
{\it Proof}: ($\Rightarrow$): If $\lambda \ne 0$, one shows by simple induction that 
$${\rm ad}^n({\lambda X + \mu Y + \nu H})(q)= n! \lambda^n a^n q^{n+1}+ h_n(q)$$
where $h_n(q)$ is
a polynomial in $q$ of degree at most $n$. It then follows that \\
$({\rm ad}^n({\lambda X + \mu Y + \nu H))}(q))_{n\in\NN}$ spans an infinite-dimensional vector space and
hence $\lambda X + \mu Y + \nu H\notin \Delta_1\cup \Delta_3$, by Corollary $6.6$ of \cite{Dixmier}. 

($\Leftarrow$): If $\lambda = 0$ the result follows from Lemma $8.6$ of \cite{Dixmier}. QED QED

\end{enumerate}

\begin{corollary}
\label{inequiv-1}
The realisation $f_I^1$ is not equivalent  to any of the realisations $f_{IIA}^{1,b}$ ($b\in\CC$) under the action of the group $\A (A_1)$.
\end{corollary}
{\it Proof:} Recall that the Dixmier partition is stable under the action of the group $\A (A_1)$ (see subsection $3.1.1$). The proof is then immediate by comparing  properties (ii) or (iii) of Propositions \ref{propI} and \ref{propIIA}. QED


\begin{proposition}
\label{propIIB}
Let $f_{IIB}^{1,b}:\s\to A_1$ be defined by 
$f(e_+) = - q, \
f(e_-) = p(b+pq), \
f(e_0) = 2pq +b$. 
Then:
\begin{enumerate}
\item [(i)] The set of eigenvalues of
${{\rm ad}}({f_{IIB}^{1,b}(e_0)})$ is $2\ZZ$. 
\item [(ii)]
There exists a nilpotent $n\in\s$ such that
$f_{IIB}^{1,b}(n)\in\Delta_2$.
\item [(iii)]
There exists a semi-simple $s\in\s$ such that
$f_{IIB}^{1,b}(s)\in\Delta_4$. 
\end{enumerate}
\end{proposition}
{\it Proof:} Similar to Proposition \ref{propIIA}, but using this time

\begin{enumerate}
\item[]
\begin{lemma}
\label{alta2}
$ \lambda f_{IIB}^{1,b}(e_-) + \mu f_{IIA}^{1,b}(e_+) + \nu f_{IIA}^{1,b}(e_0)\in \Delta_1 \cup \Delta_3$ iff $\lambda =0$.
\end{lemma}
{\it Proof:} Analogous to  Lemma \ref{alta}. QED QED 

\end{enumerate}

\noi
We conclude this subsection by

\begin{corollary}
\label{inequiv-2}
(i) The realisation $f_I^1$ is not equivalent  to any of the realisations $f_{IIB}^{1,b}$ ($b\in\CC$) under the action of the group $\A (A_1)$.\\
(ii) No realisation $f_{IIA}^{1,b}$ ($b\in\CC$) is equivalent  to any of the realisations $f_{IIB}^{1,b'}$ ($b'\in\CC$) under the action of the group $\A (A_1)$.
\end{corollary}
{\it Proof:} (i) Similar to  Corollary \ref{inequiv-1}.\\
(ii) Immediate since $f_{IIB}^{1,b}(e_+)\in\Delta_1$ and $f_{IIA}^{1,b'}(e_+)\in\Delta_2$. QED

\subsection{Equivalence of members of ${\cal F}$ under $\A (A_1)$}

%

We show that no two distinct realisations of ${\cal F}$ are equivalent under the action of the group $\A (A_1)$.

\begin{proposition}
\label{labareala}
 Let $b,b'\in  {\mathbb C}$.
 Then 
 $ f_{IIA}^{1,b}$ (resp. $f_{IIB}^{1,b}$) is  equivalent to 
$f^{1,b'}_{IIA}$ (resp. $f_{IIB}^{1,b'}$)
 under  the action of the group $\A (A_1)$ iff $b=b'$.
\end{proposition}
{\it Proof}: 
The $\s$ triplets corresponding to  $f_{IIA}^{1,b}$ and
$f_{IIA}^{1,b'}$ are
\beqa
\label{tablou}
\begin{array}{ll}
X=(b+pq)q \quad &  X'=(b'+pq)q \cr
Y=-p & Y'=-p \cr
H=2pq+b & H'=2pq+b'.
\end{array}
\eeqa
If $f_{IIA}^{1,b}$ and $f_{IIA}^{1,b'}$ are equivalent under the
action of $\A (A_1)$,  there exist  $p',q'\in A_1$ such
that $[p',q']=1$ and
\beqa
\label{tablou2}
\begin{array}{ll}
(b'+p'q')q'&=  (b+pq)q  \cr
-p'&=-p  \cr
2p'q'+b'&= 2pq+b. 
\end{array}
\eeqa
\noi
From this it follows easily that $p'=p$, $q'=q$, and $b'=b$. 

Arguing along the same lines one proves that $f_{IIB}^{1,b}$ is equivalent to 
 $f_{IIB}^{1,b'}$ iff $b=b'$. QED

\begin{corollary}
\label{muistii}
No two distinct realisations of  the family $(f_I^1,
f_{IIA}^{1,b},f_{IIB}^{1,b'})_{\tiny{ b, b'\in\CC}}$ are equivalent
under the action of $\A (A_1)$.
\end{corollary}
{\it Proof}: As pointed out in the previous subsection, the realisation $f_I^1$ is not equivalent to a
realisation of type $IIA$ (Corollary \ref{inequiv-1}) or of type $IIB$ (part (i) of Corollary \ref{inequiv-2}) and a  realisation of type $IIA$
is not equivalent to a realisation of type $IIB$  (part (ii) of Corollary \ref{inequiv-2}). The result now follows from the previous proposition. QED

\subsection{Equivalence of members of $\cal F$ under $\A (A_1)\times \A(\s)$}


Recall the notation: ${\cal N}=(f_I^1,f_{IIA}^{1,b})_{b\in\CC}$.


\begin{remark}
\label{AB}
 (also mentioned in \cite{joseph})
The realisations $f_{IIA}^{1,-(b+2)}$ and $ f_{IIB}^{1,b}$ are equivalent under the action of $\A (A_1)\times \A(\s)$ since
$$ f_{IIB}^{1,b}=\alpha\circ f_{IIA}^{1,-(b+2)} \circ \tau,$$
where $\tau\in\A(\s)$ and $\alpha\in \A(A_1)$ are given by $\tau(x)=y,\
\tau(y)=x$   and $\alpha(p)=-q, \ \alpha(q)=p$.
\end{remark}




\begin{proposition}
\label{memeCasimir}
The pairs (i) ($f_{IIA}^{1,b}$, $f_{IIA}^{1,b'}$), ($b\ne b'$),
(ii) ($f_{IIA}^{1,b}$, $f_{I}^{1}$) 
are not equivalent under the action of the group $\A (A_1)\times \A (\s)$.
\end{proposition}
\textbf{Proof.} (i): 
The $\s$ triplets corresponding to  $f_{IIA}^{1,b}$ are
$f_{IIA}^{1,b'}$ are
\beqa
\label{tablou222}
\begin{array}{ll}
X=(b+pq)q \quad &  X'=(b'+pq)q \cr
Y=-p & Y'=-p \cr
H=2pq+b & H'=2pq+b'.
\end{array}
\eeqa

If $f_{IIA}^{1,b}$ and $f_{IIA}^{1,b'}$ are equivalent under the
action of $\A (A_1)\times \A (\s)$, then there exist an automorphism $w={\rm Ad}(\begin{pmatrix} a_1 & a_2 \\ a_3 & a_4 \end{pmatrix})$ of $\s$ (see Proposition \ref{apendice2} for the notation)
and $p',q'\in A_1$ satisfying $[p',q']=1$ such that
\beqa
\label{tablou22}
\begin{array}{lll}
(b'+p'q')q'&=f_{IIA}^{1,b}\circ w(e_+)&= a_1^2 (b+pq)q  - a_3^2
(-p) - a_1a_3 (2pq+b) \cr
 -p'&= f_{IIA}^{1,b}\circ w(e_-) &=-a_2^2 (b+pq)q  + a_4^2
(-p) + a_2a_4(2pq+b)  \cr
 2p'q'+b'&=f_{IIA}^{1,b}\circ w(e_0) &=-2 a_1a_2(b+pq)q  + 2a_3a_4 (-p) + (a_1 a_4 +a_2 a_3) (2pq+b). 
\end{array}
\eeqa
\noi
Substituting the second equation in the third equation, we obtain
\beqa
\label{identificare22}
-2[-a_2^2 (b+pq)q  + a_4^2
(-p) + a_2a_4(2pq+b)]q'+b'= \nonumber \\
= -2 a_1a_2(b+pq)q  + 2a_3a_4 (-p) + (a_1 a_4 +a_2 a_3) (2pq+b). 
\eeqa
\noi
If $a_2\ne 0$ then the expansion of $q'$ in the standard basis
must consist of only a constant term, otherwise the term $-a_2^2 (b+pq)q q'$ on the
LHS contains terms which are not present in the RHS; but then
$[p',q']=0$ which is a contradiction and hence $a_2 = 0$. 
Thus, one also has $a_1a_4=1$.

Equation \eqref{identificare22} now reduces to
\beqa
\label{identificare222}
-2[ a_4^2 (-p)]q'+b'= 2a_3a_4 (-p) + (2pq+b). 
\eeqa
\noi
Equating the constant term on both sides of (\ref{identificare222}) gives $b=b'$.

In conclusion we have shown that if $b\ne b'$, $f_{IIA}^{1,b}$ is not
equivalent to  $f_{IIA}^{1,b'}$ under the action of $ \A (A_1)\times \A (\s)$ which proves part (i). Part (ii) is proved in a similar but easier way. QED


\begin{proposition}
\label{largire1}
a) If $f\in \cal F$ there exists $(\alpha,
w)\in\A(A_1)\times \A(\s)$ such that $(\alpha, w).f \in {\cal N}$.

\smallskip \noi
b) No two distinct realisations in ${\cal N}$ are equivalent
under the action of $\A (A_1)\times \A (\s)$.
\end{proposition}
{\it Proof}: Immediate from Remark \ref{AB}
and Proposition \ref{memeCasimir}. QED

\subsection{Characterisation of the orbit of ${\cal F}$ under $\A (A_1) \times \A (\s)$ }
\label{orbit-f}


In this subsection we analyse the orbit ${\cal O}_{\cal F}$ of ${\cal F}$ under the action of the group $\A (A_1) \times \A (\s)$. In terms of the Dixmier partition we define several {\it a priori} different subsets of $A_1^\s$ which all turn out to be identical with ${\cal O}_{\cal F}$. 
This means that $\cal N$ is the family of normal forms for the action of the group $\A(A_1)\times \A(\s)$ on any of these subsets.

\begin{definition}
\label{partitii}
\beqa
{\cal D}_3&=&\{ f\in A_1^\s : \exists z\in \s \setminus\{0\} \mbox{ s.t }
f(s)\in\Delta_3\} \nonumber \\
{\cal D}^\prime_3&=&\{ f\in A_1^\s : \exists z\in \s\setminus\{0\} \mbox{
s.t }
{{\rm ad}}(f(z)) \mbox{ admits an eigenvector in } \Delta_3 \}
 \nonumber \\
{\cal D}_1&=&\{ f\in A_1^\s : \exists z\in \s\setminus\{0\} \mbox{ s.t }
f(n)\in\Delta_1\} \nonumber \\
{\cal D}^\prime_1&=&\{ f\in A_1^\s : \exists z\in \s\setminus\{0\}  \mbox{ 
s.t }
\mathrm{ad}({f(z)}) \mbox{ has an eigenvector in } \Delta_1 \}
 \nonumber \\
   {\cal D}&=&\{ f\in A_1^\s : \exists (\alpha,
   w)\in\A(A_1)\times\A(\s) \mbox{ s.t } (\alpha, w).f\in {\cal N} \} \nonumber \\
{\cal O}_{\cal F}&=&\{ f\in A_1^\s : \exists (\alpha,
   w)\in\A(A_1)\times\A(\s) \mbox{ s.t } (\alpha, w).f\in {\cal F} \}. \nonumber
\eeqa
\end{definition} 

\noi
It follows from Proposition \ref{largire1} that ${\cal D}={\cal O}_{\cal F}$.
We now show that in fact all of the above  sets are identical.

\begin{theorem}
\label{TEOREMA}
${\cal D}_3={\cal D}^\prime_3={\cal D}_1={\cal D}'_1={\cal D}$.
\end{theorem}
\textit{Proof:} First note that ${\cal D}_3,{\cal D}^\prime_3,{\cal
D}_1, {\cal D}'_1$ and ${\cal D}$ are stable  under the action of  $\A (A_1)\times \A (\s)$ and that the inclusions ${\cal D}_3\subseteq {\cal D}'_3$ and ${\cal D}_1\subseteq {\cal D}'_1$ are obvious. Furthermore, 
 the inclusions ${\cal D}\subseteq {\cal D}_3$ and  ${\cal D}\subseteq {\cal D}_1$ 
follow from subsection \ref{Dixmier-F}.
Hence, to
prove the theorems it is sufficient to show that  
\begin{enumerate}
\item ${\cal D}_3\subseteq {\cal D}$,
\item  ${\cal D}'_3\subseteq {\cal D}_3$,
\item ${\cal D}'_1\subseteq {\cal D}_3$.
\end{enumerate}

\noi
$1.$ {\bf{$\mathbf{{\cal D}_3\subseteq{\cal D}}$}}: Let $f\in {\cal D}_3$.
By hypothesis there exists an $s\in\s$ such
that $f(s)\in \Delta_3$ and $s$ is semi-simple 
by Remark \ref{simplificari} because in $\s$ an element is either nilpotent or semi-simple.
By rescaling we can always suppose that
the eigenvalues (in $\s$) of ${\rm ad}(s)$ are $-2,0$ and $2$.
Then there exists $w\in\A(\s)$ such that $w^{-1}(e_0)=s$.
By Theorem $3.1.3$ ($2$), there exist $\alpha\in\A
 (A_1), \mu \in \CC^*,\ \nu\in\CC$ such that 
$\alpha \circ f\circ w^{-1} (e_0)=\mu pq + \nu$. 

\noi
By Theorem $3.4.3$ and Corollary \ref{reducereF}
\beqa
\label{rez-joseph}
(\alpha, w).f\in {\cal F}
\eeqa
\noi
Hence ${\cal D}_3\subseteq {\cal D}$ by Proposition \ref{largire1}.



\medskip

\noi
$2.$ {\bf {${\cal D}^\prime_3\subseteq{\cal D}_3$}}: We need the following lemma

\begin{itemize} \item []
\begin{lemma}
Let $S\in \Delta_3$. Then 
\beqa
\label{delta55}
\begin{array}{llll}
(i)&C(S)&\subset& \CC \cup \Delta_3 \cup \Delta_5;\\
(ii)& C(S)\cap \Delta_3 &=&\{ \mu S + \nu : \mu\in\CC^*, \nu\in\CC\}.
\end{array}
\eeqa
\end{lemma}
\textbf{Proof:} Without loss of generality we can suppose that $S=pq$ since
any element of $\Delta_3$ is equivalent under $\A (A_1)$ to $\mu' pq + \nu'$
(Theorem $9.2$ of \cite{Dixmier}) and since $C(\mu' pq + \nu')= C(pq)$.
First note that $C(pq)=\CC [pq]$ (see Proposition $5.3$ of
\cite{Dixmier}). Let $Z=a_k (pq)^k + \dots+a_0$ be a polynomial of degree $k$ in
$pq$. Then a simple induction shows that 
$$[Z,p^m q^n]= k (n-m) a_kp^{m+k-1} q^{n+k-1} + \mbox{ terms of lower degree in
$p$}. $$
From this it follows that if $k>1$ the only eigenvalue of ${\rm ad}(Z)$ is $0$
and so $D(Z)= C(Z)$. By iteration of this formula it also follows that that if
$k>1$, ${\rm Ker} \, {\rm ad}^\ell (Z) = {\rm Ker} \, {\rm ad}(Z) $ and so $N(Z)=C(Z)$. 
This means that if $k>1$, $Z\in\Delta_5$ (see Theorem $2.2$).  By Theorem $3.1.3 (2)$, $Z\in \Delta_3$ if $k=1$ and the lemma is proved. QED

\end{itemize}

Let  $f \in {\cal D}^\prime_3$.  There exists $z\in \s \setminus\{0\}$,
$S\in \Delta_3$ and $\lambda\in \CC$ such that $[f(z),S]=\lambda S$. 
This implies that ${\rm ad}^2 (S)(f(z))=0$, that is $f(z)\in N (S)$ (see subsection $3.1.1$ for notation).
Since $S\in\Delta_3$, $N(S)=C(S)$ (see Theorem $3.1.1$) and thus $f(z)\in C(S)$. 
By Lemma $3.4.21$, $f(z)\in \Delta_3 \cup \Delta_5$; but $f(z)\notin
\Delta_5$ (see Remark $3.1.4$) and so $f(z)\in \Delta_3$ and  $f\in{\cal D}_3$.

\medskip

\noi
$3.$ {\bf{${\cal D}'_1\subseteq{\cal D}_3$}}:  We need the following two lemmas

\begin{itemize} \item []
\begin{lemma}
\label{lema22}
Let $a\in A_1$, $\mu \in \mathbb C^*$ be such that
$[a,p]=-\mu p $. There exists  $\alpha_3\in \A (A_1)$
such that $\alpha_3^{-1}(a)=\mu pq+\nu$, for some $\nu\in\CC$.
\end{lemma}
\textbf{Proof:}
Let $a=\sum_{ij}h_{ij} p^iq^j$. 
Since 
\beq
\label{condHmax}
[a,p]=-\mu p \ \Leftrightarrow\ [a-\mu pq,p]=0,
\eeq
\noi and $[p,p^iq^j]=jp^iq^{j-1}$, one has
\beq
\label{muie}
a=\mu pq + \sum_{i=0}^N a_i p^i,  
\eeq
where $a_i \in \mathbb C$ and $N\in \mathbb N$.
One can then write 
$$a=\mu p(q+\sum_{i=1}^N \frac{a_i}{\mu}p^{i-1})+\nu. $$

But $[p,q']=1$ where $q'=q+\sum_{i=1}^N \frac{a_i}{\mu} p^{i-1}$ so that there exist an  unique automorphism $\alpha_3$ of $A_1$
such that
$\alpha_3(p)=p,\ \alpha_3(q)=q'$.
Hence
$$\alpha_3^{-1}(a)=\mu pq + \nu. $$
{\flushright QED}

\begin{lemma}
\label{lema12}
Let $a\in A_1$, $\lambda\in\CC^*$ and let $g(p)= \sum_{k=0}^n
a_k p^k$ be a polynomial of degree $n$.
If 
$[a, g(p)]=\lambda g(p)$ then there
exists $\alpha_2\in \A(A_1)$ such that 
$[\alpha_2^{-1}(a),
p]=\frac{\lambda}{n} p$.
\end{lemma}
\noi
\textbf{Proof:} Since $<p^iq^j>$ is a basis of $A_1$, we can write
$$[a,p^k]=\sum_{i=0}^{N_k} f_{i,k}(p)\ q^i, $$
where $k\in {\mathbb N}^*, N_k\in \mathbb N$ and $f_{N_k,k}$ is a non-zero polynomial of degree $M_k$ in
$p$. Let $c_k\ne 0$ be the coefficient of $p^{M_k}q^{N_k}$ in $[a,p^k]$.
A straightforward induction argument shows that $N_k=N_1,\ M_k=M_1+k-1$ and $c_k=k c_1$.

Thus, one has
$$[a,\sum_{k=0}^n a_k p^k]=\sum_{k=0}^n \left(\sum_{i=0}^{N_k} a_k f_{i,k}(p) q^i \right), $$
 But since $[a, g(p)]=\lambda g(p)$ by hypothesis, this means that
 $a_n c_n=a_n n c_1= a_n \lambda$, 
 $N_n=N_1=0,\ M_n=n$ and so $c_1=\frac{\lambda}{n}$ and $M_1=1$. Therefore
$$[a,p]=\frac{\lambda}{n} p + \nu$$
for some constant $\nu \in \mathbb C$.
But, $[\frac{\lambda}{n} p + \nu, \frac{n}{\lambda} q]=1$ 
there exist an unique  automorphism $\alpha_2$ of $A_1$
such that
$\alpha_2(p)=\frac{\lambda}{n} p + \nu,\ \alpha_3(q)=\frac{n}{\lambda} q$.
Hence
$$[\alpha_2^{-1}(a),p)]=\frac{\lambda}{n} p $$
{\flushright {QED}}
\end{itemize}


We now prove that ${\cal D}'_1\subseteq {\cal
D}_3$. Let $f\in {\cal D}'_1$.
By hypothesis there exist  $z\in\s$,
$N\in \Delta_1$ and $\lambda\in\CC$ such that
\beqa
\label{muita}
 [f(z), N]=\lambda N.
\eeqa
\noi
By Theorem $9.1$ of 
\cite{Dixmier} there exist $\alpha_1\in\A(A_1)$ such that
$\alpha_1(N)\in\CC[p]$. If  $\lambda\ne 0$,
$\alpha_1\circ f(z) $ satisfies the hypothesis of Lemma \ref{lema1}. 
and, by  Lemmas \ref{lema1} and  \ref{lema2},
 there exist
$\alpha_2, \alpha_3\in\A (A_1)$, $\mu\in \CC^*$ and $a_0\in\CC$ such that $\alpha_3^{-1}\circ
\alpha_2^{-1}\circ\alpha_1\circ f(z) = \mu pq + a_0$. 
Since $\mu pq + a_0$ is in $\Delta_3$,  this means that $\alpha_3^{-1}\circ
\alpha_2^{-1}\circ\alpha_1\circ f\in {\cal D}_3$ and hence $f\in {\cal D}_3$.
To complete the proof we show that the $\lambda=0$ case reduces to the 
$\lambda\ne 0$ case as follows.
If $\lambda=0$, then $f(z)\in C(N)$. But $C(N)\subseteq\Delta_1$ (by 
Theorem $9.1$ of 
\cite{Dixmier}) so that $f(z)\in\Delta_1$ and, by Remark \ref{simplificari},
 $z$ is nilpotent. There exists $s\in\s$ semi-simple such that
 $[s,z]=2z$ and then $[f(s),f(z)]=2f(z)$. Now,
replacing 
$z$ by $s$, $N$ by $f(z)$ and  $\lambda$ by $2$  in \eqref{muita}, 
we can conclude that $f\in {\cal D}_3$ as above since $\lambda\ne 0$. QED



\begin{corollary}
Let $f\in A_1^\s$. The following are equivalent:
\begin{enumerate}
\item
$f(\s)\subseteq \Delta_1 \cup \Delta_3$.
\item 
There exists $(\alpha, w)\in \A (A_1)\times \A(\s)$ such that $(\alpha, w).f=f_I^1$.
\end{enumerate}
\end{corollary}
{\it Proof:} $2 \Rightarrow 1$: By Proposition $3.4.7$ parts $(ii)$ and $(iii)$, $f_I^1 (\s)\subseteq \Delta_1 \cup \Delta_3$ and this property is invariant under the action of $\A(A_1)\times \A (\s)$.

\noi
$1 \Rightarrow 2$: If $f(\s)\subseteq \Delta_1 \cup \Delta_3$ then 
$f\in {\cal D}_1$ and by Theorem \ref{TEOREMA} there  
exists  $(\alpha, w)\in \A (A_1)\times \A(\s)$ such that
$(\alpha, w). f\in\cal F$. By Propositions $3.4.7$, $3.4.8$ and $3.4.11$ we must have  $(\alpha, w).f=f_I^1$. QED

\subsection{The isotropy groups in $\A (A_1) \times \A (\s)$}
\label{iso}

Recall that if a group $G$ acts on a set $ X$, the isotropy of $x\in X$ in $G$ is by definition the subgroup $\{g\in G : g.x=x\}$.
In this subsection we calculate the isotropy groups of $f_I^1$ and $f_{IIA}^{1,b}$ in  $\A (A_1) \times \A (\s)$. We  denote these groups respectively  by $I_I$ and  $I_{IIA}$.

\begin{proposition}
\label{iso-I}
$ I_{I}=\{ (\alpha_g, {\rm Ad}(g))\ : \ g\in SL(2,\CC)\} $, 
where $\alpha_g\in \A (A_1)$ is given by
$\alpha_g (p)=a_4 p + a_2 q, \ \alpha_g (q)=a_3 p + a_1 q$.
\end{proposition}
\begin{remark}
In fact the group $I_{I}$ is isomorphic to $SL(2,\CC)$.
\end{remark}
{\it Proof:} Let $(\alpha,w)\in I_I$. This is equivalent to $\alpha\circ f_{I}^{1} = f_{I}^{1}\circ w$. Let $p'=\alpha(p),\ q'=\alpha(q)$ and $w={\rm Ad}(\begin{pmatrix} a_1 & a_2 \\ a_3 & a_4 \end{pmatrix})$ (see Proposition \ref{apendice2} for notations).

The $\s$ triplets corresponding to $f_{I}^{1}$ and  $\alpha\circ f_{I}^{1}$  are
\beqa
\label{tablou2222}
\begin{array}{ll}
X=-\frac12 q^2 \quad &  X'= -\frac12 q^2\cr
Y=\frac12 p^2 & Y'= \frac12 p'^2\cr
H=pq-\frac12 & H'=p'q'-\frac12.
\end{array}
\eeqa


One has
\beqa
\label{tablou222-bis}
\begin{array}{llll}
\alpha\circ f_I^1 (e_+)&=-\frac12 q'^2&=f_{I}^{1}\circ w(e_+)&= -\frac 12 a_1^2 q^2  -\frac 12 a_3^2 p^2 - a_1a_3 (pq-\frac 12) \cr
\alpha\circ f_I^1 (e_-)&=\frac12 p'^2 &=f_{I}^{1}\circ  w(e_-) &=\frac 12 a_2^2 q^2   + \frac 12 a_4^2 p^2 + a_2a_4(pq-\frac12)  \cr
\alpha\circ f_I^1 (e_0)&=p'q'-\frac 12 &=f_{I}^{1}\circ w(e_0) &= a_1a_2 q^2 + a_3a_4 p^2 + (a_1 a_4 +a_2 a_3) (pq-\frac12). 
\end{array}
\eeqa
\noi
The first equation writes
\beqa
\label{eq-q'}
\alpha\circ f_I^1 (e_+)=\frac 12 q'^2=\frac12 (a_3p+a_1q)^2
\eeqa
\noi
Let $\beta\in \A (A_1)$ such that $\beta (q)= a_3 p +a_1 q$. 
By acting with $\beta^{-1}$ on \eqref{eq-q'}, one has $\beta^{-1}\circ\alpha \circ f_I^1 (e_+)=\frac 12 (\beta^{-1} (q'))^2=\frac12 q^2$.  It is then easy to see that  $\beta^{-1} (q')=\pm q$ with implies $q'=\pm \beta (q)=\pm(a_3 p +a_1 q)$.

Similarly, using the second equation of \eqref{tablou222-bis} one finds that $p'=\pm (a_4p+a_2q)$. Since $[p',q']=1$  we have either $p'=a_4p+a_2q, q'= a_3 p +a_1 q)$ or $p'=-(a_4p+a_2q),  q'= -(a_3 p +a_1 q)$. Furthermore, the last equation of \eqref{tablou222-bis} is verified by any of the two solutions found. 
Since $w={\rm Ad}(g)={\rm Ad}(-g)$ this completes our proof.
QED

\medskip

To calculate $I_{II}$ we need the following definition:

\begin{definition}
\label{luminita}
Let $\hat B = \left\{ \begin{pmatrix} a_1 & 0 \\ a_3 & \frac{1}{a_1} \end{pmatrix} : a_1\in\CC^*, a_3\in\CC \right\}$ and let $B$ be the subgroup ${\rm Ad}(\hat B)\subset \A (\s)$.
\end{definition}

\begin{proposition}
\label{iso-IIA} 
$ I_{IIA}=\{ (\alpha_w, w)\ : \ w\in B\} $
where $\alpha_w\in\A(A_1)$ is given by 
$\alpha_w(p)=\frac{1}{a_1^2}p,\ \alpha_w(q)=a_1^2 (q-\frac{a_3}{a_1}).$
\end{proposition}
\begin{remark}
In fact the group $I_{IIA}$ is isomorphic to $B$, a Borel subgroup of $\A (\s)$.
\end{remark}
{\it Proof:} 
Let $(\alpha_w,w)\in I_I$. This is equivalent to
 $\alpha_w\circ f_{IIA}^{1,b}\circ w = f_{IIA}^{1,b}$. 
Let $p'=\alpha_w(p),\ q'=\alpha_w(q)$
and $w={\rm Ad}(\begin{pmatrix} a_1 & a_2 \\ a_3 & a_4 \end{pmatrix})$ (see Proposition \ref{apendice2} for notations).
 Calculation as in the proof of Proposition \ref{memeCasimir}   up to  equation \eqref{identificare222}, as in the proof of Proposition \ref{memeCasimir},  (replacing $b'$ by $b$ in the LHS of the Equations \ref{tablou22}, \ref{identificare22}, \ref{identificare222}). Thus, one has $a_2=0$ and  $a_1a_4=1$. This  means that $w\in B$. Furthermore, equation \ref{identificare222} can be written
\beqa
\label{identificare2222}
-2[ a_4^2 (-p)]q'+b= 2a_3a_4 (-p) + (2pq+b). 
\eeqa
\noi
which gives
\beqa
\label{identificare22222}
p(a_4^2 q'+ a_3a_4 -q)=0. 
\eeqa
\noi
Since in $A_1$ there are no zero divisors (see \cite{Dixmier}), one has
\beqa
\label{q'}
q' = \alpha_w(q)= \frac{1}{a_4^2}(q- a_3a_4)=a_1^2(q- a_3a_4)= a_1^2(q- \frac{a_3}{a_1})
\eeqa
\noi
Imposing $[\alpha_w (p), \alpha_w(q)]=1$ one gets $\alpha_w (p)= a_4^2 p +k=\frac{1}{a_1^2}p+k$, with $k\in\CC$, which, inserted in the first or the second equation of (\ref{tablou22}) gives $k=0$. QED

\subsection{Other examples}

Let $ {\cal U} (\s)$ be the universal enveloping algebra of $\s$. Recall that  $ {\cal U} (\s)$ is an associative algebra containing $\s$ which has the following universal property: if $f:\s\to B$ is a Lie homomorphism from $\s$ to 
an associative algebra $B$ there exists a unique associative algebra homomorphism $\hat f:{\cal U} (\s)\to B$ such that $\hat f\vert_\s=f$.
This means  that we can define  a left action of $\A ({\cal U} (\s))$ on $A_1^\s$ as follows:
$$ (\alpha,w)\cdot f = \alpha \circ f \circ w^{-1}\vert_\s $$
where $\alpha \in \A (A_1)$ and $w\in \A ({\cal U} (\s))$.

One can construct realisations of $\s$ which are not into the orbit of ${\cal F}$ under $\A (A_1)\times \A (\s)$ (see fig. $3.1$). 
This can be done  by 
letting elements of $\A ({\cal U}(\s))\setminus \A(\s)$ act on
 $\cal N$.  Note that the group $\A(\s)$ is naturally included in the group $\A({\cal
U} (\s))$. In this section, for the sake of completeness, we
give an explicit example (see also page $127$ of \cite{joseph}).

\epsfxsize = 3cm     
$$
\epsffile{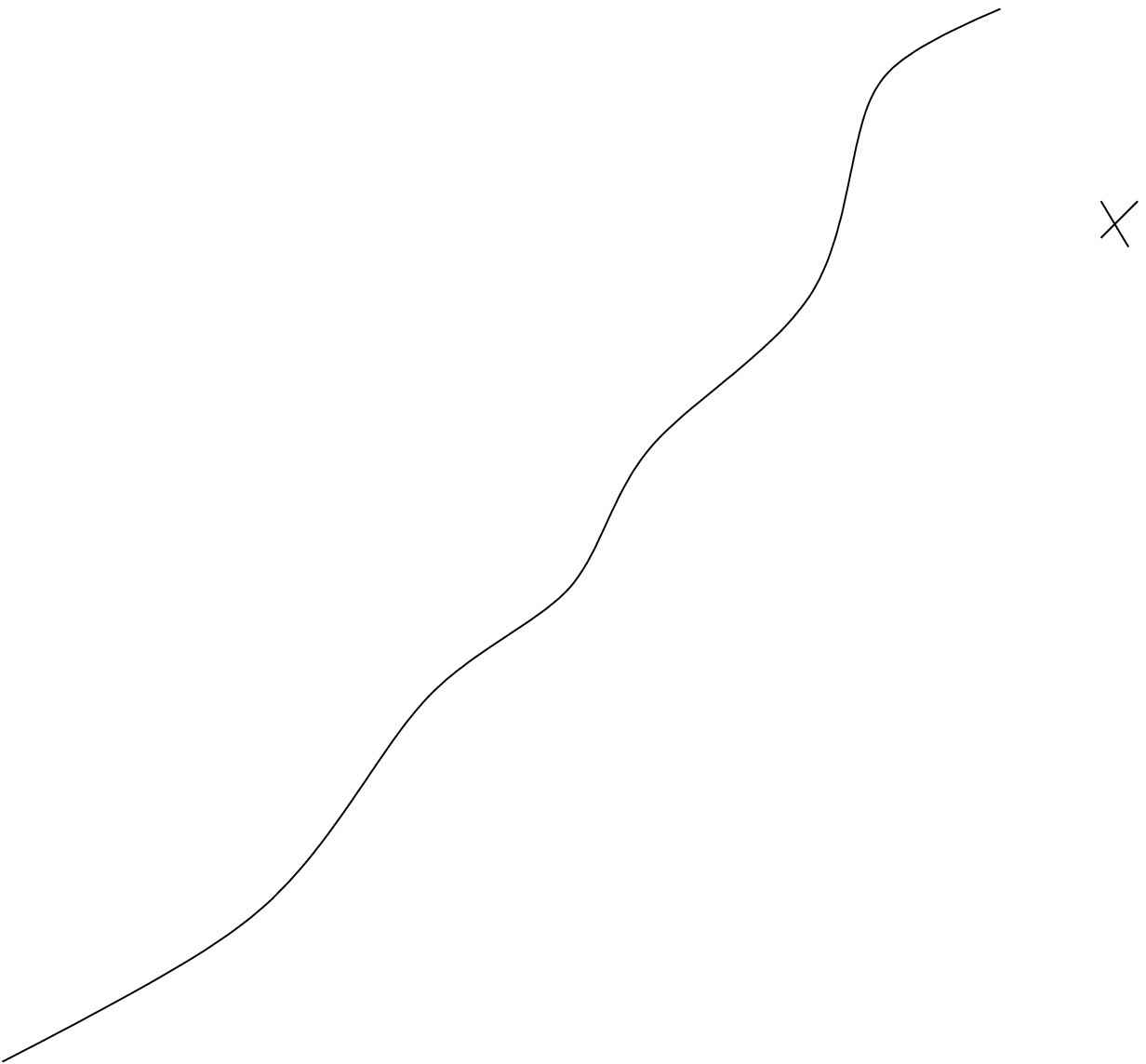}
$$
\begin{center}
{\bf Fig. 3.1:} Realisation of $\s$ which is not into the orbit of $\cal F$ under $\A(A_1)\times \A(\s)$
\end{center}


Let $x,y,h$ denote the images of 
 $e_+, e_-,e_0$ under the natural inclusion 
$\s\subset {\cal U}(\s)$.



Define now $g\in A_1^\s$ by $g=\hat f_{IIA}^{1,1}\circ w\vert_\s$, where 
$\hat f_{IIA}^{1,1}: {\cal U}(\s)\to A_1$ is the natural extension of $f_{IIA}^{1,1}$ to $ {\cal U}(\s)\to A_1$ and 
$w=\exp({\rm ad}(x^2))\in\A ({\cal U}(\s))$ is given by:
\beqa
\label{nou}
w(x)&=&x,\nonumber\\
w(y)&=&y+hx+xh-4x^3,\nonumber\\
w(h)&=&h-4x^2.
\eeqa 
\noi
Explicitly, this gives
\beqa
\label{exotic}
g(e_+)&=&(1+pq)q\nonumber \\
g(e_-)&=&-p+4p^2q^3-4p^3q^6+12p^2q^5\nonumber \\
g(e_0)&=&2pq-4p^2q^4+1.
\eeqa

\begin{proposition}
\label{compliteness}
$g\notin {\cal O}_{\cal F}$.
\end{proposition}
\textit{Proof:} To show this it is enough to prove that there does not
exist $(\alpha,w) \in \A(A_1)\times \A(\s)$ such that $(\alpha,w).g \in
\cal N$. 
Suppose for contradiction that there exist  $(\alpha,w) \in
\A(A_1)\times \A(\s)$ such that $(\alpha,w).g=f_{IIA}^{1,b}$
{\it i.e.} $\alpha\circ f_{IIA}^{1,b}=f_{IIA}^{1,b}\circ w$.

The $\s$ triplets corresponding to $g$ and $f_{IIA}^{1,b}$
are
\beqa
\label{tablou-nou}
\begin{array}{ll}
X= (1+pq)q &  X'=(b+pq)q \cr
Y=-p+4p^2q^3-4p^3q^6+12p^2q^5  & Y'=-p \cr
H=2pq+1-4p^4q^2 & H'=2pq+b.
\end{array}
\eeqa
Then writing
 $p'=\alpha(p),\ q'=\alpha(q)$ and $w={\rm Ad}(\begin{pmatrix} a_1 & a_2 \\ a_3 & a_4 \end{pmatrix})$ (see Proposition \ref{apendice2} for notations) one has
\beqa
\label{tablou-nou2}
(b+p'q')q'&=&-a_1a_3 X -a_3^2 Y - a_1a_3 H\nonumber \\
&=& -a_1a_3 ((1+pq)q)-a_3^2 (-p+4p^2q^3-4p^3q^6+12p^2q^5) -a_1a_3 (2pq+1-4p^4q^2) \nonumber \\
-p'&=& -a_2^2 X  +a_4^2 Y + a_2a_4 H \nonumber \\
&=& -a_2^2(1+pq)q  +a_4^2 (-p+4p^2q^3-4p^3q^6+12p^2q^5) + a_2a_4 (2pq+1-4p^4q^2)\nonumber  \\
2p'q'+b&=& -2a_1a_2 X + 2a_3a_4 Y + (a_1a_4+a_2a_3) H\nonumber \\
&=&-2a_1a_2 (1+pq)q + 2a_3a_4 (-p+4p^2q^3-4p^3q^6+12p^2q^5)  \nonumber\\
&&+ (a_1a_4+a_2a_3)( 2pq+1-4p^4q^2) .\nonumber
\eeqa
\noi
Substituting the second equation in the third equation, we obtain
\beqa
\label{master}
-2a_1a_2 (1+pq)q + 2a_3a_4 (-p+4p^2q^3-4p^3q^6+12p^2q^5) + (a_1a_4+a_2a_3)( 2pq+1-4p^4q^2)\nonumber\\
=-2\big( -a_2^2(1+pq)q  +a_4^2 (-p+4p^2q^3-4p^3q^6+12p^2q^5) + a_2a_4 (2pq+1-4p^4q^2)\big)q' +b.\nonumber
\eeqa

If $a_4\ne 0$ then the expansion of $q'$ in the standard basis
consists of only a scalar term, otherwise the term $-2 a_4^2 p^3q^6 q'$ on the RHS contain terms which are not present on the LHS; but then $[p',q']=0$ which is a contradiction and hence $a_4=0$.

Thus the equation above reduces to
\beqa
\label{master2}
-2a_1a_2 (q+pq^2)  - (2pq+1-4p^4q^2)
= 2a_2^2(q+pq^2) q' +b.
\eeqa

Let  $k$ be the highest power of $q$ appearing in the expansion of $q'$
in the standard basis. Then the highest power appearing in the
expansion of the RHS in the standard basis is $k+2$. Comparing with
the LHS gives $k=0$ and $q'=f(p)$ where $f$ is a polynomial in $p$. By similar arguments of degree counting on the two sides of Equation $6.7$, one shows that
 the degree of $f$ is equal to  $3$, that is
\beqa
\label{master22}
 q'=K_3 p^3 + K_2 p^2 + K_1 p + K_0 
\eeqa
\noi
where  $K_0,K_1,K_2\in\CC$ and $K_3\in\CC^*$. Inserting \eqref{master22} in \eqref{master2} and expanding explicitly the RHS in the standard basis, one sees that the term in $p^3q$ is $-10 a_2^2 K_3 p^3q$ whilst in the LHS there is no term in $p^3q$. Thus $K_3=0$ which is a contradiction.

We  have therefore proved that  there does not exist  $\alpha \in
\A(A_1)$ and $w\in \A(\s)$ such that $(\alpha,w) . g=
f_{IIA}^{1,b}$. QED

\medskip

This means that none of the elements of $g(\s)$ are strictly nilpotent or strictly semi-simple.

\begin{corollary}
With the notations above, $g(\s)\subseteq \Delta_2\cup\Delta_4$.
\end{corollary}
{\it Proof:} Immediate from Theorem \ref{TEOREMA}. QED

\begin{remark}
If we set
$${\cal D}^{\prime\prime}_1=\{ f\in A_1^\s : \exists n\in \s \mbox{ nilpotent
s.t }
{f(n)} \mbox{ is an eigenvector of some element in } \Delta_3 \}$$ 
then $g\in {\cal D}^\prime_1$
since $g(e_+)$ is an eigenvector of $pq$. Hence $ {\cal
D}^\prime_1\ne  {\cal D}$.
\end{remark}

Before ending this subsection, we relate this example to the characterisation of $A_1^\s$  given by A. Joseph in \cite{joseph}. He showed that $A_1^\s$ is a disjoint union $S_1 \cup S_2$, where $S_1$ (resp. $S_2$) is the set of realisations for which the spectrum of the standard semi-simple element $H$ is equal to $2\ZZ$ (resp. $\ZZ$). Furthermore, he showed that $S_1$ (resp. $S_2$) can be subdivided  into a disjoint union $S_{11}\cup S_{12} \cup \dots \cup S_{1\infty}$ (resp. $S_{21}\cup S_{22}\cup \dots $). 
Finally, he showed that $S_1, S_{21},S_{22}\dots$ and $S_{12},S_{13}\dots, S_{1r}$ with $r<\infty$ are stable under the action of $\A (A_1)\times \A ({\cal U}(\s)$.

With respect to 
 this decomposition, $f_I^1\in S_{21}$ and $f_{IIA}^{1,b}\in S_{11}$ (see Propositions \ref{propI} and \ref{propIIA}). Hence
$$ {\cal N}\subseteq S_{21}\cup S_{11}.$$
We will now prove that $g\in S_{1\infty}.$ For this we need the definition of $S_{11}$:

\begin{definition}
\label{def-S11}
Let $f\in A_1^\s, m\in 2\ZZ, H=f(e_+), X=f(e_+), Y=f(e_-)$ and  
$D\big(H,m\big)=\left\{ z\in A_1 :\ [H,z]=mz\right\}.$
By Lemma $3.1$ of \cite{joseph}, there exists $y_m\in D\big(H,m\big)$ such that 
\beqa
\label{vom-utiliza}
D\big(H,m\big)=y_m \CC[H].
\eeqa
\noi
The set $S_{11}$ is defined to be the set of realisations $f\in A_1^\s$ for which 
\beqa
\label{ce-nu-trebuie}
D\big(H,2\big)\ne X \CC[H] \mbox{ or } D\big(H,-2\big)\ne Y \CC[H].
\eeqa
\noi
\end{definition}

We can now prove

\begin{proposition}
\label{S1infty}
Let $g\in A_1^\s$ be defined by equations \eqref{exotic}. Then $g\in S_{1\infty}$.
\end{proposition}
{\it Proof:} 
Recall that  $g$ was obtained by acting on $f_{II}^1\in S_{11}$ with a specific element of $\A ({\cal U}(\s)$. Hence, by the properties of the Joseph decomposition to  prove that $g\in S_{1\infty}$
it is enough to show that $g\notin S_{11}$. 
From \eqref{ce-nu-trebuie}, one sees that this comes down  to proving that
$$D\big(H,2\big)= X \CC[H] \mbox{ and } D\big(H,-2\big)= Y \CC[H].$$

We now treat the first of these conditions.
By \eqref{vom-utiliza}, there  exists $y_2\in D\big(H,2\big)$ such that $D\big(H,2\big)=y_2 \CC[H]$. Since $X\in D\big(H,2\big)$ there exists a polynomial $a_n H^n+\dots+a_0$ such that
$$ X=y_2(a_n H^n+\dots+a_0),$$
which gives
$$ q+pq^2= y_2 \left(a_n (2pq-4p^2q^4+1)^n+\dots+a_0\right).$$
Comparing highest powers of $p$, one has $n=0$, $y_2=\frac{1}{a_0} X$ and thus $D\big(H,2\big)=X \CC[H]$.
A similar but more tedious argument shows that $ D\big(H,-2\big)=Y \CC[H]$. QED

\section{Perspectives}
\label{3-perspective}
 
We end this chapter by overviewing some possible perspectives of this work. We considered the problem of the construction of finite-dimensional Lie algebras from the fundamental observables $p$ and $q$ of quantum mechanics, that is the Lie algebras that can be realised in $A_1$. We  considered here only complex Lie algebras and the Weyl algebra $A_1$ was taken over the field of complex numbers. Thus, a first question one might ask is what will be the situation when considering real Lie algebras realisable in the Weyl algebra over the real numbers. This problem should be approached by an analysis of the real forms of the complex Lie algebras obtained in Theorem \ref{toate-subalg}. Moreover one could address the question of what  all this would become over an arbitrary field (for example not of characteristic $0$).

Another natural generalisation of this study would be to consider the Lie algebras that can be realised in $A_n$, the complex associative algebra defined in ($3.1$). A somewhat connected issue is to study 
the problem analogous to the problem considered in this chapter
not in the context of the quantum Lie bracket but in the context of the classical Poisson bracket. Indeed, consider $A_0$ the set of polynomials over some field in the {\it commuting} variables $x$ and $y$ and define the Poisson bracket
$$\{ f, g \}= \frac{\partial f}{\partial x}\frac{\partial g}{\partial y}-
\frac{\partial f}{\partial y}\frac{\partial g}{\partial x}, \, \forall f,g\in A_0.$$
This bracket makes $A_0$ into a Lie algebra and 
 one may thus investigate what are the Lie subalgebras of $A_0$.
 Moreover, some comparison between the obtained symmetries at the classical ($ A_0$) and the quantum ($A_1$) level are then possible.
Finally, an in some sense opposite problem is the study of infinite-dimensional subalgebras of $A_n$.

\chapter{Lie algebras of order $F$}

In the previous chapter we got familiarised with fundamental algebraic structures, the Lie (super)algebras. 
However, even if these structures are nowadays practically the only used to describe symmetries of physical systems, one may think  to define more exotic structures.
In this chapter we introduce such an example, namely  the {\it Lie algebras of order $F$} and in the next two chapters, starting from a particular Lie algebra of 
order $3$ we construct field theoretical realisations of it.
Lie algebras of order $F$ can be seen as generalisations of Lie algebras and superalgebras (which are obtained for the particular cases of $F=1$ and resp. $F=2$, as we will see in the sequel). Here 
we set  the basis of an algebraic theory of contractions and deformations of this algebraic structure.

\medskip

The chapter is structured as follows. In the first section we recall the definition of Lie algebras of order $F$. We focus on the case $F=3$ and we give some known examples. Furthermore we study the case of Lie algebras of order $3$ based on the $\s$ Lie algebra. In the second section we recall some  results on representations of these algebraic structures. In the next sections we set the basis of the study of deformations and contractions. 
An explicit non-trivial example is given.
In the fifth section we initiate a study of non-trivial extensions of the Poincar\'e algebra.
In the sixth section a different, binary approach  is proposed for Lie algebras of order $3$. 
The last section presents briefly some further perspectives.

\medskip

If in the beginning of this chapter some results are recalled from \cite{Michel-finit} and \cite{Michel-infinit}, the rest presents new results which constitute the subject of a forthcoming publication.

\section{Definition and examples of Lie algebras of order $F$}

In this section we recall the definition and some basic properties of Lie algebras of order $F$ (for a more detailed analyse see \cite{Michel-finit} and \cite{Michel-infinit}).
We then focus on the case $F=3$ and give some examples. Finally, we study the case of Lie algebras of order $3$ based on  $\s$.

\subsection{Definition}
\label{3-def}


\begin{definition}
\label{elementary}
Let $F\in\mathbb{N}^*$, $q=e^{\frac{2\pi i}{F}}$, a ${\mathbb Z}_F$ graded ${\mathbb C}-$vector space  ${\mathfrak{g}}= {\mathfrak{g}}_0 \oplus
{\mathfrak{g}}_1\oplus
{\mathfrak{g}}_2 \dots \oplus
{\mathfrak{g}}_{F-1}$ 
and $\e$ an automorphism of $\g$ satisfying $\e^F=1$, $\g_i$ being the eigenspace corresponding to the eigenvalue $q^i$ of $\e$. Then $\g$
is called a Lie algebra of order $F$ if
\begin{enumerate}
\item $\mathfrak{g}_0$ is a Lie algebra.
\item $\mathfrak{g}_i$ is a representation of $\mathfrak{g}_0$, $i=1,..,F-1$; if $[.,.]$ denotes the bracket on $\g_0$ and the action of $\g_0$ on $\g$ it is clear that $[\e  (X), \e (Y)]= \e ([X,Y])$, for all $X\in \g_0$ and $Y\in \g$.
\item  For all $i=1,..,F-1$ there exists  a $F-$linear,  $\mathfrak{g}_0-$equivariant  map
 $\mu_i : {\cal S}^F\left(\mathfrak{g}_i\right)
\rightarrow \mathfrak{g}_0$, where  ${ \cal S}^F(\mathfrak{g}_i)$ denotes
the $F-$fold symmetric product of $\mathfrak{g}_i$, map that satisfy the following (Jacobi) identity:
\beqa
\label{eq:J}
&&\sum\limits_{j=1}^{F+1} \left[ Y_j,\mu_i ( Y_1,\dots,
Y_{j-1},
Y_{j+1},\dots,Y_{F+1}) \right] =0, \nonumber
\eeqa
for any $i=1,\dots,F-1$ and for all $Y_j \in \mathfrak{g}_i$, $j=1,\dots,F+1$. One can also see that $\{ \e(Y_1), \dots, \e (Y_F) \}= \e (\{ Y_1, \dots, Y_F \})$.
\end{enumerate}
\end{definition}

\begin{remark}
If $F=1$, by definition $\mathfrak{g}=\mathfrak{g}_0$ and a Lie algebra of order $1$ is a Lie algebra. If $F=2$, then $\g$ is a Lie superalgebra. Therefore, the Lie algebras of order $F$ appear as possible generalisations of Lie algebras and superalgebras.
\end{remark}

Note that the following identities are satisfied by a Lie algebra of order $F$; we call them  {\it Jacobi identities for Lie algebras of order $F$}:\\
\textbf{1.}For any $ X,X',X'' \in \mathfrak{g}_0$, 
$$\left[\left[X,X'\right],X''\right] +
\left[\left[X',X''\right],X\right] +
\left[\left[X'',X\right],X'\right] =0, \ {(\mbox{\textbf{I.1}})}$$
which we will refer to as the ``1st Jacobi identity''. This relation expresses the fact that $\mathfrak{g}_0$ is a Lie algebra.\\
\textbf{2.} For any $X,X' \in {\mathfrak{g}}_0 \mbox{ and } Y \in {\mathfrak{g}}_i, i=1,..,F-1$,
$$ \left[\left[X,X'\right],Y\right] +
\left[\left[X',Y\right],X\right] +
\left[\left[Y,X\right],X'\right]=0, \ (\mbox{\textbf{I.2}})$$
(``2nd Jacobi identity'') since ${\mathfrak{g}}_i$ is a representation of $\mathfrak{g}_0$.\\
\textbf{3.} For $X\in {\mathfrak{g}}_0 \mbox{ and } Y_j \in {\mathfrak{g}}_i$ (with $j=1,\dots,F$, $i=1,\dots,F-1$)
$$\left[X,\mu_i (Y_1,\dots,Y_F)] =
\mu_i (\left[X_0,Y_1 \right],\dots,Y_F)  +
\dots +
\mu_i (Y_1,\dots,\left[X,Y_F\right] \right), \ (\mbox{\textbf{I.3}}) .$$ 
(`` 3rd Jacobi identity'') results from  the $\g_0$-equivariance of the map $\mu_i$.\\
\textbf{4.} For all $Y_j \in \mathfrak{g}_i$, $j=1,...,F+1,\  i=1,...,F-1$, the identity (the 4th Jacobi identity)
$$\sum\limits_{j=1}^{F+1} \left[ Y_j,\mu_i ( Y_1,\dots,
Y_{j-1},
Y_{j+1},\dots,Y_{F+1}) \right] =0 \ (\mbox{\textbf{I.4}})$$
corresponds to condition $3$ of the definition.

\begin{remark}
Let ${\mathfrak{g}}={\mathfrak{g}}_0\oplus{\mathfrak{g}}_1\dots{\mathfrak{g}}_{F-1}$ be a Lie algebra of order $F$, with $F>1$. For any $i=1,\ldots,F-1$, the graded vector spaces ${\mathfrak{g}}={\mathfrak{g}}_0\oplus{\mathfrak{g}}_i$ satisfy all the properties of  a Lie algebra of order $F$ structure. We  call these type of algebras \textit{elementary Lie algebras of order $F$}.
\end{remark}

\medskip

From now on, we consider  elementary Lie algebras of order $F$.

\medskip

An inductive construction of finite-dimensional Lie algebras of order $F$ on a given vector space (which already posses a Lie algebra of order $F_1$ structure, $1<F_1<F$) was given in \cite{Michel-finit}. We give here the main elements of this construction.
Let $\mathfrak{g}=\mathfrak{g}_0\oplus\mathfrak{g}_1$ be the elementary algebra of order $F_1$ (with the application $\mu_1$) and $F_2\in \mathbb{N}^*$. If there exists a $\mathfrak{g}_0-$equivariant, $F_2-$linear symmetric form on $\mathfrak{g}_1$,
$$ \mu_2 : {\cal S}^F\left(\mathfrak{g}_1\right)
\rightarrow {\mathbb C},$$
then $\mathfrak{g}$ admits an elementary Lie algebra of order $F_1+F_2$ structure.\\
We just give here an outline of the proof (one can check \cite{Michel-finit} for further details).
The $\g_0$-equivariant map that allows us to define the elementary Lie algebra of order $F_1+F_2$ structure is
$$\mu: {\cal S}^{F_1+F_2}(\g_1)\to \g_0 \otimes {\mathbb C} \cong \g_0, $$
with
$$ \mu (Y_1,\ldots,Y_{F_1+F_2})=\frac{1}{F_1!}\frac{1}{F_2!}\sum_{\sigma\in S_{F_1+F_2}}\mu_1(Y_{\sigma (1)},\dots Y_{\sigma (F_1)}) \otimes \mu_2 (Y_{\sigma (F_1+1)},\ldots,Y_{\sigma (F_1+F_2)})$$
where $Y_1,\ldots,Y_{F_1+F_2}\in\g_1$.
One then checks that this map satisfies the Jacobi  identities {\bf I.1-I.4}.

The particular case $F_1=1$ was also treated in \cite{Michel-finit}.

\medskip

After this general introduction,  in the rest of this thesis we consider the case $F=3$, that is of Lie algebras of order $3$. We denote the $3-$linear map by the three entries bracket $\{.,.,.\}$.

\subsection{Some examples of elementary Lie algebras of order $3$}
\label{3-ex}

So far in literature there exists non-trivial examples of Lie algebras of order $F$, both finite and resp. infinite-dimensional (see \cite{Michel-finit} and resp. \cite{Michel-infinit}). We give now some examples of finite-dimensional elementary Lie algebras of order $3$, which will be relevant in the sequel.

\begin{example}
\label{so23}
Let $\g_0={\mathfrak{so}}(2,3)$ and $\g_1$ its adjoint representation.  Let $\{J_a\}$ be a basis of $\g_0$ and $\{A_a\}$ be the corresponding basis of $\g_1$, with $a=1,\ldots,10$. Thus, one has $[J_a, A_b]={\rm ad}([J_a, J_b])$. Let $g_{ab}=Tr(A_aA_b)$ be the Killing form. Then, one can endow $\g_0\oplus\g_1$ with a 
Lie algebra of order $3$ structure given by
\beq
\{ A_a,A_b,A_c \}=g_{ab}J_c+g_{ac}J_b+g_{bc}J_a. \nonumber
\eeq
\end{example}

Notice that when one considers this type of Lie algebras of order $3$ one can obviously consider the complex structure as well as real forms of it.

\begin{example}
Let $\g_0$ be the Poincar\'e algebra and  $\{ L_{mn}, P_m:\ L_{mn}=-L_{nm},\ m<n, m,n,=0,\dots,3 \} $  be a basis of $\g_0$ with the non-zero brackets
\beqa
 \left[L_{mn}, L_{pq}\right]&=&
\eta_{nq} L_{pm}-\eta_{mq} L_{pn} + \eta_{np}L_{mq}-\eta_{mp} L_{nq},\nonumber \\
\left[L_{mn}, P_p \right]&=& \eta_{np} P_m -\eta_{mp} P_n.
\eeqa
\noi
(see subsection $2.3.2$). Let now $\g_1$ be a $4-$dimensional representation of $\g_0$ and $\{ V_m:\ m=0,\dots,3\}$ be a basis of $\g_1$. The action of $\g_0$ on $\g_1$ is given by
$$\left[L_{mn}, V_p \right]= \eta_{np} V_m -\eta_{mp} V_n, \ \
\left[P_{m}, V_n \right]= 0. $$
One can now construct a Lie algebra of order $3$ structure on $\g_0\oplus\g_1$, with the additional bracket
\beqa
\label{algebra0}
&&\{ V_m, V_n, V_r \}=
\eta_{m n} P_r +  \eta_{m r} P_n + \eta_{r n} P_m, 
\eeqa
\noi
with the metric $\eta_{mn}$ taken to be $\rm{diag}(1,-1,-1,-1)$.
\end{example}

Note that $P^2$ is a Casimir operator of this Lie algebra of order $3$.

Let us remark at this point that this example is of crucial importance for the last two chapters of this thesis. Indeed, it is exactly this algebraic structure that will be used to construct a field theoretical model in four and resp. in arbitrary  dimensions (see chapters $6$ and resp. $7$).

An additional remark to be done here is that a In\"on\"u-Wigner contraction of the algebra of Example \ref{so23} contains the algebra of Example \ref{algebra0} as a subalgebra; furthermore this provides also an example of a non-trivial deformation of Lie algebra of order $3$. We will come back to this point in subsection \ref{ex-contractie}.

\subsection{Study of elementary Lie algebras of order $3$ based on $\s$}
\label{3-sl2}

In this subsection we study the elementary Lie algebras of order $3$ $\g_0\oplus \g_1$ for which $\g_0\cong \s$. Recalling from subsection \ref{3-def} that $\g_1$ is a representation of $\g_0$, we will first treat the case when $\g_1$ is an irreducible representation of $\s$ and then the case when $\g_1$ is a reducible representation.

In this subsection we denote by $X_+, X_-, X_0$ a standard basis of $\g_0$, with the following commutation relations
\beqa
\label{3-g0}
[X_0, X_+]=2X_+,\ [X_0,X_-]=-2X_-,\ [X_+, X_-]=X_0. 
\eeqa

\begin{theorem}
\label{sl2-irrep}
Let $\g=\s\oplus \g_1$ be a Lie algebra of order $3$ such that $\g_1$ is an irreducible representation of $\g_0$.
If the bracket on $\g_1$ is not trivial (that is it exists $Y_1, Y_2, Y_3\in\g_1$ such that $\{Y_1, Y_2, Y_3 \}\ne 0$) then 
$\g$ is of dimension $6$ and 
 $\g$ is isomorphic to the Lie algebra of order $3$ given by the following non-zero relations
\beqa \begin{array}{llllll}
\label{sl2+adj}
\{ Y_{+1}, Y_{-1}, Y_0\}&=& X_0,\ &\{ Y_0, Y_0, Y_0\}&=& 6X_0\nonumber \\
\{Y_{+1}, Y_{-1}, Y_{+1}\}&=& 2 X_+,\ &\{Y_{+1}, Y_0, Y_0 \}&=& 2 X_+,\nonumber\\
\{Y_{-1}, Y_{+1}, Y_{-1}\}&=& 2 X_-,\ &\{Y_{-1}, Y_0, Y_0 \}&=& 2 X_-.
\end{array}
\eeqa
(where $Y_{+1}, Y_{-1}, Y_0$ is a standard basis of $\g_1$)
\end{theorem}
{\it Proof:} I. Firstly remark that if one considers representations of $\s$ of even dimension, then the weights will be odd numbers. Thus, calculating any bracket $\{Y_1, Y_2, Y_3 \}$ one has an odd weight which cannot be found in $\s$.

\medskip

\noi
II. The representation of dimension $1$ of $\s$, $<Y>$, is trivial: $[X_\pm, Y]=0, [X_0, Y]=0$. One necessarily has $\{ Y, Y, Y \} =\alpha X_0$ and imposing for example {\bf I.3} on the set $X_+, Y, Y, Y$ one gets $\alpha =0$.

\medskip

\noi
III. Consider now ${\rm dim}\, \g_1=3$ (that is $\g_1$ is the adjoint representation of $\s$). Hence one has the following non-zero commutation relations
\beqa
\begin{array}{llllll}
\label{3-adj}
\left[X_+, Y_{-1}\right]&=&Y_0,\ &\left[X_+,Y_0\right]&=&-2Y_{+1},\\ 
\left[X_-, Y_{+1}\right]&=&-Y_0,\ &\left[X_-,Y_0\right]&=&2Y_{-1},\\
\left[X_0, Y_{+1}\right]&=&2Y_{+1},\ &\left[X_0,Y_{-1}\right]&=&-2Y_{-1}.
\end{array} 
\eeqa
\noi
By a simple calculus of weights, one has
\beqa \begin{array}{llllll}
\label{sl2+adj-1}
\{ Y_{+1}, Y_{-1}, Y_0\}&=& \alpha_1 X_0, & & & \\
\{Y_{+1}, Y_{-1}, Y_{+1}\}&=& \alpha_2 X_+,\ &\{Y_{+1}, Y_0, Y_0 \}&=& \alpha_4 X_+,\\
\{Y_{-1}, Y_{+1}, Y_{-1}\}&=& \alpha_3 X_-,\ &\{Y_{-1}, Y_0, Y_0 \}&=& \alpha_5 X_-.\\
\{ Y_0, Y_0, Y_0\}&=& \alpha_6 X_0 & & &.
\end{array}
\eeqa
\noi
where $\alpha_i\in \CC$ (with $i=1,\dots,6$).
From symmetry considerations one has that $\alpha_2=\alpha_3$ and $\alpha_4=\alpha_5$. 
Identity {\bf I.4} on the set $Y_{+1},Y_{+1}, Y_{-1}, Y_0$ gives
$$ 2[Y_+, \{ Y_{+1}, Y_{-1}, Y_0 \}]+ [Y_{-1}, \{ Y_{+1}, Y_{+1}, Y_0 \}]+[Y_0, \{ Y_{+1}, Y_{+1}, Y_{-1} \}]=0.$$
Using now \eqref{3-g0}, \eqref{3-adj} and \eqref{sl2+adj-1} to express the different brackets, one gets 
\beqa
\label{6-1}
\alpha_2=2\alpha_1.
\eeqa
\noi
Imposing {\bf I.4} on the set $Y_0,Y_0, Y_0, Y_{+1}$ leads to
\beqa
\label{6-3}
3\alpha_4=\alpha_6.
\eeqa
\noi
Similarly, identity {\bf I.3} with respect to $X_-, Y_{+1}, Y_{-1}, Y_0$ gives
\beqa
\label{6-4}
2\alpha_1=\alpha_5.
\eeqa
\noi
Thus $(\alpha_1, \alpha_2, \alpha_3, \alpha_4, \alpha_5, \alpha_6)=(\alpha_1, 2\alpha_1, 2\alpha_1, 2\alpha_1, 2\alpha_1, 6\alpha_1)$.


\medskip

\noi
IV.  Consider now ${\rm dim}\, \g_1=5$
and let  $<Y_2, Y_1,\dots, Y_{-2}>$ be a basis of $\g_1$ on witch $\g_0$ acts in the following manner
\beqa
\begin{array}{llllllllllll}
\label{X-Y-dim5}
\left[ X_+, Y_{+1}\right]=Y_2 & \left[ X_+, Y_0\right]=2Y_1 & \left[ X_+, Y_{-1}\right]=3Y_0 & \left[ X_+, Y_{-2}\right]=4Y_{-1} \\
\left[ X_-, Y_{+2}\right]=4Y_{+1} &\left[ X_-, Y_{+1}\right]=3Y_0 & \left[ X_-, Y_0\right]=2Y_{-1} & \left[ X_-, Y_{-1}\right]=Y_{-2}  \\
\left[ X_0, Y_{+2}\right]=4Y_{+2} &\left[ X_0, Y_{+1}\right]=2Y_{+1} & \left[ X_0, Y_{+1}\right]=-2Y_{-1} & \left[ X_0, Y_{-2}\right]=-4Y_{-2}  
\end{array}
\eeqa
\noi
On the set $Y_{\pm 1}, Y_0$ the Jacobi identities  lead, as shown in part III of this proof, to
\beqa \begin{array}{llllll}
\label{dim5-1}
\{ Y_{1}, Y_{-1}, Y_0\}&=& \alpha X_0,\ &\{ Y_0, Y_0, Y_0\}&=& 6\alpha X_0\nonumber \\
\{Y_{1}, Y_{-1}, Y_{1}\}&=& 2\alpha X_+,\ &\{Y_1, Y_0, Y_0 \}&=& 2\alpha X_+,\nonumber\\
\{Y_{-1}, Y_1, Y_{-1}\}&=& 2\alpha X_-,\ &\{Y_{-1}, Y_0, Y_0 \}&=& 2\alpha X_-.
\end{array}
\eeqa
\noi
where $\alpha\in\{0,1\}$.

 Now, by the same type of weight considerations, one completes \eqref{dim5-1} with
\beqa \begin{array}{llllll}
\label{dim5-2}
\{ Y_{-2}, Y_0, Y_{+1}\}&=& \beta_1 X_-,\ & \{ Y_{-2}, Y_{+1}, Y_{+1}\}&=& \beta_2 X_0,\\
\{ Y_{+2}, Y_0, Y_{-1}\}&=& \beta_3 X_+,\ & \{ Y_{+2}, Y_{-1}, Y_{-1}\}&=& \beta_4 X_0,\\
\{ Y_{-2}, Y_{+2}, Y_{-1}\}&=& \gamma_1 X_-,\ & \{ Y_{-2}, Y_{+2}, Y_{+1}\}&=& \gamma_2 X_0,\\
\{ Y_{+2}, Y_{-2}, Y_{+1}\}&=& \gamma_2 X_-. & & & 
\end{array}
\eeqa
\noi
From symmetry considerations one has $\beta_1=\beta_3,\beta_2=\beta_4$ and $\gamma_1=\gamma_3$.
Imposing now {\bf I.3} identities on these brackets one has the following system of linear equations:
\beqa
\gamma_1 &=& 4 \beta_2 + 3 \gamma_2 \nonumber\\
\gamma_1 &=& - \gamma_2 -2 \beta_1 \nonumber\\
\beta_1&=&4\alpha + 2 \beta_2 + 2 \gamma_2\nonumber\\
\gamma_1 &=& - \beta_2 - 4 \alpha \nonumber\\
\beta_2 &=& 3\beta_1.
\eeqa
\noi 
This system implies that $\alpha =0$ and all $\beta$ and $\gamma$ constants are also zero.

\medskip

\noi
V. We now prove by induction that a representation $\g_1$ of $\s$ of even dimension $2k+1$ leads to trivial bracket on $\g_1$. We have proven that this statement is true for $k\le 2$. 
This means that 
\beqa
\label{nulitate}
\{ Y_a, Y_b, Y_c \} = 0 \ \forall a,b,c\in\{-(k-1),\dots, +(k-1)\}.
\eeqa
\noi
The only {\it a priori} non-zero brackets are those involving at least one $Y_{\pm k}$, that is of type $\{ Y_{-k}, Y_a, Y_b\}$ (with $a,b,c\in\{-(k-1),\dots, +(k-1)\}$) and symmetric brackets, of type $\{ Y_{+k}, Y_{-a}, Y_{-b}\}$
and then $\{Y_{-k}, Y_{+k}, Y_{\pm 1}\}$ or $\{Y_{-k}, Y_{+k}, Y_0\}$. 
The first type of  these  brackets write explicitly as:
\beqa \begin{array}{llllll}
\label{dimimp1}
\{ Y_{-k}, Y_{+(k-1)}, Y_{0}\}&,\ \{ Y_{-k}, Y_{+(k-1)}, Y_{+1}\}&,\ \{ Y_{-k}, Y_{+(k-1)}, Y_{+2}\}\\
\{ Y_{-k}, Y_{+(k-2)}, Y_{1}\}&,\ \{ Y_{-k}, Y_{+(k-2)}, Y_{+2}\}&,\ \{ Y_{-k}, Y_{+(k-2)}, Y_{+3}\}\\
\dots\\
\{ Y_{-k}, Y_{1}, Y_{+(k-2)}\}&,\ \{ Y_{-k}, Y_{1}, Y_{+(k-1)}\}
\end{array}
\eeqa
\noi
and symmetric ones. 
We now show that all the brackets \eqref{dimimp1} are zero. For this let us explicitly consider the brackets which are proportional to $X_{-1}$ (the first column in \eqref{dimimp1}), the rest of the terms being treated analogously. Indeed, all these brackets can be written as
$$\{Y_{-k}, Y_{a}, Y_{k-a-1}\}$$
with $a=0,\dots, k-2$. Let us  apply {\bf I.4} on  the set $(Y_{0}, Y_{-k}, Y_{a}, Y_{k-a-1}\}$ which gives
\beqa
\label{totul}
[Y_0, \{\{Y_{-k}, Y_{a}, Y_{k-a-1}\}]=-[Y_{-k},\{Y_0, Y_a, Y_{k-a-1}\}]-[Y_a,\{ Y_{-k}, Y_0, Y_{k-a-1}\}] - [ Y_{k-a-1}, \{ Y_0, Y_{-k}, Y_a\}]\nonumber\\
\eeqa
\noi
The first term in the RHS is equal to $0$ because the bracket $\{Y_0, Y_a, Y_{k-a-1}\}$ does not involve any $Y_{\pm k}$ (the induction hypothesis). The second term in the RHS is also equal to $0$ for $a\ne 0$ by weight considerations and finally the last term is also $0$ again by weight considerations. In the LHS, $\{\{Y_{-k}, Y_{a}, Y_{k-a-1}\}\propto X_{-1}$ and $[Y_0,X_{-1}]\ne 0$. Hence this analyses provides $\{\{Y_{-k}, Y_{a}, Y_{k-a-1}\}=0$ for all 
$a\ne 0$. To complete this proof one now has to treat the last remaining case, namely $\{ Y_{-k}, Y_{+(k-1)}, Y_{0}\}$. Applying {\bf I.4} on the set $Y_{-1}, Y_{-k}, Y_{+(k-1)}, Y_{0}$ one obtains the required result.

Finally the last type of brackets, $\{Y_{-k}, Y_{+k}, Y_{\pm 1}\}$ or $\{Y_{-k}, Y_{+k}, Y_0\}$, are treated similarly. For example, for the set $Y_{-k}, Y_{+k}, Y_{-1}$ one uses {\bf I.4} on the set $(Y_{-1}, Y_{-k}, Y_{+k}, Y_{- 1})$. QED




\bigskip

We now consider the case when $\g_1$ is a reducible representation of $\s$.

\begin{theorem}
\label{sl2-rep}
Let $\g=\g_0\oplus \g_1$ be a Lie algebra of order $3$, where $\g_0\cong\s$ and $\g_1$ is a reducible representation of $\g_0$.
Then the $3-$entries bracket $\{.,.,.\}$ of $\g$ is trivial.
\end{theorem}
{\it Proof:} We begin by considering the case where $\g_1$ represents two distinct copies of the adjoint representation. Let $<Y_\pm^i, Y_0^i>$ (with $i=1,2$) constitute a standard basis of $\g_1$ (that is, for each $i=1,2$, $<Y_\pm^i, Y_0^i>$ spans an adjoint representation of $\s$). Thus, from Theorem \ref{sl2-irrep}, one has 
\beqa \begin{array}{llllll}
\label{2ls2-0}
\{ Y_{+1}^i, Y_{-1}^i, Y_0^i\}&=& \alpha_i X_0,\ &\{ Y_0^i, Y_0^i, Y_0^i\}&=& 6\alpha_i X_0\nonumber \\
\{Y_{+1}^i, Y_{-1}^i, Y_{+1}^i\}&=& 2\alpha_i X_+,\ &\{Y_{+1}^i, Y_0^i, Y_0^i \}&=& 2\alpha_i X_+,\nonumber\\
\{Y_{-1}^i, Y_{+1}^i, Y_{-1}^i\}&=& 2\alpha_i X_-,\ &\{Y_{-1}^i, Y_0^i, Y_0^i \}&=& 2\alpha_i X_-.
\end{array}
\eeqa
\noi
where $i=1,2$ and $\alpha_i\in\{0,1\}$. We now list, performing again weight computations, the non-zero $3-$brackets involving $Y^1_\alpha, Y^1_\beta$ and $Y^2_\gamma$ (where $\alpha,\beta,\gamma\in\{\pm,0\}$)
\beqa
\begin{array}{llllll}
\label{2sl2-1}
\{Y_{+1}^1, Y_{-1}^1, Y_0^2\}=\beta_1 X_0 \\
\{Y_{+1}^1, Y_{-1}^1, Y_{+1}^2\}=\beta_2 X_+ & \{Y_{+1}^1, Y_0^1, Y_0^2\}=\beta_4 X_+ \\
\{Y_{-1}^1, Y_{+1}^1, Y_{-1}^2\}=\beta_3 X_- & \{Y_{-1}^1, Y_0^1, Y_0^2\}=\beta_5 X_- \\
\{Y_0^1, Y_0^1, Y_0^2\}=\beta_6 X_0 \\
\{Y_0^1, Y_0^1, Y_{+1}^2\}=\beta_7 X_+ & \{Y_{-1}^1, Y_0^1, Y_{+1}^2\}=\beta_9 X_+ \\
\{Y_0^1, Y_0^1, Y_{-1}^2\}=\beta_8 X_- & \{Y_{+1}^1, Y_0^1, Y_{-1}^2\}=\beta_{10} X_- \\
\{Y_{-1}^1, Y_{-1}^1, Y_{+1}^2\}=\beta_{11} X_- & \{Y_{+1}^1, Y_{+1}^1, Y_{-1}^2\}=\beta_{12} X_+.\\
\end{array}
\eeqa
\noi
where $\beta_j\in\CC$, $j=1,\dots,12$. From symmetry reasons (which become clear when imposing identity {\bf I.3} on these triplets) one has $\beta_2=\beta_3,\ \beta_4=\beta_5,\ \beta_7=\beta_8,\ \beta_9=\beta_{10}$ and $\beta_{11}=\beta_{12}$. Analogously to \eqref{2sl2-1} one has the $3-$brackets involving $Y^1_\alpha, Y^2_\beta$ and $Y^2_\gamma$ (where $\alpha,\beta,\gamma\in\{\pm,0\}$).

As before, one has now to impose the Jacobi identities {\bf I.3} and {\bf I.4} which involve the $3-$brackets \eqref{2sl2-1}; we proceed by imposing  {\bf I.3} on the set $X_+, Y_{+1}^1, Y_{-1}^1, Y_0^2$. This leads to 
\beqa
\label{2sl2-2}
-2 \beta_1 = \beta_4 - 2 \beta_2.
\eeqa
\noi
Similarly, from the set $X_-, Y_{+1}^1, Y_0^1, Y_0^2$ one gets
\beqa
\label{2sl2-3}
-2 \beta_1 = \beta_4 -  \beta_6.
\eeqa
\noi
Furthermore, the set $X_+, Y_0^1, Y_0^1, Y_0^2$ leads to
\beqa
\label{2sl2-4}
 \beta_6 = 2 \beta_4 +  \beta_7.
\eeqa
\noi
Arguing along the same line, the rest of independent equations one gets in this manner are
\beqa
\label{2sl2-5}
 \beta_7 &=& \beta_6 - 4  \beta_9\nonumber\\
 \beta_7 &=& 2 \beta_2 - 2 \beta_9  \nonumber\\
2\beta_9 &=& 2 \beta_{11} -  \beta_5 \nonumber\\
\beta_{11} &=& 2 \beta_9.
\eeqa
\noi
Solving the system of linear equations \eqref{2sl2-2}, \eqref{2sl2-3}, \eqref{2sl2-4} and \eqref{2sl2-5} leads to the solution
\beqa
\label{2sl2-sol}
\beta_1=\beta_2 = \beta,\ \beta_4=0,\ \beta_6=\beta_7=2\beta,\ \beta_9=\beta_{11}=0.
\eeqa
\noi

On the sets of type
$Y^1_\alpha, Y^2_\beta, Y^2_\gamma$ (where $\alpha,\beta,\gamma\in\{\pm,0\}$), following exactly the steps above,  the identity {\bf I.3} leads to the same type of solution \eqref{2sl2-sol} for the structure constants.

We now impose the Jacobi identity {\bf I.4}. Doing this for the set $Y_0^1, Y_{+1}^1,Y_{-1}^1,Y_0^2$, one gets
$$\alpha_1=0.$$
Now, doing this for the set $Y_{+1}^1, Y_{+1}^1,Y_{-1}^1,Y_0^2$ one gets
$$ \beta =0. $$
Similarly, one has that the $3-$bracket is trivial.

\medskip


Let us now treat the general case where $\g_1$ is the sum of an adjoint representation ${\cal D}_1$ and a representation ${\cal D}_2$ of dimension $2k+1$. We have seen that for $k=1$ this leads to a trivial $3-$bracket; we now prove by induction that this statement
remains true for any $k\in \NN^*$. Indeed, the only non-zero brackets will be the brackets 
\beqa
\begin{array}{llllll}
\label{1sl2-0}
\{ Y_{+1}^{(1)}, Y_{-1}^{(1)}, Y_0^{(1)}\}&=& \alpha X_0,\ &\{ Y_0^{(1)}, Y_0^{(1)}, Y_0^{(1)}\}&=& 6\alpha_i X_0\nonumber \\
\{Y_{+1}^{(1)}, Y_{-1}^{(1)}, Y_{+1}^{(1)}\}&=& 2\alpha X_+,\ &\{Y_{+1}^{(1)}, Y_0^{(1)}, Y_0^ {(1)}\}&=& 2\alpha X_+,\nonumber\\
\{Y_{-1}^{(1)}, Y_{+1}^{(1)}, Y_{-1}^{(1)}\}&=& 2\alpha X_-,\ &\{Y_{-1}^{(1)}, Y_0^{(1)}, Y_0^{(1)} \}&=& 2\alpha X_-.
\end{array}
\eeqa
\noi
with $\alpha\in \{0,1\}$ and the brackets
involving at least once $Y_{\pm k}$. If $k=2$ or $3$ one can write $\{ Y_{-1}^{(1)}, Y_{-1}^{(1)}, Y_{+k}^{(2)}\}$ (and symmetrically $\{ Y_{+1}^{(1)}, Y_{+1}^{(1)}, Y_{-k}^{(2)}\}$); in any case one writes
\beqa
\label{bra-noi}
\{ Y_{+1}^{(1)}, Y_{-k}^{(2)}, Y_{+k}^{(2)}\}&=&  \gamma_2 X_+\nonumber\\
\{ Y_{0}^{(1)}, Y_{-k}^{(2)}, Y_{+k}^{(2)}\}&=&  \gamma_3 X_0\nonumber\\
\{ Y_{0}^{(1)}, Y_{-k+1}^{(2)}, Y_{+k}^{(2)}\}&=&  \gamma_5 X_+\nonumber\\
\{ Y_{-1}^{(1)}, Y_{-k+1}^{(2)}, Y_{+k}^{(2)}\}&=&  \gamma_6 X_0\nonumber\\
\{ Y_{-1}^{(1)}, Y_{-k+2}^{(2)}, Y_{+k}^{(2)}\}&=&  \gamma_7 X_+\nonumber
\eeqa
\noi
(with $\gamma_2,\dots,\gamma_7\in \CC$) and similar brackets (involving $Y_{-k}^{(2)}$) whose coefficients are equal to the coefficients of \eqref{bra-noi}. As before, we now impose the Jacobi identities. Thus, imposing {\bf I.3} on the set $(X_-,Y_{-1}^{(1)}, Y_{-1}^{(1)}, Y_{+k}^{(2)})$ one gets $\{ Y_{-1}^{(1)}, Y_{-1}^{(1)}, Y_{+k}^{(2)}\}=0$. Furthermore, imposing {\bf I.4} on the sets $(Y_{-k}^{(2)},  Y_{+1}^{(1)}, Y_{-k}^{(2)}, Y_{+k}^{(2)})$, $(Y_{-k}^{(2)}, Y_{0}^{(1)}, Y_{-k}^{(2)}, Y_{+k}^{(2)})$, $(Y_{-k+1}^{(2)}, Y_{0}^{(1)}, Y_{-k+1}^{(2)}, Y_{+k}^{(2)})$, $(Y_{+k}^{(2)}, Y_{-1}^{(1)}, Y_{-k+1}^{(2)}, Y_{+k}^{(2)})$ and resp.  $(Y_{-k+2}, Y_{-1}^{(1)}, Y_{-k+2}^{(2)}, Y_{+k}^{(2)}$ one gets $\gamma_2=0$, $\gamma_3=0$, $\gamma_5=0$,  $\gamma_6=0$ and resp.  $\gamma_7=0$.
 Finally, imposing {\bf I.4} on the set $(Y^{(2)}_{+1}, Y_{+1}^{(1)}, Y_{-1}^{(1)}, Y_0^{(1)})$ gives $\alpha=0$ which completes this proof.

\medskip 

The remaining possibilities of direct sum of reducible representations are treated analogously leading to the same result. QED




\section{Representations  of elementary Lie algebras of order $3$}
\label{3-rep}

In this section we recall the definition of a representation of a Lie algebra of order $3$ (see \cite{Michel-finit} for more details). 

\begin{definition}
\label{def-3rep}
A representation of a Lie algebra of order $3$ $\g$ is a linear map $\rho:\g\to {\rm End} (H)$ (where $H$ is the representation space) and an automorphism $\hat \e$ such that $\hat \e^3=1$ satisfying
\begin{enumerate}
\item $\rho ([X,Y])= \rho (X) \rho (Y)-\rho (Y) \rho (X)$,
\item $\rho (\{Y_1, Y_2, Y_3 \})= \sum_{\sigma\in S_3} \rho (Y_{\sigma (1)}) \rho (Y_{\sigma (2)})\rho (Y_{\sigma (3)})$,
\item $\hat \e \rho (s) \hat \e^{-1}= \rho (\e(s))$
\end{enumerate}
\end{definition}

In \cite{Michel-finit} it was also exhibited that one has
$$ H =\bigoplus_{k=0}^2 H_k $$
where $H_k=\{ a\in H : \hat \e (a) = q^k a \}$.  Furthermore one has 
$$ \rho (Y) H_k \subseteq H_{k+1 (\rm{mod} 3)}.$$

\section{Deformations of elementary Lie algebras of order $3$}
\label{defo-elementary}

In this section we study the deformations of Lie algebras of order $3$, expressed in their elementary form. We give a possible general framework for this purpose. Furthermore, in the next section we proceed with contractions for elementary Lie algebras of order $3$.

Let $\g=\g_0\oplus\g_1$ be an elementary Lie algebra of order $3$ and let $A=(\g_0\otimes \g_0) \oplus (\g_0 \otimes \g_1) \oplus {\cal S}^3 (\g_1)$.
The multiplication of Lie algebra of order $3$  is given by the linear map
$$\varphi : A \to \g$$
satisfying the conditions {\bf I.1-I.4}.

Let $\varphi_1, \varphi_2, \varphi_3$ be the restrictions of $\varphi$ to each of the terms of $A$, that is
$$\varphi_1: \g_0 \otimes \g_0 \to \g_0, $$
$$\varphi_2: \g_0  \otimes \g_1 \to \g_1, $$
$$\varphi_3:  {\cal S}^3 (\g_1)\to \g_0, $$
The identities {\bf I.1-I.4} write
\beqa
\label{defo-Jacobi}
&&\varphi_1 (\p_1(X_1, X_2), X_3)+\varphi_1 (\p_1(X_3, X_1), X_2)+\varphi_1 (\p_1(X_2, X_3), X_1)=0\nonumber \\
&&\varphi_2 (\varphi_1 (X_1, X_2), Y)+\p_2 (\p_2 (X_2, Y), X_1) + \p_2 (\p_2 (Y, X_1), X_2) = 0  \\
&& \p_1 (X, \p_3 (Y_1, Y_2, Y_3)) - \p_3 (\p_2 ( X, Y_1), Y_2, Y_3) - \p_3 (Y_1, \p_2 (X, Y_2), Y_3) - \p_3 (Y_1,Y_2,\p_2(X,Y_3)) = 0\nonumber \\
&& \p_2 (Y_1, \p_3 (Y_2, Y_3, Y_4) + \p_2 (Y_2, \p_3 (Y_1, Y_3, Y_4)+\p_2 (Y_3, \p_3 (Y_1, Y_2, Y_4)+\p_2 (Y_4, \p_3 (Y_1, Y_2, Y_3)=0.\nonumber
\eeqa

Let now $\p$ and $\p'$ be two multiplications $\p$ and $\p'$. We can thus define the following maps 
\beqa
\label{def-criptic}
&&\p \circ_1 \p' : (\g_0\otimes \g_0\otimes\g_0)\to \g_0 \nonumber\\
&& \ \  \ X_1 \otimes X_2 \otimes X_3 \mapsto \varphi_1 (\p'_1(X_1, X_2), X_3)+\varphi_1 (\p'_1(X_3, X_1), X_2)+\varphi_1 (\p'_1(X_2, X_3), X_1), \nonumber \\
&&\p \circ_2 \p' : (\g_0\otimes \g_0\otimes\g_1)\to  \g_1 \nonumber\\
&&\ \ \ X_1\otimes X_2 \otimes Y \mapsto \varphi_2 (\varphi'_1 (X_1, X_2), Y)+\p_2 (\p'_2 (X_2, Y), X_1) + \p_2 (\p'_2 (Y, X_1), X_2),\nonumber\\
&&\p \circ_3 \p': (\g_0\otimes {\cal S}^3 (\g_1))\to \g_0 \nonumber\\
&& \ \ \ X\otimes (Y_1,Y_2,Y_3)\mapsto \p_1 (X, \p'_3 (Y_1, Y_2, Y_3)) - \p_3 (\p'_2 ( X, Y_1), Y_2, Y_3) - \p_3 (Y_1, \p'_2 (X, Y_2), Y_3) \nonumber \\
&& \hskip 11truecm - \p_3 (Y_1,Y_2,\p'_2(X,Y_3)),\nonumber \\
&&\p \circ_4 \p : (\g_1\otimes {\cal S}^3 (\g_1))\to \g_1 \nonumber \\
&&\ \ \ Y_1\otimes (Y_2,Y_3,Y_4)\mapsto \p_2 (Y_1, \p'_3 (Y_2, Y_3, Y_4) + \p_2 (Y_2, \p'_3 (Y_1, Y_3, Y_4)+\p_2 (Y_3, \p'_3 (Y_1, Y_2, Y_4)\nonumber \\
&& \hskip 9truecm
+\p_2 (Y_4, \p'_3 (Y_1, Y_2, Y_3).
\eeqa


\begin{corollary}
The map $\p$ endows $\g$ with a structure of  elementary Lie algebra of order $3$ iff
\beqa
\label{cond-defo}
\p \circ_i \p =0 \mbox{ with }i=1,\dots,4.
\eeqa
\end{corollary}

For the following definition we denote by $ {\CC}[[t]]$ the ring of formal series.

\begin{definition}
A deformation of an elementary Lie algebra of order $3$ is a multiplication
\beqa
\label{def-defo}
\p_t : A \otimes {\CC}[[t]]\to (\g_0 \oplus \g_1)\otimes  {\CC}[[t]]
\eeqa
\noi
which verifies identities \eqref{cond-defo}.
\end{definition}

Thus, a deformation of an elementary Lie algebra of order $3$ writes as
\beqa
\label{defo-gen}
\p_t=\p + t \psi^{(1)} + t^2 \psi^{(2)}+\dots +t^n \psi^{(n)}, \mbox{ with } n\in\NN^*,
\eeqa
\noi 
which verifies identities \eqref{cond-defo}.

\begin{proposition}
\label{defo-general}
Considering a deformation \eqref{defo-gen}, the maps $\psi^{(p)}$ (with $p\in\NN$) satisfy the equations
\beqa
\label{sol-gen}
\sum_{p+q=r} \psi^{(p)} \circ_i \psi^{(r)}=0, \mbox{ for any }i=1,\dots,4,\ r\in\NN
\eeqa
\noi
where we put $\psi^{(0)}=\p$.
\end{proposition}
{\it Proof:} For $i=1$, equation \eqref{sol-gen} is just the condition \eqref{rez-g} of the deformations of Gerstenhaber for Lie algebras.
We explicitly prove \eqref{sol-gen} for $i=2$ the two remaining cases being similar.

If one checks only the terms in $t^2$, then only the terms $\p + t \psi^{(1)} + t^2 \psi^{(2)}$ will matter. Inserting
\beqa
\p_{t\, 1}=\p_1 + t \psi^{(1)}_1 + t^2 \psi^{(2)}_1\nonumber\\
\p_{t\, 2}=\p_2 + t \psi^{(1)}_2 + t^2 \psi^{(2)}_2\nonumber\\
\p_{t\, 3}=\p_1 + t \psi^{(1)}_3 + t^2 \psi^{(2)}_3
\eeqa
\noi
in \eqref{defo-Jacobi}, one gets, in $t$
\beqa
\label{J2-t}
&&\psi_2^{(1)} (\p_1 (X_1,X_2), Y) + \p_2 (\psi_1^{(1)} (X_1, X_2), Y) +\p_2 (\psi_2^{(1)} (X_2,Y),X_1)\nonumber \\
&&\ \ \ \  +\psi_2^{(1)} (\p_2 (X_2,Y),X_1)+ \p_2 (\psi_2^{(1)} (Y, X_1), X_2) + \psi_2^{(1)} (\p_2 (Y, X_1), X_2)=0
\eeqa
\noi
and respectively, in $t^2$
\beqa
\label{J2-t^2}
&&\p_2 (\psi_1^{(2)} (X_1,X_2), Y) + \psi_2^{(1)} (\psi_1^{(1)} (X_1, X_2), Y) +\psi_2^{(2)} (\p_1 (X_1,X_2),Y)\nonumber \\
&&\ \ \ \  +\p_2 (\psi_2^{(2)} (X_2,Y),X_1)+ \psi_2^{(1)} (\psi_2^{(1)} (X_2, Y), X_1) + \psi_2^{(2)} (\p_2 (X_2, Y), X_1) \nonumber\\
&&\ \ \ \  +\p_2 (\psi_2^{(2)} (Y, X_1),X_2)+ \psi_2^{(1)} (\psi_2^{(1)}(Y, X_1),X_2) + \psi_2^{(2)} (Y, X_1),X_2)=0.
\eeqa
\noi
One sees that equation \eqref{J2-t} is \eqref{sol-gen} for $r=1$ and \eqref{J2-t^2} is \eqref{sol-gen} for $r=2$ (and obviously $i=2$ for both cases). Similarly, one proves \eqref{sol-gen} for any $r\in \NN^*$. QED

\bigskip

We now focus on a more particular case of the above definition, namely the infinitesimal deformations.

\subsection{Infinitesimal deformations}
\label{defo-infi}

\begin{definition}
An infinitesimal deformation is given by a multiplication $\p_t$ of the form
$$\p_t=\p+t\psi^{(1)}$$
which verifies \eqref{cond-defo}.
\end{definition}

Let $\p_t=\sum_{i=1}^3 (\p_i+t\psi_i^{(1)})$. Identities \eqref{cond-defo} lead to
\beqa
\label{defo-t}
&&\psi_1^{(1)} (\p_1 (X_1, X_2), X_3) + \p_1 (\psi_1^{(1)} (X_1, X_2), X_3) +\psi_1^{(1)} (\p_1 (X_3, X_1), X_2) \nonumber \\
&&\ \ \ \ + \p_1 (\psi_1^{(1)} (X_3, X_1), X_2)+\psi_1^{(1)} (\p_1 (X_2, X_3), X_1) + \p_1 (\psi_1^{(1)} (X_2, X_3), X_1)=0,\nonumber \\
\bigskip
&&\psi_2^{(1)} (\p_1 (X_1,X_2), Y) + \p_2 (\psi_1^{(1)} (X_1, X_2), Y) +\p_2 (\psi_2^{(1)} (X_2,Y),X_1)\nonumber \\
&&\ \ \ \  +\psi_2^{(1)} (\p_2 (X_2,Y),X_1)+ \p_2 (\psi_2^{(1)} (Y, X_1), X_2) + \psi_2^{(1)} (\p_2 (Y, X_1), X_2)=0, \nonumber \\
&&\psi_1 ^{(1)}(X, \p_3 (Y_1, Y_2, Y_3)) + \p_1 (X, \psi_3^{(1)} (Y_1, Y_2, Y_3))\nonumber\\ \bigskip
&&\ \ \ \  -\p_3 (\psi_2 ^{(1)}(X, Y_1), Y_2, Y_3) - \psi_3^{(1)} (\p_2 (X, Y_1), Y_2, Y_3)
-\p_3 (Y_1, \psi_2^{(1)} (X,Y_2),Y_3) \nonumber \\
&&\ \ \ \ -\psi_3^{(1)} (Y_1, \p_2 (X,Y_2),Y_3)
-\p_3 (Y_1, Y_2, \psi_2 ^{(1)}(X, Y_3)) -\psi_3^{(1)} (Y_1, Y_2, \psi_2^{(1)} (X, Y_3)) = 0,\nonumber\\
&&\p_2 (Y_1, \psi_3 ^{(1)}(Y_2, Y_3,Y_4)) + \psi_2 ^{(1)}(Y_1, \p_3 (Y_2, Y_3,Y_4))\nonumber\\ \bigskip
&&\ \ \ \ +\p_2 (Y_2, \psi_3^{(1)} (Y_1, Y_3,Y_4)) + \psi_2^{(1)} (Y_2, \p_3 (Y_1, Y_3,Y_4))\nonumber \\
&&\ \ \ \ +\p_2 (Y_3, \psi_3^{(1)}(Y_1, Y_2,Y_4)) + \psi_2^{(1)} (Y_3, \p_3 (Y_1, Y_2,Y_4))\nonumber \\
&&\ \ \ \ +\p_2 (Y_4, \psi_3^{(1)} (Y_1, Y_2,Y_3)) + \psi_2 ^{(1)}(Y_4, \p_3 (Y_1, Y_2,Y_3))=0.
\eeqa
\noi
Using \eqref{def-criptic} these equations write
\beqa
\label{defo-t-criptic}
\p\circ_i \psi +\psi \circ_i \p =0, \mbox{ with }i=1,\dots,4. 
\eeqa
\noi
which is just equation \eqref{sol-gen} for $r=1$.

Furthermore, the terms in $t^2$ obtained from \eqref{cond-defo} give
\beqa
&&\psi_1^{(1)} (\psi_1^{(1)} (X_1, X_2),X_3)+\psi_1^{(1)} (\psi_1^{(1)} (X_3, X_1),X_2)+\psi_1^{(1)} (\psi_1^{(1)} (X_2, X_3),X_1)=0\nonumber\\
&&\psi_2^{(1)} (\psi_1^{(1)} (X_1, X_2), Y)+\psi_2^{(1)} (\psi_2^{(1)} (X_2, Y), X_1)+\psi_2^{(1)} (\psi_2^{(1)} (Y,X_1), X_2)=0,\nonumber\\
&&\psi_1^{(1)} (X, \psi_3^{(1)} (Y_1, Y_2, Y_3))-\psi_3^{(1)}(\psi_2^{(1)} (X, Y_1),Y_2,Y_3) - \psi_3^{(1)} (Y_1,\psi_2^{(1)} (X, Y_2), Y_3)\nonumber\\
&& \hskip 7truecm -\psi_3^{(1)} (Y_1, Y_2, \psi_2^{(1)} (X,Y_3))=0\nonumber\\
&&\psi_2^{(1)} (Y_1, \psi_3^{(1)} (Y_2, Y_3,Y_4)) + \psi_2^{(1)} (Y_2, \psi_3^{(1)} (Y_1, Y_2,Y_4))+\psi_2^{(1)} (Y_3, \psi_3^{(1)} (Y_1, Y_2,Y_4))+\nonumber\\
&& \ \ \ \ \hskip 7truecm \psi_2^{(1)} (Y_4, \psi_3^{(1)} (Y_1, Y_2,Y_3))=0.
\eeqa
\noi
which write
\beqa
\psi^{(1)}\circ_i \psi^{(1)} = 0, \mbox{ with }i=1,\dots,4.
\eeqa

\begin{definition}
Denote by
$$Z(A)=\{ \psi_1+\psi_2 +\psi_3 : A\to \g \},$$
where $\psi_i$ ($i=1,2,3$) satisfy \eqref{defo-t}. The vector space $Z(A)$ is called the infinitesimal deformation space of $A$.
\end{definition}

\subsection{Isomorphic deformations}

Let us now consider a multiplication law $\p$ and perform a formal change of basis on the Lie algebra of order $3$. We then want to express the deformed law $\p_t$ as a function of the law $\p$.

\begin{proposition}
Let $\g=\g_0 \oplus \g_1$ be an elementary Lie algebra of order $F$. If one considers a formal change of basis given by  $f=(f_1,f_2)$ such that $g=(g_0, g_1)$, $g_0= 1+tf_0\in GL(\g_0)$, $g_1= 1+tf_1\in GL(\g_1)$, then the isomorphic multiplication $\p_t$ writes as the deformation
$$ \p_t = \p + t \psi + {\cal O} (t^2), $$ 
where $\psi = \psi_1+\psi_2 + \psi_3$ is given by
\beqa
\label{deltele}
&&\psi_1(X_1, X_2)=\p_1 (f_0(X_1), X_2)+\p_1 (X_1,f_0 (X_2))-f_0 (\p_1 (X_1, X_2)) \nonumber \\
&&\psi_2(X, Y)=\p_2 (f_0(X), Y)+\p_2 (X,f_1 (Y))-f_1 (\p_2 (X,Y)) \nonumber \\
&&\psi_3(Y_1, Y_2, Y_3)=\p_3 (f_1(Y_1), Y_2, Y_3)+\p_3 (Y_1,f_1 (Y_2), Y_3)+\p_3 (Y_1, Y_2, f_0 (Y_3))-f_0 (\p_3 (Y_1, Y_2, Y_3)).\nonumber \\
\eeqa
\end{proposition}
\begin{remark}
One denotes  $\psi$ by $\delta_\p f$, that is 
\beqa
\psi_1 (X_1, X_2)&=&(\delta_{\p_1} f)(X_1, X_2),\nonumber\\
\psi_2 (X,Y)&=&(\delta_{\p_2} f)(X, Y),\nonumber\\
\psi_3 (Y_1, Y_2, Y_3)&=&(\delta_{\p_3} f)(Y_1, Y_2, Y_3).
\eeqa
\end{remark}
{\it Proof:} 
Let $f\in{\rm End}_\g (\g_0\oplus \g_1)$ ($\g_0$ of basis $\{ X_i:\ i=1,\dots,m\} $ and $\g_1$ of basis $\{ Y_j : j=1,\dots,n\}$) be a graded homomorphism $f=(f_0,f_1)$. Consider a formal change of basis defined by the isomorphism $g=({\rm Id}+tf_0, {\rm Id}+tf_1)\in GL (\g)$ which gives
\beqa
\label{noua-baza}
&&\tilde X_i = g_0 (X_i)= ({\rm Id}+tf_0) (X_i)= X_i + t f_0 (X_i),\nonumber\\
&&\tilde Y_j = g_1 (Y_j)= ({\rm Id}+tf_1) (Y_j)= Y_j + t f_1 (Y_j).
\eeqa

One has
\beqa
\p_1 (\tilde X_1, \tilde X_2)&=&g_0^{-1} \p_1 (g_0 (X_1), g_0 (X_2)) \nonumber \\
\p_2 (\tilde X_1, \tilde Y_2)&=&g_1^{-1} \p_2 (g_0 (X_1), g_1 (Y_2)) \nonumber \\
\p_3 (\tilde Y_1, \tilde Y_2, \tilde Y_3) &=& g_0^{-1} \p_3 ( g_1 (Y_1), g_1 (Y_2), g_1 (Y_3)).
\eeqa
\noi
This can be written as
\beqa
\p_1 (\tilde X_1, \tilde X_2)&=&\p_1 ( X_1, X_2)+t (\delta_{\p_1} f)(X_1, X_2) + {\cal O}(t^2) \nonumber \\
\p_2 (\tilde X, \tilde Y)&=&\p_2 ( X, Y)+t (\delta_{\p_2} f)(X,Y) + {\cal O}(t^2) \nonumber \\
\p_3 (\tilde Y_1, \tilde Y_2, \tilde Y_3)&=&\p_3 ( Y_1, Y_2, Y_3)+t (\delta_{\p_3} f)(Y_1, Y_2, Y_3) + {\cal O}(t^2),
\eeqa
\noi
where, by a tedious but straightforward calculation, one has \eqref{deltele}. QED

\begin{definition}
Let
$$ B(A)=\{ \psi \in Z(A): \ \psi = \delta_\p f \}. $$
\end{definition}

\begin{remark}
$B(A)\subset Z (A)$.
\end{remark}

\bigskip

We can now conclude this subsection with

\begin{theorem}
The non-trivial infinitesimal deformations of $A$ are parametrised by the quotient space $Z(A)/B(A)$.
\end{theorem}




\subsection{Rigid elementary Lie algebra of order $3$}

\begin{definition}
An elementary Lie algebra of order $3$ $\g=\g_0\oplus\g_1$ is called rigid if all deformations of $\g$ are  isomorphic to $\g$.
\end{definition}


As an example of rigid Lie algebra of order $3$ one has $\s \oplus {\rm ad}\, \s$ with $\alpha=1$ (see Theorem \ref{sl2-irrep} for notations). Other examples are given in subsections \ref{ex-contractie} and \ref{another}. 
An example of non-rigid Lie algebra of order $3$ is also exhibited in subsection \ref{another}.

\subsection{The variety of elementary Lie algebras of order $3$}

Let $\g=\g_0\oplus \g_1$ be an elementary Lie algebra of order $3$ and $X_1,\dots,X_m,Y_1,\dots,Y_n$ be a given basis of $\g$. The maps $\p_i$ ($i=1,2,3$)  are given by their structure constants
\beqa
\label{3-constante}
\p_1 (X_i, X_j)= C_{ij}^k X_k,\ \
 \p_2 (X_i, Y_j)=D_{ij}^k Y_k\mbox{ and } \p_3 (Y_i, Y_j, Y_k)=E_{ijk}^l X_l.
\eeqa
\noi
Since the basis is fixed, one can identify $\p=\p_1+\p_2+\p_3$ to its structure constants $(C_{ij}^k, E_{ij}^k, D_{ijk}^l)$, who verify also the following (anti)symmetry conditions
\beqa
&&C_{ij}^k=-C_{ji}^k,\ \nonumber \\ 
&&E_{ijk}^l=E_{ikj}^l=E_{kij}^l=E_{kji}^l=E_{jik}^l=E_{jki}^l.
\eeqa
\noi
Let $N=mC_m^2+mn^2+m(C_{n+1}^2+ C_n^2 + \dots + C_2^2)$ be the number of these structure constants. Hence $\p$ corresponds to a product of the vector space ${\CC}^N$.

The identities {\bf I.1-I.4} write
\beqa
\label{jacobi-constante}
&&C_{ij}^l C_{lk}^s + C_{jk}^lC_{li}^s+C_{ki}^l C_{lj}^s=0 \nonumber\\
&&-D_{jk}^l D_{il}^s + D_{ik}^l D_{jl}^s + C_{ij}^l D_{lk}^s=0 \nonumber\\
&&E_{jkl}^m C_{im}^s - D_{ij}^m E_{mkl}^s - D_{ik}^m E_{jml}^s - D_{il}^m E_{jkm}^s=0 \nonumber \\
&&E_{jkl}^m D_{im}^s+ E_{ikl}^m D_{jm}^s + E_{ijl}^m D_{km}^s + E_{ijk}^m D_{lm}^s=0.
\eeqa
\noi
These identities define an algebraic variety structure $F_{mn}$  in ${\CC}^N$. 


Let us now consider the action of the group $GL({m,n})\cong GL(m)\times GL(n)$ on  $F_{mn}$. For any $(g_0,g_1)\in GL_{m,n}$, this action is defined by
$$ (g_0, g_1) \cdot (\p_1 +\p_2 +\p_3) \to (\p'_1 + \p'_2 + \p'_3) $$
where 
\beqa
\p'_1 (X_1, X_2)&=&g_0^{-1} \p_1 (g_0 (X_1), g_0 (X_2)) \nonumber \\
\p'_2 (X_1, Y_2)&=&g_1^{-1} \p_2 (g_0 (X_1), g_1 (Y_2)) \nonumber \\
\p'_3 (Y_1, Y_2, Y_3) &=& g_0^{-1} \p_3 ( g_1 (Y_1), g_1 (Y_2), g_1 (Y_3)).
\eeqa
\noi
Denote by ${\cal O}_\p$ the orbit of $\p=\p_1+\p_2+\p_3$ with respect to this action. 

\section{Contractions of elementary Lie algebras of order $3$}
\label{3-contractions}


Using the notions introduced above, we can now give the general definition

\begin{definition}
A contraction of $\p$ is a point $\p'\in F_{mn}$ such that $\p'\in \bar {{\cal O}_\p}$, with adherence in the Zariski sense.
\end{definition}

Let us recall here that the Zariski topology is a topology well-suited for the study of polynomial equations, since a Zariski topology has more open sets that the usual metric topology; the only closed sets are defined by polynomial equations.

As was the case for contractions of Lie algebra (see subsection $2.5.1$) this general notion is not useful from the point of view of physical applications. Thus, we particularise it, in the sense of the definition of subsection \ref{contractions}.

\bigskip

Let $\p=(\p_1,\p_2,\p_3)$ be a given multiplication of elementary Lie algebras of order $3$ and let $(g_p)_{n\in\NN}$ (with $g_p=(g_{0,p},g_{1,p})\in GL({m,n})$) be a series of isomorphisms. Define
$$\p_{1,p}=\p_{1,p}+\p_{2,p}+\p_{3,p}$$ 
by
\beqa
\label{fiurile}
\p_{1,p} (X_1, X_2)&=&g_{0,p}^{-1} \p_1 (g_{0,p} (X_1), g_{0,p} (X_2)) \nonumber \\
\p_{2,p} (X_1, Y_2)&=&g_{1,p}^{-1} \p_2 (g_{0,p} (X_1), g_{1,p} (Y_2)) \nonumber \\
\p_{3,p} (Y_1, Y_2, Y_3) &=& g_{0,p}^{-1} \p_3 ( g_{1,p} (Y_1), g_{1,p} (Y_2), g_{1,p} (Y_3)).
\eeqa
\noi
If 
the limit 
\beqa
\label{limita}
{\rm lim}_{p\to\infty} \p_p
\eeqa
exists, then this limit is a {\it contraction}.

As we have mentioned above we are interested in this particular case of contractions. For the sake of completeness let us recall again that in the literature there exists contractions which are not of this type \cite{Burde}.


\bigskip

We now particularise furthermore the definition above by giving a specific form to the automorphism $g$. We do this here inspired from the Weimar-Woods construction (see subsection \ref{ww}). Thus we take 
$$g_\e={\rm diag} (\e^{a_1},\dots,\e^{a_m},\e^{b_1},\dots, \e^{b_n})$$
with $a_i,b_j\in\ZZ$ ($i=1,\dots,m,\ j=1,\dots,n$). Hence $g_{0,\e} (X_i)=\e^{a_i} X_i$, $g_{0,\e}^{-1} (Y_j)=\e^{a_j} Y_j$, $g_{0,\e}^{-1} (X_i)=\e^{-a_i} X_i$ and $g_{0,\e}^{-1} (Y_j)=\e^{-a_j} Y_j$. Thus, equations \eqref{fiurile} become
\beqa
\p_{1,\e} (X_i, X_j)&=&\e^{a_i+a_j-a_k}C_{ij}^k X_k\nonumber\\
\p_{2,\e} (X_i, Y_j)&=&\e^{a_i+b_j-b_k}D_{ij}^k Y_k\nonumber\\
\p_{3,\e} (Y_i, Y_j,Y_k)&=&\e^{b_i+b_j+b_k-a_l}E_{ijk}^l X_l
\eeqa
\noi
As already stated, one can define a contraction if the limit $\e\to 0$  exists, that is if $a_i+a_j-a_k\ge 0$, $a_i+b_j-b_k\ge 0$ and $b_i+b_j+b_k-a_l\ge 0$ for any $a$ and $b$.

\subsection{Fundamental example: the variety $F_{11}$}
\label{ex-contractie}

In this subsection we 
obtain all contractions of elementary Lie algebras of order $3$ of dimension $2$. 

As already mentioned in section \ref{contractions-deformations}, the most frequently used in physics calculations are the In\"on\"u-Wigner contractions (see subsection \ref{iw}). Following this line of reasoning, for the situation of Lie algebras of order $3$ this implies to have automorphisms $g_\e=(g_{0,\e}, g_{1,\e})$ of the form $\g_{0,\e}=g_{0}^1 + \e g_0^2$ and $\g_{1,\e}=g_{1}^1 + \e g_1^2$ with $g_0^1, g_1^1$ singular, $g_0^2, g_1^2$ regular
 and $\e\approx 0$.

\begin{proposition}
\label{dim3-ele}
Any  Lie algebra of order $3$ $\g=\g_0\oplus\g_1$ of dimension $2$, with $\g_0$ spanned by $X$ and  $\g_1$ spanned by $Y$
is isomorphic to one of the following Lie algebras of order $3$ (we give here only the non-zero brackets):\\
i) $\g^3_1$, $\{Y,Y,Y\}=X$\\
ii) $\g^3_2$, $[X,Y]=Y$,\\
iii) $\g^3_3$, the trivial Lie algebra of order $3$.
\end{proposition}
\textit{Proof:} Considering the most general possibility for the structure constants of a $3-$dimensional binary Lie algebra of order $3$:
\beqa
\label{genu-ele}
\pard X,Y \pari= - \pard Y,X \pari = \alpha_1 Y, 
 \{Y,Y,Y\}=\alpha_2 X
\eeqa
(again we give only the non-zero brackets).
Imposing the Jacobi identities {\bf I.1-I.4} one has the following Lie algebras of order $3$:
\beqa
\label{sol3-ele}
&&\g^3_1,  \alpha_1=0, \alpha_2=1, \nonumber \\
&&\g^3_2,  \alpha_1=1, \alpha_2=0; \nonumber \\
&&\g^3_3,  \alpha_1=0, \alpha_2=0. \nonumber
\eeqa
QED

\begin{corollary}
\label{corolar-ele}
The variety of $2-$dimensional elementary  Lie algebras of order $3$, $F_{1,1}$ is the reunion of two components $U_1$ and $U_2$ with
$$ U_1=\bar{\g}^3_1 \mbox{ and }  U_2=\bar{\g}^3_2.$$
\end{corollary}
\textit{Proof:} 
One has the following contraction scheme


$$\xymatrix{
 \g^3_2 \ar@/_/[dr] & & \g^3_1 \ar@/^/[dl]\\
& \g^3_3 & }
$$
where by $A\to B$ we denote a contraction of the algebra $A$ to the algebra $B$ ($B$ is a contraction of $A$). QED

\subsection{Another example of contraction}
\label{another}

As already mentioned in subsection \ref{3-ex}, we now give an example of contraction of the elementary Lie algebra  of order $3$ ${\mathfrak {so}}(2,3)\oplus{\rm ad}\, {\mathfrak {so}}(2,3)$ of Example \ref{so23} to the \3 algebra of Example \ref{algebra0} (see \cite{Michel-finit}); the latter will be the starting point  for the construction of the theoretical field model of the next chapters. This mechanism can be seen as the analogous of the contraction to the SUSY algebra of certain Lie superalgebras. 

Let $M_{mn}, M_{m4}$ (with $m,n=0,\dots,3$ and $m<n$) a basis of ${\mathfrak {so}}(2,3)$ and  $J_{mn}, J_{m4}$ (with $m,n=0,\dots,3$ and $m<n$)  a basis of ${\rm ad}\, {\mathfrak {so}}(2,3)$. To simplify the next formulae, we put for  $m\ge n$,  $M_{mn}=-M_{nm}$, $J_{mn}=-M_{nm}$, $M_{4m}=-M_{m4}$ and $J_{4m}=-J_{m4}$. The multiplication law $\p$ writes
\beqa
\label{p-so23}
\p_1(M_{mn}, M_{pq})&=&-\eta_{nq} M_{mp}-\eta_{mp} M_{nq}+\eta_{mq} M_{np}+\eta_{np} M_{mq}\nonumber\\
\p_1(M_{mn}, M_{p4})&=&-\eta_{mp} M_{n4} +\eta_{np} M_{m4}\nonumber\\
\p_1 (M_{m4}, M_{p4})&=&-M_{mp},\nonumber\\
\bigskip
\p_2(M_{mn}, J_{pq})&=&-\eta_{nq} J_{mp}-\eta_{mp} J_{nq}+\eta_{mq} J_{np}+\eta_{np} J_{mq}\nonumber\\
\p_2(M_{mn}, J_{p4})&=&-\eta_{mp} J_{n4} +\eta_{np} J_{m4}\nonumber\\
\p_2(M_{m4}, J_{pq})&=&-\eta_{mp} J_{4q} +\eta_{mq} J_{4p}\nonumber\\
\p_2 (M_{m4}, J_{p4})&=&-J_{mp},\nonumber\\
\bigskip
\p_3 (J_{mn},J_{pq}, J_{rs})&=&g_{(mn)(pq)} M_{rs}+g_{(mn)(rs)} M_{pq}+g_{(pq)(rs)} M_{mn},\nonumber\\
\p_3 (J_{mn},J_{pq}, J_{r4})&=&g_{(mn)(pq)}M_{r4},\nonumber\\
\p_3 (J_{mn},J_{p4}, J_{r4})&=&g_{(p4)(r4)} M_{mn},\nonumber\\
\p_3 (J_{m4},J_{p4}, J_{r4})&=&g_{(m4)(p4)} M_{r4}+g_{(m4)(r4)} M_{p4}+g_{(p4)(r4)} M_{m4},\nonumber
\eeqa
\noi
where $\eta_{mn}$ extends  with $\eta_{m4}=\eta_{4m}=0$, $\eta_{44}=1$ (the anti-de Sitter metric) and $g_{(MN)(PQ)}=\eta_{MP}\eta_{NQ}-\eta_{MQ}\eta_{NP}$ (for any $M,N,P,Q=0,\dots , 4$) which gives  $g_{(m4)(p4)}=\eta_{mp}$, $g_{(mn)(44)}=0$, $g_{(mn)(r4)}=0$ {\it etc.} 

Define now
\beqa
\label{F-contractia}
L_{mn}=g_0 (M_{mn})= M_{mn}\nonumber \\
P_m= g_0 (M_{m4})=\e M_{m4} 
\eeqa
\noi
Note that taking the limit $\e \to 0$ one has the well-known In\"on\"u-Wigner contraction from the anti-de Sitter algebra ${\mathfrak {so}}(2,3)$ to the Poincar\'e algebra.
Define also 
\beqa
V_{mn}=g_1 (J_{mn})= \sqrt[3]{\e} J_{mn}\nonumber\\
V_m =g_1 (J_{m4})=  \sqrt[3]{\e}  J_{m4}
\eeqa
\noi
The limit $\e \to 0$ realises the contraction; the contracted Lie algebra of order $3$ has $\g_0$ isomorphic to the Poincar\'e algebra and $\g_1$ is generated by $V_{mn}$ and $V_m$ which lie in the adjoint and resp. vectorial representation of ${\mathfrak {so}}(1,3)$. The multiplication law becomes
\beqa
\label{lege-mica}
\p_1(L_{mn}, L_{pq})&=&-\eta_{nq} L_{mp}-\eta_{mp} L_{nq}+\eta_{mq} L_{np}+\eta_{np} L_{mq}\nonumber\\
\p_1(L_{mn}, P_{p})&=&-\eta_{mp} P_{n} +\eta_{np} P_{m}\nonumber\\
\p_1 (P_m, P_p)&=&0,\nonumber\\
\bigskip
\p_2(L_{mn}, V_{pq})&=&-\eta_{nq} V_{mp}-\eta_{mp} V_{nq}+\eta_{mq} V_{np}+\eta_{np} V_{mq}\nonumber\\
\p_2(L_{mn}, V_{p})&=&-\eta_{mp} V_{n} +\eta_{np} V_{m}\nonumber\\
\p_2(P_{m}, V_{pq})&=&0\nonumber\\
\p_2(P_{m}, V_{p})&=&0,\nonumber\\
\bigskip
\p_3 (V_{mn},V_{pq}, V_{rs})&=&0,\nonumber\\
\p_3 (V_{mn},V_{pq}, V_{r})&=&g_{(mn)(pq)} P_{r},\nonumber\\
\p_3 (V_{mn},V_{p}, V_{r})&=&0,\nonumber\\
\p_3 (V_{m},V_{p}, V_{r})&=&\eta_{mp} P_{r}+\eta_{mr} P_{p}+\eta_{pr} P_{m}.
\eeqa
\noi

\begin{remark}
In this contraction, the subspace generated by $M_{mn}, M_{m4}$ and $J_{p4}$ (which is not a Lie algebra of order $3$) contracts on the \3 algebra of Example \ref{algebra0}.
\end{remark}

\bigskip

This endows with an example of a deformation, which is defined in the following manner. Let $\p=\p_1+\p_2+\p_3$ the law defined by \eqref{lege-mica}. The deformation $\p_{t}=\p_{t1}+\p_{t\, 2}+\p_{t\, 3}$ is given by
\beqa
\p_{t\, 1}(L_{mn}, L_{pq})&=& \p_1 (L_{mn}, L_{pq})\nonumber\\
\p_{t\, 1}(L_{mn}, P_{p})&=& \p_1 (L_{mn}, P_{p})\nonumber\\
\p_{t\, 1}(P_{m}, P_{p})&=& -t^6 L_{mp}\nonumber\\
\p_{t\, 2}(L_{mn}, V_{pq})&=& \p_2 (L_{mn}, V_{pq})\nonumber\\
\p_{t\, 2}(L_{mn}, V_{p})&=& \p_2 (L_{mn}, V_{p})\nonumber\\
\p_{t\, 2}(P_{m}, V_{pq})&=& t^3(\eta_{mp} V_q - \eta_{mq} V_p) \nonumber\\
\p_{t\, 2}(P_{m}, V_{p})&=& t^3 V_{mp} \nonumber\\
\p_3 (V_{mn},V_{pq}, V_{rs})&=&t^3(g_{(mn)(pq)} L_{rs}+g_{(mn)(rs)} L_{pq}+g_{(pq)(rs)} L_{mn}),\nonumber\\
\p_3 (V_{mn},V_{pq}, V_{r})&=&\p_3 (V_{mn},V_{pq}, V_{r}),\nonumber\\
\p_3 (V_{mn},V_{p}, V_{r})&=&t^3g_{(p4)(r4)} L_{mn},\nonumber\\
\p_3 (V_{m},V_{p}, V_{r})&=&\p_3 (V_m, V_p, V_r).\nonumber
\eeqa

\section{Towards a classification of Lie algebras of order $3$ based on the Poincar\'e algebra}

In this section, guided by the extension \eqref{algebra0} of the Poincar\'e algebra, we investigate the $14-$dimensional Lie algebras of order $3$ based on the Poincar\'e algebra. 
Recall that the Poincar\'e algebra has as semi-simple part the Lorentz algebra ${\mathfrak {so}}(1,3)$ and to this one adds the momentums $P_m$, lying in some representation of ${\mathfrak {so}}(1,3)$ (see subsection $2.3.2$). Now, 
to extend it {\it via} a structure of Lie algebra of order $3$, we place ourselves in the same representation of ${\mathfrak {so}}(1,3)$ and we systematically investigate all the possibilities to construct a Lie algebras of order $3$. The result we obtain is that one finds only one such non-trivial structure, which will correspond to \eqref{algebra0}. 

Recall that in particle physics, using the superalgebras framework, one extends the Poincar\'e algebra by placing himself in the spinor representation and thus finding the SUSY algebra (the Haag-Lopuszanski-Sohnius no-go theorem, see subsection $5.1.2$). Thus our approach here may be seen as some kind of a similar tentative. Nevertheless, to obtain a complete result, one should study all the extensions, by analysing all the possibilities of $4-$dimensional representations of ${\mathfrak {so}}(1,3)$, as we will show in the sequel.

\medskip

We write here the Poincar\'e algebra as (note the difference of convention here, difference useful for our calculations)
\beqa
\label{Poincare}
&&\left[M_{mn}, M_{pq}\right]=
i(\eta_{nq} M_{pm}-\eta_{mq} M_{pn} + \eta_{np}M_{mq}-\eta_{mp} M_{nq}),
\ \left[M_{mn}, P_p \right]= i(\eta_{np} P_m -\eta_{mp} P_n), \nonumber \\
&&\left[P_m, P_n \right]= 0, \ \
\eeqa
\noindent
where $\eta_{mn}$ is the Minkowski metric.

Let $\mathcal{P}_\mathbb{C}=\mathcal{P}\otimes_\mathbb{R} \mathbb{C}$ be the complexified of $\mathcal{P}.$ 
It is well-know
that its semi-simple Levi part is  isomorphic to $\s \oplus \s$. 


In \eqref{Poincare}, let us now consider the following change of basis
$$
\begin{array}{ll}
U_1=iM_{01}-M_{23} & \\
U_2=-\frac{1}{2}(M_{02}-M_{12}+iM_{03}-iM_{13}) & \\
U_3=\frac{1}{2}(M_{02}+M_{12}-iM_{03}-iM_{13}) & \\
V_1=iM_{01}+M_{23} & \ \ \  \\
V_2=-\frac{1}{2}(M_{02}-M_{12}-iM_{03}+iM_{13}) & \\
V_3=\frac{1}{2}(M_{02}+M_{12}+iM_{03}+iM_{13}). & 
\end{array}
$$
The semi-simple Levi part is then written
$$
\begin{array}{ll}
\left[U_1,U_2\right]=-2U_2,& \ \ \left[V_1,V_2\right]=-2V_2,\\
\left[U_1,U_3\right]=2U_3, & \ \ \left[V_1,V_3\right]=2V_3,\\
\left[U_3,U_2\right]=U_1 , &\ \ \left[V_3,V_2\right]=V_1.
\end{array}
$$
The momentum $(P_0,P_1,P_2,P_3)$ generates an irreducible representation.
If we put
$$
\begin{array}{ll}
p_1=P_3-iP_2, & \ \ p_2=P_3+iP_2,\\
p_3= P_0-P_1,& \ \ p_4=P_0+P_1
\end{array}
$$
then we have the following description of this representation
\beqa
\label{rel1}
\begin{array}{llll}
\left[U_1,p_1\right]=-p_1,& \ \left[U_1,p_2\right]=p_2,& \ \left[U_1,p_3\right]=-p_3,& \ \left[U_1,p_4\right]=p_4 \\
\left[U_2,p_2\right]=-p_3,& \ \left[U_2,p_4\right]=-p_1,& &  \\
\left[U_3,p_1\right]=-p_4,& \ \left[U_3,p_3\right]=-p_2 & & \\
\left[V_1,p_1\right]=p_1,& \ \left[V_1,p_2\right]=-p_2,& \ \left[V_1,p_3\right]=-p_3,& \ \left[V_1,p_4\right]=p_4 \\
\left[V_2,p_1\right]=p_3,& \ \left[V_2,p_4\right]=p_2,& & \\
\left[V_3,p_2\right]=p_4,& \ \left[V_3,p_3\right]=p_2. & &
\end{array}
\eeqa
\noi
The roots $\alpha _i$ of each one of the vectors $p_i$ are respectively:
\beqa
\label{roots}
\alpha _1=(-1,1), \ \alpha _2=(1,-1), \ \alpha _3=(-1,-1), \ \alpha _4=(1,1).
\eeqa
\noi
The Poincar\'e algebra corresponds to the products
$$\left[p_i,p_j\right]=0.$$
 Let us suppose that $(q_1,q_2,q_3, q_4)$ is a basis of the $4$-dimensional representation of $\frak{g}_0$ we consider here, 
and these vectors are eigenvectors of ${\rm ad}(U_1)$ and ${\rm ad}(V_1)$. We denote by $\alpha_i=(n_i,m_i)$ the roots associated to $q_i$ that is $[U_1,q_i]=n_iq_i$ and 
$[V_1,q_i]=m_iq_i$. 
As already stated, we place ourselves in the representation \eqref{roots}, that is
$\alpha _1=(-1,1), \ \alpha _2=(1,-1), \ \alpha _3=(-1,-1), \ \alpha _4=(1,1)$, one has
\beqa
\label{rel2}
\left\{
\begin{array}{llll}
\left[U_1,q_1\right]=-q_1 , & \left[U_1,q_2\right]=q_2 , & \left[U_1,q_3\right]=-q_3 , & \left[U_1,q_4\right]=q_4  \\
\left[U_2,q_1\right]=0 , & \left[U_2,q_2\right]=q_3 , & \left[U_2,q_3\right]=0 , & \left[U_2,q_4\right]=q_1  \\
\left[U_3,q_1\right]=q_4, & \left[U_3,q_2\right]=0, & \left[U_3,q_3\right]=q_2 , & \left[U_3,q_4\right]=0  \\
\end{array}
\right.
\\
\left\{
\begin{array}{llll}
\left[V_1,q_1\right]=q_1, & \left[V_1,q_2\right]=-q_2 , & \left[V_1,q_3\right]=-q_3 , & \left[V_1,q_4\right]=q_4  \\
\left[V_2,q_1\right]=q_3 , & \left[V_2,q_2\right]=0 , & \left[V_2,q_3\right]=0, & \left[V_2,q_4\right]=q_2  \\
\left[V_3,q_1\right]=0, & \left[V_3,q_2\right]=q_4, & \left[V_3,q_3\right]=q_1, & \left[V_3,q_4\right]=0.  \\
\end{array}
\right.
\eeqa
A simple examination of the weights gives
\beqa
\label{rel3}
[p_i, q_j]=0\ \forall i,j=1,\dots 4.
\eeqa

\medskip

What misses from the picture at this point is the $3-$bracket. Again by weight considerations the non-zero brackets are
\beqa
\label{rel4}
\begin{array}{lllllllll}
\{ q_1, q_1, q_2 \}= \beta_1 p_1 &\
\{ q_1, q_2, q_2 \}= \beta_2 p_2 &\
\{ q_1, q_2, q_3 \}= \beta_3 p_3 & \
\{ q_1, q_2, q_4 \}= \beta_4 p_4 \\
\{ q_1, q_3, q_4 \}= \beta_5 p_1 &\
\{ q_2, q_3, q_4 \}= \beta_6 p_2 &\
\{ q_3, q_3, q_4 \}= \beta_7 p_3 &\
\{ q_3, q_4, q_4 \}= \beta_8 p_4.
\end{array}
\eeqa
\noi
with $\beta_i\in \CC$. To determine these structure constants $\beta_i$ one has to impose the identities {\bf I.3-I.4}. 

Imposing {\bf I.4} one obtains no information for these structure constants. Indeed, this is the case because $\{q_i, q_j, q_k\} \propto p_\ell$ and by \eqref{rel3} $[p_\ell, q_m]= 0$.

Now, imposing {\bf I.3} for the sets 
$(U_3, q_1,q_1,q_2)$, 
$(V_2, q_1,q_1,q_2)$, 
$(U_2, q_1,q_2,q_2)$, 
$(U_3, q_1,q_2,q_3)$, 
$(V_3, q_1,q_2,q_3)$, 
$(U_3, q_1,q_3,q_4)$, 
$(V_2, q_1,q_3,q_4)$, 
one gets the following system of linear equations
\beqa
-\beta_1 &=& 2\beta_4\nonumber\\
\beta_1 &=& 2\beta_4\nonumber\\
-\beta_2& =& 2 \beta_3\nonumber\\
-\beta_3& =& \beta_6 +\beta_2\nonumber\\
\beta_3& =& \beta_5 +\beta_1\nonumber\\
-\beta_5& =& \beta_8 +\beta_4\nonumber\\
\beta_5& =& \beta_7 +\beta_3.
\eeqa
\noi
Thus $(\beta_1,\beta_2, \beta_3,\beta_4, \beta_5,  \beta_6,\beta_7,\beta_8)= (-2 \beta_4, 2\beta_4, - \beta_4, \beta_4, \beta_4, -\beta_4, 2\beta_4, -2\beta_4)$.

Therefore, the requested Lie algebras of order $3$ are isomorphic to 
\beqa
\label{urma}
\begin{array}{lllllllll}
\{ q_1, q_1, q_2 \}= -2\beta_4 p_1 &\
\{ q_1, q_2, q_2 \}= 2\beta_4 p_2 &\
\{ q_1, q_2, q_3 \}= -\beta_4 p_3 & \
\{ q_1, q_2, q_4 \}= \beta_4 p_4 \\
\{ q_1, q_3, q_4 \}= \beta_4 p_1 &\
\{ q_2, q_3, q_4 \}= -\beta_4 p_2 &\
\{ q_3, q_3, q_4 \}= 2\beta_4 p_3 &\
\{ q_3, q_4, q_4 \}= -2\beta_4 p_4.
\end{array}
\eeqa
\noi


Thus, placing ourselves in the particular representation \eqref{roots} the only Lie algebra of order $3$ extending the Poincar\'e algebra is isomorphic to 
 \eqref{urma}.

\medskip

Before ending this section let us recall that for the sake of completeness, one has to check in a similar manner all the possibilities for the roots $\alpha_i$.

\section{A binary approach}

One can see that the Lie algebras of order $3$ have an inhomogeneous structure. Indeed, as one notices from Definition \ref{elementary}, the algebra has bilinear composition laws ($\g_0 \times \g_0 \to \g_1, \g_0 \times \g_1 \to \g_1$) but also a trilinear composition law (${\cal S}^3 (\g_1) \to \g_0$). Thus, in order to have an homogeneous structure, we now define a completely quadratic composition law. 
This is done starting from an elementary Lie algebra of order $3$. 
Thus a Lie algebra of order $3$ may appear as an ordinary algebra.

\begin{definition}
\label{binary}
Let $\g_0\oplus\g_1$ be an elementary Lie algebra of order $3$.
Then $\g =\mathfrak{g}_0\oplus\mathfrak{g}_1\oplus\mathfrak{g}_2$, with $ \g_2 = \g_1 \tilde \otimes \g_1$ is a binary Lie algebra of order $3$ if, for any $X,X',X''\in \g_0$, $Y,Y',Y'',Y_1,\dots,Y_4\in\g_2$ and $Z,Z'\in\g_2$ one has\\
1. $\pard X, X' \pari = [X,X']$,\\
2. $\pard X, Y \pari = [X,Y]$,\\
3. $\pard X, Z \pari = \pard X, Y\tilde{\otimes}Y' \pari = [X,Y]\tilde{\otimes}Y' + Y\tilde{\otimes}[X,Y']$, where $Z=Y\tilde{\otimes}Y'$,\\
4. $\pard Y, Y' \pari = Y\tilde{\otimes}Y',$\\
5. $\pard Y, Z \pari =\{Y,Y',Y''\}$, where $Z=Y'\tilde{\otimes}Y''$,\\
6. $\pard Z, Z' \pari =  \pard Y_1 \tilde{\otimes} Y_2, Y_3\tilde{\otimes}Y_4 \pari = \pard Y_1, \pard Y_2, Y_3 \tilde \otimes Y_4 \pari \pari + \pard Y_2, \pard Y_1, Y_3 \tilde \otimes Y_4 \pari \pari + \pard Y_3, \pard Y_1, Y_2 \tilde \otimes Y_4 \pari \pari + \pard Y_4, \pard Y_1, Y_2 \tilde \otimes Y_3 \pari \pari$, 
where $Z=Y_1\tilde{\otimes}Y^1_2, Z'=Y_3\tilde{\otimes}Y_4$, for indecomposable vector (with the symbol $\tilde \otimes$ for the symmetrised tensor product).
\end{definition}


\begin{remark}
\label{indecomposable}
Actually, when referring to  a generic $Z \in \mathfrak{g}_2$, one must take linear combinations of the symmetrised tensor products $Y_1\tilde{\otimes}Y_2$, with $Y_1,Y_2 \in \mathfrak{g}_1$.
\end{remark}

\medskip

Nevertheless, this type of approach leads to a different problematic; these problems are related to finding appropriate representations for this structure or, when making a similar deformation study, the issue of the stability of different conditions of the definition, {\it etc.} Further investigations are thus required for these specific purposes.

\section{Concluding remarks and perspectives}

In this chapter, after recalling the definition and some basic examples, we have set the basis of the study of deformation and contractions of Lie algebras of order $3$; explicit examples were also given.

A further perspective is  the study of extensions of the Poincar\'e algebra, study initiated in section $4.5$.
Furthermore, questions related to this algebraic study (like for example the issue of representations of binary Lie algebras of order $3$) are to be answered.

Such an algebraic study may give valuable insights on the eventual applications of this structures in theoretical physics models. As already stated, in chapters $5$ and $6$ of this thesis we will implement a field theoretical model on a particular Lie algebra of order $3$ extending non-trivially the Poincar\'e symmetry.

\chapter{Cubic supersymmetry}

In the first chapters we got  familiarised with algebraic aspects of more conventional (Lie (super)algebras) or more exotic mathematical structures (Lie algebras of order $3$). In the rest of this thesis we will focus on their physical applications.

We will use the Lie algebra of order $3$ \eqref{algebra0} introduced previously. This algebra is a non-trivial extension of the Poincar\'e algebra, different of the supersymmetric extension and it will be used to construct a field theoretical model, the cubic supersymmetry or \3.  We obtain different bosonic multiplets which we use to
construct free invariant Lagrangians. We also analyse the compatibility between this new symmetry and the abelian gauge invariance.
Finally we prove that no self-interacting terms between the considered bosonic multiplets
are allowed.

This chapter is structured as follows. In the first section, which can be seen as some preamble for the rest, we recall the most important no-go theorems of particle physics and some basic features of supersymmetry. We then
 briefly overview existing extensions of the Poincar\'e algebra and we explain the motivations for this type of approach.
We then make a short discussion about the compatibility between the assumption of analicity of the Coleman-Mandula theorem and the possibility of interactions for our model. We then continue with the explicit construction of \3.
In the third section we first recall the algebraic structure; we then exhibit matrix representations and bosonic multiplets one can obtained from these representations.
We end this section by a discussion about generators of symmetries and some issues related to the Noether theorem.
In the fourth section we construct invariant free Lagrangians and diagonalise them. Furthermore, the  compatibility with abelian gauge invariance is investigated. Finally, we analyse the interaction possibilities for these bosonic multiplets and conclude that \3 does not allow such terms.

If in sections $5.3$, beginning of section $5.4$ and subsection $5.4.5.$ we mostly exhibit results of \cite{articol}, in the rest of the chapter we show new results published in \cite{io1}.

\section{Preamble: no-go theorems and supersymmetry}

We begin by overviewing the reasons for which particle physics today exceeds to go beyond its Standard Model (SM). We then introduce  the no-go theorems of Coleman-Mandula \cite{cm} and Haag-Lopuszanski-Sohnius \cite{hls}. We will thus be able to see what are the open gates these no-go theorems leave; it is these open gates that will make the connection with the last part of this thesis, the cubic SUSY. 
As general references for supersymmetry, one can consult the review article \cite{sohnius} of M. Sohnius, the book \cite{wessbagger} of J. Wess and J. Bagger or the course \cite{derendinger} of J-P. Derendinger.

\subsection{Necessities of going beyond  the Standard Model}

 In nowadays physics, the Standard Model is one of the theories with theoretical predictions verified with an extremely high accuracy by experiments: for example, the fine structure constant $\alpha$ is measured with an impressive precision, giving though a very good test for Quantum Electrodynamics (QED).

 But, in the present status, it is clear for a physicist that the Standard Model, with all its virtues, cannot be an Ultimate Theory. One of the most important theoretical reason is the  problem of hierarchy. One of the aspects of this stringiest problem is that the mass of the Higgs boson is not protected from radiative corrections, i.e. the loop corrections are of the order: $\delta M^2_\Phi=10^{30} M_\Phi^2$!, where $M_\Phi$ is the mass of the Higgs boson $\Phi$. A connected aspect of this problem is the small number problem, the fact that we have a huge difference of scale between the weak ($M_{W}$) and gravitational ($M_G$) scale: $\frac{M_{W}}{M_G}\sim 10^{-14}$, which seems a very unnatural situation that the gravitational force needs such a huge fundamental scale.

 Other important theoretical reasons is the big number of free parameters (masses and mixing angles) which, being free, can be (fine) tuned, which is not a very elegant property for a theoretical model. The choice of gauge groups also, doesn't have any theoretical explanation. Another important issue is that SM predictions do not achieve a unification of the  coupling constants of the three fundamental forces: the electromagnetic, the weak and the strong force).
  Finally, one shouldn't forget that gravity, the fourth fundamental force is not taken into consideration within the frames of the Standard Model.

 All these were strong theoretical reasons, but to them one can also add \textbf{facts}, that is experimental results: dark matter and energy are present in the Universe (more than $90\%$) and the Standard Model doesn't have any candidate for it. Also, for example, the baryon asymmetry in the Universe is much bigger than the SM predicts.

 All these examples show us that going beyond the SM is becoming more and more of a necessity today. We will see now how this can be allowed by  no-go theorems and finally, one of the most appealing candidate of new physics, SUSY.

\subsection{Symmetries of nature; no-go theorems}

 In this section we show how the no-go theorems (the Coleman-Mandula theorem and the Haag-Lopuszanski-Sohnius theorem)  restrict the symmetries of our theory.


\subsubsection{The Coleman-Mandula theorem; symmetries in the SM frame}  
\label{CM}

\textbf{The Coleman-Mandula no-go theorem:} this theorem shows what are the possible symmetries (within a Lie group) of a QFT model.

A detailed presentation of the assumptions of an interacting relativistic QFT (Poincar\'e invariance, non-trivial S-matrix etc.) does not lie in the purposed of this work (the interested reader can consult for examples chapters 1-8 of the book \cite{lopuszanski} of J. Lopuszanski). 
One can also recall the axiomatic approach to QFT (for example the Wightman formalism \cite{uit}). 
The naturalness of  these type of assumptions can hardly be questioned from the point of view of today's fundamental physics.

However, one of the corner stones of the Coleman-Mandula theorem is the restriction (absolutely natural at that moment, in $1967$) to bosonic symmetries, grouped in a Lie group of symmetries. 
As we will see, it is exactly this assumption that will leave the door open for other types of symmetries.

Let us now exhibit the assumptions
made in \cite{cm} for these bosonic symmetries: 
the symmetry group contains the Poincar\'e group as a subgroup, 
the S-matrix is based on a local, interacting relativistic quantum field theory in $4$ dimensions (thus one has for example analytic dependence of the center-of-mass energy $s$ and invariant momentum transfer $t$ of the elastic-scattering amplitude); 
there are only a finite-number of different particles associated with one-particle states of a given mass and there is an energy gap between the vacuum and the one-particle states. 
The theorem then states 
that  the demanded symmetry group is the direct product of an internal symmetry group and the Poincar\'e group.

Note that by \textit{internal symmetry group} one understands a group composed of symmetry transformations which commute with the Poincar\'e group; they are thus Poincar\'e scalars. This means that these internal symmetry transformations act only on particle-type indices and have no  matrix elements between particles of different four-momentum or different spin.


Notice that maybe the most questionable assumption from a physical point of view is the fact that the possibility of having all one-particle states as massless is not allowed by the assumptions of the Coleman-Mandula theorem. Dropping up this assumption and considering only massless one-particle states give rise to the {\it conformal symmetry}, which is thus a symmetry group going beyond the Poincar\'e group.

Let us also mention that the Coleman-Mandula theorem is the guerdon of previous results, amongst which one can recall the O'Raifeartaigh theorem \cite{o}. This theorem states that if one considers Lie algebras of a group of finite order (which contains the Poincar\'e group as a subgroup of it) then the particles appearing in some multiplet will all have the same mass.


{\bf Symmetries of Nature in the SM frame:}  We have thus seen, if one restricts to bosonic symmetries, {\it i.e.} to the SM frame, that one has two types of symmetries: space-time symmetries and internal symmetries.

 The \textit{space-time symmetries} are regrouped in a Lie algebra - the Poincar\'e algebra, already recalled in  \eqref{Poincare-algebra}
\beqa
\label{Poincare-algebra2}
&&\left[L_{mn}, L_{pq}\right]=
\eta_{nq} L_{pm}-\eta_{mq} L_{pn} + \eta_{np}L_{mq}-\eta_{mp} L_{nq},\nonumber\\
&&\left[L_{mn}, P_p \right]= \eta_{np} P_m -\eta_{mp} P_n, \nonumber \\
&&\left[P_m, P_n \right]= 0, \ \
\eeqa
\noindent
where  $P_m,m=1,..,4$ are the momentums, 
$ L_{mn}$ are the generators of rotations and Lorentz boosts and $\eta_{mn}$ is the Minkowski metric ($\mathrm{diag}(1,-1,-1,-1)$).

  If the space-time symmetries are completely determined by the Coleman-Mandula theorem, this is not the case for the internal symmetries. The only theoretical informations are, as already stated above, that the internal symmetries are scalars of the Poincar\'e transformations and that their group must be the direct product of a compact semi-simple group with $U(1)$ factors.
This group is called the \textit{gauge group}, and, for the SM, phenomenological reasons make it to be
$$ SU(3)_C \times SU(2)_L \times U(1)_Y .$$
This group corresponds to the gauge groups of the strong, weak and electromagnetic interaction.

\subsubsection{The Haag-Lopuszanski-Sohnius theorem; supersymmetry}

One might think that the Coleman-Mandula no-go theorem forbids other possibilities of physics beyond the SM. This is not the case, because, as already stated here the no-go theorems do leave open gates for new theoretical models. This means that, under different assumptions, one can evade the no-go theorems, thus obtaining other interesting physical results. We will see this for SUSY case and, in a more exotic manner in the following chapters, for the 3SUSY case.

\bigskip

The Haag-Lopuszanski-Sohnius no-go theorem is in some way an analogous of the Coleman-Mandula theorem to the SUSY frame. This theorem is obtained by enlarging the allowed type of symmetries with  fermionic symmetries. 
Such generators obey anticommutation relations with each others; thus they do not generate any kind of Lie groups and are therefore not ruled out by the Coleman-Mandula no-go theorem.

The adapted algebraic structure at the infinitesimal level will not be a Lie algebra anymore, but a Lie superalgebra. Thus, in the same framework of interacting relativistic QFT, the Haag-Lopuszanski-Sohnius theorem reads:
 amongst all Lie superalgebras only the SUSY algebras can generate symmetries of the S-matrix which are coherent with a relativistic QFT.

\medskip

\noi 
For details of the proof, the interested reader can check for example \cite{sohnius}. Let us give here a short sketch of it. Since the underlying algebraic structure is a Lie superalgebra which has the Poincar\'e algebra as bosonic sector, the fermionic generators $Q$ must lie in a representation of the Poincar\'e algebra. By an examination of weights, in \cite{hls} it is proven that these fermionic generators lie in  the $2$-dimensional representations $(\frac 12 , 0)$ and $(0, \frac 12 )$ of the Lorentz algebra, thus being left-handed (LH) and right-handed (RH) Weyl spinors. 
Imposing the super Jacobi identities (\ref{super-Jacobi2}) one gets all of these constants. The result will be the well-known SUSY algebra (given here for $N=1$, as in   \eqref{SUSYalgebra}):
\beqa
\label{SUSYalgebra2}
&&\left[L_{mn}, L_{pq}\right]=
\eta_{nq} L_{pm}-\eta_{mq} L_{pn} + \eta_{np}L_{mq}-\eta_{mp} L_{nq},
\ \left[L_{mn}, P_p \right]= \eta_{np} P_m -\eta_{mp} P_n, \nonumber \\
&&\left[Q_\alpha, P_m\right]=0=[\bar Q_{\dot \alpha},P_m], \nonumber \\
&&\left[Q_\alpha, L_{mn}\right]=\frac{1}{2} (\sigma_{mn})_\alpha^{\mbox {    } \beta} Q_\beta \ \left[\bar Q_{\dot \alpha}, L_{mn}\right]=\frac{1}{2} \bar Q_{\dot \beta}(\bar \sigma_{mn})^{\dot \beta}_{\mbox{    } \dot \alpha}, \nonumber \\
&& \{Q_\alpha,Q_\beta\}=0=\{\bar Q_{\dot \alpha},\bar Q_{\dot \beta}\} \nonumber \\
&& \{Q_\alpha,\bar Q_{\dot \beta}\}=2(\sigma^m)_{\alpha \dot \beta} P_m 
\eeqa
\noindent
where, as before,  the supercharges $Q_\alpha$, a two-component LH Majorana spinor, and $\bar Q_{\dot \alpha}$, a two-component RH Majorana spinor (the undotted - $ \alpha,\beta$ and the dotted $ \dot \alpha, \dot \beta$ indices can take the values $1$ and $2$ -\textit{the two component notation)} and
$\sigma_m=(1,\sigma_i)$ are the usual Pauli  matrices, $\bar \sigma_m=(1,-\sigma_i)$  and $\sigma_{mn}=\frac 12 (\sigma_m \bar \sigma_n- \sigma_n \bar \sigma_m)$, $\bar \sigma_{mn}=\frac 12 (\bar \sigma_m  \sigma_n- \bar \sigma_n \sigma_m)$.

A crucial observation is that, looking at the last anticommutation relation, one can interpret the supercharges $Q$ as {\bf 'square roots' of translations}.
We will see in the next sections what this interesting feature becomes in the case of 3SUSY.

Note also that the SUSY transformation writes
\begin{eqnarray}
\label{SUSYtransf}
\delta_\e (\Phi)=(\e Q + \bar \e \bar Q) \Phi
\end{eqnarray}
where $\e,\bar \e$ are two-component LH (resp. RH) anticommuting Majorana spinor and $\Phi$ any SUSY field.

Recall that if for the Poincar\'e algebra $[P^2,{\rm anything}]=0$, this identity holds also for the supercharges, $[P^2, Q]=0$. Thus, for the the SUSY algebra also one has
$$[P^2, {\rm anything}]=0.$$
Thus all the states in an unbroken SUSY multiplet will have  equal mass.

\section{Extensions of the Poincar\'e symmetry and motivations for such approaches}
\label{extensions}

We give here some of the extensions of the Poincar\'e symmetry, \3 being a particular case in this larger class.  We then exhibit some possible motivations for this type of approach. We end this section by discussing to what would lead assumption of analicity of the Coleman-Mandula theorem (see subsection $5.1.2$)
when applied to the \3 construction.

\subsection{Non-trivial extensions of the Poincar\'e algebra}
  
In today's literature, the most known such extensions, different of the supersymmetric extension, are parasupersymmetry  and fractional supersymmetry. An additional interest one should also give to ternary algebraic structures.

\medskip

Parasupersymmetry was originally introduced by V. A. Rubakov and V. P. Spiridonov in \cite{paraSUSY}.  In a non-equivalent way, J. Beckers and N. Debergh \cite{paraSUSY2} define a new type of algebraic structure, Lie parasuperalgebras, in relation with parafermions (see for example \cite{parastat}, where different type of canonical commutations are proposed, leading to parastatistics).
 To construct a quantum field theoretical model, such a particular Lie parasuperalgebra is used, 
 the Poincar\'e parasuperalgebra. The Poincar\'e algebra is extended by spinor charges $Q_\alpha, \bar Q_{\dot \alpha}$ which do not close with anticommutators (as was the case for the SUSY algebra \eqref{SUSYalgebra2}) but with double commutators which write
\beqa
[Q_\alpha, [Q_\beta, Q_\gamma]]&=&0,\ [\bar Q_{\dot\alpha}, [\bar Q_{\dot \beta}, \bar Q_{\dot \gamma}]]=0,\nonumber\\
\left[Q_\alpha, \left[Q_\beta, \bar Q_{\dot \gamma}\right]\right]&=&-4Q_\beta (\sigma_m)_{\alpha \gamma}P^m,\nonumber\\
\left[\bar Q_{\dot \alpha}, \left[Q_\beta, \bar Q_{\dot \gamma}\right]\right]&=&4\bar Q_\gamma (\sigma_m)_{\beta \alpha}P^m.
\eeqa 
\noi
The theoretical field  model constructed from this algebraic structure allows interacting terms. Thus, a possible connexion with our approach may be of interest as a possibility of interactions for \3.

\medskip

The latter of the extensions mentioned above, the fractional supersymmetry (FSUSY) \cite{FSUSY,FSUSY2,FSUSY3, FSUSY4,FSUSY5,FSUSY6,FSUSY7,FSUSY8, FSUSY9} is based on the idea of extracting roots of order $F$ of the translation generators $P$, $Q^F\propto P$. 
Let us recall here this algebraic structure in the simplest case of $1$ dimension. FSUSY is generated by $H$, the Hamiltonian and $Q$, the generator of FSUSY transformation. One has
$$ [Q, H]=0,\ Q^F=H.$$
If in the literature this idea is used to construct quantum mechanics models, our approach  constructs a field theoretical model 
 which is based, as  mentioned above on a particular Lie algebra of order $3$ \eqref{algebra0}.

\medskip

A different approach is proposed in \cite{Kerner}, where
R. Kerner extracts cubic roots of  translations. 
This is achieved defining generalised covariant derivations (in analogy with SUSY). Indeed, defining three different kinds of variables $\theta^A, \hat \theta^{\hat B}$ and $\dot \theta^{\dot C}$ (in connexion with the three roots of unity $1,j$ and $j^2$), one defines the three operators
\beqa
\label{ker}
{\cal D}_A&=&\rho^m_{A\hat B \dot C} \hat \theta^{\hat B} \dot \theta^{\dot C} \partial_m + j \partial_A,\nonumber\\
{\cal D}_{\hat B}&=&\rho^n_{A\hat B \dot C}  \dot \theta^{\dot C} \theta^A \partial_n + j \partial_{\hat B},\nonumber\\
{\cal D}_{\dot C}&=&\rho^r_{A\hat B \dot C}  \theta^A \hat \theta^{\hat B}\partial_r + j \partial_{\dot C},
\eeqa
\noi
where the derivations $\partial_A,\partial_{\hat B},\partial_{\dot C} $ are understood to be with respect to the variables $\theta^A, \hat \theta^{\hat B},\dot \theta^{\dot C}$ and the $\rho$ matrices are some cubic generalisations of Pauli's matrices (see \cite{Kerner} for explicit conventions). Cubic roots of translations are obtained, in the following sense
$$ {\cal D}_A {\cal D}_{\hat B} {\cal D}_{\dot C} +\mbox{ all permutations } = 6 \rho^m_{A\hat B \dot C} \partial_m.$$
The algebraic realisation we propose here is a different from the realisation \eqref{ker}.

\medskip

Let us end this subsection by remarking that 
 a systematic study of the possible connexions between all these algebraic structures may be of interest from both mathematical and theoretical physics point of view.

\subsection{Motivations}

We have seen in subsection $5.1.2$ that one of the hypothesis of the Coleman-Mandula theorem was to group the generators of symmetries in  { Lie groups} (resp. at infinitesimal level in a Lie algebra) and for the Haag-Lopuszanski-Sohnius theorem in  {Lie supergroups} (resp. at infinitesimal level in a Lie superalgebra).
However, 3SUSY (and also the rest of the algebraic structures overviewed in the previous subsection) is  {\it a priori} {\bf allowed by no-go theorems} because of its algebraic underlying structure, the Lie algebra of order $3$ (\ref{algebra0}). Indeed, the additional generators $V_m$ of the 3SUSY algebra (see equation (\ref{algebra0})) make possible for this  construction to evade the hypothesis of the no-go theorems: Lie algebras of order $3$ do not form Lie groups and obviously they are not Lie superalgebras. Thus the no-go  theorems do not apply and the coast is clear for the possibility of new theoretical models.

Obviously, the Lie algebra of order $3$ we have chosen to study here is not the only Lie algebra of order $3$ which extends in a non-trivial way the Poincar\'e algebra. Other algebraic structures of this type may lead to interesting physical results.

   

We now make ask the legitimate question of {\it why} we  bother with the construction of such type of model. The motivation for our approach has to consider two options.

$\bullet$ First, as for any physical model, the ideal purpose would be  to obtain a self-coherent theoretical approach which enlightens new symmetries of nature.

$\bullet$ Otherwise, if this purpose proves to be unreachable for different reasons, this might be an indication for more powerful no-go theorems. Indeed, if \3 is proven not to work, then one may hope to prove more powerful no-go theorem by further studying all the class of Lie algebras of order $3$.






\subsection{The assumption of analicity}
\label{analicitate}

We have seen in subsection \ref{CM} that one of the assumptions of the Coleman-Mandula theorem was {\it analicity}. In this section we will look to this issue in more detail and give an illustration of why additional exotic symmetries in the Standard Model frame would violate this assumption. We do this explicitly on a simple example of two-body scattering. We then discuss what this becomes for the case of SUSY and for \3. This short discussion is drawn upon E. Witten's analyse in \cite{witten}.

\medskip

The assumption of analicity means that the elastic-scattering amplitudes are analytic functions of their momentum and spin. As S. Coleman and J. Mandula say in their original paper \cite{cm}, the naturalness of this assumption is above any doubt, being ``something that most physicist believe to be a property of the real world''.

Let us now consider the basic case of two-particle scattering:

\epsfxsize = 4cm     
$$
\epsffile{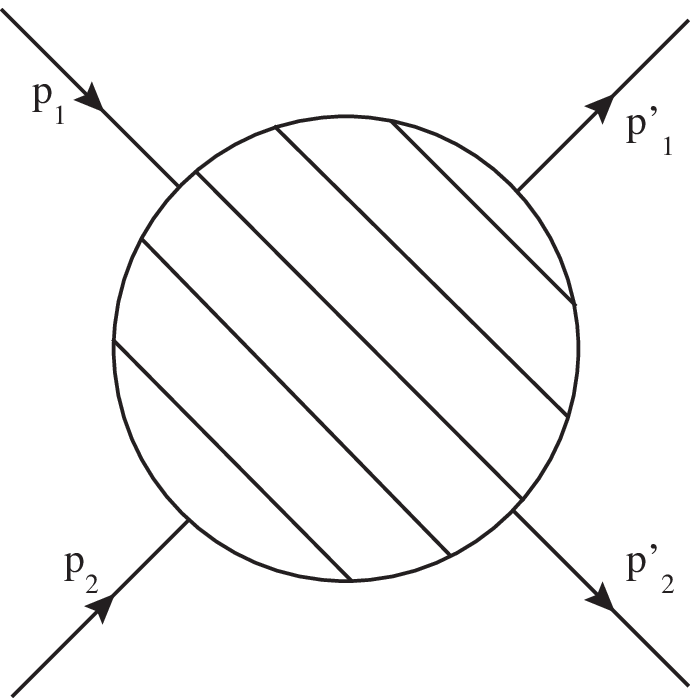}
$$
\begin{center}
{\bf Fig. 6.1:} Two-particle scattering
\end{center}

Considering the momentum and angular momentum as the only conserved charges, one has the cross section depending only on the scattering angle $\theta$. This can be seen in the center of mass frame, where the problem is replaced by the diffusion of the relative particle in a central potential. At classical level, imposing the conservation laws mentioned above, one has the Rutherford formulae $\frac{d\sigma}{d\Omega} \propto {\rm sin}^{-4} \frac{\theta}{2}$. At quantum level, considering the same conserved quantities, one needs more elaborate methods, namely the phase shift analysis, which  leads to more complicated dependencies $\sigma (\theta)$ on the scaterring angle $\theta$.
Obviously, these dependencies are  here analytical.
 
Let us now consider that one has an additional exotic conserved charge, say a  symmetric, traceless tensor $Z_{mn}$, which closes with the rest of the Poincar\'e generators by commutation relations. These commutation relations would not be trivial  and hence $Z_{mn}$ has no trivial matrix elements between particles of different four-momentum and spin.

For simplicity, we consider here only spinless particle states \footnote{All the hypothesis made now for simplificity reasons will not change the final conclusion, as we will see at the end of this reasoning.}. By Lorentz invariance, one takes for the matrix element of $Z_{mn}$ in a one-particle state of momentum $p$
\beqa
\label{matrix-element}
<p|Z_{mn}|p>=p_mp_n-\frac 14 \eta_{mn} p^2.
\eeqa
\noi
One has thus expressed the new matrix element with the help of the momentum four-vector.
Now, for the two-particle scattering above, assume that the matrix element in the two-particle state $|p_1p_2>$ is the sum of the matrix elements in the states $|p_1>$ and $|p_2>$, given by \eqref{matrix-element} above. Hence, the conservation of $Z$, $<p_1p_2|Z_{mn}|p_1p_2>=<p'_1p'_2|Z_{mn}|p'_1p'_2>$, leads  to
\beqa
\label{contrainte}
p_{1\, m}p_{1\, n}+p_{2\, m}p_{2\, n}=p'_{1\, m}p'_{1\, n}+p'_{2\, m}p'_{2\, n}.
\eeqa
\noi
This is a supplementary conservation law, which further implies  
the independence of the scattering amplitude on the last remaining parameter, the scattering angle $\theta$.

We see also that the hypothesis we made to simplify the equation \eqref{contrainte} will not lead to a different conclusion. Indeed, if one considers for example particles with non-zero spin, the explicit form of \eqref{contrainte} will be more complicated, but this does not change the fact that we are dealing with a further constraint that will violate the analicity assumption.

If one considers now the case of SUSY, one cannot make the same reasoning. Indeed, since $Q$ is a spinor, one cannot construct a  matrix element $<p|Q_\alpha|p>$ (the analogous of \eqref{matrix-element}) with the help of momentum and spin variables.

To conclude this section, let us now shortly discuss the case of \3. As we have already mentioned the additional symmetries introduced $V_m$ are Lorentz vectors and close within a structure of Lie algebra of order $3$. Hence, one can {\it a priori} construct a matrix element 
\beqa
\label{3-matrix-element}
<p|V_m|p>= C\, p_m,
\eeqa
\noi
which will lead to a certain constraint. If one considers the simpler case of spinless particle, then $C=C(p^2)= C(m^2)$.
Moreover, considering for example more elaborate situations, with non-zero spin, more complicated matrix elements \eqref{3-matrix-element}  can be written (or the same but with $C$ depending not only of the momentum but also of the spin) thus giving rise to a new, exotic conservation law, which, as above would contradict analicity and thus the theory constructed upon in $4$ dimensions would have to be non-interacting.


\medskip

This short remark is obviously not a proof of the impossibility of interacting \3, but rather an indication towards this conclusion. This indication goes along the theorem we prove in section \ref{interactiuni}.  
Furthermore one can obviously imagine more complicated situations, like for example $n$ particle scaterring, or to consider different types of interactions between these particles. A complete analysis of all possible cases is not of interest here since, already in this simple case, we have indicated this discrepancy.

Nevertheless, when up-lifting to $D$ dimensions, the above analysis has to change (considering for example the no-go theorems in extra-dimension \cite{Dcm}). It is tempting to consider also the possibility of interactions with extended objects, hence the interest of studying these new algebraic structures in extra-dimensions (see chapter $6$).

\bigskip

We now proceed with the construction of our model in $4$ dimensions.

\section{3SUSY algebra and multiplets}

We start this section by recalling the Lie algebra of order $3$ that we use to construct \3. Matrix representations are then given. We then obtain bosonic multiplets associated to these representations and their transformation laws under \3. Further properties of these multiplets, useful for the sequel are then shown.

\subsection{3SUSY algebra}
\label{prima}

Recall \eqref{algebra0} the following particular Lie algebra of order $3$, to which we refer as the {\it 3SUSY algebra}
\beqa
\label{algebra} 
&&\left[L_{mn}, L_{pq}\right]=
\eta_{nq} L_{pm}-\eta_{mq} L_{pn} + \eta_{np}L_{mq}-\eta_{mp} L_{nq},
\ \left[L_{mn}, P_p \right]= \eta_{np} P_m -\eta_{mp} P_n, \nonumber \\
&&\left[L_{mn}, V_p \right]= \eta_{np} V_m -\eta_{mp} V_n, \ \
\left[P_{m}, V_n \right]= 0, \\
&&\left\{V_m, V_n, V_r \right \}=
\eta_{m n} P_r +  \eta_{m r} P_n + \eta_{r n} P_m, \nonumber
\eeqa

 \noindent
where 
$$\{V_m,V_n,V_r \}=
V_m V_n V_r + V_m V_r V_n + V_n V_m V_r + V_n V_r V_m + V_r V_m V_n +
V_r V_n V_m $$
  stands for the symmetric product of order $3$ and
$\eta_{mn} = \mathrm{diag}\left(1,-1,-1,-1\right)$  is the Minkowski
metric. 

{\it Comparison with the SUSY construction:} this construction is a SUSY inspired construction, extending the Poincar\'e symmetries with other types of symmetries. Nevertheless, as already noticed a main difference with the SUSY case is that the new generators $V$ (which one may call ``3charges'') lie in the vector representation of the Lorentz algebra. This is not the case for SUSY, where the supercharges $Q$ lie in the spinor representation of the Lorentz algebra (as already remarked in subsection $5.1.2$)
An important consequence of this fact is that the 3SUSY multiplets
contain fields of the same statistics (the $V$ generators do not mix
bosons and fermions). This is obviously a crucial difference if one
compares the  SUSY and \3 constructions.

Another important remark is that, if in the SUSY case, one speaks about the supercharges $Q$ as ``square roots of translations'' (since $\{ Q, Q\}\propto P$, see subsection $5.1.2$), we can now speak of {\it ``cubic roots of translations''} (since $\{ V, V,V\}\propto P$); hence the terminology of {\it cubic} SUSY.


\subsection{Irreducible representations}

In order to proceed with the implementation of this algebraic structure at field theoretical level, the next step is to have irreducible matrix representations. The 3SUSY algebra (\ref{algebra}) has a twelve-dimensional representation 
\beqa
\label{12}
V_m = \begin{pmatrix}0&\Lambda^{1/3}\gamma_m&0 \cr
                 0&0& \Lambda^{1/3} \gamma_m \cr
                 \Lambda^{-2/3} \partial_m&0&0 \end{pmatrix} , 
\eeqa
\noi
with $\gamma^m=(\gamma^0,\gamma^i)$ the Dirac matrices
and $\Lambda$ a 
parameter with mass dimension   that we take  equal to $1$ 
(in appropriate units). 
This representation is actually obtained 
firstly by writing the 3-entries bracket of \eqref{algebra} as 
$$ \{ V_m, V_n, V_r \}= f_{mnr}=f_{mnr}^sP_s$$
with $f_{mnr}^s= \eta_{mn} \delta_r^s + \eta_{mr} \delta_n^s + \eta_{rn}\delta_m^s$. This means that to the symmetric tensor $f_{mnr}$ one associated the cubic polynomial $f(v)=f(v^0,\dots, v^3)$ in the variables $v^m$, $m=0,\dots,3$ defined by $f(v)=f_{mnr}v^mv^nv^r=3(v\cdot P)(v\cdot v)$. Thus, the algebra \eqref{algebra} writes now $f(v)=(v^m V_m)^3$ (this relation can actually be verified by expanding the cube and identifying each term with the help of the $3-$entries bracket, see \cite{articol} for more details). Thus one has some extension of the Clifford algebra called the Clifford algebra of polynomials \cite{Clifford-pol,Clifford-pol2, Clifford-pol3, Clifford-pol4} and the study of representations of \eqref{algebra} reduces to the study of the representations of this Clifford algebra of polynomials.
(We just recall here that these representations  are not classified and only some special matrix representations are known).

Representation \eqref{12} is reducible and leads to two inequivalent $6-$dimensional representations, denoted  by the indices $+$ and $-$
\begin{eqnarray}
\label{matirred}
V_{+}{}_m=\begin{pmatrix} 0&\sigma_m& 0 \cr
                           0&0&\bar \sigma_m \cr
                           \partial_m&0&0
\end{pmatrix}, \mbox{ or }
V_{-}{}_m=\begin{pmatrix} 0& \bar \sigma_m& 0 \cr
                           0&0&\sigma_m \cr
                           \partial_m&0&0
\end{pmatrix}
\end{eqnarray}

\noi
with   $\sigma^m=(\sigma^0=1,\sigma^i)$, and 
$\bar \sigma^m =(\bar \sigma^0=1,-\sigma^i)$,  $\sigma^i$ 
the Pauli matrices.
These two representations are referred to as conjugated to each other and they will give rise to different types of multiplets, as we will see later on.

One may also notice here that these representations are not proven to be the only irreducible representations. 
Indeed, in \cite{articol}, a different matrix realisation of the Clifford algebra of polynomials was given (involving dimension $9$ matrices); however, this realisation breaks down Lorentz invariance.
Nevertheless, if other  representations exist (which should also obviously respect the Lorentz invariance), then they may lead to different result theoretical approaches.

\medskip


Let us now make the remark that these matrix representations do not involve the notion of massive or massless multiplets. As we will see later on in this chapter, these representations allow one to construct invariant massless or massive terms in the Lagrangians. This aspect is different from the SUSY case, where one considers either massless or massive representations.


\subsection{3SUSY multiplets}
\label{sec-multi}

The matrices $V_+$ and resp. $V_-$ (\ref{matirred}) act on $\Psi_+$ and resp. $\Psi_-$, 
\beqa
\label{spinor}
\Psi_+= \begin{pmatrix} \psi_{1 +} \cr  \psi_{2-}
\cr  \psi_{3+} \end{pmatrix},\mbox{ or }  \Psi_-=\begin{pmatrix}  \psi_{1-}  \cr  \psi_{2+}
\cr   \psi_{3-} \end{pmatrix},
\eeqa

\noi
where $\psi_{1 +}$ is a left-handed (LH) $2-$component Weyl spinor, $\bar \psi_{2-}$ is a right-handed (RH) $2-$component Weyl spinor {\it etc.}

The 3SUSY transformation is
\beqa
\label{3-transf}
\delta_v \Psi_\pm=v^m V_{\pm m} \Psi_\pm
\eeqa
where $v$ is a {\it commuting Lorentz vector}. One should notice that $v$ plays the role of the anticommuting Majorana spinor $\e$ in the case of SUSY transformation (\ref{SUSYtransf}).

The transformation \eqref{3-transf} gives
\beqa
\label{actiune}
\delta_v \psi_{1+}=v^n \sigma_n  \psi_{2-}, \nonumber \\
\delta_v  \psi_{2-}=v^n \bar \sigma_n \psi_{3+}, \nonumber \\
\delta_v \psi_{3+}=v^n \partial_n  \psi_{1+}.
\eeqa


To further use, we call the states of $\psi_1$ states of gradation $-1$, the states of $\psi_2$ states of gradation $0$ and the states of $\psi_3$ states of gradation $1$. (Recall from section \ref{3-rep} that since the algebra \eqref{algebra0} has a ${\mathbb Z}_3$-graded structure, if one now has a representation $\rho$ of a Lie algebra of order $3$ on a vector space $H$ then this vector space can be decomposed as $ H =\bigoplus_{k=0}^2 H_k$. 
Furthermore one has 
$ \rho (V) H_k \subseteq H_{k+1\, (\rm{mod} 3)}$ for some $V\in\g_1$.)

One has further similitudes with SUSY, in the sense already mentioned of ``cubic roots of translations'': one has $\psi_1\to\psi_2\to \psi_3\to\psi_1$ (That is, acting with the generator $V$ on a state of gradation $-1$ one has a state of gradation $0$, acting again with a generator $V$ one has a state of gradation $1$ and finally, acting one more time with a generator $V$, one has a state of gradation $-1$). One could have reached the same final state just by acting with some translation generator on the initial state (recall that a similar phenomena happened in the case of SUSY). This structure has been studied in \cite{articol}.

In representation ($6.9$) the vacuum, denoted by $\Omega$,  is taken to be a $\3$ and Lorentz singlet. 
However it is possible to consider the vacuum as a \3 singlet but
 lying  in a different  representation of the Lorentz algebra. One may consider  for example  the vacuum to be a Lorentz  spinor (see below). Another possibility, when the vacuum is a Lorentz vector was also treated in \cite{articol}.
(Recall here (see subsection $5.3.1$) that, since the generators $V$ lie in the vector representation of the Lorentz algebra, one will not have fields of different statistics within the same multiplet. Thus the \3 multiplets will be either bosonic or fermionic.)




When the vacuum $\Omega$ is a 
Lorentz scalar, then one is able to obtain fermionic multiplets, as saw in ($6.9$).
One may also obtain {\it bosonic multiplets} if considering the
vacuum in the spinor representation of the Lorentz algebra. Thus one
has to consider the possibilities $\Omega_+$ and $ \Omega_-$, a LH and
RH Weyl spinor. Therefore one has four possibilities $\Xi_{\pm \pm}$ for the tensor
product $\Psi_\pm\otimes\Omega$, with $\Psi_\pm$ given in \eqref{spinor} (that lead to four bosonic multiplets)
\beqa
\label{4-produse}
{\mathbf \Xi}_{++}&=&\Psi_+ \otimes \Omega_+ = \begin{pmatrix} \Xi_{1++} \\ 
  \Xi_{2-+} \\ \Xi_{3++} \end{pmatrix} \nonumber \\
{\mathbf \Xi}_{--}&=&\Psi_- \otimes  \Omega_- =
\begin{pmatrix} \bar \Xi_1{}_{--} \cr 
 \Xi_2{}_{+-} \cr \bar 
\Xi_3{}_{--} 
\end{pmatrix}\nonumber \\
{\mathbf \Xi}_{-+}&=&\Psi_- \otimes  \Omega_+ =
\begin{pmatrix} \Xi_1{}_{-+} \cr 
  \Xi_2{}_{++} \cr \Xi_3{}_{-+} 
 \end{pmatrix}\nonumber \\
{\mathbf \Xi}_{+-}&=&\Psi_+ \otimes  \Omega_- =
\begin{pmatrix}  \Xi_1{}_{+-} \cr 
 \Xi_2{}_{--} \cr 
\Xi_3{}_{+-} 
\end{pmatrix}.
\eeqa

The following step is the decomposition of these products of spinors on
$p-$forms (for definitions and basic properties of $p-$forms see subsection $6.1$ and Appendix A).
For this purpose consider  the Clifford algebra in $4$ dimensions
\beqa
\label{4-clifford}
\{ \gamma_m, \gamma_n \}= 2\eta_{mn}.
\eeqa
\noi
This algebra has a $4-$dimensional complex representation on the Dirac spinor space (which we denote by ${\cal S}$).
One can choose a representation  were the Dirac matrices $\gamma_m$ write 
\beqa
\label{Dirac}
\gamma_m=\begin{pmatrix} 0 & \sigma_{m} \\ \bar \sigma_m & 0 \end{pmatrix}.
\eeqa
\noi
A basis for the representation space is defined by the $16$ antisymmetric matrices
\beqa
\label{gamele}
\gamma^{(\ell)}_{m_1,\dots,m_l}=\frac{1}{\ell !} \sum_{\sigma\in S_\ell} \e(\sigma) \gamma_{m_{\sigma (1)}}\dots \gamma_{m_{\sigma(\ell)}}, \mbox{ with }\ell=0,\dots,4.
\eeqa
\noi
Amongst these matrices of special importance are the Lorentz generators (note the difference of convention $\gamma_{mn}=\frac12 \gamma_{mn}^{(2)}$ )
\beqa
\label{gmn}
\gamma_{mn}=\frac14 [\gamma_{m},\gamma_{n}],
\eeqa
\noi
where, inserting \eqref{Dirac}, one gets
\beqa
\label{gmn-2}
\gamma_{mn}=\frac14 \begin{pmatrix} \sigma_m \bar \sigma_n - \sigma_n \bar \sigma_m & 0 \\
0 & \bar \sigma_m \sigma_n - \bar \sigma_n \sigma_m \end{pmatrix} = \frac 12 \begin{pmatrix} \sigma_{mn} & 0 \\ 0 & \bar \sigma_{mn} \end{pmatrix},
\eeqa
\noi
which allows to define the matrices $\sigma_{mn}=\frac 12 (\sigma_m \bar \sigma_n- \sigma_n \bar \sigma_m)$ and $\bar \sigma_{mn}=\frac 12 (\bar \sigma_m \sigma_n - \bar \sigma_n \sigma_m)$. This definition and the $\sigma$-matrix identity $\sigma_m\bar\sigma_n + \sigma_n \bar \sigma_m=2\eta_{mn}$  leads to the identity
\beqa
\label{utila}
\sigma_m \bar \sigma_n = \sigma_{mn}+ \eta_{mn}.
\eeqa



Denote now by ${\cal S}^*$ the dual representation of ${\cal S}$ (on which the matrices $-\gamma^{t}_{mn}$, ${}^t$ being the transpose operation, act in the same way the matrices $\gamma_{mn}$ act on $\cal S$). One can find an element
\beqa
\label{calC}
{\cal C}=i\gamma_2 \gamma_0
\eeqa
 of ${\rm End}({\cal S})$ ($\cal C$ being call the charge conjugation matrix) such that ${\cal C} \gamma_{m}{\cal C}^{-1}=-\gamma_{m}^t$ and 
\beqa
\label{dual-spinor}
{\cal C} \gamma_{mn}{\cal C}^{-1}=-\gamma_{mn}^t.
\eeqa
\noi
Hence, these representations ($\cal S$ and $\cal S^*$) are in fact equivalent and
$\psi^t\cal C\in {\cal S}^*$ for any Dirac spinor $\psi\in \cal S$. One can see the charge conjugation matrix $\cal C$ as some intertwining operator.

Furthermore, if a spinor $\psi$ transforms under a Lorentz transformation by $S(\Lambda)=e^{\frac 12 \Lambda^{mn}\Gamma_{mn}}$ (with $\Lambda^{mn}$ the parameter of the transformation) then $\Xi^t{\cal C}$  transforms under a Lorentz transformation by $S(\Lambda)^{-1}$. Hence
\beqa
\label{sc}
\Xi^t{\cal C}\psi
\eeqa
\noi
is a spinor invariant and 
\beqa
\label{L-TENSOR}
\Xi^t{\cal C}\gamma^{(\ell)}_{m_1\dots m_\ell}\psi
\eeqa
\noi
transforms as an antisymmetric tensor of order $\ell$.

\medskip

The $\gamma^{(\ell)}$ matrices act on Dirac spinor and thus define a linear application on $\cal S$; one can thus say that they belong to ${\rm End}(\cal S)$.
Now, since ${\rm End}({\cal S})\cong {\cal S}\otimes {\cal S}^*$, one can decompose any product of Dirac spinors $\psi\otimes \psi'^t {\cal C}$ on the basis \eqref{gamele} of the set of  $\gamma^{(\ell)}_{m_1,\dots,m_\ell}$ matrices.
Since the coefficient of this development are 
  the antisymmetric tensors of order $\ell$ (with $\ell=0,\dots,4$),  this can be written schematically as
\beqa
\label{4-mare}
\psi \otimes \psi'^t{\cal C}=[0]\oplus[1]\oplus[2]\oplus[3]\oplus [4].
\eeqa
\noi
where $[\ell]$ denotes an $\ell-$form. 
Now, using the Hogde equivalence $[\ell]\equiv[4-\ell]$, one can choose to write \eqref{4-mare} as
\beqa
\label{4-mic}
\psi \otimes \psi'^t{\cal C}=[0]^2\oplus[1]^2\oplus[2].
\eeqa
\noi

\medskip

Let us now exhibit one final (anti-)self-dual property useful for the sequel. For this, denote first the Levi-Civita tensors $\e_{m_1\dots m_4}$ and $\e^{m_1\dots m_4}=\e_{n_1\dots n_4}\eta^{m_1n_1}\dots \eta^{m_4n_4}$ defined by $\e_{m_1\dots m_4}=1$ and $\e^{m_1\dots m_4}= -1$. Furthermore, introduce $\gamma^m=\eta^{mn}\gamma_n$ and define $\gamma=\gamma_0\dots\gamma_3$. This definition implies that
\beqa
\gamma^{(2)n_1n_2}\gamma=-\frac 12 \e^{n_1n_2m_1m_2}\gamma^{(2)}_{m_1,m_2}
\eeqa
\noi
Thus one gets
\beqa
\label{mister}
{}^*(\gamma^{(2)}\pm i \gamma^{(2)}\gamma)=\pm i (\gamma^{(2)}\pm i \gamma^{(2)}\gamma),
\eeqa
\noi
where we have denoted by ${}^*$ the dual operation (${}^* X_{mn}=\frac12 \e_{mnpq}X^{pq}$, see \eqref{hodge}). This means that \eqref{mister} is (anti-)self-dual (see \eqref{self}).

\medskip

Define now the chirality matrix $\gamma_5$ as
\beqa
\label{gama5}
\gamma_5=-i\gamma_0\dots\gamma_3.
\eeqa
\noi
From this definition and \eqref{4-clifford} one proves
\beqa
\label{prop-gama5-1}
\{\gamma_m,\gamma_5\}=0
\eeqa
\noi
which, using the definition \eqref{gamele}, gives
\beqa
\label{prop-gama5-2}
\gamma^{(\ell)}\gamma_5=(-1)^{\ell}\gamma_5 \gamma^{(\ell)}.
\eeqa
\noi
Moreover, in our representation, by \eqref{Dirac} and \eqref{gama5}, one has that  $\gamma_5$ is block-diagonal 
\beqa
\label{gama5-Weyl}
\gamma_5=\begin{pmatrix} -1 & 0 \\ 0 & 1 \end{pmatrix}.
\eeqa
Furthermore one has
\beqa
\label{prop-gama5-3}
\gamma_5^t=\gamma_5.
\eeqa
\noi
From \eqref{calC} one also has
\beqa
\label{prop-C}
{\cal C}\gamma_5 {\cal C}^{-1}=\gamma_5.
\eeqa
\noi

\medskip

The chirality matrix is used to define the irreducible $2-$dimensional LH and RH Weyl spinors and  a $4-$dimensional Dirac spinor decomposes as 
$$ \psi_D= \begin{pmatrix} \psi_+ \\  \psi'_- \end{pmatrix} $$
where $\psi_+$ (and resp. $\psi'_-$) is a  LH (resp. RH) Weyl spinor. 
From \eqref{gama5-Weyl} one sees that a Dirac spinor with only LH (resp. RH) components is an eigenstate of $\gamma_5$ with eigenvalue $-1$ (resp. $+1$); this writes
\beqa
\label{adevaru}
\gamma_5 \psi_{D\e} = -\e \psi_{D \e}, \mbox{ with } \e=\pm.
\eeqa


\medskip

After seeing these general properties of the  $\gamma_5$ and $\cal C$ matrices,
 we now obtain the products of such two distinct Weyl spinors. Consider
\beqa
\label{tensor}
T_{m_1,\dots,m_\ell }= \psi'^t_{D \e_2} {\cal C} \gamma^\ell_{m_1,\dots,m_\ell}\psi_{D \e_1}.
\eeqa
\noi
(Recall that by \eqref{L-TENSOR}, \eqref{tensor} transforms like a tensor of order $\ell$). 
Using \eqref{adevaru} one has
\beqa
T_{m_1,\dots,m_\ell }=-\e_1 \psi'^t_{D \e_2} {\cal C} \gamma^\ell_{m_1,\dots,m_\ell}\gamma_5\psi_{D \e_1}.
\eeqa
\noi
Using now \eqref{prop-gama5-2} one gets
\beqa
T_{m_1,\dots,m_\ell }=-(-1)^\ell\e_1 \psi'^t_{D \e_2} {\cal C} \gamma_5\gamma^\ell_{m_1,\dots,m_\ell}\psi_{D\e_1}.
\eeqa
\noi
We now make use of \eqref{prop-C} to write
\beqa
T_{m_1,\dots,m_\ell}=-(-1)^\ell\e_1 \psi'^t_{D \e_2} \gamma_5{\cal C}\gamma^\ell_{m_1,\dots,m_\ell}\psi_{D\e_1}.
\eeqa
\noi
Using \eqref{prop-gama5-3} one has
\beqa
T_{m_1,\dots,m_\ell}=-(-1)^\ell\e_1 (\gamma_5 \psi'_{D \e_2})^t {\cal C}\gamma^\ell_{m_1,\dots,m_\ell}\psi_{D \e_1}.
\eeqa
\noi
Finally, making use again of \eqref{adevaru} one finds
\beqa
T_{m_1,\dots,m_\ell}=(-1)^\ell\e_1 \e_2\psi'^t_{D \e_2} {\cal C}\gamma^\ell_{m_1,\dots,m_\ell}\psi_{D\e_1}.
\eeqa
\noi
Hence $T_{m_1,\dots,m_\ell}$ is vanishing if $(-1)^\ell\e_1 \e_2=-1$, which explicitly gives
\begin{itemize}
\item for $\ell = 0$ one has  $\psi'^t_{D\pm} {\cal C} \psi_{D\mp}=0$,
\item for $\ell =1$ one has $\psi'^t_{ D\pm} {\cal C} \gamma_m \psi_{D \pm}=0$,
\item for $\ell = 2$ one has  $\psi'^t_{D \pm} {\cal C} \gamma_{mn}\psi_{D \mp}=0$.
\end{itemize}

Hence, the decomposition \eqref{4-mic} reduces, at the level of Weyl spinor products at
\beqa
\label{4-decompozitie}
{\psi}_+ \otimes  {\psi}_+^{\prime t}\cal C &=&
 [0] \oplus  [2]^{(+)}
\nonumber \\
{\psi}_- \otimes  {\psi}_-^{\prime t}\cal C &=&
 [0] \oplus [2]^{(-)} \\
{\psi}_+ \otimes  {\psi}_-^{\prime t}\cal C &=&  
 [1]  \nonumber 
\eeqa
\noindent
where 
 $[2]^{(\pm)}$ represents an (anti-)self-dual
$2-$form. Indeed, the $2$-form in $4$ dimensions is reducible. By relation \eqref{mister} one can decompose this as
\beqa
\label{reductibil-Lorentz}
[2]=[2]^{(+)}\oplus [2]^{(-)}.
\eeqa
\noi
This equation is also correct from the point of view of the dimension ($6$ in the LHS and $3+3$ in the RHS). The same argument of dimension counting in \eqref{4-decompozitie} shows that one must have considered the (anti-)self-dual $2-$forms. The choice we have made (self-dual for ${\psi}_+ \otimes  {\psi}_+^{\prime t}\cal C$ and resp. anti-self-dual ${\psi}_- \otimes  {\psi}_-^{\prime t}\cal C $) is compatible with our conventions, as it will become clear from \eqref{muie1832}.


We now apply the decomposition \eqref{4-decompozitie} for the case of the ${\mathbf \Xi}_{++}=\Psi_+ \otimes \Omega_+ = \begin{pmatrix} \Xi_{1++} \\ \bar
  \Xi_{2-+} \\ \Xi_{3++} \end{pmatrix}$ (see \eqref{4-produse}). Using \eqref{Dirac} and \eqref{gmn-2} this decomposition writes explicitly
\beqa
\label{muie1832}
\Xi_{1++}&=&\p+\frac14 B_{mn}\sigma^{mn},\nonumber \\
\Xi_{2-+}&=&\tA_m \bar \sigma^m,\nonumber \\
\Xi_{3++}&=&\ttP+\frac14 \ttB_{mn}\sigma^{mn},
\eeqa
\noi
where we denote by $\p,\ttP$ two scalar fields, $\tA_m$ a vector and $B_{mn},\ttB_{mn}$ two self-dual $2-$forms (because with our convention \eqref{gmn-2} ${}^*\sigma_{mn}=i \sigma_{mn}$; this can be seen for example by components; thus $i\sigma_{01}=\frac12 \e_{01pq}\sigma^{pq}$ {\it etc.}).

We can now apply this analysis for the four product of spinors \eqref{4-produse} which will thus lead to the four multiplets
$\Xi_{\pm \pm}$ with the following field content
\beqa
\label{4-decomposition}
{\mathbf \Xi}_{++}=
 \begin{pmatrix} 
\varphi, B_{mn} \cr \tA_m \cr \ttP, \ttB_{mn}
\end{pmatrix}&& \ \ \ 
{\mathbf \Xi}_{+-}=
\begin{pmatrix}    A'_m \cr \tP', \tB'_{mn} \cr  \ttA'_m
\end{pmatrix}
\nonumber 
\\
{\mathbf \Xi}_{--}=
\begin{pmatrix}
\varphi', B^\prime_{mn} \cr \tA'_m \cr \ttP', \ttB'_{mn}\end{pmatrix}
&& \ \ \ 
{\mathbf \Xi}_{-+}=
\begin{pmatrix} A_m \cr \tP, \tB_{mn} \cr  \ttA_m \end{pmatrix},
\eeqa
\noi
where  $\varphi, \ttP, \varphi', \ttP',
\tP, \tP'$ are scalars fields, $\tA, \tA', A, \ttA, A',\ttA'$ are vector 
fields, $B, \tB, \ttB$, $B',\tB', \ttB'$ are $2-$forms.
As we haved mentioned above, these $2-$forms, namely $B, \tB, \ttB$  are self-dual ({\it i.e.}
${}^*B=iB$, where by ${}^*B$ we mean the dual of $B$) and resp. $B',\tB', \ttB'$ are anti-self-dual ({\it i.e.}
 ${}^*B'=iB'$); thus these $2-$forms must be complex. 
As we will
see from the transformation laws (\ref{transfo2}), this implies that all
the \3 fields are complex.
To have minimum field content, one takes ${\mathbf \Xi}_{++}={\mathbf \Xi}_{--}^*$ and ${\mathbf \Xi}_{+-}={\mathbf \Xi}_{-+}^*$ (that is $\varphi=\varphi^{'*}, B=B^{'*}$, {\it etc.})\footnote{One should pay attention at the notation used, that is ${}^*B$ denotes the dual of $B$ whereas $B^{*}$ denotes the complex conjugated of $B$.}. 
 This choice is proven furthermore to be compatible with the transformation laws of the fields \eqref{transfo2}.  
We call the couples ${\mathbf \Xi}_{++}-{\mathbf \Xi}_{--}$, ${\mathbf \Xi}_{+-}- {\mathbf \Xi}_{-+}$ {\it conjugated multiplets} and the couples ${\mathbf \Xi}_{++}-{\mathbf \Xi}_{+-}$, ${\mathbf \Xi}_{--}- {\mathbf \Xi}_{-+}$ {\it interlaced multiplets}.

Following the convention defined earlier, one can say, for example for
the multiplet ${\mathbf \Xi}_{++}$ that the fields $\varphi, B$ are of gradation
$-1$, the field $\tA$ is of gradation $0$ and the fields $\ttP,\ttB$ are
of gradation $1$.

Before ending this subsection let us write down the converse formulae. Using trace properties of the $\sigma$ matrices, one can express the fields as functions of the original product of spinors. We do this for the multiplet ${\mathbf \Xi}_{++}$, similar formulae existing for the other three multiplets. \beqa
\label{++-conv}
\p&=&\frac12 {\rm Tr}(\Xi_{1++}),\nonumber\\
B_{mn}&=&\frac12{\rm Tr}(\sigma_{mn}\Xi_{1++}),\nonumber\\
\tA_m&=&\frac12{\rm Tr}(\sigma_{m}\Xi_{2-+}),\nonumber\\
\ttP&=&\frac12 {\rm Tr}(\Xi_{3++}),\nonumber\\
\ttB_{mn}&=&\frac12{\rm Tr}(\sigma_{mn}\Xi_{3++}).
\eeqa
\noi
To prove these formulae one can reinsert the original decompositions \eqref{muie1832}. For example, for the first of the formulae \eqref{++-conv}, one has
$$\p=\frac12 {\rm Tr} (\p + \frac 14 B_{mn}\sigma^{mn})$$
which is verified, since ${\rm Tr}(\sigma^{mn})=0$.


\subsection{Transformation laws of the fields}
\label{4-transf}

The transformation laws of the fields are obtained from the 3SUSY
transformation law (\ref{3-transf}), remembering that the vacuum $\Omega$
is a \3 scalar, {\it i.e.} it 
is not transforming under \3; for example, for the ${\mathbf \Xi}_{++}$ multiplet one has
\beqa
\label{lege-iar}
\delta_v {\mathbf \Xi}_{++}= (\delta_v {\mathbf \Psi}_{+})\otimes \Omega_+ =(v^m V_{+ m} \Psi_+)\otimes  \Omega_+
\eeqa
\noi
Now, inserting the form \eqref{matirred} of $V_{+m}$ one has
\beqa
\label{4-desfacut}
\delta_v \Xi_{1 ++}&=&v^m \sigma_m \Xi_{2-+},\nonumber \\
\delta_v \Xi_{2 -+}&=&v^m \bar \sigma_m \Xi_{3-+},\nonumber \\
\delta_v \Xi_{3 ++}&=&v^m \partial_m \Xi_{1-+}.
\eeqa
\noi
We first find the transformation law of the unique field of gradation $0$ of ${\mathbf \Xi}_{++}$, namely $\tA_m$. For this we start with its expression \eqref{++-conv} which gives
\beqa
\label{muie1437}
\delta_v \tA_m=\frac12 {\rm Tr}(\sigma_m \delta_v \Xi_{2-+}).
\eeqa
\noi
Inserting the second equation of \eqref{4-desfacut} one has
\beqa
\label{muie1438}
\delta_v \tA_m=\frac{v^n}{2} {\rm Tr}(\sigma_m \bar \sigma_n \Xi_{3++}).
\eeqa
\noi
We make now use of \eqref{utila}, thus obtaining
\beqa
\label{aproape}
\delta_v \tA_m=\frac{v^n}{2} {\rm Tr}(\sigma_{mn}\Xi_{3++})+\frac{v_m}{2} {\rm Tr}(\Xi_{3++}).
\eeqa
\noi
We now use again \eqref{++-conv}, this time to extract the fields $\ttP$ and $\ttB_{mn}$. Hence, one finally gets
\beqa
\delta_v \tA_m=v_m\ttP + v^n \ttB_{mn}.
\eeqa

Let us now find the transformation laws of the fields of gradation $1$, namely $\ttP$ and $\ttB_{mn}$. For the case of $\ttP$, we proceed as above, starting with its expression \eqref{++-conv} which gives
\beqa
\label{muie1355}
\delta_v \ttP=\frac 12 {\rm Tr}(\delta_v \Xi_{3++})
\eeqa
\noi
The last equation of \eqref{4-desfacut} gives
\beqa
\label{muie1356}
\delta_v \ttP=\frac{v^m}{2} \partial_m {\rm Tr}(\delta_v \Xi_{1++})
\eeqa
\noi
Now, equation \eqref{++-conv} allows us to write
\beqa
\delta_v \ttP=v^m \partial_m \p.
\eeqa
\noi
Similarly, one gets
\beqa
\delta_v \ttB_{mn}^{(+)}=v^p \partial_p B_{mn}^{(+)}.\nonumber\\
\eeqa
\noi
Arguing along the same lines one has
$$ \delta_v \p=v^m A_m.$$

The last remaining transformation law for the ${\mathbf \Xi}_{++}$ multiplet,  namely for  $B_{mn}$ is more tricky. As before, we start with its expression \eqref{++-conv} which gives
\beqa
\delta_v B_{mn}^{(+)}=\frac12 {\rm Tr} (\sigma_{mn} \delta_v \Xi_{1++}). 
\eeqa
\noi
The first equation of \eqref{4-desfacut} leads to
\beqa
\label{muie14372}
\delta_v B_{mn}^{(+)}=\frac{v^p}{2} {\rm Tr} (\sigma_{mn} \sigma_p \Xi_{2-+}). 
\eeqa
\noi
This time we first compute the quantity  $\sigma_p \Xi_{2-+}$. Using \eqref{muie1832} one has
\beqa
\label{muie14382}
\sigma_p \Xi_{2-+}=\sigma_p \tA^r \bar \sigma_r = \tA^r (\sigma_{pr}+\eta_{pr})
\eeqa
\noi
(where we have used again \eqref{utila}). Now, inserting \eqref{muie14382} in \eqref{muie14372}, and recalling that, by definition ${\rm Tr}(\sigma_{mn})=0$, one has
\beqa
\label{chiarultima}
\delta_v B_{mn}^{(+)}=\frac{v^p}{2} \tA^r{\rm Tr} (\sigma_{mn} \sigma_{pr}).
\eeqa
\noi
To conclude, one uses the trace identity $\frac12 {\rm Tr}(\sigma_{mn} \sigma_{pr})= -(\eta_{mp}\eta_{nr} - \eta_{mr}\eta_{np})+i\e_{mnpr}$ (one can easily check this by components: for example, for $(m,n,p,r)=(0,1,0,1)$ one has $-1$ in the LHS and by a direct application of the definition of the Pauli matrices, the same in the LHS; for $(m,n,p,r)=(0,1,0,1)$ one has $i$ in the RHS and the same in the LHS by a direct application of the definition of the Pauli matrices). Inserting this in \eqref{chiarultima} gives
\beqa
\delta_v B_{mn}^{(+)}=- (
v_m \tA_n - v_n \tA_m )
+ i \varepsilon_{mnpq} v^p \tA^q{}.
\eeqa
\noi
Note that the last term prevents the self-dual character of the $2-$form.


\bigskip

Similar reasonings for the other multiplets complete the following table:
\beqa
\label{transfo2}
 \begin{array}{ll}
~~~~~~~~~~~~~~~~~~~~~~~~~{\mathbf \Xi}_{++} & ~~~~~~~~~~~~~~~~~~~~~~~~~{\mathbf \Xi}_{+-} \cr
 \left\{\begin{array}{l}
\delta_v \varphi =  
v^m \tA_m   \cr
\delta_v B_{mn} = - (
v_m \tA_n - v_n \tA_m )
+ i \varepsilon_{mnpq} v^p \tA^q{} \cr
\delta_v \tA_m =  (
v_m \ttP +
v^n  \ttB_{mn} ) \cr
\delta_v \ttP = v^m
\partial_m \varphi \ \ \  
\delta_v  \ttB_{mn} = v^p \partial_p
B_{mn} \end{array} \right. &
\left\{ \begin{array}{l}
\delta_v A_m^\prime = (v^n \tB^\prime_{mn}+
v_m \tP^\prime) \cr
\delta_v \tP^\prime =  v^m 
\ttA_m^\prime \cr
\delta_v \tB^\prime _{mn} = - (
v_m \ttA^\prime_n - v_n \ttA^\prime_m ) 
- i \varepsilon_{mnpq} v^p \ttA^\prime{}^q{}
\cr
\delta \ttA^\prime_m =  v^n \partial_n A^\prime_m
\end{array}\right. \end{array} 
\nonumber \\
\\
\begin{array}{ll}
~~~~~~~~~~~~~~~~~~~~~~~~~{\mathbf \Xi}_{--} & ~~~~~~~~~~~~~~~~~~~~~~~~~{\mathbf \Xi}_{-+} \cr
\left\{\begin{array}{l}
\delta_v \varphi' =  
v^m \tA'_m   \cr
\delta_v B'_{mn} = - (
v_m \tA'_n - v_n \tA'_m )
- i \varepsilon_{mnpq} v^p \tA'^q{} \cr
\delta_v \tA'_m =  (
v_m \ttP' +
v^n  \ttB'_{mn} ) \cr
\delta_v \ttP' = v^m
\partial_m \varphi' \ ,  \ \ \  
\delta_v  \ttB'_{mn} = v^p \partial_p
B'_{mn} \end{array} \right. &
\left\{ \begin{array}{l}
\delta_v A_m = (v^n \tB_{mn}+
v_m \tP) \cr
\delta_v \tP =  v^m 
\ttA_m \cr
\delta_v \tB _{mn} = - (
v_m \ttA_n - v_n \ttA_m ) 
+ i \varepsilon_{mnpq} v^p \ttA{}^q{}
\cr 
\delta \ttA_m =  v^n \partial_n A_m
\end{array}\right. \end{array} \nonumber 
\eeqa
\noi
As we have pointed out in the previous subsection,
 these transformations laws are compatible with our choice of complex conjugation of multiplets, ${\mathbf \Xi}_{++}^*={\mathbf \Xi}_{--},\ {\mathbf \Xi}_{+-}={\mathbf \Xi}_{-+}^*$.

Finally, 
let us mention that in the next chapter we will generalise this in arbitrary dimensions.

\subsection{Derivation of a multiplet}
\label{deriv-mult}


After having found the transformation laws \eqref{transfo2} of the \3 multiplets considered here, we now make use of them to obtain an interesting property of these multiplets.

From now on, let us denote by  $X_{[mn]\pm}$  the  (anti-)self-dualisation of any second rank tensor $X_{mn}$, {\it i.e.}
\begin{equation} \label{dualisation}
X_{[m n]_\pm} = X_{m n} - X_{n m} \mp i \varepsilon_{mnpq} X^{p q},
\end{equation}

Let us now consider the fields of a ${\mathbf \Xi}_{+-}$ multiplet, that is
$$ A'_m;\, \tP', \tB'_{mn};\, \ttA'_m$$
to construct a different type of multiplet using partial derivatives $\partial_m$. Thus, to construct a ${\mathbf \Xi}_{++}$ multiplet, one has to have expressions for any field of the ${\mathbf \Xi}_{++}$ multiplet (respecting the self-dual character of the $2-$forms). Saturating the Lorentz indices, one possible solution is
\beqa
\label{partialXipm}
{\cal D} \,{\mathbf \Xi}_{+-} = 
\Big(\psi, \psi_{mn}, \; \tpsi_m, \; \ttpsi, \ttpsi_{mn} \Big) 
 &\equiv&  \Big(\partial_m A'^m, \partial_{[m}A'_{n]_+}; \; \partial_m \tP'+ 
\partial^n \tB'^{}_{nm}; \; \partial_m \ttA'^m, \partial_{[m}\ttA'_{n]_+}\Big). \nonumber \\
\eeqa
\noi
One has now to see if the set ${\cal D} \,{\mathbf \Xi}_{+-} = \Big(\psi, \psi_{mn}, \; \tpsi_m, \; \ttpsi, \ttpsi_{mn})$ transforms like a \3 multiplet.
To check this, we directly apply \eqref{transfo2} on \eqref{partialXipm}. 
Let us illustrate this by obtaining firstly the \3 transformation law of $\psi=\partial_m A^m$. From \eqref{transfo2} one has $\delta_v \psi =\partial_m \delta_v A'^m=v_n \tB'^{mn}+v^m \tP'=v^m (\partial^n \tB'^{}_{nm}+ \partial_m \tP')$ and 
$$\delta_v \psi= v^m \tpsi_m.$$
Let us now give the proof for the most difficult case above, the transformation law $\delta_v \psi_{mn}$. Since $\psi_{mn}=\partial_{[m}A'_{n]_+}$ and $\delta_v A'_m= v^n \tB'_{mn}+v_m \tP'$ (see \eqref{transfo2}) one will have 
$$\delta_v \psi_{mn}=v_n \partial_m \tP' - v_m \partial_n \tP' + i \e_{mnpq} v^p \partial^q \tP'+v^p (\partial_m \tB'_{np} - \partial_n \tB'_{mp} - i \e_{mnp'q}\partial^{p'}\tB'^q_p).$$
Using now the definition \eqref{partialXipm} of $\tpsi_m=\partial_m \tP'+ 
\partial^n \tB'^{}_{nm}$ and \eqref{dualisation} (to  recombine the component in $\tB'$ of $\delta_v \psi_{mn}$ above) the transformation law of $\psi_{mn}$ can be written
$$ \delta_v \psi_{mn}= v_n \tpsi_m - v_m \tpsi_n + i \e_{mnpq} v^p \tpsi^q 
- (v^r \partial_{[n} \tB^{\prime}_{m]_+ r} - v_{[n}\partial^r \tB^{\prime}_{m]_+ r} )$$
By (\ref{prop}) we have  $v^r \partial_{[n} \tB^{\prime}_{m]_+ r} - v_{[n}\partial^r \tB^{\prime}_{m]_+ r} =0$ and hence
$$ \delta_v \psi_{mn}= v_n \tpsi_m - v_m \tpsi_n + i \e_{mnpq} v^p \tpsi^q 
$$
\noi
Similarly one finds
\beqa
\delta_v \ttpsi&= &v^m \partial_m \psi, \nonumber \\ 
\delta_v \tpsi^m &=& v^m \ttpsi + v_n \ttpsi^{mn} \nonumber \\
\delta_v \ttpsi_{mn}&=& v^r \partial_r \psi_{mn}.  
 \nonumber 
\eeqa
\noi
Therefore one notices that \eqref{partialXipm} transforms like a $\Xi_{\pm \pm}$ multiplet. We have thus showed a mechanism of obtaining a multiplet of a certain type (here ${\mathbf \Xi}_{++}$) by ``deriving'' a multiplet of another type (here ${\mathbf \Xi}_{+-}$). We call ${\cal D} {\mathbf \Xi}_{+-}$ a {\it derivative multiplet}.


One can actually define such a ``derivation'' for every  \3 multiplet ${\mathbf \Xi}_{\pm\pm}$. For example, defining as ${\cal D}{\mathbf \Xi}_{-+}$ in \eqref{partialXipm}, one has ${\cal D}{\mathbf \Xi}_{-+}\ne {\mathbf \Xi}_{++}$ but ${\cal D}{\mathbf \Xi}_{-+} ={\mathbf \Xi}_{--}$.
This type of property will be used when analysing the compatibility of our model with abelian gauge invariance (see subsection \ref{joaja-abeliana}) and when treating the possibilities of interaction (see section $5.5$).

\section{Free theory}
\label{4-liber}

In this section we construct free Lagrangians  invariant under the \3 transformations \eqref{transfo2}. 
From now on we use the field strengths associated to the fields, $F_{mn}=\partial_m A_n-\partial_n A_m$ for any vector field $A_m$ and $H_{mnp}=\partial_m B_{np}+\partial_p B_{mn}+\partial_n B_{pm}$ for any $2-$form $B_{mn}$. Definitions and basic operations of $p-$forms are given in  section $6.1$ and Appendix \ref{a-pforme}).

\subsection{Coupling between conjugated multiplets}
\label{detoate}

If we consider the quadratic couplings between conjugated multiplets, as denoted in subsection \ref{sec-multi}, one can construct two real Lagrangians, one for each pair
${\mathbf \Xi}_{++}-{\mathbf \Xi}_{--}$ and ${\mathbf \Xi}_{+-}-{\mathbf \Xi}_{-+}$ 
\beqa
\label{free4}
{\cal L}_0&=&{\cal L}_0({\mathbf \Xi}_{++}) +{\cal L}_0({\mathbf \Xi}_{--} ) \nonumber \\
&=&\partial_m \varphi \partial^m \ttP + \frac{1}{12} H_{mnp}
{\tilde {\tilde H}}^{mnp} +\frac12 {}^\star H_m {}^\star {\tilde {\tilde
H}}^m -\frac14 \tilde F_{mn} \tilde F^{mn} 
-\frac12 \left(\partial_m \tilde A^m\right)^2 \nonumber \\
&+& \partial_m \varphi' \partial^m \ttP' + \frac{1}{12} H'_{mnp}
{\tilde {\tilde H}}^{\prime mnp} +\frac12 {}^\star H'_m {}^\star {\tilde {\tilde
H}}^{\prime m} -\frac14 \tilde F'_{mn} \tilde F'^{mn} 
-\frac12 \left(\partial_m \tilde A'^m\right)^2
\eeqa
and
\beqa
\label{4free2}
{\cal L'}_0&=&{\cal L}_0({\mathbf \Xi}_{-+}) +{\cal L}_0({\mathbf \Xi}_{+-} ) \nonumber \\
&=&\frac12 \partial_m \tP \partial^m \tP
+\frac{1}{24} \tilde H_{mnp} \tilde H^{mnp} -
\frac14  {}^\star \tilde H_m {}^\star \tilde H^m 
-\frac12 F_{mn} {\tilde {\tilde F}}^{mn} - 
(\partial_m A^m)(\partial_n \ttA^n)  \nonumber \\
&+&\frac12 \partial_m \tP' \partial^m \tP'
+\frac{1}{24} \tilde H'_{mnp} \tilde H'^{mnp} -
\frac14  {}^\star \tilde H'_m {}^\star \tilde H'^m 
-\frac12 F'_{mn} {\tilde {\tilde F'}}^{mn} - 
(\partial_m A'^m)(\partial_n \ttA'^n).
\eeqa
\noi
We consider here ${\cal L}_0$, ${\cal L}'_0$ being similar. 
Since we use complex conjugated terms, the Lagrangian \eqref{free4} is real; furthermore the Lagrangian is
of gradation $0$. 
Here we have denoted ${}^*H_m=\frac16 \e_{mnpq} H^{npq}=\partial^n B_{mn}$ the dual of the field strength $H$ (see \eqref{hodge} for the general definition).

To prove its invariance,  we first notice by a simple overview of \eqref{transfo2}, that the two lines of the Lagrangian \eqref{free4} 
do not mix under \3 transformation laws. Hence, we only prove the invariance of the first line of  \eqref{free4}, that is of ${\cal L}_0({\mathbf \Xi}_{++})$.
One can explicitly check the following general identities (up to total derivatives) for any $1-$form $\tA_m$ and for any $2-$form self-dual $B_{mn}$ and $\ttB_{mn}$ 
\beqa
\label{ajutor}
\frac 14 \tilde F_{mn}\tilde F^{mn}+\frac 12 (\partial_m \tA_m)^2&=&\frac 12 (\partial_m \tA_n)^2,\nonumber\\
  \frac{1}{12} H_{mnp}
{\tilde {\tilde H}}^{mnp} +\frac12 {}^\star H_m {}^\star {\tilde {\tilde
H}}^m &=& \frac 14 (\partial_m B_{np})(\partial^m \ttB^{np}).
\eeqa
\noi
Using these relations, one can rewrite ${\cal L}_0({\mathbf \Xi}_{++})$ using Fermi-like terms 
\beqa
\label{inv1}
{\cal L}_0({\mathbf \Xi}_{++})=\partial_m \varphi \partial^m \ttP-\frac 12 (\partial_m \tA_n)^2+\frac 14 (\partial_m B_{np})(\partial^m\ttB^{np}).
\eeqa
\noi
This is a more suited form to explicitly apply the \3 transformation laws \eqref{transfo2}. One thus obtains
\beqa
\label{inv2}
\delta_v {\cal L}_0({\mathbf \Xi}_{++})&=& \partial_m (v^p \tA_p)\partial^m \ttP + \partial_m \varphi (\partial^m v^p \partial_p \varphi) 
- \partial_m (v_n \ttP + v^p \ttB_{np}) \partial^m \tA^n \nonumber\\
&-&v_n\partial_m \tA_p \partial^m \ttB^{np}
+\frac 14 \partial_m B_{np} (\partial^mv^r\partial_r B^{np})
\eeqa
\noi
(where we have used the last identity of \eqref{dualprop1} to write $\frac 14 (\delta_v \partial_m B_{np})(\partial^m\ttB^{np})=-v_n\partial_m \tA_p \partial^m \ttB^{np}$). Since $\partial_m \varphi (\partial^m v^p \partial_p \varphi)=\frac 12 v^p\partial_p ((\partial_m \varphi)^2)$ and $\frac 14 \partial_m B_np (\partial^mv^r\partial_r B^{np})=\frac 18 v^r \partial_r ((\partial_m B_{np})^2)$ one has
$$\delta_v {\cal L}_0({\mathbf \Xi}_{++})=\mbox{total derivative},$$
and hence the Lagrangian \eqref{free4} is \3 invariant.

\medskip

Let us now perform  in \eqref{free4} the following change of variables 
\beqa
\label{real}
\tA_1=\frac{\tA+\tA^\prime}{\sqrt{2}} &,&
\tA_2=  i\frac{\tA-\tA^\prime}{\sqrt{2}}, \nonumber \\
B_1= \frac{B+B^\prime}{\sqrt{2}} &,& 
B_2= i\frac{B-B^\prime}{\sqrt{2}}, \nonumber \\
\ttB_1=  \frac{\ttB+\ttB^\prime}{\sqrt{2}} &,& 
\ttB_2=  i\frac{\ttB- \ttB^\prime}{\sqrt{2}}, \\
\varphi_1= \frac{\varphi+\varphi^\prime}{\sqrt{2}} &,& 
\varphi_2= i\frac{\varphi-\varphi^\prime}{\sqrt{2}}, \nonumber \\
\ttP_1=  \frac{\ttP+\ttP^\prime}{\sqrt{2}} &,& 
\ttP_2=  i\frac{\ttP- \ttP^\prime}{\sqrt{2}}. \nonumber
\eeqa 
\noi
The Lagrangian writes now
\beqa
\label{free4_2-0}
{\cal L}_0
&=&\partial_m \varphi_1 \partial^m \ttP_1 -
 \partial_m \varphi_2 \partial^m \ttP_2\nonumber\\ 
&+& \frac{1}{6} H_1{}_{mnp} {\tilde {\tilde H}}_1^{mnp} 
+ \partial^n B_1{}_{nm} \partial_p \ttB_1{}^{pm} 
- \frac{1}{6} H_2{}_{mnp} {\tilde {\tilde H}}_2^{mnp} 
- \partial^n B_1{}_{nm} \partial_p \ttB_1{}^{pm}
\nonumber \\
&-&\frac14 F_1{}_{mn}  F_1{}^{mn} 
+\frac14 {\tilde F}_2{}_{mn} {\tilde F}_2{}^{mn} 
-\frac12 \left(\partial_m \tilde A_1{}^m\right)^2 
+\frac12 \left(\partial_m \tilde A_2{}^m\right)^2.
\eeqa
\noi
Notice at this point that by the redefinition \eqref{real}, we find ourselves with $2-$forms $B_1, B_2, \ttB_1$ and $\ttB_2$ which are neither self-dual nor anti-self-dual. 
Indeed, it is in the theory of representations of the Lorentz group $SO(1,3)$ that the $2-$forms are (anti-)self-duals. In the case of the Poincar\'e group, whose little group is $SO(2)$, it is the $1-$forms which are (anti-)self-duals. 

Moreover, using the redefinition \eqref{real} and the definition \eqref{self} of
the (anti-)self-duality of  $2-$forms, one observes
\beqa
\label{B1-B2}
{}^\star B_1 = B_2, {}^\star \ttB_1 = \ttB_2
\eeqa
\noindent
Therefore, using the identities \eqref{4dd} for these $2-$forms, 
 one can eliminate two of them, for example $B_2$ and $\ttB_2$; thus ${\cal L}_0$  now becomes
\beqa
\label{free4_2}
{\cal L}_0
&=&\partial_m \varphi_1 \partial^m \ttP_1 -
 \partial_m \varphi_2 \partial^m \ttP_2 
+ \frac{1}{6} H_1{}_{mnp} {\tilde {\tilde H}}_1^{mnp} 
+ \partial^n B_1{}_{nm} \partial_p \ttB_1{}^{pm} 
\nonumber \\
&-&\frac14 F_1{}_{mn}  F_1{}^{mn} 
+\frac14 {\tilde F}_2{}_{mn} {\tilde F}_2{}^{mn} 
-\frac12 \left(\partial_m \tilde A_1{}^m\right)^2 
+\frac12 \left(\partial_m \tilde A_2{}^m\right)^2.
\eeqa

\noi

Proceeding with the analyses of the Lagrangian, one notices that the terms in the first line of (\ref{free4_2})
are not diagonal. For this purpose, we now define
\beqa
\label{diag}
\hat \varphi_1 = \frac{\varphi_1 + \ttP_1}{\sqrt{2}},
{\hat{ \hat \varphi}}_1 &=& \frac{\varphi_1 - \ttP_1}{\sqrt{2}},
\hat \varphi_2 = \frac{\varphi_2 + \ttP_2}{\sqrt{2}},
{\hat{ \hat \varphi}}_2 = \frac{\varphi_2 - \ttP_2}{\sqrt{2}}, \nonumber \\
\hat B_1 &=& \frac{B_1 + \ttB_1}{\sqrt{2}},
{\hat{ \hat B}}_1 = \frac{B_1 - \ttB_1}{\sqrt{2}},
\eeqa
\noi
and, with the new fields,  ${\cal L}_0$ writes
\beqa
\label{fermi2}
 {\cal L}_0 &=&
\frac12 \partial_m \hat \varphi_1 \partial^m  \hat \varphi_1
- \frac12 \partial_m {\hat {\hat \varphi}}_1 
\partial^m  {\hat {\hat \varphi}}_1 -
\frac12 \partial_m \hat \varphi_2 \partial^m  \hat \varphi_2
+ \frac12 \partial_m {\hat {\hat \varphi}}_2 
\partial^m  {\hat {\hat \varphi}}_2 \nonumber \\
&-&\frac14 \tilde F_1{}_{mn} \tilde F_1{}^{mn} 
+\frac14 \tilde F_2{}_{mn} \tilde F_2{}^{mn} 
-\frac12 \left(\partial_m \tilde A_1{}^m\right)^2 
+\frac12 \left(\partial_m \tilde A_2{}^m\right)^2 \nonumber \\
&+&\frac16 \hat H_1{}_{mnp} \hat H_1^{mnp} + \partial^n \hat B_1{}_{nm}\partial_p \hat B_1^{pm}
-\frac16 {\hat {\hat H}}_1{}_{mnp} {\hat {\hat H}}_1^{mnp} - \partial^n \hhB_1{}_{nm}\partial_p \hhB_1^{pm}.
\eeqa

Using again identities of type \eqref{ajutor}, one  writes the above Lagrangian as 
\beqa
\label{fermi}
 {\cal L}_0 &=&
\frac12 \partial_m \hat \varphi_1 \partial^m  \hat \varphi_1
- \frac12 \partial_m {\hat {\hat \varphi}}_1 
\partial^m  {\hat {\hat \varphi}}_1 -
\frac12 \partial_m \hat \varphi_2 \partial^m  \hat \varphi_2
+ \frac12 \partial_m {\hat {\hat \varphi}}_2 
\partial^m  {\hat {\hat \varphi}}_2 \nonumber \\
&-&\frac12 \partial_m \tA_1{}_n \partial^m \tA_1{}^n 
+\frac12 \partial_m \tA_2{}_n \partial^m \tA_2{}^n \\
&+&\frac14 \partial_m \hat B_1{}_{np} \partial^m \hat B_1{}^{np}
 -\frac14 \partial_m {\hat {\hat B}}_1{}_{np} \partial^m {\hat {\hat B}}_1{}^{np}.   \nonumber 
\eeqa
\noi

Also, using the differential form notations  (see \eqref{ext} for the definition of the exterior derivative $d$ and resp. \eqref{extdag} for its adjoint $d^\dag$)
one can write the Lagrangian as
\beqa
\label{lag-dif}
{\cal L}_0 &=&\frac 12  d \hat \varphi_1 d \hat \varphi_1- \frac 12  d{\hat {\hat \varphi}}_1 d{\hat {\hat \varphi}}_1 
 -\frac 12  d \hat \varphi_2 d \hat \varphi_2 +  \frac 12  d{\hat {\hat \varphi}}_2 d{\hat {\hat \varphi}}_2 \nonumber \\
&-& \frac 14 d  \tA_1  d  \tA_1 - \frac 12 d^\dag \tA_1 d^\dag \tA_1 +  \frac 14 d  \tA_2  d  \tA_2 + \frac 12 d^\dag \tA_2 d^\dag \tA_2 \nonumber\\
&+& 2(\frac{1}{12} d { {\hat B}}_1 d { {\hat B}}_1 +\frac12 d^\dag { {\hat B}}_1 d^\dag { {\hat B}}_1 - \frac{1}{12} d { \hat {\hat B}}_1 d {\hat {\hat B}}_1 - \frac12 d^\dag { \hat {\hat B}}_1 d^\dag {\hat {\hat B}}_1).
\eeqa

\medskip

Let us now consider the general gauge transformations
\beqa
\label{joaja-normala}
A_m&\to& A_m + \partial_m \chi,\nonumber\\
B_{mn}&\to&B_{mn}+\partial_m \chi_n-\partial_n \chi_m
\eeqa
\noi
where $\chi$ and $\chi_m$ are the gauge parameters. 
Hence one sees in the Lagrangian  \eqref{fermi2} the presence of kinetic terms and Feynman gauge fixing terms (of type $-\frac12 \left(\partial_m A^m\right)^2$ for a generic vector field $A$ or of type $\partial^n B_{nm}\partial_p B^{nm}$ for a generic $2-$form $B$). These gauge fixing terms are not just a {\it choice} of gauge, but they are {\it required by \3 invariance}. One can thus affirm, that {\it one symmetry, \3, fixes another, the gauge symmetry}.



The gauge fixing terms above (of type $-\frac12 \left(\partial_m A^m\right)^2$ and resp. $\partial^n B_{nm}\partial_p B^{pr}$)  imply some constraints on the gauge parameters defined in \eqref{joaja-normala}, namely
\beqa
\label{contrainte-joaja}
\partial^m \partial_m \chi&=& 0\nonumber\\
\partial^m (\partial_m \chi_n-\partial_n \chi_m)&=&0.
\eeqa
\noi
 We will come back on this issue of gauge transformation in subsection \ref{joaja-abeliana}.
The presence of this gauge fixing terms has a lot of consequences on different aspects of our models, as we will see in the rest of this thesis.

The first of them is related to the number of degrees of freedom of our fields. A $p-$form in $D$ dimensions has 
\beqa
\label{p-comp}
C_D^p
\eeqa
\noi
independent components. If one deals with a generic {\it free} $p-$form $\omega_{[p]}$, than the gauge transformation is
\beqa
\label{p-gauge}
 \omega_{[p]}\to \omega_{[p]} + d \chi_{[p-1]},
\eeqa
\noi
where the gauge parameter $\chi_{[p-1]}$ is a $(p-1)-$form. One can thus eliminate $C_D^{p-1}$ of the degrees of freedom of $\omega_{[p]}$. 
However, in addition, one can write a gauge transformation on  $\chi_{[p-1]}$ (reducible gauge transformation)
\beqa
\label{p-reductive}
 \chi_{[p-1]}\to \chi_{[p-1]} + d \chi'_{[p-2]}.
\eeqa
\noi
Thus, one needs now to add $C_D^{p-2}$ degrees of freedom to the count. The process continues and one has, for such a free $p-$form $ \omega_{[p]}$
\beqa
\label{p-off-shell}
C_{D}^p-C_D^{p-1}+C_D^{p-2}-\ldots +(-1)^p C_D^0 = C_{D-1}^p
\eeqa
\noi
degrees of freedom
 (see for example \cite{gomis}).  
Thus, such a free off-shell $p-$form has $C_{D-1}^p$ degrees of freedom.

Consider for example the well-known case of a photon, a $1-$form, in four dimensions, we have $C_3^1=3$ degrees of freedom. To find the well-known number of $2$ degrees of freedom for a {\it physical} photon, one uses Ward identities. Indeed, consider for example a simple physical process involving an external photon with momentum $k^m=(k,0,0,k)$. Denote the amplitude $i{\cal M}(k)=i{\cal M}^m (k)\e^*_m(k)$, where $\e^*_m(k)$ is the polarisation vector of the photon (the amplitude always contains the factor $\e^*_m(k)$, so we have just extracted it to define ${\cal M}^m (k)$). Using the classical equations of motion one is able to prove the Ward identity $k_m {\cal M}^m (k)=0$. This identity is used to simplify the square of the amplitude $|{\cal M}|^2$ which will depend only on the $2$ transverse polarisation states. (For a detailed analysis see for example \cite{peskin}).

For the general case of a on-shell $p-$form ($p\le D-2$) in $D$ dimensions, similar Ward identities lead to 
\beqa
\label{p-on-shell}
C_{D-2}^p
\eeqa
\noi
 physical degrees of freedom.

Let us also mentioned that $(D-1)-$ and $D-$forms in $D$ dimensions are known to be non-propagating forms.

\medskip

This is not the case for the \3 fields;  the gauge parameters 
are subject to constraints of type \eqref{contrainte-joaja}; thus one cannot eliminate anymore degrees of freedom of the $p-$form as was the case before. Let us illustrate this with the case of the $2-$form present here in $4$ dimensions, which has $6$ independent components. Its gauge parameter $\chi_m$ has $4$ components which must satisfy $4$ independent constraints
$$ \Box \chi_n -\partial_m \partial_n \chi^m=0.$$
Thus, we see that the components of $\chi_m$ are themselves constraint hence they will not be able to eliminate degrees of freedom of the $2-$form.
Generally in our case, the $p-$form has 
$$C_D^p$$
 degrees of freedom ($4$ for a vector field and $6$ for a $2-$form).
Related to this one might address here the issue of placing oneself in the Lorentz gauge. However this cannot be done. (Indeed, considering for example that we put $\partial_m \ttA^m=0$. Applying the \3 transformation, one has $\partial_m \delta_v \tA^m=0$ which, by \eqref{transfo2} gives $\partial_n A_m=0$ which is obviously not satisfactory.)

Hence, related to this subject of degrees of freedom counting, one may consider the problem of
other \3 compatible mechanisms of elimination of unphysical degrees of freedom.
A more elaborate analysis involving presence of ghosts (related to quantification issues) would be required (see section $6.4$).




\medskip

Another important aspect of the Lagrangian (\ref{fermi}) is that the fields ${\hat {\hat \varphi}}_1,{\hat { \varphi}}_2, \tA_2, {\hat {\hat B}}_1$ have wrong sign for their kinetic term. This implies {\it a priori} a problem of unboundedness from below of their potentials. 
To illustrate this consider the simplest case of a scalar $\p$.  
We use the definition $T_{00}=\frac{\partial {\cal L}}{\partial \partial^0 \p} \partial_0 \p - \eta_{00}{\cal L}$ of the energy density as component of the energy-momentum tensor. If the Lagrangian writes $\frac12 \partial_m \p \partial^m \p$, then $T_{00}=\frac12 [(\partial_0 \p)^2 + (\nabla \p)^2]$ which is bounded from below. If one considers an opposite sign in the Lagrangian, namely $-\frac12 \partial_m \p \partial^m \p$,, then $T_{00}=-\frac12 [(\partial_0 \p)^2 + (\nabla \p)^2]$
which is now unbounded from below.
This problem might by corrected by suited interaction terms. 
However, we will see in section \ref{interactiuni} that no self-interacting terms are allowed for these bosonic multiplets.
  
A possible solution to the problem of unboundness from below is based on Hodge duality of $p-$forms. 

\bigskip

{\bf Dualisation:} 
Recall that for the situation of free forms, dualisation is performed at the level of the field strength; this implies an equivalence of the theories of a $p-$form and a $(D-2-p)-$form (in $D$ dimensions). 
Indeed, starting from a generic $p-$form $\omega_{[p]}$, one considers its field strength, $d\omega_{[p]}$ which is $(p+1)-$form. Considering its Hodge dual, we have a $(D-p-1)-$form, which is the field strength of $(D-p-2)-$form.
One can see that these theories are equivalent, having the same number of physical degrees of freedom
$$C_{D-2}^p=C_{D-2}^{D-p-2}.$$
One can write this schematically as
$$\xymatrix{
   F_{[p+1]}=d\omega_{[p]} & \cong & ({}^*F)_{[D-p-1]}\ar[d]\\
   \omega_{[p]} \ar[u] & & \omega'_{[D-p-2]}
}              
$$
Obviously, this is not true  in the case of {\it interacting} $p-$forms.

Let us also mention here some issues concerning the electric-magnetic duality. Thus, the Bianchi identity writes $d F=0$, whereas the equation of motion involving  its dual ${}^*F$ writes $d{}^*F={}^*J$, where $J$ is the current $J_m=(\rho, \overrightarrow j)$. Notice that the asymmetry between the equations for $F$ and for ${}^*F$ corresponds physically to the absence of magnetic monopoles. When dualising in this way, the place of the Bianchi identity is taken by the equation of motion and {\it viceversa} and one also reverses the role of particles and solitons.

\medskip

Differently, the dualisation we propose for the \3 case  is not performed at the level of the field strength as above but at the level of the potential itself, {\it i.e.} we  replace the potential by its Hodge dual. 
Indeed,
as we have argued before, the gauge fixation present in the \3 model freezes the number of degrees of freedom of a $p-$form at $C_{D}^p$. 

To simplify notations we use here differential forms notations (see subsection $6.1$ and Appendix C) giving the following identities
\beqa
\label{4dd}
\frac{1}{(p+1)!} d A_{[p]}  d A_{[p]} 
&=&- \frac{1}{(4-p-1)!} d^\dag  B_{[4-p]}  
d^\dag   B_{[4-p]}, \nonumber\\ 
\frac{1}{(p-1)!} d^\dag A_{[p]}  
d^\dag A_{[p]} &=& - \frac{1}{(4-p+ 1)!} d B_{[4-p]}  d  B_{[4-p]} 
\eeqa 
\noi
with $B_{[4-p]}= {}^\star A_{[p]}$; this is actually a particular case, in $4$ dimensions, of identities \eqref{dd}. We thus see that, if we pass to the Hodge dual ($A_{[p]} \to B_{[4-p]}$), the kinetic term of $A_{[p]}$ becomes the gauge fixing term of its Hodge dual and {\it viceversa}. Moreover, one has a global change of sign, property witch remains true in arbitrary dimension. Thus one is tempted to replace the fields with a wrong sign in the Lagrangian (\ref{lag-dif}) by their Hodge duals, using the identities above.
The new fields are
\beqa
\label{dual}
\begin{array}{ll}
{\hat {\hat D}}_1= {}^\star {\hat {\hat \varphi}}_1,
{\hat { D}}_2= {}^\star  {\hat { \varphi}}_2, 
&\mbox{$4-$forms,}\cr
 \tilde C_2={}^\star \tA_2, & \mbox{$3-$form,} \cr
{\hat {\hat {\cal B}}}_1={}^\star{\hat {\hat B}}_1,
& \mbox{$2-$form,} 
\end{array} 
\eeqa

\noi
These new fields have an appropriate sign of their kinetic term.

Thus, at the very end, the free Lagrangian for the conjugated multiplets ${\mathbf \Xi}_{++}-{\mathbf \Xi}_{--}$ becomes
\beqa
\label{Lagrangien_libre}
{\cal L}_0&=&
\frac12 d \hat \varphi_1 d \hat \varphi_1  
+\frac12  d {\hat {\hat \varphi}}_2 d {\hat {\hat \varphi}}_2 \nonumber \\
&-& \frac14 d \tA_1 d \tA_1 -\frac12 d^\dag \tA_1 d^\dag \tA_1 \nonumber \\
&+&\frac{1}{12} d {\hat  B}_1 d {\hat  B}_1 + \frac12 d^\dag {\hat B}_1 
d^\dag {\hat B}_1 +
\frac{1}{12} d {\hat {\hat {\cal B}}}_1 d {\hat {\hat {\cal B}}}_1 
+ \frac12 
d^\dag {\hat {\hat {\cal B}}}_1 d^\dag {\hat {\hat {\cal B}}}_1
 \\
&-& \frac{1}{48} d \tilde C_2 d \tilde C_2  - \frac{1}{4} 
d^\dag \tilde C_2 d^\dag \tilde C_2  \nonumber \\
&+&\frac{1}{12} d^\dag {\hat {\hat D}}_1  d^\dag {\hat {\hat D}}_1
+\frac{1}{12} d^\dag { {\hat D}}_2  d^\dag { {\hat D}}_2
\nonumber
\eeqa

The field content is now the following: in the sector of gradation $-1$ and $1$ two
$0-$forms ($\hat \varphi_1, {\hat {\hat \varphi}}_2$), 
two neither self-dual, nor anti-self-dual $2-$forms
(${\hat B}_1, {\hat {\hat {\cal B}}}_1$) and  two $4-$forms 
(${\hat {\hat D}}_1, \hat D_2$); in the zero-graded sector one
$1-$form $\tA_1$ and one $3-$form $\tilde C_2$. As one can see from the transformations performed, (\ref{real}) and  (\ref{diag}), these fields are mixtures of states belonging
to two conjugate multiplets  and 
also  mixtures of the
graded $(-1)-$ and the graded $1-$sectors. Let us also observe that, if for the $1-$, $2-$ and $3-$forms above we have kinetic and gauge fixing terms, in  the particular cases of the $0-$forms (resp. $4-$forms) we have only a kinetic (resp. gauge fixing) term. 
In \cite{siegel}, W. Siegel remarked that for a $4-$form, the gauge fixing term takes the form of the Lagrangian of a free massless scalar field, which is also the case in our model. Furthermore, the $3-$ and $4-$forms in $4$ dimensions are referred to  as non-propagating forms which is not the case in our model, as a consequence of the gauge fixation. Several classical or quantum properties of these objects had been further investigated (see for example \cite{34forme}). One can see this approach as some kind of generalisation of the study of $2-$forms in $4$ dimensions (see for example \cite{2forme, 2forme2}).
Finally, before ending this subsection recall that, as already stated above, for our \3 model some deeper investigation of  the presence of ghosts in connexion with unitarity issues and possible elimination of degrees of freedom in different \3 sectors may be further required.

\subsection{Coupling between interlaced multiplets}
\label{embedded}

So far we have analysed \3 invariant terms that arise from couplings of conjugated multiplets, ${\mathbf \Xi}_{++}-{\mathbf \Xi}_{--}$ and ${\mathbf \Xi}_{+-}-{\mathbf \Xi}_{-+}$. We now look closer to couplings between the pairs of interlaced multiplets. 
We prove that \3 allows quadratic coupling terms between these pairs.
 
 Starting the calculations with the fields given in (\ref{4-decomposition}), one can write the following $0-$graded, real coupling Lagrangian
\beqa
\label{coup}
&{\cal L}_{\mathrm{c}}& = {\cal L}_{\mathrm{c}}({\mathbf \Xi}_{++},{\mathbf \Xi}_{+-}) + {\cal L}_{\mathrm{c}}({\mathbf \Xi}_{--},{\mathbf \Xi}_{-+})
\nonumber \\
&=&\lambda \left(
\partial_m \varphi \ttA^\prime{}^m 
+\partial_m \ttP  A^\prime{}^m
-\partial_m \tA^m \tP^\prime - \partial_m \tA_n \tB^\prime{}^{mn}
+\partial^m  B_{mn}  \ttA^\prime{}^n
 +\partial^m \ttB_{mn} A^\prime{}^n
\right)  \nonumber \\
&+&
\lambda^\star \left(
\partial_m \varphi' \ttA{}^m 
+\partial_m \ttP'  A^m
-\partial_m \tA'^m \tP - \partial_m \tA'_n \tB^{mn}
+\partial^m  B'_{mn}  \ttA^n
 +\partial^m \ttB'_{mn} A^n
\right),  \nonumber \\
\eeqa

\noi
with $\lambda=\lambda_1 + i \lambda_2$ a complex  coupling constant
with mass dimension.

To study the \3 invariance of (\ref{coup}) one may study separately the invariance of ${\cal L}_{\mathrm{c}}({\mathbf \Xi}_{++},{\mathbf \Xi}_{+-})$ and ${\cal L}_{\mathrm{c}}({\mathbf \Xi}_{--},{\mathbf \Xi}_{-+})$ because they do not mix under \3 transformations (\ref{transfo2}). Up to total derivative, one has
\beqa
\label{var}
 \delta_v{\cal L}_{\mathrm{c}}({\mathbf \Xi}_{++},{\mathbf \Xi}_{+-}) =
-\frac14 \lambda \tB'^{mn}
 \left(v^r \partial_{[m} \ttB_{n]_-r} -v_{[m} \partial^r \ttB_{n]_-r}\right)
\eeqa

By (\ref{prop}), one can conclude that the Lagrangian (\ref{coup}) is \3 invariant.

\medskip

We now give a different method of proving this invariance, method that uses the properties of the derivative multiplets of subsection $6.3.5$. One writes the Lagrangian ${\cal L}_c ({\mathbf \Xi}_{++}, {\mathbf \Xi}_{+-})$ as
\beqa
\label{muie-michel}
-\frac{1}{\lambda} {\cal L}_c ({\mathbf \Xi}_{++}, {\mathbf \Xi}_{+-}) &=& \p \partial_m \ttA'^{m} + B_{mn}\partial^m \ttA'^n - \tA^m (\partial_m \tP' + \partial_n \tB'^{nm})\nonumber\\
&+&\ttP \partial_m A'^m + \ttB_{mn} \partial^m A'^n.
\eeqa
\noi
This Lagrangian is just the one expressed in \eqref{X++coup}, where the multiplet $\psi$ is identical to ${\cal D}{\mathbf \Xi}_{+-}$.  
We prove in subsection \ref{cuplaje-posibile} that, if the $\psi$ fields transform as a ${\mathbf \Xi}_{++}$ multiplet, then \eqref{muie-michel} is invariant. This is indeed the case, since, as already stated the role of the fields $\psi$ is played here by ${\cal D}{\mathbf \Xi}_{+-}$, which, as we have proven in subsection $6.3.5$, does transform as ${\mathbf \Xi}_{++}$ multiplet.

\subsection{Diagonalisation of the total Lagrangian}
\label{diagonalizarea-mare}

We have thus constructed two types of free Lagrangians, firstly by coupling conjugated multiplets and then by coupling interlaced multiplets. Since all these terms are allowed by \3 invariance, the total Lagrangian to be considered is 
\beqa
\label{free}
{\cal L}= {\cal L}_0 + {\cal L}^\prime_0 + {\cal L}_c,
\eeqa

\noi
where ${\cal L}_0$ and ${\cal L}'_0$ are given in (\ref{free4}) and resp. \eqref{4free2} and  
${\cal L}_c$ is given in (\ref{coup}). Since ${\cal L}$ is quadratic in the fields, we deal with a non-interacting theory and it should be possible, by field redefinitions to write the Lagrangian in a diagonal form.

In order to do this, we first perform the changes of variable (\ref{real}) and (\ref{diag}) that make  ${\cal L}_0$ explicitly real and diagonal. Obviously, the same redefinitions (keeping the same type of notations for the redefined fields) must be made for ${\cal L}'_0$. After all this, the field content is:\\
$6$ scalar fields,   $\hP_1,\hhP_1, \hP_2,\hhP_2$ 
(in ${\cal L}_0$), $\tP_1,\ttP_2$ (in ${\cal L}'_0$);\\
$6$ vector fields,   $\tA_1,\ttA_2$ (in ${\cal L}_0$), 
$\hA_1,\hhA_1, \hA_2,\hhA_2$ (in ${\cal L}'_0$);\\
$3$ two-forms        
$\hB_1, \hhB_1$ (in ${\cal L}_0$) $\tB_2$ (in ${\cal L}'_0$).
\noi

We thus have a total of $15$ independent fields. Expressed with these new fields, ${\cal L}$ decouples into $3$ distinct pieces, each of them having the exact same dependence on a set of $5$ fields (two scalars, two vectors and one $2-$forms, denoted generically by $\varphi_1,\varphi_2,A_1,A_2$ and $B$). This Lagrangian writes
\beqa
\label{L1/3}
{\cal L}(\varphi_1, \varphi_2, A_1, A_2, B)&=&
\frac12 (\partial_m \varphi_1)^2- \frac12 (\partial_m \varphi_2)^2- \frac12 
(\partial_m A_1{}_n)^2 +\frac12 (\partial_m A_2{}_n)^2
+\frac14 (\partial_m B_{np})^2 \nonumber \\
&+&\lambda_1\left(A_1{}^m \partial_m \varphi_1 + A_2{}^m \partial_m \varphi_2
 - B^{mn} \partial_m A_1{}_n - {}^\star B^{mn} \partial_m A_2{}_n
\right)\\
&+&\lambda_2\left(-A_2{}^m \partial_m \varphi_1 + A_1{}^m \partial_m \varphi_2
 + B^{mn} \partial_m A_2{}_n - {}^\star B^{mn} \partial_m A_1{}_n
\right). \nonumber 
\eeqa


Thus, for diagonalising ${\cal L}$ it is enough to work on ${\cal L}(\varphi_1, \varphi_2, A_1, A_2, B)$. Before proceeding, a few remarks about the terms appearing in the Lagrangian are in order to be done. Firstly, one can check that the gauge fixation of ${\cal L}_0$ and ${\cal L}'_0$ is still demanded by the terms of ${\cal L}_c$ (the last two lines of Eq. \ref{L1/3}). Indeed, if one looks at the $2-$form $B$, then terms of type $\frac12 B^{mn} F{}_{mn}$ fix the gauge, while terms of type ${}^\star B^{mn} F{}_{mn}$
are  gauge invariant. Indeed,  the gauge transformation writes (see \eqref{joaja-normala})
$$ B\to B + d \chi_{[1]}$$
with $\chi_{[1]}$ a general $1-$form (the gauge parameter). Hence, for the dual field ${}^* B$, the gauge transformations writes
$$ {}^* B \to {}^* B + {}^* d\chi_{[1]}$$
which since, in $4$ dimensions ${}^{**}\chi_{[1]}=\chi_{[1]}$ (see \eqref{star2}), writes further as
$$ {}^* B \to {}^* B + {}^* d{}^{**}\chi_{[1]}.$$
Using now the definition of the derivative $d^\dag$ (see section $6.1$) which states in $4$ dimensions that $d^\dag = -{}^* d{}^{*}$ and denoting $\chi_{[3]}=-{}^* \chi_{[1]}$, one has
$$ {}^* B \to {}^* B + d^\dag \chi_{[3]}.$$
We can now use this to check the gauge transformation of ${}^\star B^{mn} F{}_{mn}$. Since the field strength $F= d A$ is gauge invariant, one has
$$ {}^* B dA \to ({}^* B + d^\dag \chi_{[3]}) dA .$$
For this to be gauge invariant, one needs
$$ d^\dag \chi_{[3]} dA = 0 $$
which is equivalent to
$$ d^\dag d^\dag \chi_{[3]}=0 .$$
Since $d^\dag d^\dag = 0$, this identity is trivially satisfied, thus completing our proof.

\medskip

These terms, known as $BF-$terms are related to topological theories \cite{BF, BF2,BF3,BF4}. Nevertheless, this line of work is not the one used here (for example we have never been concerned with surface terms in any of our invariance calculations).

A last thing to notice here is that couplings like $A^m \partial_m \varphi$ present in \eqref{L1/3} are of Goldstone type. Usually, they are gauged away and are responsible for  appearance of mass. For illustration, consider an abelian example  with a complex scalar field $\p$ coupled both to itself and to a vector field $A_m$. The complex scalar field $\p$ is decomposed as $\p=\p_0 + \frac{1}{\sqrt{2}}(\p_1+i\p_2)$ where $\p_0$ is its nonvanishing vacuum expectation value. When one expands the Lagrangian about this vacuum states, he finds terms of type $A^m\partial_m \p_2$. One then makes a particular choice of gauge, namely the {\it unitary gauge}, where the scalar field $\p(x)$ becomes real-valued at every point $x$. With this choice, the unwanted coupling terms are eliminated from the theory. 
However, one sees that this mechanism cannot be applied in the case of our model since as we have already stated above, the gauge is partially fixed.

\medskip

After these remarks we now return to our diagonalisation calculus. We first express the Lagrangian in the Fourier space
\beqa
\label{lag-fourier}
\tilde {\cal L}&=&\frac12 p^2 \tP_1 (p) \tP_1 (-p)
-\frac12 p^2 \tP_2 (p) \varphi_2 (-p) \nonumber \\
&-&\frac12 p^2 \tA_1 (p) A_1 (-p) + \frac12 p^2 \tA_2 (p) \tA_2 (-p)
+\frac14 p^2 \tB (p) \tB (-p) \nonumber\\
&+&\frac 12 i \lambda_1 \left(\tA_1^m(p) p_m \tP_1(-p)-\tA_1^m (-p)p_m \varphi_1(p)\right)\nonumber\\
&+&\frac 12 i \lambda_1 \left(\tA_2^m(p) p_m \tP_2(-p)-\tA_2^m (-p)p_m \tP_2(p)\right)\nonumber\\
&-&\frac 12 i \lambda_1 \left(\tB^{mn}(p) p_m \tA_1{}_n(-p)- \tB^{mn}(-p) p_m \tA_1{}_n(p)\right)\nonumber\\
&-&\frac 12 i \lambda_1 \left({}^*\tB^{mn}(p) p_m \tA_2{}_n(-p)- {}^* \tB^{mn}(-p) p_m \tA_2{}_n(p)\right)\nonumber\\
&-&\frac 12 i \lambda_2 \left(\tA_2^m(p) p_m \varphi_1(-p)-\tA_2^m (-p)p_m \tP_1(p)\right)\nonumber\\
&+&\frac 12 i \lambda_2 \left(\tA_1^m(p) p_m \tP_2(-p)-\tA_1^m (-p)p_m \tP_2(p)\right)\nonumber\\
&+&\frac 12 i \lambda_2 \left(\tB^{mn}(p) p_m \tA_1{}_n(-p)- \tB^{mn}(-p) p_m \tA_1{}_n(p)\right)\nonumber\\
&-&\frac 12 i \lambda_2 \left({}^*\tB^{mn}(p) p_m \tA_2{}_n(-p)- {}^* \tB^{mn}(-p) p_m \tA_2{}_n(p)\right).
\eeqa
\noi
where the tilde denotes the Fourier transform (not to be confused with the tilde in
the fields we had until  (\ref{L1/3})).

The first step now is to complete a perfect square for the terms involving $\tA_1$. We have thus to define 
\beqa
\tA'_1{}_m(p)= \tA_1{}_m(p)  +
 \frac{ \lambda_1}{p^2}i p_m \tP_1(p) + \frac{  \lambda_2}{p^2}i p_m \tP_2(p)
+  \frac{ \lambda_1}{p^2}i p^r\tB_{rm}(p) + \frac{ \lambda_2}{p^2}ip^r
({}^\star \tB_{rm}(p)).
\eeqa
\noi
The next step is to complete a perfect square for the terms involving $\tA_2$; now we define
\beqa
\tA'_2{}_m(p)= \tA_2{}_m(p)  - \frac{ \lambda_1}{p^2}ip_m \tP_2(p) + 
\frac{ \lambda_2}{p^2}ip_m \tP_1(p)
+  \frac{ \lambda_2}{p^2}ip^r \tB_{rm}(p) - \frac{ \lambda_1}{p^2}ip^r
 ({}^\star \tB_{rm}(p)). 
\eeqa

\noi
The Lagrangian writes
\beqa
\label{L-diag}
\tilde {\cal L} &=&\frac12\left(p^2-(\lambda_2^2-\lambda_1^2)\right)
 \tP_1(p)\tP_1(-p) 
               - \frac12\left(p^2-(\lambda_2^2-\lambda_1^2)\right) 
\tP_2(p)\tP_2(-p)   \nonumber \\
&+& \lambda_1 \lambda_2   \left( \tP_1(p) \tP_2(-p) + \tP_2(p) \tP_1(-p)
 \right) \nonumber \\
&-& \frac12 p^2 \tA'_1{}_m(p)  \tA'_1{}^m(-p)    + \frac12 p^2 \tA'_2{}_m(p) 
 \tA'_2{}^m(-p) +
\frac14    p^2 \tB_{mn}(p)  \tB^{mn}(-p)  \\
&+&\frac12 \frac{1}{p^2} p^r p_s(\lambda_1^2 -\lambda_2^2)\left(
\tB_{rm}(p) \tB^{sm}(-p)- {}^\star \tB_{rm}(p) {}^\star \tB^{sm}(-p)\right)
 \nonumber \\
&+&\frac{\lambda_1 \lambda_2}{p^2}  p^r p_s
\left(\tB_{rm}(p) {}^\star \tB^{sm}(-p)+ {} \tB^{s m}(p) {}^\star 
\tB_{rm}(-p)\right), \nonumber
\eeqa

\noi

A final diagonalisation can be written on the $\tP$ part of (\ref{L-diag}). Thus, defining
\beqa
\tP'(p)=\tP(p)+\frac{\lambda_1 \lambda_2}{\frac12 (p^2-(\lambda_2^2-\lambda_1^2))} \tP_2(p),
\eeqa
the Lagrangian finally writes as
\beqa
\label{L-diag2}
\tilde {\cal L} &=&\frac12\left(p^2-(\lambda_2^2-\lambda_1^2)\right)
 \tP'_1(p)\tP'_1(-p) 
               - \frac12\left(p^2-(\lambda_2^2-\lambda_1^2)+\frac{\lambda_1^2 \lambda_2^2}{\frac12 (p^2-(\lambda_2^2-\lambda_1^2))} 
\right) \tP_2(p)\tP_2(-p)   \nonumber \\
&-& \frac12 p^2 \tA'_1{}_m(p)  \tA'_1{}^m(-p)    + \frac12 p^2 \tA'_2{}_m(p) 
 \tA'_2{}^m(-p) +
\frac14    p^2 \tB_{mn}(p)  \tB^{mn}(-p)  \\
&+&\frac12 \frac{1}{p^2} p^r p_s(\lambda_1^2 -\lambda_2^2)\left(
\tB_{rm}(p) \tB^{sm}(-p)- {}^\star \tB_{rm}(p) {}^\star \tB^{sm}(-p)\right)
 \nonumber \\
&+&\frac{\lambda_1 \lambda_2}{p^2}  p^r p_s
\left(\tB_{rm}(p) {}^\star \tB^{sm}(-p)+ {} \tB^{s m}(p) {}^\star 
\tB_{rm}(-p)\right), \nonumber
\eeqa

One can now remark that not all values of the  parameters $\lambda_1$ and $\lambda_2$ are allowed if we do not want  tachyons to be present. Some allowed values (which simplify considerably the Lagrangian (\ref{L-diag2})) are $\lambda_1=\lambda_2$ or $\lambda_1=0$. However one remarks a non conventional form of the kinetic term for the $2-$form $B$.
One final remark is that we have done this diagonalisation on the Lagrangian without the dualisation; thus the sign of the kinetic terms will not change; however the same calculation may be performed for the dualised Lagrangian \eqref{Lagrangien_libre}.

\bigskip

One more remark is to be done here. 
We have so far considered non-massive fields. However, invariant mass terms can be explicitly added to our Lagrangian, since, as we have already noticed, the representations we have been working with do not depend whether we deal with massless or massive multiplets. Moreover, as it has already been noticed in subsection \ref{3-ex}, $P^2$ is a Casimir operator and therefore all states in an irreducible representation must have the same mass. For example, for the ${\mathbf \Xi}_{++}$ multiplet, one has
\beqa
\label{mass-4}
{\cal L}[{\mathbf \Xi}_{++}]_{\mathrm{mass}}=
m^2(\varphi {\tilde {\tilde \varphi}} +\frac14 B^{mn} \ttB_{mn}
-\frac12 \tA_m \tA^m),
\eeqa

The different analysis done in these section for the massless multiplets do not  drastically change. For example, in the massive case one has no gauge invariance and the number of degrees of freedom does not change (one does not have gauge parameters to eliminate any degree of freedom).

Finally, looking at the transformation laws (\ref{transfo2}) one can easily see that a linear term
\beqa
\label{tadpole}
{\cal L}_\varphi= g \ttP
\eeqa
\noi
is invariant on its own.

\subsection{Abelian gauge invariance}
\label{joaja-abeliana}

We now look closer to the problematic of the compatibility of \3 and abelian gauge symmetry. We have seen that these two symmetries are intimately connected, in the sense that the \3 symmetry fixes the gauge symmetry, by the Feynman gauge fixing terms required in the Lagrangian (see subsections $5.4.1$ and $5.4.3$). 

Nevertheless, another question is entitled at this level. If one acts with the gauge transformation on a \3 multiplet will the result be a \3 multiplet? Or, schematically,
$${\mathbf \Xi} \overset{\mbox{gauge}}\longrightarrow {\mathbf \Xi'}?$$
So what one has to check is whether or not ${\mathbf \Xi'}$ is a \3 multiplet. Moreover we also find in what conditions the gauge parameters may form a \3 multiplet. Recall that this is the case for SUSY, where a gauge transformation sends a vector superfield $V$ to $V+\Phi+\Phi^\dag$, with $\Phi$ a chiral superfield (see for example \cite{sohnius}).


\medskip

{\it {\underline{I}}}. 
Let us firstly write the general gauge transformation one uses for the physical fields $\hat \p_1, \hat \p_2$, $\hhP_1, \hhP_2, \tA_1, \tA_2, \hat B_1, \hhB_1$ (with whom for example the Lagrangian \eqref{lag-dif} was expressed)
\beqa
\label{jauge-normale}
\hP_1 &\to& \hP_1 + \hat k_1\nonumber\\
\tA_{1m}&\to &\tA_{1m}+\partial_m \tilde \chi_1\nonumber\\
\hB_{1mn} &\to& {\hB}_{1mn}+\partial_m \hat \chi_{1n} - \partial_n \hat \chi_{1m}.
\eeqa
\noi
and similarly for the rest of the fields ($\hat k_1$ being some constant). 
Recall that these physical fields were obtained from linear redefinitions of the original fields (\ref{4-decomposition}). Hence one can write down gauge transformations of the  fields (\ref{4-decomposition}) also. For example, the vector field $\tA_1$ was obtained from the fields $\tA$ and $\tA'$ by the redefinition \eqref{real}, $\tA_1=\frac{1}{\sqrt 2}(\tA + \tA')$. Hence the gauge parameter $\tilde \chi_1$ of \eqref{jauge-normale} is written as $\tilde \chi_1=\frac{1}{\sqrt 2}(\tilde \chi + \tilde \chi')$ which will thus allow us to obtain the gauge parameters $\tilde \chi$ and $\tilde \chi'$ of the field $\tA$ and resp. $\tA'$.


Hence the gauge transformations for the original fields (\ref{4-decomposition}) thus write
\beqa \label{g-trans}
\phi &\to& \phi + k,  \nonumber \\
\tA_m &\to& \tA_m + \partial_m \tilde \chi,   \\
{ B}^{(\pm)}_{m n} &\to& { B}^{(\pm)}_{m n} + 
\partial_m \chi_n - \partial_n \chi_m  \mp i \varepsilon_{mnpq} \partial^p \chi^q\nonumber
\eeqa
\noindent
and similar for the rest of the fields. 
Note however that for the case of $2-$forms, the compatibility between the  transformations \eqref{jauge-normale} and \eqref{g-trans} needs a closer look.
Indeed, write in $p-$form notation the transformation \eqref{g-trans} for the self-dual $2-$form $B_{mn}$
\beqa
\label{g-trans-forme}
{B}^{} &\to& {B}^{} + 
d \chi_{[1]}S - i {}^* (d\chi_{[1]}).
\eeqa
\noi
Recall now that $B'=B^*$; one has
\beqa
\label{ttb'}
{B'} &\to& {B'} + 
d \chi^*_{[1]}+ i {}^* (d\chi^*_{[1]}).
\eeqa
\noi
(One should pay attention at the notations used for the dual and complex conjugation, that is ${}^*B$ denotes the dual of $B$ whereas $B^{*}$ denotes the complex conjugated of $B$.) 
Eq. \eqref{real} combined these two $2-$ forms in $B_1=\frac{1}{\sqrt 2}(B+B')$. Hence, for the real $2-$ form $B_1$ one obtains
\beqa
\label{b1}
B_1\to B_1+ \frac{1}{\sqrt 2} \left(d(\chi_{[1]}+\chi^*_{[1]})-i{}^* d(\chi_{[1]} -\chi^*_{[1]})\right)
\eeqa
\noi
Thus, one immediate solution for \eqref{b1} to be the gauge transformation \eqref{jauge-normale} for a $2-$ form, is to impose that the $1-$form $\chi_{[1]}$ is real.

However, we now prove that this compatibility can still be achieved even if  the $1-$form $\chi_{[1]}$ is complex. For this denote by 
$\lambda_{[1]}=\frac{1}{\sqrt 2} (\chi_{[1]}+\chi^*_{[1]})$ and 
$\lambda_{[3]}=-i\frac{1}{\sqrt 2}{}^* (\chi_{[1]} -\chi^*_{[1]})$. Notice that  $\lambda_{[1]}$ and $\lambda_{[3]}$ are real. By \eqref{star2} one has  ${}^{**}(\chi_{[1]} -\chi^*_{[1]})=\chi_{[1]} -\chi^*_{[1]}$ and using also  the definition $d^\dag={}^*d^*$ (see section $6.1$), \eqref{b1} writes
\beqa
\label{b11}
B_1\to B_1+ d\lambda_{[1]} + d^\dag\lambda_{[3]}.
\eeqa
\noi
We now prove that for a $2-$form $B_1$ one can write a gauge transformation as $B_1\to B_1 + d\lambda_{[1]}$, with $d^\dag d\lambda_{[1]}=0$ but also as  $B_1\to B_1 +d^\dag\lambda_{[3]}$ with $d d^\dag \lambda_{[3]}=0$ . Indeed, since $d^\dag d\lambda_{[1]}=0$ this means (by the Poincar\'e lemma) that there exists a $3-$form $\lambda_{[3]}$ such that $d\lambda_{[1]}=d^\dag\lambda_{[3]}$. Thus, the gauge transformation can be written as $B_1\to B_1 +d^\dag\lambda_{[3]}$. Moreover, adding this two equivalent types of gauge transformations one can write a ``general'' gauge transformation for $B_1$ as in \eqref{b11}. This means that \eqref{jauge-normale} and \eqref{g-trans} are compatible for the $2-$forms also even if the gauge parameter $\chi_{[1]}$ is complex.

\medskip

Let us now argue about the set of gauge parameters ${\mathbf \Lambda}=(k, \chi_m; \tilde \chi; \tilde {\tilde k}, \tilde {\tilde \chi}_m)$ (defined in \eqref{g-trans})
and the possibility to form a \3 multiplet. Suppose that ${\cal D}{\mathbf \Lambda}=(k,\partial_{[m} \chi_{n]_+}; \partial_m \tilde \chi; \tilde{\tilde k}, \partial_{[m} \tilde {\tilde \chi}_{n]+})$ transforms like a ${\mathbf \Xi}_{++}$ multiplet.
Then one has  $\delta_v \partial_m \tilde \chi=v_m \tilde {\tilde k} + v^n \partial_{[m}\tilde {\tilde \chi}_{n]+}$; writing in components one has  $\partial_0 \delta_v \tilde \chi=v_0 \tilde {\tilde k} + v^n \partial_{[0}\tilde {\tilde \chi}_{n]+}$, $\partial_1 \delta_v \tilde \chi=v_1 \tilde {\tilde k} + v^n \partial_{[1}\tilde {\tilde \chi}_{n]+}$ {\it etc.} and integrating on $x^0$ or $x^1$ {\it etc.} one obtains four distinct expressions for $\delta_v \tilde \chi$.

\medskip

We now show what are the constraints on the gauge parameters $k,\tilde \chi$ and $\chi_m$ of the the gauge transformations \eqref{g-trans} (constraints  imposed by the gauge fixing terms of Lagrangians (\ref{free4}) and \eqref{coup}).
As before, the case of the scalar and vector gauge parameter is simple, leading to 
$$\partial_m k = 0 \mbox{ and }\Box \tilde \chi = 0$$ 
For the case of a $2-$form ${ B}^{(\pm)}$, one checks separately the invariance under \eqref{g-trans-forme} of $(d{ B}^{(\pm)})^2$ and $(d^\dag { B}^{(\pm)})^2$. 
\footnote{Note that for the case of a real $2-$form the gauge parameter $\chi'_{[1]}$ defined in \eqref{jauge-normale} is from the same reason subject to the constraint  $d^\dag d \hat \chi_{[1]}=0$.}
For $(d{ B}^{(\pm)})^2$ to be invariant,  one needs 
\beqa
d^*d\chi_{[1]}=0.
\eeqa
\noi
This is equivalent to
\beqa
{}^*d^*d\chi_{[1]}=0
\eeqa
\noi
Using now the definition of $d^\dag$ in $4$ dimensions (see section $6.1$), namely $d^\dag={}^*d^*$ one sees that the gauge fixation condition writes
\beqa
d^\dag d \chi_{[1]}= 0.
\eeqa
\noi
Now, the invariance of $(d^\dag { B}^{(\pm)})^2$ leads furthermore to
$$d^\dag {}^* d \chi_{[1]}=0$$
which, using again the definition of $d^\dag$ leads to
$$d{}^{**} d \chi_{[1]}=0.$$
This is equivalent to
$$dd\chi_{[1]}=0$$
which is trivially satisfied (recall that $d^2=0$).

Hence, these constraints on a general gauge parameter $\chi_{[p]}$ write 
\beqa
\label{gen}
d^\dag d \chi_{[p]}= 0.
\eeqa
\noi
In our particular case, using component notations, one has
\begin{equation} \label{g-fix}
\partial_m k = 0, \;\; \Box \tilde \chi = 0, \;\; \Box \chi_n - \partial_n  \partial_m \chi^m = 0.
\end{equation}


The strategy we adopt here is to find explicit forms of the gauge parameters defined in \eqref{g-trans} and subject to the constraints \eqref{g-fix}. 

\medskip

{\it {\underline{II}}}. The second step of our analysis is to have some transformations of a \3 multiplet into a \3 multiplet of same type, transformation that can then be matched with the gauge transformation (\ref{g-trans}). Since in subsection \ref{deriv-mult} we have introduced the derivative multiplets (that transform like \3 multiplets), we can use them to do the job:
\beqa
\label{pe-plac}
{\mathbf \Xi}_{++}\to {\mathbf \Xi}_{++}+{\cal D}\,{\mathbf \Xi}_{+-}.
\eeqa
\noi
For instance,  one can use the derivative of a multiplet ${\mathbf \Xi}_{+-}=\big(\lambda_m, \tl, \tl_{mn}, \ttl_m \big)$, writing thus \eqref{pe-plac} as
\beqa  \label{g-trans1}
 \varphi &\to& \varphi + \partial_m \lambda^m  \nonumber \\
 B_{m n} &\to&  B_{m n} + \partial_m \lambda_n - \partial_n \lambda_m 
- i \varepsilon_{mnpq} \partial^p \lambda^q \nonumber \\
 \tA_m &\to& \tA_m + \partial_m \tl + \partial^n \tl_{nm} \\
 \ttP &\to& \ttP + \partial^m \ttl_m \nonumber \\
  \ttB_{m n} &\to& \ttB_{m n} + \partial_m \ttl_n - \partial_n \ttl_m 
- i \varepsilon_{mnpq} \partial^p \ttl^q\nonumber
\eeqa

\medskip

{\it {\underline{III}}}. The last step of this programme is to make (\ref{g-trans1}) a gauge transformation, that is to match it with (\ref{g-trans}) and the conditions (\ref{g-fix}).  Since these equations have practically the same form, it now become clear why we have chosen to work with gauge transformations of type \eqref{g-trans} and not the gauge transformations \eqref{jauge-normale} of the real fields since we work directly on the ${\mathbf \Xi}_{++}$.

First, remark that the  actual matching of these transformations  implies a non-trivial condition for the parameters of the transformation of the vector field, namely

\begin{equation} \label{g-conditions4}
\partial^n \tl_{nm} = \partial_m \chi
\end{equation} 

\noi
and, since  $\tl_{nm}$ is antisymmetric, one has
\begin{equation} \label{g-conditions4pp}
\Box \chi =0.
\end{equation}
\noi
Note that obviously \eqref{g-conditions4pp} does not imply \eqref{g-conditions4}. Nevertheless, as we will see in the sequel here we will first find solutions for $\chi$ satisfying \eqref{g-conditions4pp} and then we will find solution for $\tl_{nm}$ satisfying \eqref{g-conditions4}.


Now, imposing the conditions $(\ref{g-fix})$ on the set of gauge parameters $\big(\lambda_m; \tl, \tl_{mn}; \ttl_m \big)$, one has 
\beqa
&& \partial_m \; (\partial \cdot \lambda) = \partial_m \; 
(\partial \cdot \ttl) = 0  \label{g-conditions1}\\
&& \Box \lambda_m = \Box \ttl_m =0
 \label{g-conditions2}\\
&& \Box \tl + \partial^m \partial^n \tl_{nm} \equiv \Box \tl = 0.
\label{g-conditions3} 
\eeqa 

   Now we have to explicitly find solutions of the gauge parameters which are compatible with all these constraints. In order to do this, 
\begin{enumerate}
\item we determine the
solutions for $\tl, \chi$ satisfying the constraints 
 (\ref{g-conditions3}, \ref{g-conditions4pp}). 
\item knowing $\chi$, 
we then construct   an anti-self-dual $2-$form $\tilde \lambda_{mn}$ 
satisfying (\ref{g-conditions4}).
\item
we finally find explicit solutions for  $\lambda_m, \ttl_m$
satisfying (\ref{g-conditions1}, \ref{g-conditions2}).  
\end{enumerate}

The existence of solutions to these constrains  would prove at this step the compatibility between \3 and the abelian gauge invariance. 

\medskip

\noi
$1.$ Now, if  the scalar functions 
$\tl$ and $\chi$ depend only on the space-time Lorentz  invariant $x_m x^m$,
then the conditions (\ref{g-conditions3}, \ref{g-conditions4pp}) determine
uniquely their form, $\tl(x^2) \propto \chi(x^2) \propto 1/x^2$ up to
some additive constants.
In the context of \3, whose generators and transformation parameters are $4-$vectors, it is somewhat natural to include dependence of  
a $4-$vector $\xi_m$. Moreover, by analysing  solutions of \eqref{g-conditions3} and \eqref{g-conditions4pp} when $\xi^2$ is equal or different of $0$, we find more general allowed configurations for $\lambda$ and $\chi$ when $\xi^2=0$ (see \cite{io1}). Hence we explicitly treat this case in this subsection.


We thus have to treat an equation of type
\beqa
\label{ec-scalar}
\Box f (x^2, x \cdot \xi)= 0
\eeqa
\noi
where $f$ denotes generically $\tilde \lambda$ or $\chi$. 
To solve equation \eqref{ec-scalar} let us firstly denote by
\beqa
\hat f = \frac {\partial f}{\partial x^2},\ f' = \frac {\partial f}{\partial x\cdot \xi}
\eeqa
\noi
the first derivatives of $f$ with respect to its variables. Hence one has
\beqa
\label{prima-der}
\partial^n f = \xi^n f' + 2 x^n \hat f.
\eeqa
\noi
Now, acting with $\partial_n$ on \eqref{prima-der} one obtains
\beqa
\label{eqI}
\Box f = 4 {x^2} {\hat {\hat f}} + 4 x \cdot \xi \hat f' + 8 \hat f =0.
\eeqa
\noi
Acting again with $\partial^n$ on \eqref{eqI} one has
\beqa
\label{eqII-III}
(4 x\cdot \xi \hat f''  + 4 x^2 \hat {\hat f}' + 12 \hat f')\xi^n 
+ ( 8 \xi \cdot x \hat {\hat f}'' + 24 \hat {\hat f} + 8 x^2 \hat {\hat {\hat f}} ) x^n=0.
\eeqa
\noi
Since $\xi$ and $x$ are independent variables, one has
\beqa
\label{eqII}
x\cdot \xi  \hat f'' + x^2 {\hat {\hat f}}' + 3 \hat f'=0\\
\label{eqIII}
\xi \cdot x \hat {\hat f}'' + 3\hat {\hat f} + x^2 \hat {\hat {\hat f}}=0.
\eeqa
\noi
Now, multiplying equations \eqref{eqII} and \eqref{eqIII} with $x\cdot\xi$ one has, together with \eqref{eqI} a system of three equations in the variables $(x\cdot \xi) \hat f', x^2 {\hat {\hat f}}, x^2 (x\cdot \xi) {\hat {\hat f}}'$. ``Solving'' this system one gets 
\beqa
\label{ABC}
(x\cdot \xi) \hat f'&=&-\frac16 (6 \hat f - (x^2)^2 {\hat {\hat {\hat f}}}+ (x \cdot \xi)^2 {\hat f}'')\\
x^2 {\hat {\hat f}}&=&- \frac16 ( 6 \hat f + (x^2)^2 {\hat {\hat {\hat f}}}-  (x\cdot \xi)^2 {\hat f}''  \\
x^2 (x\cdot \xi) {\hat {\hat f}}'&=&-\frac12 (-6\hat f +(x^2)^2 {\hat {\hat {\hat f}}}+ (x\cdot \xi)^2 {\hat f}''. 
\eeqa
\noi
Integrate now the first equation of \eqref{ABC} with respect to the variable $x^2$.  One gets
\beqa
\label{A1}
(x\cdot \xi) f'&=&- f +\frac16\int (x^2)^2 {\hat {\hat {\hat f}}} d(x^2)-\frac16(x \cdot \xi)^2 {f}' + K_1 (x\cdot \xi)
\eeqa
\noi
where $K_1$ is an arbitrary function of $x\cdot \xi$.
By a double integration by parts  one gets $\frac16\int (x^2)^2 {\hat {\hat {\hat f}}} d(x^2)= \frac 16 (x^2)^2 - \frac13 x^2 \hat f +\frac13 f.$ 
Hence \eqref{A1} writes
\beqa
\label{A'}
f=-\frac32 (x\cdot\xi) f' - \frac14 (x\cdot \xi)^2 f'' - \frac12 x^2 {\hat f}  + \frac14 (x^2)^2 {\hat {\hat f}} + \frac 32 K_1 (x\cdot \xi).
\eeqa
\noi
Applying the same type of treatment to the second equation of \eqref{ABC}, one gets
\beqa
\label{B'}
f= - 2 x^2 \hat f -\frac12 (x^2)^2 {\hat {\hat f}} + \frac12 (x\cdot \xi)^2 f'' +  K_2 (x\cdot \xi)
\eeqa
where $K_2$ is an arbitrary function of $x\cdot \xi$.

Now multiply \eqref{A'} by $2$ and add this to \eqref{B'}; one gets
\beqa
\label{separabila}
f=-(x\cdot \xi)f' - x^2 \hat f + K (x\cdot \xi)
\eeqa
\noi
where $K= 3K_1+ K_2$. If in this equation one considers $K=0$, then the solution is given by $f=(\xi \cdot x)^{-1} 
H(\frac{x^2}{(\xi \cdot x)})$ where $H$ is an arbitrary function. Furthermore considering $K\ne 0$, one may add to this solution a general term $ G(\xi \cdot x)$.

Thus, a solution of \eqref{ec-scalar} is given by
\beqa
\label{sol-scalar}
f (x^2, \; \xi \cdot x) = G(\xi \cdot x) + (\xi \cdot x)^{-1} 
H(\frac{x^2}{(\xi \cdot x)})
\eeqa
\noi
where G and H are arbitrary functions. 

To check that \eqref{sol-scalar} satisfies the requested constraint one makes use of partial derivatives formulae of type
$$ \Box \frac{1}{(x^2)^k}=4k (k-1)\frac{1}{(x^2)^{k+1}},$$
where $k\in \NN^*$.

Thus, solution \eqref{sol-scalar} provides us with an explicit form for the parameters $\tilde \lambda$ or $\chi$
\begin{equation} \label{eqtl}
\tl(\xi \cdot x, x^2) = G_1(\xi \cdot x) + 
(\xi \cdot x)^{-1} H_1(\frac{x^2}{(\xi \cdot x)}),
\end{equation}

\begin{equation} \label{eqchi}
\chi(\xi \cdot x, x^2) = G_2(\xi \cdot x) + 
(\xi \cdot x)^{-1} H_3(\frac{x^2}{(\xi \cdot x)}), 
\end{equation}
thus, completing step $1$ of the programme.

 
\medskip

\noi  
$2.$ As already stated, we now have to find a form of $\tl_{mn}$ which satisfies the constraint (\ref{g-conditions4}), with $\chi$ given by (\ref{eqchi}) above. A possible solution is 

\begin{equation} \label{eqtlmn}
\tl_{m n}(\xi \cdot x, x^2) = x_{[m} \xi_{n]_-} F(\xi \cdot x,  x^2) 
\end{equation}

\noi
where the function $F$ can be expressed in terms of $G_2, H_3$
appearing in (\ref{eqchi}). One has 
\begin{equation}
F(\xi \cdot x, x^2) = - (\xi \cdot x)^{-2} H_3(\frac{x^2}{(\xi \cdot x)})
+ (\xi \cdot x)^{-1} G_2(\xi \cdot x)
 -2(\xi \cdot x)^{-3} \int \limits_0^{\xi \cdot x} G_2(t)\; t \; dt 
\end{equation}

\medskip

\noi
$3.$ Similarly to step $1$, we  now investigate possible solutions for 
the gauge parameters $\lambda_m, \ttl_m$, which satisfies equations \eqref{g-conditions1} and \eqref{g-conditions2}. We consider them as functions of the vectors $x$ and $\xi$ and, as before, we assume $\xi^2=0$.
Hence, the problem is reduced to finding explicit solutions of equations of type
\beqa
\label{ec-vector}
\partial_m \partial_p {\cal A}^p (x,\xi)&=&0 \nonumber\\
\Box {\cal A}_m (x,\xi)&=&0
\eeqa
\noi
(which are just equations \ref{g-conditions1}) and (\ref{g-conditions2} and $\cal A$ stands for $\lambda_m$ or ${\tilde {\tilde \lambda}}_m$). As before, some solution of these equation is given by
\beqa
\label{sol-vector}
 {\cal A}_m (x, \xi)= g(\xi \cdot x) \xi_m + \alpha x_m + 
(\frac{1}{(x^2)^2} \alpha_{m r} + \beta_{m r}) x^r +
\kappa (\frac{x^2}{(\xi \cdot x)^3} \xi_m - \frac{x_m}{(\xi \cdot x)^2})
\eeqa
\noi
where $g$ is an arbitrary function, $\kappa, \alpha, \beta_{m n}$ arbitrary
constants and $\alpha_{m n}$ an arbitrary anti-symmetric tensor.

Thus one can now write the following expressions for the last gauge parameters $\lambda_m$ or ${\tilde {\tilde \lambda}}_m$
\beqa \label{eqlambdam}
\lambda_m (\xi \cdot x,  x^2) &=& g_1(\xi \cdot x) \xi_m + 
 \alpha x_m + (\frac{1}{(x^2)^2} \alpha_{m r} + \beta_{m r}) x^r \nonumber \\
&&
 +\kappa_1 (\frac{x^2}{(\xi \cdot x)^3} \xi_m - \frac{x_m}{(\xi \cdot x)^2}), 
\eeqa

\beqa \label{eqttlm}
\ttl_m (\xi \cdot x, x^2)& =& \tilde{\tilde g}_1(\xi \cdot x)
 \xi_m  + 
\tilde{\tilde \alpha} x_m + (\frac{1}{(x^2)^2} \tilde{\tilde \alpha}_{m r} +
 \tilde{\tilde \beta}_{m r}) x^r \nonumber \\
&& + \tilde{\tilde \kappa}_1 (\frac{x^2}{(\xi \cdot x)^3} \xi_m - 
\frac{x_m}{(\xi \cdot x)^2}). 
\eeqa
\noindent

  We have thus obtained the prove of existence of gauge transformations which are compatible with \3 symmetry. Nevertheless, we have found special forms of our gauge parameters $\lambda_m$, $\tl$, $\tl_{mn}$ and resp. $\ttl_m$ (equations \eqref{eqlambdam}, \eqref{eqtl}, \eqref{eqtlmn} and resp. \eqref{eqttlm}).

\subsection{Generators of symmetries and Noether currents}
\label{noether}

Before going further to the analysis of possibilities of interaction we make here a short discussion related to generators of symmetries and Noether currents for \3.

In \cite{sohnius}, M. Sohnius argues about generators of symmetries. Such generators $G$ can be written as the product of an annihilation operator $a$ (which annihilates a particle of momentum $\overrightarrow q$) and a creator operator $a^\dag$ (which creates a particle of momentum $\overrightarrow p$)
\beqa
\label{generator}
G=\sum_{ij}\int d^3 p d^3 q \, a_i^\dag (\overrightarrow p)K_{ij} (\overrightarrow p, \overrightarrow q) a_j (\overrightarrow q)
\eeqa
\noi
where 
$K_{ij}$ is the integral kernel. Equation \eqref{generator} writes symbolically
\beqa
\label{gen-sim}
G=a^\dag * K * a.
\eeqa
\noi
Such generators can replace a boson with a boson or  a fermion and a fermion with a fermion or  a boson. Hence, $G$ can be decomposed into an even part $B$ (containing the terms which replace bosons by bosons and fermions by fermions) and an odd part $F$ (containing the terms which replace bosons by fermions and fermions by bosons)
$$G=B+F$$
with
\beqa
\label{gen-decompo}
B&=&b^\dag * K_{bb} * b + f^\dag * K_{ff}*f,\nonumber\\
F&=&f^\dag * K_{fb} * b + b^\dag * K_{bf}*f.
\eeqa
\noi
where $b$ (resp. $f$) denote generically annihilation operators for bosons (resp. fermions).

Finally, assuming canonical commutation relations for the particle operators $b$ and $f$, one finds that the symmetry generators $B$ and $F$ obey to the (anti)commutation laws of a Lie superalgebra.

\medskip

However, in the case of \3 this interpretation is lost. In the model considered here we have only bosonic generators (the Poincar\'e generators and the \3 generators $V$) and furthermore only bosonic fields (or only fermionic  fields). Thus, considering canonical commutation relations for the particle operators $b$, one cannot express the bracket $\{V,V,V\}$ as a proper symmetry generator as above. Non conventional aspects are also present at the level of the algebra of conserved charges (see below); nevertheless, one is still able to realise the algebra \eqref{algebra} as in \eqref{realizare-caca}, as we will further explain.

\medskip

Let us now consider some issues related to  Noether theorem; we begin by recalling some existing results in field theory.  At classical level, one has a general symmetry transformation generated by some $Q_m$ (whose conserved charge is denoted by $\hat Q_m$) and parametrised for example by a vector parameter $v_m$
\beqa
\label{sym-clasic}
\delta_v \Phi = \{v^m \hat Q_m, \Phi\}_{P.B.}
\eeqa
\noi 
where $\Phi$ is any field present in the model and $\{.,.\}_{P.B.}$ is the  Poisson bracket.

Recall that the symmetry transformations  form a Lie algebra whose law is the usual commutator
\beqa
\label{interme}
[\delta_a, \delta_b]\Phi\equiv\delta_a(\delta_b\Phi)-\delta_a(\delta_b\Phi)=f_{ab}^c \delta_c \Phi
\eeqa
\noi
where $f_{ab}^c$ are the structure constant of the Lie algebra which was associated with the generators $V$.
Inserting \eqref{sym-clasic} in \eqref{interme} one gets
\beqa
\label{se-poate-cl}
\{ \hat Q_a, \{\hat Q_b, \Phi \}_{P.B.}\}_{P.B.}- \{ \hat Q_b, \{\hat Q_a, \Phi \}_{P.B.}\}_{P.B.}=f_{ab}^c \{\hat Q_c, \Phi\}_{P.B.} 
\eeqa
\noi
Using now the Jacobi identity, one has
\beqa
\label{nu-se-poate-1}
\{\{\hat Q_a, \hat Q_b\}_{P.B.},\Phi\}_{P.B.}=f_{ab}^c\{\hat Q_c, \Phi\}_{P.B.}
\eeqa
\noi
which further writes
\beqa
\label{rez-clasic}
\{ \hat Q_a, \hat Q_b \}_{P.B.} = f_{ab}^c \hat Q_c.
\eeqa
\noi
Hence, the conserved charges form a Lie algebra whose law is represented by the Poisson bracket.  
Moreover, the algebra of transformations \eqref{interme} is isomorphic to the algebra of conserved charges \eqref{rez-clasic}.

\medskip

The transition to the quantum case is performed by replacing the Poisson brackets 
 with the usual commutator, following the canonical quantisation procedure.

Thus, instead of \eqref{sym-clasic} one has
\beqa
\label{sym-q}
\delta_v \Phi = [v^m \hat Q_m, \Phi]
\eeqa
\noi 
As in the classical case,  the symmetry transformations  form a Lie algebra \eqref{interme}. Now, inserting \eqref{sym-q}  one gets
\beqa
\label{se-poate-q}
[ \hat Q_a, [\hat Q_b, \Phi ]]- [ \hat Q_b, [\hat Q_a, \Phi]]=f_{ab}^c [\hat Q_c, \Phi] 
\eeqa
\noi
which, using  as above the Jacobi identity, leads to
\beqa
\label{nu-se-poate-2}
[[\hat Q_a, \hat Q_b],\Phi]=f_{ab}^c[\hat Q_c, \Phi].
\eeqa
\noi
This gives
\beqa
\label{rez-q}
[ \hat Q_a, \hat Q_b] = f_{ab}^c \hat Q_c
\eeqa
\noi
which is the quantum correspondent of the result \eqref{rez-clasic}.

\medskip

In the \3 case the situation is however different (see \cite{articol}). Thus, even though Lie algebras of order $3$ do have their own Jacobi identity, they do not allow to obtain conventional results like \eqref{rez-clasic} (resp. \eqref{rez-q}). Nevertheless, the original \3 algebra is realised in a different manner. Indeed, 
classically one has
$$\left\{ \hat V_m \left\{\hat V_n, \left\{ \hat V_r, \Phi \right\}_{P.B.}\right\}_{P.B.}\right\}_{P.B.}+\mbox{ all perm.}=\eta_{mn}\left\{ \hat P_r+\eta_{mr} \hat P_n+\eta_{rn} \hat P_m, \Phi\right\}_{P.B.}$$
However, one does not have the analogous of \eqref{rez-clasic}.
Furthermore, writing $\delta_m\Phi ={\rm ad}(\hat V_m) (\Phi)= [\hat V_m, \Phi]$ (thus placing in the adjoint representation) one has relations of type
\beqa
\label{realizare-caca}
(\delta_m \delta_n \delta_r +\mbox{all perm.})\Phi&=&\nonumber\\
\left[ \hat V_m \left[ \hat V_n, \left[ \hat V_r, \Phi \right]\right]\right]+\mbox{ all perm.}&=&\eta_{mn}\left[ \hat P_r, \Phi\right] +\eta_{mr}\left[ \hat P_n, \Phi\right]+\eta_{rn}\left[ \hat P_m, \Phi\right]
\eeqa
\noi
which is in fact the correspondent of relation \eqref{se-poate-cl} (resp. \eqref{se-poate-q}).

However,  one cannot write down the corresponding of \eqref{nu-se-poate-2}
\beqa
\label{caca-mare}
[\{\hat V_m, \hat V_n,\hat V_r\},\Phi]=[\eta_{mn} \hat P_r +\eta_{mr} \hat P_n+\eta_{rn} \hat P_m]
\eeqa

This means that one cannot have the $\Phi$ independent relations witch appear 
when doing physics with Lie (super)algebras. This may not be so surprising since the algebraic structure we use, a Lie algebra of order $3$, is not a conventional one, thus making us think of what a physical symmetry might look like when implemented on such structures.

In \cite{articol} it is also argued that one may however obtain $\Phi$ independent relations but on one-particle states; 
hence on acting on multi-particle Fock states, one needs to be more careful because these states will not be obtained by the usual tensor product of one-particle states, this being different from the conventional algebraic structures (see \cite{articol} for more details).

\bigskip

Recall now that generally, comparing $\delta_v {\cal L}$ with the result obtained by varying the fields, one has
\beqa
\label{cand}
\delta_v {\cal L}=\frac{\partial {\cal L}}{\partial \Phi}\delta_v \Phi + \frac{\partial {\cal L}}{\partial \partial_m \Phi}\partial_m \delta_v \Phi.
\eeqa
\noi
Now, writing the last term  as 
$$ \frac{\partial {\cal L}}{\partial \partial_m \Phi}\partial_m \delta_v \Phi= \partial_m \left(\frac{\partial {\cal L}}{\partial \partial_m \Phi}\delta_v \Phi\right) -\left( \partial_m \frac{\partial {\cal L}}{\partial \partial_m \Phi}\right) \delta_v \Phi $$
and using the equation of motion, equation \eqref{cand} simplifies to
\beqa
\label{cand2}
\delta_v{\cal L}=\partial_m \left(\frac{\partial {\cal L}}{\partial \partial_m \Phi}\delta_v X\right).
\eeqa
\noi
Now, if $\delta_v {\cal L} = v_n \partial^n K = v_n \partial_m \eta^{mn} K $ and $\delta_v \Phi = v_n \Phi'^n$ equation \eqref{cand2} becomes
\beqa
\label{conse}
v^m \partial^n J_{mn}=0
\eeqa
\noi
where the conserved current writes
\beqa
\label{conserved}
J_{mn}=\frac{\partial {\cal L}}{\partial \partial^m \Phi}\phi'_n -\eta_{mn}K.
\eeqa
\noi
Thus \eqref{conse} implies that $\partial^n J_{mn}=0$; furthermore the conserved charges write
 $\hat Q_m=\int d^3x J_{0m}$.

\medskip

Let us now look at the \3 situation in particular. One has the physical fields $\hat \Phi =(\hat \p_1, \hat \p_2$, $\hhP_1, \hhP_2, \tA_1, \tA_2, \hat B_1, \hhB_1)$ (with whom for example the Lagrangian \eqref{lag-dif} was expressed). 
 Recall that the canonical momentum for any such field $\hat \Phi$ is written
\beqa
\label{hcanonic}
\pi^{\hat\Phi}=\frac{\partial {\cal L}}{\partial \partial_0 \hat \Phi}.
\eeqa
\noi
Moreover one has the equal-time commutation relations
\beqa
\label{p-comm-ph}
[\pi^{\hat \Phi}(t, \overrightarrow y), \hat \Phi (t, \overrightarrow x)]= \delta^3 (\overrightarrow x- \overrightarrow y).
\eeqa
\noi
Let us now show that one can however impose this type of commutation relations also for the original fields $\Phi\in\{\p, B_{mn},\tA_m, \ttP, \ttB_{mn}\}$ (with whom for example the Lagrangian \eqref{inv1} was expressed).
 Indeed, the physical fields $\hat \Phi$ were obtained from the fields $\Phi$ above by the linear transformations \eqref{real} and \eqref{diag}. Let us generically write
\beqa
\label{legatura}
\Phi^i=U^i_j \hat \Phi^j.
\eeqa
\noi
Now, the associated canonical momentum for such a field $\Phi$ writes
\beqa
\label{canonic}
\pi^{\phi}=\frac{\partial {\cal L}}{\partial \partial_0  \Phi}
\eeqa
\noi
Inserting \eqref{legatura}, one has
\beqa
\label{hcanonic2}
\pi^{\phi}_i=\frac {\partial \hat \Phi^j}{\partial \Phi^i}\frac{\partial {\cal L}}{\partial \partial_0 \hat \Phi^j}=(U^{i}_j)^{-1}\pi^{\hat \Phi^j}.
\eeqa
\noi
Using again \eqref{legatura} one proves the requested formulae
\beqa
\label{p-comm}
[\pi^{ \Phi}(y),  \Phi (x)]= \delta^3 (\overrightarrow x- \overrightarrow y)
\eeqa
\noi

Thus we can work with the fields $\Phi\in\{\p, B_{mn},\tA_m, \ttP, \ttB_{mn}\}$ also; for them we will prove that
\beqa
\label{maretie}
\delta_v \Phi (x) = v^m[\hat V_m (y), \Phi(x) ].
\eeqa
\noi

Recall that the Lagrangian \eqref{inv1} writes
\beqa
\label{inv1-bis}
{\cal L}=\partial_m \varphi \partial^m \ttP-\frac 12 (\partial_m \tA_n)^2+\frac 14 (\partial_m B_{np})(\partial^m\ttB^{np}).
\eeqa
\noi
which gives under \3 (see subsection \ref{detoate})
\beqa
\label{var-l}
\delta_v {\cal L} = v^\rho \partial_\rho K
\eeqa
\noi
where
\beqa
\label{K}
K= \frac12 (\partial_m \p)^2 + \frac 18 (\partial_m B_{nr})^2.
\eeqa
\noi
Thus, inserting \eqref{K} into the expression of the conserved current \eqref{conserved} one obtains
\beqa
\label{curent}
J_{mn}&=&(\partial_m \ttP) \tA_n + (\partial_m \p) (\partial_n \p)
-(\partial_m \tA^r)(\eta_{rn} \ttP + \ttB_{rn})\nonumber\\
&+&\frac 14 (\partial_m B_{rp})(\partial_n B^{rp})
+\frac14 \partial_m \ttB_{rs}(-\eta^r_n\tA^s+\eta^s_n \tA^r + i\e^{rsab}\eta_{an} \tA_b) \nonumber\\
&-&\eta_{mn}\left(\frac12 (\partial_s\p)^2 + \frac 18 (\partial_r B_{st})^2)\right).
\eeqa

As usual the conserved charges are written as
\beqa
\label{usual}
\hat V_m (y)= \int d^3 y J_{0m}.
\eeqa

Obviously, a similar part of the conserved current can be written for ${\cal L}({\mathbf \Xi}_{--})$, the complex conjugated multiplet. The conserved current will have the same form as \eqref{curent} where one has to substitute $\p\to\p', A\to A'$ {\it etc.}

Let us thus prove \eqref{maretie}  firstly for the simpler case $\Phi=\p$. 
Moreover, it is subject to the equal-time commutation relation of type
\beqa
\label{eeqa}
[\pi^\Phi(t, \overrightarrow y), \Phi (t, \overrightarrow x)]= \delta^3 (\overrightarrow x- \overrightarrow y).
\eeqa
For $\Phi=\p$ this gives
\beqa
\label{p-phi}
\pi^\p (y)= \partial_0 \ttP
\eeqa
\noi
which obeys to
\beqa
\label{p-comm-p}
[\pi^\p  (y), \p (x)]= \delta^3 (\overrightarrow x - \overrightarrow y).
\eeqa
\noi
We now calculate 
\beqa
\label{interm0}
v^m[\hat V_m (y), \varphi (x)]
\eeqa
\noi
Inserting \eqref{usual} and \eqref{p-phi}, equation \eqref{interm0} becomes
\beqa
\label{interm}
v^m \int d^3 y \left[ (\pi^\p (y) \tA_m + \dots \, , \,\p (x) \right]
\eeqa
\noi
where the rest of the terms (not explicitly written here) involve terms in the canonical momentums associated to the rest of the fields, and which {\it have zero commutator with $\p$} (hence not being useful for this calculus). Making now use of \eqref{p-comm}, equation \eqref{interm} gives
$$ v^m \tA_m (x)$$
which completes the proof of  \eqref{maretie}.

Let us now look at the situation of the vector $\tA$. One has
$$ \pi^{\tA}_r= - \partial_0 \tA_r. $$
Notice that, because of our gauge fixation we have $\pi^{\tA}_0\ne 0$ which is differt from the usual, not gauge-fixed situation, which appears for example in the case of electromagnetism. As above, computing 
$$ v^n [\hat V_n (y), \tA_s (y)] $$
one obtains
$$ v^n \int d^3 y \left[ (\pi^{\tA}_r (y) (\eta_{rn} \ttP+\ttB_{rn}) + \dots \, , \,\tA_s (x) \right] = v_s \ttP (x) + v^n \ttB_{sn} (x) $$
which is indeed $\delta_v \tA_s (x)$.

Finally let us check this for $B_{mn}$. The momentum writes
$$ \pi^{B}_{rs}=\frac 14 \partial_0\ttB_{rs}$$
As above one has
$$ v^n \int d^3 y \left[ (\pi^{B}_{rs} (y) (-\eta^r_n\tA^s+\eta^s_n \tA^r + i\e^{rsab}\eta_{an} \tA_b) + \dots \, , \, B_{pq} (x) \right] = -v_p\tA_q+v_q \tA_p + i\e_{pqab}v^a \tA^b $$
which is equal to as $\delta_v B_{mn}$ (as expected). For the remaining fields, $\ttP$ and $\ttB$, the situation is analogous, {\it i.e.} equation \eqref{maretie} is verified similarly.




\medskip

Before ending this section, a final comment is to be done.
The additional symmetries $V$ even though they lie in the vector representation of the Lorentz algebra do not close with classical (anti)commutation relations. 
Usually in physics literature (for example in the case of superalgebras), one denotes generators that close with commutators as bosonic generators and to generators that close with anticommutators as fermionic generators. Obviously this is not the case here 
(this is how \3 evades the no-go theorems and apparently, the price to pay). Nevertheless, the physical fields $\varphi, A, B$ are bosons (and in the case of the fermionic multiplets in \cite{articol} they are fermions), thus obeying the conventional statistics; furthermore one obtains the correct transformations laws \eqref{transfo2} through the conventional equal-time commutation relations.
Technically, we got to this situation by the decomposition (\ref{4-decompozitie}) on $p-$forms.



\section{Interaction possibilities}
\label{interactiuni}

In the previous section we treated non-interacting terms that \3 allows for the bosonic multiplets $\Xi_{\pm \pm}$. Here we investigate the possibility of allowing interacting terms, terms which must have a degree in the fields higher than $2$ (thus not being possible to diagonalise them back to kinetic terms), thus not being possible to diagonalise them as we did in the previous section. 
The main result here is that for the bosonic multiplets considered here, no self-interaction terms are allowed.

To approach this issue we first make use of a tensor calculus adapted for \3. We then make a systematic study of all interaction possibilities for our multiplets.

\subsection{Tensor calculus for \3}
\label{tensor-calculus}

The tensor calculus is a technique that was successfully used for SUSY to construct interacting terms\footnote{This technique was used before the more elegant formulation of superspace, see for example \cite{sohnius}.}. Its  basic idea  is the following: starting from two multiplets, one quadratically constructs another multiplet; then the process can be reiterated to get higher order multiplets and thus also invariant terms. In the case of \3, one cannot build such a quadratic multiplet.

To implement this process, one obviously needs to consider all pairing of multiplets to quadratically obtain a third multiplet (which can be {\it a priori} of any type also). Here we explicitly treat the case when we couple two multiplets of type ${\mathbf \Xi}_{++}$ of field content $\big(\varphi_1, B_1,\tA_1, \ttP_1,\ttB_1\big)$ and resp. $\big(\varphi_2, B_2,\tA_2, \ttP_2,\ttB_2\big)$; the third multiplet we look for is also of type  ${\mathbf \Xi}_{++}$, with field content $\Xi_{12}{}_{++}=\big(\varphi_{12}, B_{12},\tA_{12}, \ttP_{12},\ttB_{12}\big)$. The other possible pairing of multiplet types are treated similarly.

From the fields of the two initial multiplets, we begin by quadratically construct the scalar field $\ttP_{12}$ (whose transformation under \3 is a total derivative). In the following subsection we prove that for such a coupling, the only possibility is
\beqa
\label{p12}
\ttP_{12}= \varphi_1 \ttP_2 + \varphi_2 \ttP_1 + 
\frac14  B_1{}^{mn} \ttB_2{}_{mn} + \frac14  B_2{}^{mn} \ttB_1{}_{mn}
-\tA_1{}_m  \tA_2{}^m.
\eeqa

\noindent
Using (\ref{transfo2}) we now write explicitly the transformation of $\ttP_{12}$ under \3. On the other hand, since $\ttP_{12}$ should lie in a ${\mathbf \Xi}_{++}$ multiplet, it must transform like  $\delta_v \ttP_{12} = v^m \partial_m \varphi_{12}$. Thus, by identification, one gets (up to an additive constant) the expression of $\varphi_{12}$
$$\varphi_{12} = \varphi_1 \varphi_2 + \frac14 B_1{}_{mn} B_2{}^{mn}.$$
Following the same algorithm, from $\delta_v \varphi_{12}=v^m \tA_{12}{}_m$ one gets
$$\tA_{12}{}_m= \tA_1{}_m \varphi_2 + \tA_2{}_m \varphi_1 +
\tA_1{}^n B_{2}{}_{nm} + \tA_2{}^n B_{1}{}_{nm}. $$
Now, applying again the transformation laws (\ref{transfo2}) on this equation one finds
\beqa
\label{contrad} 
\delta_v \tA_{12}{}_m&=&v_m\left(\ttP_1 \varphi_2 +\ttP_2 \varphi_1 +
2 \tA_1{}_m \tA_2{}^m\right) \nonumber \\ & +&
v^n \left(\ttB_2{}_{mn} \varphi_1 +B_2{}_{mn} \ttP_1 +
\ttB_1{}_{mn} \varphi_2 +B_1{}_{mn} \ttP_2 +
\ttB_1{}_{pn} B_2{}^p{}_m+ \ttB_2{}_{pn} B_1{}^p{}_m\right).\nonumber\\
\eeqa
\noi
But the transformation law required for $\tA_{12}$ must be $\delta_v \tA_{12}{}_m = v_m \ttP_{12} + v^n \ttB_{12}{}_{mn}$, with $\ttB_{12}{}_{mn}$ a self-dual $2-$form. Hence we have a contradiction (in (\ref{contrad}) we don't find the form (\ref{p12}) for $\varphi_{12}$ and moreover, we do not find a self-dual $2-$form).

By similar calculatoric arguments, the other possibilities (by regrouping the differently multiplets) also fail.

\bigskip

After this simpler idea we now systematically study any possible interaction terms. We first find what are the fields $\Psi$ (content and transformation laws) that can couple to  \3 multiplets in an invariant quadratic way. We then express these fields $\Psi$ as a function $\Psi({\mathbf \Xi}_{++},{\mathbf \Xi}_{--},{\mathbf \Xi}_{+-},{\mathbf \Xi}_{-+})$. We find this function to be linear in the multiplets; hence the most general invariant terms to be build are quadratic and thus non-interacting.

\subsection{Possible couplings of a given multiplet}
\label{cuplaje-posibile}

We consider in this subsection the coupling of a multiplet with some generic fields $\Psi$ \cite{io1}. 
We focus on the coupling of a ${\mathbf \Xi}_{++}$ multiplet, the other cases being similar. 

Thus, considering its field content, the most general possibility of quadratic coupling with some set of unknown fields $\Psi$ is 
\beqa
\label{X++coup}
{\cal L}({{\mathbf \Xi}_{++}, \Psi})=
\varphi \ttpsi + \ttP \psi + \frac14 B^{(+)mn} \ttpsi_{mn} +
  \frac14 \ttB^{(+)}_{mn} \psi^{mn} - \tA_m \tpsi^m
\eeqa

\noi
with $\psi,\ttpsi$ two scalars, $\tpsi_m$ a vector and $\psi_{mn}, \ttpsi_{mn}$ two 
$2-$forms which, by (\ref{dualprop1}) are self-dual.
{\it A priori} some of the fields $\Psi$ can be set to zero.

To find the set of fields $\Psi$ we impose that (\ref{X++coup}) transforms under \3 as a total derivative. 
We first treat the case where the fields $\Psi$ contain no derivative terms. Thus, from (\ref{transfo2}) one has the \3 variation of (\ref{X++coup})
\beqa
\label{theo1-0}
\delta_v {\cal L}({\mathbf \Xi}_{++},\Psi)&=&
v^m \tA_m \ttpsi + \p \delta_v \ttpsi + v^m (\partial_m \p) \psi + \ttP \delta_v \psi \nonumber\\
&+& \frac14 (-v_m \tA_n + v_n \tA_m + i \e_{mnpq} v^p \tA^q)\ttpsi^{mn} + \frac14 B_{mn}^{(+)}\delta_v \ttpsi^{mn}\\
&+& \frac14 v^p \partial_p B^{(+)}_{mn})\psi^{mn} + \frac14 \ttB^{(+)}_{mn} \delta_v \psi^{mn} - (v_m \ttP + v_n \ttB_{mn})\tpsi^m - \tA_m \delta_v \tpsi^m. \nonumber
\eeqa
\noi 
Using \eqref{dualprop1-3} to ``compress'' $\frac14 (-v_m \tA_n + v_n \tA_m + i \e_{mnpq} v^p \tA^q)\ttpsi^{mn}$ and then to ``decompress'' $v^n\tpsi^m\ttB^{(+)}_{mn}$, equation \eqref{theo1-0} becomes
\beqa
\label{theo1}
\delta_v  {\cal L}({\mathbf \Xi}_{++},\Psi)&=&
\varphi \Big(\delta_v \ttpsi -v^m \partial_m \psi\Big)
+\ttP\Big(\delta_v \psi - v_m \tpsi^m\Big)
+\frac14 B_+{}_{mn} \Big(\delta_v \ttpsi^{mn} - v^p \partial_p \psi^{mn}\Big) 
\nonumber \\ 
&+&\frac14 \ttB^{(+)}_{mn} \Big(\delta_v \psi^{mn}  +
v^m \tpsi^n -v^n \tpsi^m - i \e^{mnpq}   v_p \tpsi_q\Big) 
 \\
&-&\tA_m\Big(\delta_v \tpsi^m- v_n\ttpsi^{mn}  -v^m \ttpsi\Big)
+v^p \partial_p\Big(\varphi \psi +\frac14 B^{(+)}_{mn} \psi^{mn} \Big).
\nonumber 
\eeqa
\noi
Since we have assumed that we have no presence of derivative terms in the $\Psi$ fields, one cannot have anymore total derivatives present. Thus, one has $\delta_v  {\cal L}({\mathbf \Xi}_{++},\Psi)=0$, which gives the \3 transformation laws of the fields $\Psi$. By a simple comparison with (\ref{transfo2}) one sees that the $\Psi$ fields transform like a ${\mathbf \Xi}_{++}$ multiplet (the self-dual character of the $2-$forms being also preserved)
\beqa
\label{theo2}
&&\delta_v \psi= v^m \tpsi_m, \ \
\delta_v \ttpsi= v^m \partial_m \psi, \ \ 
\delta_v \tpsi^m = v^m \ttpsi + v_n \ttpsi^{mn} \nonumber \\
&&\delta_v \psi_{mn}= v_n \tpsi_m - v_m \tpsi_n + i \e_{mnpq} e^p \tpsi^q \\
&&\delta_v \ttpsi_{mn}= v^r \partial_r \psi_{mn}  
 \nonumber 
\eeqa
\noi
Moreover, one cannot put to $0$ any of these fields. We thus have 

\begin{enumerate}
\item[]{\bf I:} {\it If the $\psi$ fields contain no derivative terms and 
(\ref{X++coup}) is invariant, then they  form a  multiplet  
of  type ${\mathbf \Xi}_{++}$.}
\end{enumerate}

We now treat the second possibility, when the $\psi$ fields contain exactly one derivative term. Then, in all generality, they can be expressed as
\beqa
\label{derivative}
&&\psi= \partial_m \lambda^m,\ \  \ttpsi = \partial_m \ttl^m  \nonumber \\
&&\psi_{mn}= \partial_m \lambda'_n -\partial_n \lambda'_m -
i \varepsilon_{mnpq} \partial^p \lambda'^q   \\
&&\ttpsi_{mn}= \partial_m \ttl'_n -\partial_n \ttl'_m -
i \varepsilon_{mnpq} \partial^p \ttl'^q,  \nonumber \\
&&\tpsi_m = \partial_m \tl + \partial^n \tl_{nm}.\nonumber
\eeqa
\noi
One can remark  that, {\it a priori} the vectors $\lambda_m$ and $\lambda'_m$ are different and that we have no information on the (anti-)self-dual character of the $2-$form $\tl_{nm}$.


We calculate the transformation laws of the fields $\psi$ (and for the moment not of their ``building blocks'', the fields $\lambda$). Hence, the calculation is identical to that made to prove {\bf I}: the variation of 
(\ref{X++coup}) gives (\ref{theo1-0}) and finally (\ref{theo1}).
Nevertheless,  
now one cannot put directly $\delta_v {\cal L}({\mathbf \Xi}_{++},\Psi)$ equal to $0$, since as we see from \eqref{derivative}, the fields $\psi$ do contain partial derivative, so the hypothesis of further total derivative presence cannot be directly casted away. However, we now show that in this case also, further total derivatives {\it cannot be present}.

Suppose for contradiction that these total derivatives can be present. They will lead to some further terms in the transformation laws \eqref{theo2} (deduced of \eqref{theo1}) which write
\beqa
\label{theo2-1}
&&\delta_v \psi= v^m \tpsi_m+X , \ \
\delta_v \ttpsi= v^m \partial_m \psi+{\tilde {\tilde X}}, \ \ 
\delta_v \tpsi^m = v^m \ttpsi + v_n \ttpsi^{mn} + \tilde X_m \nonumber \\
&&\delta_v \psi_{mn}= v_n \tpsi_m - v_m \tpsi_n + i \e_{mnpq} v^p \tpsi^q  + X_{mn}\\
&&\delta_v \ttpsi_{mn}= v^r \partial_r \psi_{mn} + {\tilde {\tilde X}}_{mn} 
 \nonumber 
\eeqa
\noi
where (since we consider the fields of the ${\mathbf \Xi}_{++}$ multiplet to be independent) one has $\ttP X$, $\p {\tilde {\tilde X}}$, $\tA_m \tilde X^m$, $\ttB_{mn} X_{mn}$ and $B_{mn} {\tilde {\tilde X}}_{mn} $ equal to some total derivatives. Obviously, one does not need to have all the fields $X$ above present (some of them, but not all, could be set to zero).
Nevertheless, the fields $X$ have to be constructed from the fields $\lambda$ considered in \eqref{derivative}. For example, for $X$ one may consider the terms: 
$$v^m \lambda_m,\ \partial^m \lambda_m,\ \tl,\ \lambda_m \lambda'^m$$
 and similar ones. By direct inspection, one can easily check that no $X$ field can be constructed to fulfil the requested condition $\ttP X$ is a total derivative. Indeed, for example the term $\p \partial^m \lambda_m$ would require a term $\partial^m \p \lambda_m$ in order to have the total derivative $\partial^m (\p \lambda_m)$. Nevertheless this term is not present and thus one cannot obtain the required total derivatives. Thus one concludes by
\beqa
\label{cemata}
X=0.
\eeqa
\noi
Similar considerations apply for the rest of $X$ fields. However, a somewhat more particular situation appears for the case of the $2-$forms $X_{mn}$ and ${\tilde {\tilde X}}_{mn} $ (which by \eqref{dualprop1} must be self-dual). The situation is more tricky because one now has at his disposal 
identities of type \eqref{prop}; one thus writes may write terms of type
$$ \ttB_{mn} (v^r \partial_{[m} \tilde \lambda_{n]+r}- v_{[m} \partial^r \tilde \lambda_{n]+r})$$
(if $\tilde \lambda_{mn}$ is anti-self-dual). But, by \eqref{prop} itself, this contribution is trivially identical to $0$, hence not leading to any complications in the transformation laws \eqref{theo2}. 


Hence, we conclude that the fields $\psi$ transform in this case also as indicated in \eqref{theo2}.

One might ask about the variation of the fields $\lambda$. Indeed, from (\ref{theo2}) and  (\ref{derivative}) one has a constraint on these variations. 
For example, 
 by \eqref{theo2} one has
$$ \delta_v \psi= v^m \tpsi_m$$
inserting the explicit expressions \eqref{derivative} for $\psi$ and $\tpsi_m$ one has
$$ \partial^m \delta_v \lambda_m = v_m \partial^m \tilde \lambda + v^n \partial^m \tilde \lambda_{mn}$$
which leads (keeping in mind that we have all the set of the $\lambda$ fields at our disposal) to
\beqa
\label{transfolambda}
\delta_v \lambda_m = v_m\tilde \lambda + v^n \tilde \lambda_{mn}+
a(v_m \partial_n \lambda^n - v_n \partial^n \lambda_m) +
a'(v_m \partial_n \lambda'^n - v_n \partial^n \lambda'_m),
\eeqa
\noi
However, 
by a direct inspection of formulae  (\ref{theo2}) and (\ref{transfolambda}) one sees that 
it is much more convenient to work with the $\psi$ fields, the ``derivatives'' of the $\lambda$ fields. One important technical reason is that, as seen from  (\ref{theo2}), the $\Psi$ fields transform like a ${\mathbf \Xi}_{++}$ multiplet. 

A final remark to be done here is that, if one takes $\lambda'_m = \lambda_m, \, \ttl'_m = \ttl_m$, the Lagrangian  (\ref{coup}) is identical to the Lagrangian \eqref{X++coup}  considered here. In this case,
the  $\psi$'s form the derivative multiplet (\ref{partialXipm})
 of the $\lambda$'s. 

\smallskip

We thus have the following result

\begin{enumerate}
\item[]{\bf II:} {\it If the $\psi$ fields are as in (\ref{derivative})  and 
the Lagrangian (\ref{X++coup})
is invariant, then they  form a    
${\mathbf \Xi}_{++}$ multiplet}.
\end{enumerate}

We now treat the most general case with one derivative, namely when the $\psi$ fields contain terms with one derivative and terms with no derivative. Their most general form is
\beqa 
\label{field3}
&&\psi=  \rho + \partial_m \lambda^m,\ \  
\ttpsi =  \ttr + \partial_m \ttl^m, \nonumber \\
&&\psi_{mn}= \rho_{mn}+ \partial_m \lambda'_n -\partial_n \lambda'_m -
i \varepsilon_{mnpq} \partial^p \lambda'^q,   \\
&&\ttpsi_{mn}= \ttr_{mn}+ \partial_m \ttl'_n -\partial_n \ttl'_m -
i \varepsilon_{mnpq} \partial^p \ttl'^q,  \nonumber \\
&&\tpsi_m = \tr_m+ \partial_m \tl + \partial^n \tl_{nm}.\nonumber
\eeqa
\noi
where, as before, the $2-$forms $\psi$ and $\rho$ are self-dual.

As before, not taking into consideration the explicit ``content'' \eqref{field3} of the $\psi$ fields,
one obtains their transformation laws, which are identical to \eqref{theo2-1}. The difference comes now at the level of the fields $X$, which can be constructed using the $\rho$ and the $\lambda$ fields. For example, considering again the field $X$ that satisfies 
$$\ttP X=\mbox{total derivative},$$
 one must also consider, except the terms already enumerated in the proof of {\bf II}, terms like 
$$\rho,\ v^m \tr_m,\ \partial^m \tr_m,\ \ttr,\ v^m \tr_m \mbox{ and }\rho \ttr.$$
 As before, by a direct inspection of all these terms, one can easily see that 
such total derivatives cannot be constructed. For example, for the term $\ttP \partial^m \tr_m$ to contribute to a total derivative $ \partial^m (\ttP \tr_m)$, one would need an additional term $\partial^m \ttP  \tr_m$, term which is not present. One can thus conclude
$$X=0.$$
Arguing along the same lines, all the fields $X$ of \eqref{theo2-1} are equal to $0$. Hence we have the same transformation laws as in \eqref{theo2}.

As for {\bf II}, working with the $\psi$ fields is
more appropriate for our study that the $\rho$ and $\lambda$ fields. However, in this case, some of the fields $\rho$ or $\lambda$ may be absent from (\ref{field3}).

We illustrate this by choosing, as above the case of the transformation law
$$ \delta_v \psi = v^m \tpsi_m $$
which, by \eqref{field3} leads to
$$ \delta_v \rho + \partial_m \delta_v \lambda^m = v^m \tr_m + v^m \partial_m \tl + v^m \partial^n \tl_{nm} $$
which gives
\beqa
\delta_v \rho &=& v^m \tr_m \nonumber\\
\delta_v \lambda_m&=&v_m\tilde \lambda + v^n \tilde \lambda_{mn}+
a(v_m \partial_n \lambda^n - v_n \partial^n \lambda_m) +
a'(v_m \partial_n \lambda'^n - v_n \partial^n \lambda'_m).
\eeqa

It is thus obvious that working with the $\psi$ fields (and not their ``building blocks'' $\lambda$ and $\rho$) is more suited for calculations.

We now conclude by

\bigskip
\begin{enumerate}
\item[]{\bf III:} {\it If the $\psi$ are as in (\ref{field3})  and 
the Lagrangian (\ref{X++coup})
is invariant, then they  transform as in (\ref{theo2})}.
\end{enumerate}

\bigskip


Similar arguments hold if one allows a higher number of derivative in
the fields $\Psi$. For example, if one considers two derivatives, one
can write fields of type 
$$ \psi= \Box \lambda,\ \psi_m=\alpha \Box \lambda_m+\beta \partial_m\partial_n
\lambda^n,\ \psi_{mn}=\alpha' \Box \lambda_{mn}+\beta'\partial^p\partial_{[m}\lambda_{n]+p}$$ 
Generally speaking, if one consider an even number $n$ of partial derivatives, the terms that can be added are
$$ \psi = \Box^{\frac n2} \lambda,\ \psi_m= \alpha \Box^{\frac n2} \lambda_m + \beta \Box^{\frac{n-2}{2}} \partial^n \partial_m \lambda_n,\ \psi_{mr}=\alpha \Box^{\frac n2} \lambda_{mr}+ \beta \Box^{\frac{n-2}{2}} \partial^p\partial_{[m}\lambda_{n]+p}$$
If $n$ is an odd number, one can construct
$$\psi=\Box^{\frac{n-1}{2}} \partial^m \lambda_m,\ \psi_m=\alpha \Box^{\frac{n-1}{2}} \partial_m \lambda + \beta \Box^{\frac{n-1}{2}} \partial^p \lambda_{pm},\ \psi_{mp}= \Box^{\frac{n-1}{2}} \partial_{[m}\lambda_{n]+}  $$
As before, the variation laws of the $\lambda$ fields are more 
complicated and they are not of interest for our study. However, one
obtains the variation laws of the $\psi$ fields, which are 
transformation laws of type \eqref{theo2-1}. The only thing to prove is that, again, the $X$ fields are not present. We do this by induction. Considering for example $X$ such that 
$$\ttP X=\mbox{total derivative}.$$ 
We have seen that for $n<2$ one cannot construct such a field $X$. The terms composing the $X$ fields and which have a number of derivatives $<n$ were shown not to lead to any total derivative.
Thus, to complete the proof one has to obtain the same conclusion for the supplementary terms, involving exactly $n$ derivatives.
These terms one can add to $X$ are
$$\Box^{\frac n2} \lambda,\ v_{m} \partial^{m} \Box^{\frac{n-2}{2}} \lambda $$
if $n$ is even, and
$$\Box^{\frac{n-1}{2}} \partial^m\lambda_m,\ v_{m} \partial^{m} \Box^{\frac{n-1}{2}} \lambda. $$
Consider for the example the term $\ttP \Box^{\frac n2} \lambda=\ttP\Box^{\frac{n-2}{2}} \partial_m \partial^m\lambda $. In order to obtain a total derivative $\partial_m(\ttP \Box^{\frac{n-2}{2}} \partial^m\lambda $ one needs a supplementary term $(\partial_m\ttP) \Box^{\frac{n-2}{2}} \partial^m\lambda $, term which, as before is not present.
Thus, one has $X=0$ for any number $n$ of derivatives and thus the transformation laws of the $\Psi$ fields are identical to (\ref{theo2}).

Note that if one considers only dependence of type $\Box^n$ then the Lagrangian obtained here is as the Lagrangian of the previous section.

\bigskip

We have thus seen which are the most general couplings
of a given multiplet. Hence we  have now the set of fields $\Psi$, its
content and \3 transformation laws. We have also seen that the $\Psi$
fields may content terms with $0,1,2,\dots$ derivatives
in terms of the $\lambda$ fields. In the rest of this section we see
what is the most general way one can construct these $\Psi$ fields out
of the original ${\mathbf \Xi}_{\pm\pm}$ multiplets.

\subsection{Generalised tensor calculus}

In subsection \ref{tensor-calculus} we have applied a tensor calculus method for \3, trying to
build a multiplet out of two other multiplets, method which proved to fail. Since we have seen that
the $\Psi$ fields (which were the most general possibility of  coupling with a \3 multiplet) transform like a \3 multiplet, what we must do now is to
try to construct such a multiplet in all generality; hence the name of
``generalised tensor calculus''.

Here also we focus on the ${\mathbf \Xi}_{++}$ multiplet, the other cases
being similar.
So, what we look for is to express the fields $\psi, \psi_{mn}, \tpsi_m,\ttpsi,\ttpsi_{mn}$ of the previous subsection as functions of the fields of the \3 multiplets
${\mathbf \Xi}_{++}, {\mathbf \Xi}_{--},{\mathbf \Xi}_{+-},{\mathbf \Xi}_{-+}$
and of their  derivatives. More over we consider here holomorphic functions.

\medskip

We first consider the simpler case when the fields $\Psi$ have no
derivative dependence on any of the \3 fields. Furthermore, we also
assume for the moment that the  $\Psi$ fields depend only of the
${\mathbf \Xi}_{++}$ multiplet. We have find out in the previous subsection how
the $\Psi$ fields transform under \3, so, for instance one writes
\beqa
\label{F1}
\delta_v \ttpsi = v^m \partial_m \psi =
\frac{\partial \ttpsi}{\partial \ttP} \delta_v \ttP +
\frac{\partial \ttpsi}{\partial \varphi} \delta_v \varphi +
 \frac{\partial \ttpsi}{\partial B_{mn}} \delta_v B_{mn}+
\frac{\partial \ttpsi}{\partial \ttB_{mn}} \delta_v \ttB_{mn} +
\frac{\partial \ttpsi}{\partial \tA_{m}} \delta_v \tA_{m}.
\eeqa
\noindent
Now, in \eqref{F1}  one can insert the explicit variation
(\ref{transfo2}) of the ${\mathbf \Xi}_{++}$ fields which gives
\beqa
v^m \partial_m \psi &=&
\frac{\partial \ttpsi}{\partial \ttP}  v^m \partial_m \varphi +
\frac{\partial \ttpsi}{\partial \varphi}  v^m \tA_m 
+ \frac{\partial \ttpsi}{\partial B_{mn}} (-v_m \tA_n+ v_n \tA_m + i\e_{mnpq}v^p\tA^q) 
\nonumber \\
&+&
 \frac{\partial \ttpsi}{\partial \ttB_{mn}}  v^p \partial_p B_{mn}+
\frac{\partial \ttpsi}{\partial \tA_{m}} (v_m \ttP + v^n \ttB_{mn}).
\eeqa
\noi
Using  \eqref{dualprop1-3} this simplifies to
\beqa
\label{F11}
 v^m \partial_m \psi &=&
\frac{\partial \ttpsi}{\partial \ttP}  v^m \partial_m \varphi +
\frac{\partial \ttpsi}{\partial \varphi}  v^m \tA_m 
-4  \frac{\partial \ttpsi}{\partial B_{mn}} v_m \tA_n 
\nonumber \\
&+&
 \frac{\partial \ttpsi}{\partial \ttB_{mn}}  v^p \partial_p B_{mn}+
\frac{\partial \ttpsi}{\partial \tA_{m}} (v_m \ttP + v^n \ttB_{mn}).
\eeqa
\noindent
Since we are treating the case where we have no derivative dependence
in $\psi$, we observe that in the LHS of the equation (\ref{F11}) we
have one spatial derivative. Recalling that $v$ is arbitrary, one has
\beqa
\label{acum1}
\frac{\partial \ttpsi}{\partial \p} \tA_m -4 \frac{\partial \ttpsi}{\partial B^{mn}}  \tA^n + \frac{\partial \ttpsi}{\partial \tA_m}\ttP + \frac{\partial \ttpsi}{\partial \tA_p}\ttB_{pm} =0
\eeqa
\noi
and
\beqa
\label{F11-bis}
  \partial_m \psi &=&
\frac{\partial \ttpsi}{\partial \ttP}   \partial_m \varphi 
+
 \frac{\partial \ttpsi}{\partial \ttB_{pq}}  \partial_m B_{pq}
\eeqa
\noindent
Then, by integration by parts of \eqref{F11-bis} one has
$$\partial_m( \psi -
\frac{\partial \ttpsi}{\partial \ttP} \varphi  
 -  \frac{\partial \ttpsi}{\partial \ttB_{mn}} B_{mn} )
= - \varphi \partial_m 
\frac{\partial \ttpsi}{\partial \ttP}
- B_{np} \partial_m  \frac{\partial \ttpsi}{\partial \ttB_{np}}.$$
\noindent
In the LHS one has a total derivative; to obtain a total derivative in the RHS also one needs to have $\ttpsi= \p  \ttP + \frac 14  B_{np}  \ttB^{np}$. However, the \3 transformation of $\ttpsi$  is a total derivative; one can now recall that we have proven in the previous subsections that the scalar involving these terms and transforming under \3 in such a way is ${\cal L} ({\mathbf \Xi}_{++})=  \p \ttP + \frac 14  B_{np}  \ttB^{np}-\frac12  \tA_n^2$.
However, in subsection $5.5.1$ we have proven by explicitly applying the \3 transformations laws \eqref{transfo2} that this type of coupling cannot lead to the whole field content of a ${\mathbf \Xi}_{++}$ multiplet.
Thus one has
 $$\partial_m \frac{\partial \ttpsi}{\partial \ttP}=0,\ \partial_m  \frac{\partial \ttpsi}{\partial \ttB_{np}}=0.$$
One can now write
\beqa
\label{salvarea-v}
\ttpsi= \alpha \ttP +  X_{mn} \ttB^{mn} + \ttpsi_R (\tA, \p, B)
\eeqa
\noi
where $\alpha$ is a scalar and $X_{mn}$ is a self-dual $2-$form, invariant under \3 and $x-$independent and $\ttpsi_R$ is an arbitrary function. Hence \eqref{acum1} becomes
\beqa
\label{acum2}
\frac{\partial \ttpsi_R}{\partial \p} \tA_m -4 \frac{\partial \ttpsi_R}{\partial B^{mn}} \tA^n + \frac{\partial \ttpsi_R}{\partial \tA^m}\ttP + \frac{\partial \ttpsi_R}{\partial \tA_p}\ttB_{pm} =0
\eeqa
\noi
Suppose now that $ \frac{\partial \ttpsi_R}{\partial \tA_m}\ne 0$; this implies that $\ttpsi_R$ contains $\ttP$ or $\ttB$ which is a contradiction. Hence
$$ \frac{\partial \ttpsi_R}{\partial \tA_m}=0.$$
Equation \eqref{acum2} can now be written
\beqa
\label{s-a}
\tA^n \left(\frac{\partial \ttpsi_R}{\partial \p} \eta_{mn} - 4 \frac{\partial \ttpsi_R}{\partial B^{mn}}\right)=0
\eeqa
\noi
which implies that 
$$\frac{\partial \ttpsi_R}{\partial \p}=0=\frac{\partial \ttpsi_R}{\partial B_{mn}}$$
because a symmetric tensor ($\eta_{mn}$) cannot be equal to an antisymmetric tensor ($\frac{\partial \ttpsi_R}{\partial B_{mn}}$).

Hence, \eqref{salvarea-v} writes
\beqa
\label{salvarea}
\ttpsi= \alpha \ttP +  X_{mn} \ttB^{mn}.
\eeqa
\noi
With this form for $\ttpsi$, applying its \3 transformation one has
$$ \delta_v \ttpsi= v^m \partial_m \psi = \alpha v^m \partial_m \p + X_{mn} v^p \partial_p B^{mn} $$ which gives 
$$\psi= \alpha \varphi +  X_{mn} B^{mn}.$$
 Furthermore, the \3
transformation of $\psi$ above writes
$$\delta_v \tpsi= v^m \tpsi_m= \alpha v^m \tA_m - 4 X_{mn} v^m \tA^n $$
and thus gives  
$$\tpsi_m= \alpha \tA_m - 4 X_{mn} \tA^n.$$ 
Now, from the transformation law for $\tpsi_m$ one gets  
\beqa
\label{muie0303}
\delta_v \tpsi_m = v_m \ttpsi + v^n \ttpsi_{mn}=\alpha v_m \ttpsi+v^n(\alpha \ttB_{mn}-4X_{mn}\ttP-4 \ttB^p_nX_{mp}).
\eeqa
\noi
Hence one gets
$$ \ttpsi_{mn}= \alpha \ttP$$
which, compared to \eqref{salvarea} gives
\beqa
\label{salvarea2}
 X_{mn}=0. 
\eeqa
\noi
Now the equations above immediately lead to 
$$\Psi({\mathbf \Xi}_{++}) = \alpha {\mathbf \Xi}_{++}$$

\medskip


Let us now consider the case $\Psi({\mathbf \Xi}_{+-})$.
Arguing along the same lines as above one firstly writes
\beqa
\label{F1bis}
\delta_v \ttpsi = v^m \partial_m \psi =
\frac{\partial \ttpsi}{\partial A'_m} \delta_v A'_m +
\frac{\partial \ttpsi}{\partial \tP'} \delta_v \tP' +
 \frac{\partial \ttpsi}{\partial \tB'_{mn}} \delta_v \tB'_{mn}+
\frac{\partial \ttpsi}{\partial \ttA'_{m}} \delta_v \ttA'_{m}.
\eeqa
\noindent
Now, in \eqref{F1bis}  one can insert the explicit variation
(\ref{transfo2}) of the ${\mathbf \Xi}_{++}$ fields which gives
\beqa
\label{interm22}
v^m \partial_m \psi =
\frac{\partial \ttpsi}{\partial A'_m}  (v^n \tB^\prime_{mn}+
v_m \tP^\prime)+
\frac{\partial \ttpsi}{\partial \tP'} v^m 
\ttA_m^\prime 
-4  \frac{\partial \ttpsi}{\partial \tB'_{mn}} v_m \ttA'_n +
\frac{\partial \ttpsi}{\partial \ttA'_{m}} v^n \partial_n A^\prime_m
\eeqa
\noi
(where we have used \eqref{dualprop1-3} to simplify $\frac{\partial \ttpsi}{\partial \tB'_{mn}} \delta_v \tB'_{mn}$). As before, since we do not allow derivative dependences of $\psi$ on the \3 fields, we have
\beqa
\label{tragere11}
\frac{\partial \ttpsi}{\partial A'_p} \tB'_{pm} + \frac{\partial \ttpsi}{\partial A'_m}\tP^\prime + \frac{\partial \ttpsi}{\partial \tP'}\ttA'_m -4 \frac{\partial \ttpsi}{\partial \tB'^{mn}}\ttA'_n=0
\eeqa
and
$$ \partial_m \psi = \frac{\partial \ttpsi}{\partial \ttA'_{n}}  \partial_m A^\prime_n$$
One can integrate by parts this last equation to obtain a total derivative
$$\partial_m (\psi -\frac{\partial \ttpsi}{\partial \ttA'_{n}} A^\prime_n)= -\left(\partial_m \frac{\partial \ttpsi}{\partial \ttA'_{n}}\right)  A^\prime_n$$
We now apply the same type of arguments as in the previous case. One needs $\ttpsi= \ttA'_n A^n$ but the condition that its transformation under \3 and the results of subsection $5.5.1$ finally lead to
 $$\partial_m \frac{\partial \ttpsi}{\partial \ttA'_{n}}=0$$
 thus one has
$$ \ttpsi=\alpha^m \ttA'_m +\ttpsi_R (A', \tP', \tB'),$$
where $\alpha^m$ is an arbitrary vector, invariant under \3 and $x$ independent and $\ttpsi_R$ is an arbitrary function. Equation \eqref{tragere11} becomes
\beqa
\label{acum2-1}
\frac{\partial \ttpsi_R}{\partial A'_p} \tB'_{pm} + \frac{\partial \ttpsi_R}{\partial A'_m}\tP^\prime + \frac{\partial \ttpsi_R}{\partial \tP'}\ttA'_m -4 \frac{\partial \ttpsi_R}{\partial \tB'^{mn}}\ttA'_n=0
\eeqa
\noi
Suppose now $\frac{\partial \ttpsi_R}{\partial \tP'}$ and $\frac{\partial \ttpsi_R}{\partial \tB'^{mn}}\ne 0$. Then, since $\ttpsi$ is not a function of $\ttA$, this leads to
$$\frac{\partial \ttpsi_R}{\partial \tP'}\ttA'_m -4 \frac{\partial \ttpsi_R}{\partial \tB'^{mn}}\ttA'_n=0$$
which, as before is not possible. Hence $\frac{\partial \ttpsi_R}{\partial \tP'}$ or $\frac{\partial \ttpsi_R}{\partial \tB'^{mn}}=0$. But, since $\ttpsi$ is not a function of $\ttA$, one has  
$$\frac{\partial \ttpsi_R}{\partial \tP'}=0 \Leftrightarrow\frac{\partial \ttpsi_R}{\partial \tB'^{mn}}=0$$
and hence
$$\frac{\partial \ttpsi_R}{\partial \tP'}=0,\ \frac{\partial \ttpsi_R}{\partial \tB'^{mn}}=0$$
Then $\ttpsi_R$ remains a function only in $A'$ and obviously \eqref{acum2-1} implies now that
$$\frac{\partial \ttpsi_R}{\partial A'_p}=0. $$
Thus one has
\beqa
\label{ttpsi-0}
\ttpsi= \alpha^m \ttA'_m
\eeqa
\noi
  We now use the transformation laws of the $\Psi$ multiplet to find in a step-by-step process
\beqa
\psi&=&\alpha^m A'_m,\nonumber\\
\tpsi_m &=&\alpha^n \tB'_{nm},\nonumber\\
\ttpsi_{mn}&=&\alpha_{[m}\ttA'_{n]+},\nonumber\\
\psi_{mn}&=&\alpha_{[m}A'_{n]+}. 
\eeqa
\noi
To conclude, one imposes the transformation law of $\psi_{mn}$, $\delta_v \psi_{mn}=-v_{[n} \tpsi_{m]+}$, witch, since we now have the explicit form of $\tpsi_{m}$ leads to 
$$\alpha^m =0.$$ Thus one has
$$\Psi ({\mathbf \Xi}_{+-})=0.$$


Let us now treat the  more complicated case of dependence on several multiplets: 
$$\Psi ({\mathbf \Xi}_{++},{\mathbf \Xi}_{+-}).$$
 As above we begin by writing
\beqa
\delta_v \ttpsi = v^m \partial_m \psi &=&
\frac{\partial \ttpsi}{\partial \ttP} \delta_v \ttP +
\frac{\partial \ttpsi}{\partial \varphi} \delta_v \varphi +
 \frac{\partial \ttpsi}{\partial B_{mn}} \delta_v B_{mn}+
\frac{\partial \ttpsi}{\partial \ttB_{mn}} \delta_v \ttB_{mn} +
\frac{\partial \ttpsi}{\partial \tA_{m}} \delta_v \tA_{m}\nonumber\\
&+&\frac{\partial \ttpsi}{\partial A'_m} \delta_v A'_m +
\frac{\partial \ttpsi}{\partial \tP'} \delta_v \tP' +
 \frac{\partial \ttpsi}{\partial \tB'_{mn}} \delta_v \tB'_{mn}+
\frac{\partial \ttpsi}{\partial \ttA'_{m}} \delta_v \ttA'_{m}.\nonumber
\eeqa
\noi
Making use of the explicit form of the transformation laws \eqref{transfo2} this becomes
\beqa
\label{panacand}
\delta_v \ttpsi = v^m \partial_m \psi &=&
\frac{\partial \ttpsi}{\partial \ttP}  v^m \partial_m \varphi +
\frac{\partial \ttpsi}{\partial \varphi}  v^m \tA_m 
-4  \frac{\partial \ttpsi}{\partial B_{mn}} v_m \tA_n 
\nonumber \\
&+&
 \frac{\partial \ttpsi}{\partial \ttB_{mn}}  v^p \partial_p B_{mn}+
\frac{\partial \ttpsi}{\partial \tA_{m}} (v_m \ttP + v^n \ttB_{mn})\\
&+&\frac{\partial \ttpsi}{\partial A'_m}  (v^n \tB^\prime_{mn}+
v_m \tP^\prime)+
\frac{\partial \ttpsi}{\partial \tP'} v^m 
\ttA_m^\prime 
-4  \frac{\partial \ttpsi}{\partial \tB'_{mn}} v_m \ttA'_n +
\frac{\partial \ttpsi}{\partial \ttA'_{m}} v^n \partial_n A^\prime_m \nonumber
\eeqa
\noi
As before, comparing the number of partial derivatives in the LHS and the RHS, one obtains that 
one has
\beqa
\label{acum3}
\frac{\partial \ttpsi}{\partial \p} \tA_m -4 \frac{\partial \ttpsi}{\partial B_{mn}} \partial_m \tA^n + \frac{\partial \ttpsi}{\partial \tA_m}\ttP + \frac{\partial \ttpsi}{\partial \tA_p}\ttB_{pm}\nonumber\\
+\frac{\partial \ttpsi}{\partial A'_p} \tB'_{pm} + \frac{\partial \ttpsi}{\partial A'_m}\tP^\prime + \frac{\partial \ttpsi}{\partial \tP'}\ttA'_m -4 \frac{\partial \ttpsi}{\partial \tB'^{mn}}\ttA'_n=0.
\eeqa
\noi
and
\beqa
 v^m \partial_m \psi &=&
\frac{\partial \ttpsi}{\partial \ttP}  v^m \partial_m \varphi +
 \frac{\partial \ttpsi}{\partial \ttB_{mn}}  v^p \partial_p B_{mn}+
\frac{\partial \ttpsi}{\partial \ttA'_{m}} v^n \partial_n A^\prime_m \nonumber
\eeqa
\noi
In the same way as before, one constructs  total derivatives in the LHS which, by grouping the arguments in the two previous cases, will allow us to obtain
\beqa
\partial_p\frac{\partial \ttpsi}{\partial \ttP}=0,\
\partial_p  \frac{\partial \ttpsi}{\partial \ttB_{mn}}=0,\
\partial_p \frac{\partial \ttpsi}{\partial \ttA'_{m}}=0.
\eeqa
\noi
Thus, one now has
$$\ttpsi=  \alpha \ttP +  X_{mn} \ttB^{mn}+ \alpha_m \ttA'_m + \ttpsi_R (\p, B, \tA, A', \tP', \tB')$$
where $\alpha, X_{mn}, \alpha_m$ are $x$ independent.
Hence, \eqref{acum3} writes
\beqa
\label{acum3-1}
\frac{\partial \ttpsi_R}{\partial \p} \tA_m -4 \frac{\partial \ttpsi_R}{\partial B_{mn}} \partial_m \tA^n + \frac{\partial \ttpsi_R}{\partial \tA_m}\ttP + \frac{\partial \ttpsi_R}{\partial \tA_p}\ttB_{pm}\nonumber\\
+\frac{\partial \ttpsi_R}{\partial A'_p} \tB'_{pm} + \frac{\partial \ttpsi_R}{\partial A'_m}\tP^\prime + \frac{\partial \ttpsi_R}{\partial \tP'}\ttA'_m -4 \frac{\partial \ttpsi_R}{\partial \tB'^{mn}}\ttA'_n=0.
\eeqa
\noi
By arguments identical to the two cases treated above one has 
$$\frac{\partial \ttpsi_R}{\partial \tA_m}=0,\ \frac{\partial \ttpsi_R}{\partial \tP'}=0,\ \frac{\partial \ttpsi_R}{\partial \tB'^{mn}}=0. $$
Thus \eqref{acum3-1} writes
\beqa
\label{acum3-2}
\frac{\partial \ttpsi_R}{\partial \p} \tA_m -4 \frac{\partial \ttpsi_R}{\partial B_{mn}} \partial_m \tA^n 
+\frac{\partial \ttpsi_R}{\partial A'_p} \tB'_{pm} + \frac{\partial \ttpsi_R}{\partial A'_m}\tP^\prime =0.
\eeqa
\noi
and furthermore $\ttpsi_R$ is a function only of $\p, B$ and $A'$. Hence its partial derivatives cannot contain $B'$ and $\tP'$ which means that
$$\frac{\partial \ttpsi_R}{\partial A'_p}=0.$$
Now what remains of \eqref{acum3-2} was already treated above and leads to
$$ \frac{\partial \ttpsi_R}{\partial \p}=0, \ \frac{\partial \ttpsi_R}{\partial B_{mn}}=0.$$
Thus one has
$$ \ttpsi=  \alpha \ttP +  X_{mn} \ttB^{mn}+ \alpha_m \ttA'_m $$
 Actually, because of the fact that $\alpha, X_{mn}, \alpha_m$ are $x$ independent and also because of the fact that the fields $ \ttP, \ttB^{mn}$ on one hand and $\ttA'$ on the other belong to different \3 multiplets (namely ${\mathbf \Xi}_{++}$ and ${\mathbf \Xi}_{+-}$) and their transformation laws \eqref{transfo2} do not mix, one can now { completely separate} the dependences on these two distinct multiplets. Hence this case reduces to the two distinct studied above.

\medskip

One can thus state that, taking into account the no-derivative dependences on all four multiplets,
$$\Psi = \alpha {\mathbf \Xi}_{++}.$$


\medskip

More generally, one must consider dependence on the \3 fields derivatives also.
The proof
follows the same line of reasoning. Thus, let us  obtain the $\Psi$ fields as fields with one derivative dependence of the ${\mathbf \Xi}_{+-}$ multiplet,
$$ \Psi = \Psi (\partial_p A'_m, \partial_p \tP', \partial_p \tB'_{mn}, \partial_p \ttA'_m).$$
Thus, arguing along the same lines as above one firstly writes
\beqa
\label{F1bis3}
\delta_v \ttpsi = v^m \partial_m \psi =
\frac{\partial \ttpsi}{\partial \partial_p A'_m} \partial_p \delta_v A'_m +
\frac{\partial \ttpsi}{\partial \partial_p \tP'} \partial_p \delta_v \tP' +
 \frac{\partial \ttpsi}{\partial \partial_p \tB'_{mn}} \partial_p \delta_v \tB'_{mn}+
\frac{\partial \ttpsi}{\partial \partial_p \ttA'_{m}} \partial_p \delta_v \ttA'_{m}.
\eeqa
\noindent
Now, in \eqref{F1bis3}  one can insert the explicit variation
(\ref{transfo2}) of the ${\mathbf \Xi}_{+-}$ fields which gives
\beqa
\label{interm223}
v^m \partial_m \psi =
\frac{\partial \ttpsi}{\partial \partial_p A'_m}  \partial_p (v^n \tB^\prime_{mn}+
v_m \tP^\prime)+
\frac{\partial \ttpsi}{\partial \partial_p \tP'} v^m \partial_p
\ttA_m^\prime 
-4  \frac{\partial \ttpsi}{\partial \partial_p  \tB'_{mn}} v_m \partial_p \ttA'_n +
\frac{\partial \ttpsi}{\partial \partial_p \ttA'_{m}} v^n \partial_p \partial_n A^\prime_m\nonumber\\
\eeqa
\noi
(where we have used \eqref{dualprop1-3} to simplify $\frac{\partial \ttpsi}{\partial \partial_p \tB'_{mn}} \delta_v \partial_p \tB'_{mn}$). 
Now one cannot put anything to zero as before because we have derivatives in all terms 
 (in the LHS the field $\psi$ can or cannot content derivative dependences)
As before, we now use integration by parts to express a total derivative on the LHS, which will allow us to obtain (since $v$ is an arbitrary vector) 
\beqa
\label{cacat}
\partial_m (\psi - \frac{\partial \ttpsi}{\partial \partial_p \ttA'_{n}} \partial_p A'_n) - \partial_p \left(\frac{\partial \ttpsi}{\partial \partial_p A'_n}\tB'_{nm} + \frac{\partial \ttpsi}{\partial \partial_p \tP'}\ttA'_m + \frac{\partial \ttpsi}{\partial \partial_p A'^m}\tP' - 4 \frac{\partial \ttpsi}{\partial \partial_p  \tB'^{mn}}\ttA'^n\right) =\nonumber\\
-\partial_p \frac{\partial \ttpsi}{\partial \partial_p A'_n}\tB'_{nm} 
- \partial_p \frac{\partial \ttpsi}{\partial \partial_p \tP'}\ttA'_m
-\partial_p \frac{\partial \ttpsi}{\partial \partial_p A'^m}\tP'
+4 \partial_p \frac{\partial \ttpsi}{\partial \partial_p  \tB'^{mn}}\ttA'^n
-\partial_m \frac{\partial \ttpsi}{\partial \partial_p \ttA'_{n}} \partial_p A'_n.
\eeqa
\noi
The last term in the RHS can be used to obtain a total derivative. Indeed consider $\ttpsi=-\partial_p A'_n \partial^p \ttA'^n $; this gives  $\partial_m \frac{\partial \ttpsi}{\partial \partial_p \ttA'_{n}} \partial_p A'_n = \frac12 \partial_m (\partial_p A'_n)^2$. 
As before, we now use \3 transformation arguments to cast away this possibility also. Indeed $\partial_p A'_n \partial^p \ttA'^n $ alone does not transform like a total derivative, but $-\partial_p A'_n \partial^p \ttA'^n+\frac12 (\partial_m \tP')^2 +\frac14 (\partial_m \tB'_{np})^2$ does. In the previous subsections we have moreover proven that this is the only such scalar which under \3 transformation gives a total derivative. Moreover in subsection $5.5.1$ it was proven that this mechanism cannot lead to a complete field content for ${\mathbf \Xi}_{++}$. 
Hence
$$\partial_n \frac{\partial \ttpsi}{\partial \partial_p A'_m}=0,\ \partial_m \frac{\partial \ttpsi}{\partial \partial_p \tP'}=0,\ \partial_r \frac{\partial \ttpsi}{\partial \partial_p  \tB'_{mn}}=0,\ \partial_n \frac{\partial \ttpsi}{\partial \partial_p \ttA'_n}=0$$
Thus one has
\beqa
\label{ttpsi-1}
\ttpsi= \beta^{pm} \partial_p A'_m + \beta^p \partial_p \tP' + \beta^{pmn} \partial_p  \tB'_{mn} + \gamma^{pm} \partial_p \ttA'_{m}.
\eeqa
\noi 
where $\beta^{pm}, \beta^p, \beta^{pmn}, \gamma^{pm}$ are arbitrary scalars.  We now use the \3 transformation laws of $\psi$ 
$$ \delta_v \ttpsi = v^m \partial_m \psi = \beta^{pm} \partial_p 
(v^n \tB^\prime_{mn}+
v_m \tP^\prime) + \beta^p v^m \partial_p \ttA'_m + \beta^{pmn} \partial_p (
- v_m \ttA^\prime_n + v_n \ttA^\prime_m - i \varepsilon_{mnpq} v^p \ttA^\prime{}^q{})
+ \gamma^{pm}v^n \partial_p \partial_n A'_m .$$
Since in the LHS we have $v\cdot \partial$, one has $\beta^{pm}=0=\beta^{pmn}=\beta^p$ and hence
\beqa
\label{prostie1}
\ttpsi= \gamma^{pm} \partial_p \ttA'_m, \ \psi= \gamma^{pm} \partial_p A'_m.
\eeqa 
\noi
We now continue this process of applying the \3 transformation laws which lead furthermore to
\beqa
\label{prostie2}
\psi&=&\gamma^{pm} \partial_p A'_m,\nonumber\\
\tpsi_m &=&\gamma^{pn} \partial_p  \tB'_{nm} + \gamma_{pm}\partial^p \tP',\nonumber\\
\ttpsi_{mn}&=&\gamma_{p[m}\partial^p \ttA'_{n]+},\nonumber\\
\psi_{mn}&=&\gamma_{p[m} \partial^p A'_{n]+}. 
\eeqa
\noi
The last remaining transformation law to impose is 
\beqa
\label{finalizare}
\delta_v \psi_{mn}= -v_{[n} \tpsi_{m]+}= v_{[m} \tpsi_{n]+}.
\eeqa
Since \eqref{prostie2} gives us the explicit expression of $\psi_{mn}$ and $\tpsi_{n}$, equation \eqref{finalizare} translates in an equation in the \3 fields $\tP'$ and $\tB'_{mn}$. Equating the most easy parts in $\tP'$ on both sides of \eqref{finalizare} one has
$$\gamma_{p[m}\partial^p v_{n]+}\tP'= \partial_{[m} v_{n]+}\tP'$$
which leads to
\beqa
\label{sfarsit}
\gamma_{pm}=\delta_{pm}.
\eeqa
\noi
This equation, one inserted in \eqref{prostie1} and \eqref{prostie2} immediately leads to
$$\Psi ({\mathbf \Xi}_{+-})= {\cal D} {\mathbf \Xi}_{+-} $$ 
(with ${\cal D} {\mathbf \Xi}_{+-}$ being defined in (\ref{partialXipm})).
The 
derivative dependence on the other multiplets (considered alone) is done analogously. 

\medskip

Let us now consider the one-derivative dependence on several the \3 multiplets: 
$$\Psi (\partial_p \p, \partial_p B_{mn}, \partial_p \tA_m, \partial_p \ttP, \partial_p \ttB_{mn}; \partial_p A'_m, \partial_p \tP', \partial_p \tB'_{mn}, \partial_p \ttA'_m).$$
As before we start by writing
\beqa
\delta_v \ttpsi = v^m \partial_m \psi &=&
\frac{\partial \ttpsi}{\partial \partial_p \ttP}\partial_p \delta_v \ttP+
 \frac{\partial \ttpsi}{\partial \partial_p p}\partial_p \delta_v \p+
\frac{\partial \ttpsi}{\partial \partial_p \tA_m}\partial_p \delta_v \tA_m+
\frac{\partial \ttpsi}{\partial \partial_p \ttB_{mn}}\partial_p \delta_v \ttB_{mn}+
\frac{\partial \ttpsi}{\partial \partial_p B_{mn}}\partial_p \delta_v B_{mn}\nonumber\\
&+&\frac{\partial \ttpsi}{\partial \partial_p A'_m} \partial_p \delta_v A'_m +
\frac{\partial \ttpsi}{\partial \partial_p \tP'} \partial_p \delta_v \tP' +
 \frac{\partial \ttpsi}{\partial \partial_p \tB'_{mn}} \partial_p \delta_v \tB'_{mn}+
\frac{\partial \ttpsi}{\partial \partial_p \ttA'_{m}} \partial_p \delta_v \ttA'_{m}.\nonumber
\eeqa
\noindent
Once more we make use of the explicit form \eqref{transfo2} of the transformation laws thus obtaining
\beqa
\delta_v \ttpsi = v^m \partial_m \psi &=&
\frac{\partial \ttpsi}{\partial \partial_p \ttP}\partial_p (v^m \partial_m \p)+
 \frac{\partial \ttpsi}{\partial \partial_p \p}\partial_p (v^m \tA_m)+
\frac{\partial \ttpsi}{\partial \partial_p \tA_m}\partial_p (v_m \ttP+ v^n \ttB_{mn})\nonumber\\
&+&\frac{\partial \ttpsi}{\partial \partial_p \ttB_{mn}}\partial_p (v^r \partial_r B_{mn})
-4\frac{\partial \ttpsi}{\partial \partial_p B_{mn}}\partial_p (v_m \tA_n)\nonumber\\
&+&\frac{\partial \ttpsi}{\partial \partial_p A'_m}  \partial_p (v^n \tB^\prime_{mn}+
v_m \tP^\prime)+
\frac{\partial \ttpsi}{\partial \partial_p \tP'} v^m \partial_p
\ttA_m^\prime 
-4  \frac{\partial \ttpsi}{\partial \partial_p  \tB'_{mn}} v_m \partial_p \ttA'_n +
\frac{\partial \ttpsi}{\partial \partial_p \ttA'_{m}} v^n \partial_p \partial_n A^\prime_m\nonumber
\eeqa
\noi
As in the previous case, by integrating by parts one can now construct total derivatives in the LHS; then applying the same arguments as in the previous case one has
\beqa
\partial_p \frac{\partial \ttpsi}{\partial \partial_p \ttP}=0,\
\partial_p \frac{\partial \ttpsi}{\partial \partial_p \p}=0,\
\partial_p \frac{\partial \ttpsi}{\partial \partial_p \tA_m}=0,\
\partial_p \frac{\partial \ttpsi}{\partial \partial_p \ttB_{mn}}=0,\
\partial_p \frac{\partial \ttpsi}{\partial \partial_p B_{mn}}=0,\nonumber\\
\partial_n \frac{\partial \ttpsi}{\partial \partial_p A'_m}=0,\ \partial_m \frac{\partial \ttpsi}{\partial \partial_p \tP'}=0,\ \partial_r \frac{\partial \ttpsi}{\partial \partial_p  \tB'_{mn}}=0,\ \partial_n \frac{\partial \ttpsi}{\partial \partial_p \ttA'_{m}}=0.\nonumber
\eeqa
\noi
This now gives
\beqa
\label{ttpsi-mix}
\ttpsi= \tilde{\tilde \alpha}^p \partial_p \ttP + \tilde \alpha^{pm} \partial_p \tA_m + \alpha^{pmn} \partial_p B_{mn}  + \tilde {\tilde \alpha}^{pmn}\partial_p \ttB_{mn} + \alpha^p \partial_p \p \nonumber\\
+\beta^{pm} \partial_p A'_m + \beta^p \partial_p \tP' + \beta^{pmn} \partial_p  \tB'_{mn} + \gamma^{pm} \partial_p \ttA'_{m}. \nonumber
\eeqa
\noi 
where $\tilde{\tilde \alpha}^p,\tilde \alpha^{pm}, {\it etc.}$, $\beta^{pm},\beta^p, \beta^{pmn}$ and $\gamma^{pm}$ are $x-$independent. For this reason and the fact that the two lines of \eqref{ttpsi-mix} do not mix under the \3 transformation laws \eqref{transfo2} (since they belong to different multiplets, namely ${\mathbf \Xi}_{++}$ and ${\mathbf \Xi}_{+-}$) the dependences of the $\Psi$ fields on the two considered multiplets can now be completely separate. Hence this case also reduces to the cases treated above.

\bigskip

One thus concludes with
\beqa
\label{lemme4}
\Psi ({\mathbf \Xi}_{++}, {\mathbf \Xi}_{+-}, {\mathbf \Xi}_{+-},{\mathbf \Xi}_{--}) = \alpha {\mathbf \Xi}_{++} + \beta
{\mathbf \Xi}_{--}^* +
\gamma{\cal D} {\mathbf \Xi}_{+-}+ \mu {\cal D} {\mathbf \Xi}_{-+}^*.
\eeqa

 We can thus state 

\bigskip
{\it {\bf IV:} The only function $\Psi$ 
with at most first order derivatives in the fields and  transforming as
a ${\mathbf \Xi}_{++}$ multiplet is
$$\Psi  = \alpha {\mathbf \Xi}_{++} + \beta
{\cal D} {\mathbf \Xi}_{+-}.$$}


Moreover, if one considers several copies of the same multiplet, the mathematical relations above will remain identical (one will just have more copies of them) and hence the conclusion does not change.

\bigskip
    
 The case of functions involving higher number of derivatives goes along the 
same lines, by comparing the number of spatial derivatives.  Let us illustrate this by treating the case $\Psi =\Psi (\partial^{(k)} {\mathbf \Xi}_{++})$, where we write $\partial^{(r)}=\partial^{r_1}\dots\partial^{r_k}$ ($r$ being a multiindex). We consider here explicitly the case $k$ even.

One writes
\beqa
\delta_v \ttpsi = v^m \partial_m \psi =
\frac{\partial \ttpsi}{\partial \partial^{(r)}\ttP} \delta_v \partial^{(r)}\ttP 
&+&\frac{\partial \ttpsi}{\partial \partial^{(r)}\varphi} \delta_v \partial^{(r)}\varphi +
 \frac{\partial \ttpsi}{\partial \partial^{(r)}B_{mn}} \delta_v \partial^{(r)}B_{mn}
+\frac{\partial \ttpsi}{\partial \partial^{(r)}\ttB_{mn}} \delta_v \partial^{(r)}\ttB_{mn} \nonumber\\
&+&\frac{\partial \ttpsi}{\partial \partial^{(r)}\tA_{m}} \delta_v \partial^{(r)}\tA_{m}.
\eeqa
\noindent
We make use again of the form \eqref{transfo2} of the transformation laws of the \3 fields. This gives
\beqa
\label{F-n}
v^m \partial_m \psi &=&
\frac{\partial \ttpsi}{\partial \partial^{(r)} \ttP}  \partial^{(r)} v^m \partial_m \varphi +
\frac{\partial \ttpsi}{\partial \partial^{(r)} \varphi}  \partial^{(r)} v^m \tA_m 
+ \frac{\partial \ttpsi}{\partial \partial^{(r)} B_{mn}} \partial^{(r)}(-v_m \tA_n+ v_n \tA_m + i\e_{mnpq}v^p\tA^q) 
\nonumber \\
&+&
 \frac{\partial \ttpsi}{\partial \partial^{(r)} \ttB_{mn}}  \partial^{(r)} v^p \partial_p B_{mn}+
\frac{\partial \ttpsi}{\partial \partial^{(r)} \tA_{m}} \partial^{(r)}(v_m \ttP + v^n \ttB_{mn}).
\eeqa
\noi


\noi
Since $v$ is an arbitrary vector and comparing the number the of spatial derivative one has some kind of a generalisation of the first of the cases treated here, namely
\beqa
\label{acum-u}
\frac{\partial \ttpsi}{\partial \partial^{(r)} \tA^m} \partial^{(r)} \p 
+ \frac{\partial \ttpsi}{\partial \partial^{(r)} \tA_n} \partial^{(r)} \ttB_{nm}+\frac{\partial \ttpsi}{\partial \partial^{(r)} \p} \partial^{(r)} \tA_m 
-4 \frac{\partial \ttpsi}{\partial \partial^{(r)} B^{mn}}\partial^{(r)} \tA^n=0
\eeqa
\noi
and
\beqa
\partial_m \psi= \frac{\partial \ttpsi}{\partial \partial^{(r)} \ttP} \partial_m \partial^{(r)} \p +  \frac{\partial \ttpsi}{\partial \partial^{(r)} \ttB_{pq}}  \partial_m \partial^{(r)} B_{pq}
\eeqa
Then, integrating by parts this last equation one has
$$\partial_m( \psi -
\frac{\partial \ttpsi}{\partial \partial^{(r)}\ttP} \partial^{(r)}\varphi  
 -  \frac{\partial \ttpsi}{\partial \partial^{(r)}\ttB_{mn}} \partial^{(r)}B_{mn} )
= - \partial^{(r)}\varphi \partial_m 
\frac{\partial \ttpsi}{\partial \partial^{(r)}\ttP}
- \partial^{(r)}B_{np} \partial_m  \frac{\partial \ttpsi}{\partial\partial^{(r)} \ttB_{np}}.$$
\noindent
In the LHS one has a total derivative; generalising the mechanism observed in the previous cases, to obtain a total derivative in the RHS also one needs to have $\ttpsi=\partial^{(r)} \p \partial_{(r)} \ttP + \frac 14 \partial^{(r)} B_{np} \partial_{(r)} \ttB^{np}$. 
Now, since we need $\ttpsi$ to be a Lorentz scalar, this means that $\ttpsi=\Box^{\frac {k}{2}} \p  \Box^{\frac k2} \ttP + \frac 14 \Box^{\frac k2} B_{np} \Box^{\frac k2} \ttB^{np}$. Since we need  the \3 transformation of $\ttpsi$  to be  a total derivative, one needs as before   ${\cal L} ({\mathbf \Xi}_{++})= \Box^{\frac k2} \p \Box^{\frac k2}\ttP + \frac 14  \Box^{\frac k2}B_{np} \Box^{\frac k2} \ttB^{np}-\frac12 ( \Box^{\frac k2}\tA_n)^2$. However, as before, it was proven in subsection $5.5.1$ by explicitly applying the \3 transformations laws \eqref{transfo2} that this type of coupling cannot lead to the whole field content of a ${\mathbf \Xi}_{++}$ multiplet.
Thus one has
 $$\partial_m \frac{\partial \ttpsi}{\partial \partial^{(r)}\ttP}=0,\ \partial_m  \frac{\partial \ttpsi}{\partial \partial^{(r)} \ttB_{np}}=0.$$
One can now write
\beqa
\label{salvarea-v-k}
\ttpsi= \alpha \Box^{\frac k2}\ttP +  X_{mn} \Box^{\frac k2}\ttB^{mn} + \ttpsi_R (\partial^{(r)} \tA, \partial^{(r)}\p, \partial^{(r)} B)
\eeqa
\noi
where $\alpha$ is a scalar and $X_{mn}$ is a self-dual $2-$form, invariant under \3 and $x-$independent and $\ttpsi_R$ is an arbitrary function. Hence \eqref{acum-u} becomes
\beqa
\label{acum2-k}
\frac{\partial \ttpsi_R}{\partial \partial^{(r)}\p} \partial^{(r)}\tA_m -4 \frac{\partial \ttpsi_R}{\partial \partial^{(r)}B^{mn}} \partial^{(r)}\tA^n + \frac{\partial \partial^{(r)}\ttpsi_R}{\partial \partial^{(r)} \tA^m}\ttP + \frac{\partial \ttpsi_R}{\partial \partial^{(r)}\tA_p}\partial^{(r)}\ttB_{pm} =0
\eeqa
\noi
Suppose now that $ \frac{\partial \ttpsi_R}{\partial \partial^{(r)}\tA_m}\ne 0$; this implies that $\ttpsi_R$ contains $\partial^{(r)}\ttP$ or $\partial^{(r)}\ttB$ which is a contradiction. Hence
$$ \frac{\partial \ttpsi_R}{\partial \tA_m}=0.$$
Using now arguments of type \eqref{s-a}, one has
$$\frac{\partial \ttpsi_R}{\partial \p}=0=\frac{\partial \ttpsi_R}{\partial B_{mn}}$$

Hence, \eqref{salvarea-v-k} writes
\beqa
\label{salvarea-k}
\ttpsi= \alpha  \Box^{\frac k2} \ttP +  X_{mn}  \Box^{\frac k2}\ttB^{mn}.
\eeqa
\noi
With this explicit form for $\ttpsi$, one can now apply in the usual step-by-step process the \3 variation laws which leads to

$$X_{mn}=0$$


\noi
and thus
$$ \Psi = \alpha \Box^{\frac k2} {\mathbf\Xi}_{++}. $$


When $\Psi$ is a function of the second type of multiplet the calculation is done analogously, leading, when $k$ is even  to the solution
$$\Psi = \alpha \Box^{[\frac k2]} {\cal D}{\mathbf \Xi}_{+-}.$$

When considering dependences on the fields and their derivatives (of a given order) the reasonment is done by applying all the situations described above, leading to the same conclusion.

\medskip

One thus concludes that the $\psi$ fields can be obtained only linearly out of the four considered \3 bosonic multiplets. Comparing this result with the one of the previous subsection (which was stating that these $\psi$ fields are the most general possibility to quadratically couple the \3 multiplets to some arbitrary fields) one concludes that no invariant terms of order higher then two in the fields can be constructed. This means that one cannot obtained invariant self-interacting terms for the bosonic multiplets ${\mathbf \Xi}_{\pm \pm}$.


\section{Concluding remarks and perspectives}
\label{4-perspective}

We have thus proved that self-interaction terms between the four bosonic multiplets considered in this model are forbidden by \3 invariance. Nevertheless this  can be compared  with the compatibility for the usual electromagnetism, where photons do not self-interact.

More general possibilities of interaction have to be investigated for a verdict to this issue. One might reconsider at this level the fermionic multiplets of \cite{articol} and investigate a possible interaction between them and  boson multiplets. Furthermore, interactions with another type of bosonic multiplets (eventually more general multiplets) may be taken into consideration.

Maybe a more promising approach consists in regarding the \3 algebra in extra-dimensions, as we will do in the next chapter. Considering this extension which leads naturally to $p-$forms, one might consider the possibility of coupling the \3 fields with extended objects (see next chapter).

As already stated throughout this chapter, some deeper analyses of possible mechanisms of elimination of unphysical degrees of freedom of the fields may be entitled. This may eventually involve some presence of ghosts, in connexion with suited quantification procedures.

In the same line of reasoning, different perspectives given in chapter $6$ are also pertinent for the case of $4$ dimensions. Amongst these ones, let us recall the eventual connexion between our extension of the Poincar\'e symmetry and the other extensions listed in subsection $5.2.1$. Also one may think of the possibility of constructing  non-abelian models. Finally let us mention 
that considering  $p-$forms with $p\ge 2$ implies 
 a high rigidity for the interaction possibilities,
see \cite{henneaux}).

\chapter{Cubic symmetry in arbitrary dimensions}
 
We have seen in the previous chapter the main features of a \3 model
in $4$ dimensions. In this chapter we extend some of these issues to arbitrary 
dimensions.

The chapter is organised as follows. 
Firstly we give a short motivation for $p-$forms in general and for our approach in particular; some basic properties of $p-$forms are also listed. 
We then introduce the algebraic construction in
extra-dimensions, matrix representations, the fields and their
transformation laws. We then present invariant Lagrangians and
conclude on future perspectives.

This chapter presents  results published in \cite{praga}.

\section{Motivation for $p-$forms and cubic supersymmetry in arbitrary dimensions}
\label{p-forme}

In this section we generally define a $p-$form; brief motivations for their use are given. We then give some motivations for this extension of \3 in arbitrary dimensions. We will end this section by recalling some basic operations on $p-$forms.

Throughout this chapter we work using components of the $p-$forms.

\subsection{$p-$forms and \3 in arbitrary dimensions}
  
  The components of a  $p-$form $A_{M_1,\dots,M_p}$ (with $M_1,\dots, M_n=0,\dots,D-1$) in $D$ dimensions represent an antisymmetric tensor field of rank $p$.

A trivial example is the $0-$form, which is a scalar field. Another well-known example is given by electromagnetism, where the vector potential $A_M$ is nothing but a $1-$form. Hence, one can see the $p-$forms as {\it generalised gauge fields}.

  A $1-$form couples to charged point particles. Indeed, considering the particle's worldline $x_M(\lambda)$  (worldline parametrised by the coordinate $\lambda$),  the current generated by this particle writes
$$ J_M(u)=e\int dx_M \delta^D (u-x(\lambda))=e\int d\lambda\dot x_M \delta^D (u-x(\lambda))$$
where $\dot x_M$ means the derivative of $x_M$ with respect to the parameter $\lambda$.
One then has a coupling 
$$\int A^M J_M d^D u= e \int d\lambda \dot x_M A^M$$
between the vector field and the current.

   Similarly, a  $2-$form $B_{mn}$ couples to some ``charged'' string, a one-dimensional extended object. The corresponding notion of the worldline for a particle is in the case of the string the worldsheet $x_m (\tau,\sigma)$ which is a surface (parametrised by the coordinates $\tau$ and $\sigma$, see fig. $7.1$) obtained from the evolution in space-time of the string. 
\epsfxsize = 5cm     
$$
\epsffile{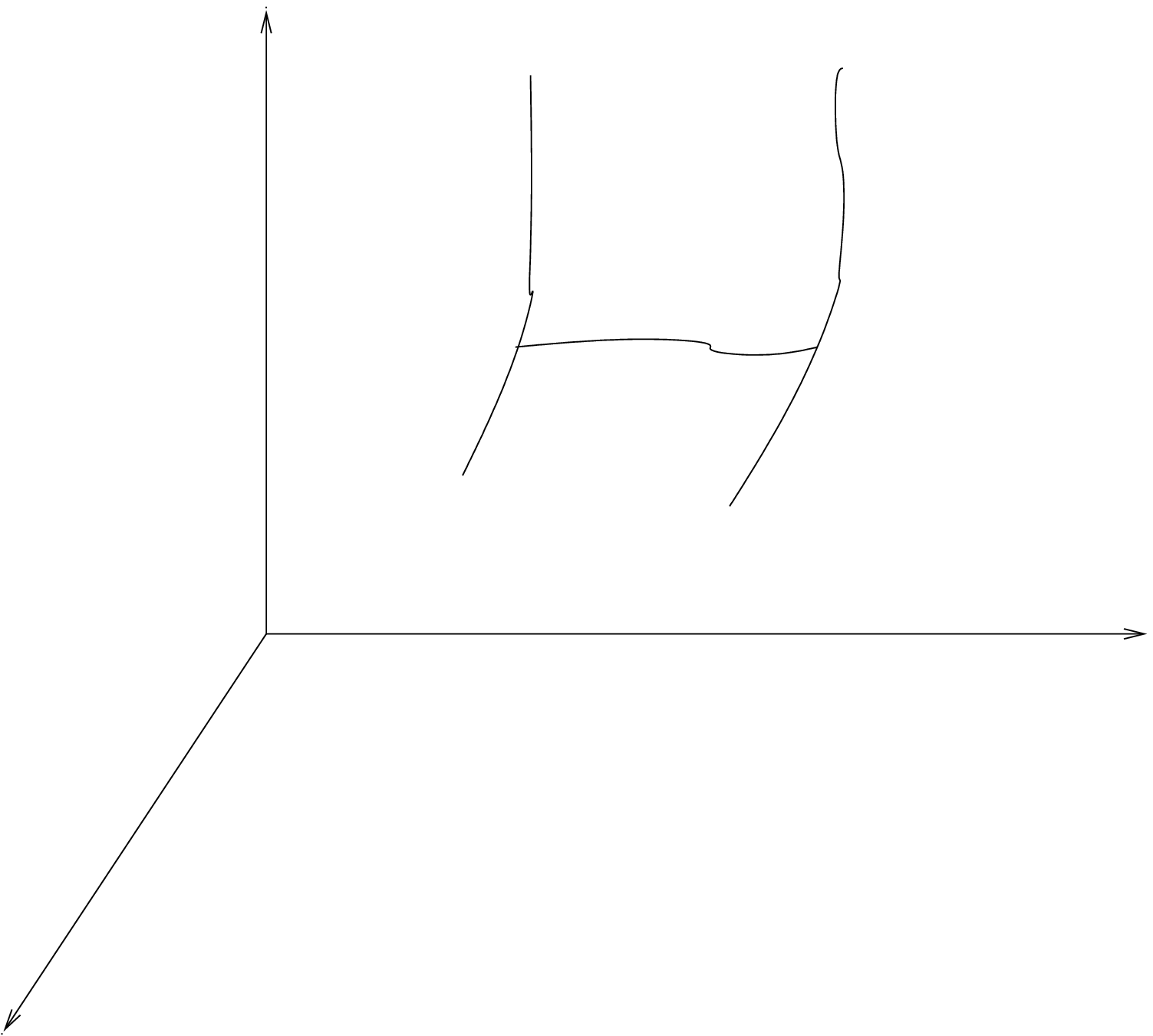}
$$
\begin{center}
{\bf Fig. 7.1:} Worldsheet parametrised by two coordinates $\tau$ and $\sigma$
\end{center}

\noi
Define now
$$\sigma_{mn}=\dot x_m x'_n - x'_m \dot x_n$$
where $\dot x_m$ (resp. $x'_m$) means derivation of $x_m$ with respect to $\tau$ (resp. $\sigma$). This can be interpreted as the exterior product of $\dot x$ and $x'$ (see next subsection for the definition).
The current of the strings writes 
$$J_{MN}(u)=
g\int d\sigma d\tau \sigma_{MN} \delta^D (u-x(\sigma,\tau))$$
and one has a coupling
$$\int A^{MN} J_{MN} d^D u=g\int d\sigma d\tau A^{MN} \sigma_{MN}.$$

      One generalises this mechanism hence concluding that  $p-$forms couple to elementary objects of spatial dimension  $p-1$ (strings, (mem)branes). Indeed, 
such a  $(p-1)$-dimensional object spans in the $D-$dimensional space a hypersurface $x_M (\sigma_1,\dots,\sigma_p)$ (hypersurface parametrised by the coordinates $\sigma_1,\dots,\sigma_p$); as above define $\sigma_{M_1\dots M_p}=\sum_{\tau\in S_p} \e(\tau)\frac{\partial x_M}{\partial \sigma_{\tau (i)}}\dots \frac{\partial x_M}{\partial \sigma_{\tau (p)}}$ (where $S_p$ is the group of permutations of $p$ elements and $\e(\tau)$ is the signature of the permutation $\tau$).  
The current associated with such a ``charged'' $(p-1)$-dimensional object is
$$J_{M_1\dots M_{p}}(u)=g \int d\sigma_1\dots d\sigma_{p} \sigma_{M_1\dots M_p} \delta^D (u-x(\sigma,\tau))$$
and the respective coupling with a $p-$form $A^{M_1\dots M_p}$ writes
$$\int A^{M_1\dots M_p} J_{M_1\dots M_p} d^D u=g\int d\sigma_1\dots d\sigma_{p} A^{M_1\dots M_p} \sigma_{M_1\dots M_p}.$$

\medskip

These antisymmetric gauge fields  play an increasingly fundamental
        role in theoretical physics today. The interest for them basically emerges from their natural appearance in theories
  like  supergravity or superstrings (see for example  \cite{wessbagger} or \cite{pol-book}). 
Thus, it seems
tempting  to attempt to transfer to these extended systems ideas and
      theoretical tools that were successful when applied to point-like objects. And what is the most powerful of these
theoretical methods, if not, obviously,  {\it symmetries}?

Considerations of arbitrary dimensions lead naturally to $p-$forms.
We  have an important motivation for our approach, approach which allows to obtain  a {new type of symmetry} at this level of antisymmetric gauge fields.
Another important
motivation for the extension of our model to arbitrary dimensions comes from the possibilities of interactions, as we have
already mentioned in the last chapter, when we argued about the eventuality  to try to obtain such interactions between the \3 fields and extended objects (by mechanisms similar to the mechanisms illustrated above).

The algebraic construction being the same as described in the previous
chapters, \3 in extra-dimensions is also {\it a priori} allowed by the no-go
theorems. In $4$ dimensions, as stated in subsection $6.1.2$ of this thesis, one has the results of the Coleman-Mandula and Haag-Lopuszanski-Sohnius theorems. Some generalisations to an arbitrary number $D$ of dimensions exists. In \cite{Dcm}, the authors provide a generalisation in this sense of the Coleman-Mandula no-go theorem.
Our construction evades this type results from the same reasons as the one exposed in the previous chapter, for the case $D=4$.

\subsection{Some basic operations on $p-$forms}

If one considers a $p-$form in $D$ dimensions, several operations can be defined. (Recall that we refer  to components of these $p-$forms.) One has the Hodge dual\footnote{In some references called the Poincar\'e dual.}, which associates to a $p-$form $A_{[p]}$ a $(D-p)-$form $B_{[D-p]}={}^*(A_{[p]}$. In components this writes
\beqa
\label{hodge}
 B_{[D-p]}{}_{M_1 \ldots M_{D-p}}=({}^*A_{[p]})_{M_1 \ldots M_{D-p}}= \frac{1}{p!} 
\varepsilon_{M_1 \ldots M_{D-p} N_1 \ldots
N_p} A_{[p]}^{N_1 \ldots N_p}.
\eeqa
\noindent
where the Levi-Civita tensor $\varepsilon_{M_1 \ldots M_{D-p} N_1 \ldots
N_p}$ is defined in \eqref{levi-civita}.

When the dimension is even, the dual of $(D/2)-$form is also a $(D/2)-$form; thus  one can define an (anti-)self-dual $(D/2)-$form by

\beqa
\label{self}
{}^\star A_{[D/2]}= \left\{
\begin{array}{ll}
\pm A_{[D/2]}&\mathrm{~when~}D/2\mathrm{~odd~}({}^{\star^2}=1) \cr
\pm i A_{[D/2]}&\mathrm{~when~}D/2\mathrm{~even~}({}^{\star^2}=-1).
\end{array}\right.
\eeqa

\noindent
One thus notices that these (anti-)self-dual $D/2-$forms are complex when $D/2$ is an even number.

\medskip

We now define the following set of operations

\begin{enumerate}
\item the exterior derivative $d$ which maps a $p-$forms to a $(p+1)-$form and which (in components) is given by

\beqa
\label{ext}
(d A_{[p]})_{M_1 \ldots M_{p+1}}= \frac{1}{p!} \delta_{M_1 \ldots 
M_{p+1}}^{N_1 
\ldots N_{p+1}} {\partial_{N_1}} A_{[p]}{}_{N_2 \ldots N_{p+1}}
\eeqa
\noindent
where
\beqa
\label{gros-delta} 
\delta_{M_1 \ldots 
M_{p+1}}^{N_1 
\ldots N_{p+1}}= \left|  
\begin{array}{lllllll}
\delta_{M_1}^{N_1} &  \dots & \delta _{M_1}^{N_{p+1}}\\
\vdots & & \vdots\\
\delta_{M_{p+1}}^{N_1}&  \dots & \delta_{M_{p+1}}^{N_{p+1}}
\end{array}
\right|.
\eeqa
\noi
The exterior derivative is used to construct the field strength out of the potentials. For example, in the case of a vector field $A_M$, taking $F=dA$, one obtains the well-known field strength $F_{MN}=\partial_M A_N - \partial_N A_M$. In the case of a $2-$form $B_{MN}$, taking $H=dB$, one obtains the field strength $H_{MNP}=\partial_M B_{NP}+\partial_P B_{MN}+\partial_N B_{PM}$.

Furthermore, the exterior derivative $d$ is subject to
$$ dd=0.$$

One can further define its adjoint $d^\dag$ which maps a $p-$form into a $(p-1)-$form; it is defined by $d^\dag=(-1)^{pD+D} {}^\star d {}^\star$ and in components writes
\beqa
\label{extdag}
(d^\dag A_{[p]})_{M_2 \ldots M_{p}}={ \partial^{M_1}} 
 A_{[p]}{}_{M_1 N_2 \ldots N_{p}}.
\eeqa

Using now a $1-$form $v$, we further define

\item the inner product $i_v$, which also maps a  $p-$form into a $(p-1)-$form and writes (in components) 
\beqa
\label{iner2}
 (i_v A_{[p]})_{M_1 \ldots M_{p-1}}= A_{[p]}{}_{M_1 \ldots M_p}
v^{M_p}.
\eeqa
\noindent
Notice the convention used here, useful for further calculations: the summation 
is done on the
last index instead of the first one.

\item the exterior product $\wedge$ maps a  $p-$form into a $(p+1)-$form and in components writes
\beqa
\label{ext2}
(A_{[p]}{}\wedge v)_{M_1 \ldots M_{p+1}}= 
\frac{1}{p!} \delta_{M_1 \ldots M_{p+1}}^{N_1\ldots N_{p+1}} A_{[p]}{}_{N_1 
\ldots N_p}
v_{N_{p+1}}.
\eeqa

\item the action of the vector field 
\beqa
\label{vector-field}
v A_{[p]}=v^M \partial_M A_{[p]}
\eeqa
\end{enumerate}

Some important features of a free $p-$form were given in subsection \ref{detoate}.  We have seen that such a $p-$form has $C_D^p$ components (see \eqref{p-comp}); however, using the gauge invariance and its reducibility (for $p>1$, see \eqref{p-gauge} and  \eqref{p-reductive}), one has $C_{D-1}^p$ degrees of freedom off-shell (see \eqref{p-off-shell}). Finally, using Ward identities one has $C_{D-2}^p$ degrees of freedom on-shell (see \eqref{p-on-shell}).

For some useful identities on general $p-$forms, one may refer to Appendix \ref{a-pforme}.

\section{Algebra and $p-$forms}
\label{D-algebra}

After this brief introduction on antisymmetric gauge fields and the motivation we have offered for our approach, in this section we proceed with the algebraic construction in $D$ dimensions, some matrix representations, the $p-$form content of the multiplets treated here and finally their transformation laws.

\subsection{Algebraic construction}

The \3 algebra \eqref{algebra} of the previous chapter generalises
naturally to an arbitrary number of dimensions. For this purpose, one uses the
Poincar\'e algebra in $D$ dimensions, generated by $L_{MN}, P_N$
($M,N=0,\ldots, D-1$). To this algebra one adds the generators $V_M$, lying
in the vector representation of ${\mathfrak{so}}(1,D-1)$. One thus
has the following Lie algebra of order $3$
 \beqa
\label{Dalgebra}
&&\left[L_{MN}, L_{PQ}\right]=
\eta_{NP} L_{PM}-\eta_{MP} L_{PN} + \eta_{NP}L_{MQ}-\eta_{MP} L_{NQ},
\nonumber \\
 &&\left[L_{MN}, P_P \right]= \eta_{NP} P_M -\eta_{MP} P_N, \nonumber \\
&&\left[L_{MN}, V_P \right]= \eta_{NP} V_M -\eta_{MP} V_N, \ \
\left[P_{M}, V_N \right]= 0, \\
&&\left\{V_M, V_N, V_R \right \}=
\eta_{MN} P_R +  \eta_{MR} P_N + \eta_{RN} P_M, \nonumber
\eeqa
 \noindent
where $\{V_M,V_N,V_P \}=
V_M V_N V_R + V_M V_R V_N + V_N V_M V_R + V_N V_R V_M + V_R V_M V_N +
V_R V_N V_M $  stands, as in the previous chapter, for the symmetric product of order $3$ and
$\eta_{MN} = \mathrm{diag}\left(1,-1,\ldots ,-1\right)$  is the $D-$dimensional
 Minkowski metric (see also Appendix E).

\subsection{Matrix representations}
\label{D-matrice}

Following the same line of reasoning as in the previous chapter one has the
following matrix representation

\begin{eqnarray}
\label{matred}
V_M=\begin{pmatrix} 0&\Lambda^{1/3} \Gamma_M& 0 \cr
                           0&0&\Lambda^{1/3}\Gamma_M \cr
                           \Lambda^{-2/3}\partial_M&0&0\end{pmatrix}
\end{eqnarray}

\noindent
with $\Gamma_M$ the $D-$dimensional $\Gamma-$matrices
(see  Appendix E).
As in the previous chapter, the mass parameter $\Lambda$ is taken to be equal to $1$.

When $D$ is an even number ($D=2k$), the $\Gamma-$matrices can be written
as 
\beqa
\label{D-Gamma}
\Gamma_M = \begin{pmatrix} 0& \Sigma_M \cr \bar \Sigma_M &0 \end{pmatrix},
\eeqa
with  $\Sigma_M=(\Sigma_0=1, \Sigma_I)$, $\bar \Sigma_M =(\bar \Sigma_0=1, \bar \Sigma_I=-\Sigma_I)$ (with $I=1,\dots,2k-1$), $\Sigma_I$ being the generators of the Clifford algebra $SO(2k-1)$, $\{\Sigma_I, \Sigma_J\}=2 \delta_{IJ}$ (generalisation
to arbitrary dimensions of the Pauli matrices, see  Appendix \ref{a-spinori}).

Thus, the matrix representation (\ref{matred}) becomes reducible and 
leads to two conjugated irreducible representations, denoted as in the
previous chapter by $+$ and $-$
\begin{eqnarray}
\label{matirred-D}
V_{+}{}_M=\begin{pmatrix} 0& \Sigma_M& 0 \cr
                           0&0&\bar \Sigma_M \cr
                           P_M&0&0 \end{pmatrix}, \nonumber \\
\\
V_{-}{}_M=\begin{pmatrix} 0& \bar \Sigma_M& 0 \cr
                           0&0&\Sigma_M \cr
                           P_M&0&0 \end{pmatrix}. \nonumber 
\end{eqnarray}

\bigskip

From now on we explicitly treat on the case 
$$D=2k$$ 
with 
$$k=2n,$$
 keeping in mind that similar analysis can be performed for different other parities of $D$ and $[\frac D2]$. 

\subsection{$p-$form content}


As in subsection \ref{sec-multi}, one considers the spinors on which these matrices act; one also needs also to consider the vacuum $\Omega$, a  \3 singlet lying in  specified representation of the Lorentz algebra. 
Considering the vacuum  in the trivial representation of the Lorentz algebra ${\mathfrak{so}}(1,D-1)$, 
 the two conjugated matrices  (\ref{matirred-D}) act on 
$$\Psi_+ =\begin{pmatrix} 
\psi_1{}_+ \cr
 \psi_2{}_- \cr \psi_3{}_+ \end{pmatrix} \mbox{ and } \Psi_- =\begin{pmatrix} 
 \psi_1{}_- \cr
\psi_2{}_+ \cr  \psi_3{}_- \end{pmatrix} $$ 
where $\psi_i{}_+$ denotes a LH Weyl spinor and $ \psi_i{}_-$ denotes a RH Weyl spinor ($i=1,2,3$), see also Appendix E.

Considering now the vacuum as a Weyl spinor (lying in the spinor representation of the Lorentz algebra), one has two possibilities: LH spinor, denoted by $\Omega_+$ or RH spinor, denoted by $ \Omega_-$. Hence, one can construct, as in the previous chapter, four product of spinors
\beqa
\label{even-odd}
\begin{array}{ll}
{\mathbf \Xi}_{++}= \Psi_+\otimes
\Omega_{+}=\begin{pmatrix}  \psi_1{}_+ \\ \psi_2{}_- \\  \psi_3{}_+ 
\end{pmatrix} \otimes
\Omega_{+}=\begin{pmatrix} \Xi_1{}_{++} \\ \Xi_2{}_{-+} \\  \Xi_3{}_{++} 
\end{pmatrix}, &
{\mathbf \Xi}_{-+}= \Psi_-\otimes
\Omega_{+}=\begin{pmatrix} \psi_1{}_- \\ \psi_2{}_+ \\ \psi_3{}_- \end{pmatrix}
 \otimes
\Omega_{+}=\begin{pmatrix} \Xi_1{}_{-+} \\ \Xi_2{}_{++} \\  \Xi_3{}_{-+} 
\end{pmatrix} \\
{\mathbf \Xi}_{--}=\Psi_- \otimes
\Omega_{-}=\begin{pmatrix} \psi_1{}_- \\  \psi_2{}_+ \\  \psi_3{}_- \end{pmatrix}
 \otimes
 \Omega_{-}=\begin{pmatrix} \Xi_1{}_{--} \\ \Xi_2{}_{+-} \\  \Xi_3{}_{--} 
\end{pmatrix},&
{\mathbf \Xi}_{+-}= \Psi_+\otimes
\Omega_{-}=\begin{pmatrix} \psi_1{}_+ \\ \psi_2{}_- \\ \psi_3{}_+ \end{pmatrix} 
\otimes
 \Omega_{-}=\begin{pmatrix} \Xi_1{}_{+-} \\ \Xi_2{}_{--} \\  \Xi_3{}_{+-} 
\end{pmatrix}
\end{array} \nonumber
\eeqa


\noi
The decomposition of these product of spinors gives the following field content for the $4$ multiplets
\beqa
\label{multiplets_even2}
\begin{array}{lllllll}
{\mathbf \Xi}_{++}=\begin{pmatrix} \Xi_1{}_{++} \\ \Xi_2{}_{-+} \\  \Xi_3{}_{++} 
\end{pmatrix}& \to&  \left\{  \begin{array}{l}
A_{[0]},  A_{[2]}\ldots, A_{[2n]}^{(+)} \cr 
\tilde A_{[1]}, A_{[3]}\ldots, \tilde A_{[2n -1]} \cr
{\tilde {\tilde A}}_{[0]}, {\tilde {\tilde A}}_{[2]},\ldots,{\tilde {\tilde  A}}_{[k]}^{(+)}
\end{array} \right.
& \ \ \ &
{\mathbf \Xi}_{--}=\begin{pmatrix} \Xi_1{}_{--} \\ \Xi_2{}_{+-} \\  \Xi_3{}_{--} 
\end{pmatrix}& \to&  \left\{  \begin{array}{l}
A^\prime{}_{[0]}, A^\prime{}_{[2]},\ldots, A^\prime{}_{[k]}^{(-)} \cr 
\tilde A^\prime{}_{[1]}, \tilde A^\prime{}_{[3]},\ldots, \tilde A^\prime{}_{[2n -1]} \cr
{\tilde {\tilde A}}^\prime{}_{[0]}, {\tilde {\tilde A}}^\prime{}_{[2]},\ldots,
{\tilde {\tilde  A}}^\prime{}_{[2n]}^{(-)}
\end{array} \right. \cr
\cr
{\mathbf \Xi}_{+-}=\begin{pmatrix} \Xi_1{}_{+-} \\ \Xi_2{}_{--} \\  \Xi_3{}_{+-} 
\end{pmatrix} & \to&  \left\{  \begin{array}{l}
A_{[1]}, A_{[3]}, \ldots, A_{[2n-1]} \cr 
\tilde A_{[0]}, \tilde A_{[1]}, \ldots, \tilde A_{[2n]}^{(-)} \cr
{\tilde {\tilde A}}_{[1]}, {\tilde {\tilde A}}_{[3]},\ldots,{\tilde {\tilde  A}}_{[2n-1]}
\end{array} \right.
& \ \ \ &
{\mathbf \Xi}_{-+}=\begin{pmatrix} \Xi_1{}_{-+} \\ \Xi_2{}_{++} \\  \Xi_3{}_{-+} 
\end{pmatrix}& \to&  \left\{  \begin{array}{l}
A^\prime{}_{[1]}, A^\prime{}_{[3]},\ldots , A^\prime{}_{[k-1]} \cr 
\tilde A^\prime{}_{[0]}, \tilde A^\prime{}_{[2]}, \ldots, \tilde A^\prime{}_{[k]}^{(+)} \cr
{\tilde {\tilde A}}^\prime{}_{[1]}, {\tilde {\tilde A}}^\prime{}_{[3]},\ldots,
{\tilde {\tilde  A}}^\prime{}_{[k-1]}.
\end{array} \right.
\end{array}
\eeqa
\noi 
For the explicit proof of this decomposition one may check Appendix \ref{a-spinori}. (This is actually a generalisation of the mechanism we have shown in subsection \ref{sec-multi} for the case of $4$ dimensions.)

As in the previous chapter, one can chose the multiplets to be complex conjugated (since $D=4n$), that is ${\mathbf \Xi}_{++}={\mathbf \Xi}_{--}^*$, ${\mathbf \Xi}_{+-}={\mathbf \Xi}_{-+}^*$; this choice of minimal field content will prove consistent with the transformation laws of the fields (see next subsection).

Moreover we denote the $p-$forms $A_{[p]}, A^\prime_{[p]}$  to be of gradation $-1$, the $p-$forms $\tilde A_{[p]}, \tilde A^\prime_{[p]}$ to be of gradation $0$ and the $p-$forms $ \tilde{ \tilde {A}}_{[p]}, \tilde{ \tilde {A}}^\prime_{[p]}$ to be of gradation $1$.

\subsection{Transformation laws of $p-$forms}
\label{D-transfo}

In this subsection we deduce the transformation laws of the sets of $p-$forms above. 
The key element of this calculation  is the relation \eqref{p-spin2}  between $p-$forms and the original product of spinors.

Recall that in the specific case we treat here ($D=2k=4n$) the transformation laws are obtained by {\it generalising the mechanism exposed for $D=4$} (see subsection \ref{4-transf}).  

As before the transformation writes
\beqa
\label{transfo++}
\delta_v {\mathbf \Xi}_{++} = (v^M V_{+M} \Psi_+)\otimes \Omega_+,
\eeqa
and similarly for the other multiplets.

 Using the form \eqref{matirred-D} of $V_{+M}$, one gets
\beqa
\label{transfo-spin}
\delta_v \Xi_{1++} &=& v^M \Sigma_M  \Xi_{2-+}, \nonumber \\
\delta_v  \Xi_{2-+} &=& v^M \bar \Sigma_M \Xi_{3++}, \\
\delta_v \Xi_{3++} &=& v^M \partial_M \Xi_{1++}. \nonumber 
\eeqa
\noi

Now, to obtain the variation of any $p-$form of \eqref{multiplets_even2} one has to make use of \eqref{p-spin2}. Thus, for any $p=1,\ldots,n-1$ one has
\beqa
\label{muie1}
\delta_v A_{[2p]}{}_{M_1\ldots M_{2p}}=\frac{1}{2^{2n-1}} \mathrm{Tr}\left(\Sigma_{M_1\ldots M_{2p}}\delta_v\Xi_{1++}\right).
\eeqa
\noi
One can now insert \eqref{transfo-spin} in \eqref{muie1} thus obtaining
\beqa
\label{muie11}
\delta_v A_{[2p]}{}_{M_1\ldots M_{2p}}=\frac{v^M}{2^{2n-1}} \mathrm{Tr}\left(\Sigma_{M_1\ldots M_{2p}}v^M\Sigma_M\Xi_{2+-}\right).
\eeqa
\noi
Furthermore, for $p=1,\ldots,n-1$ one uses \eqref{identities} to write
\beqa
\label{muie2}
\delta_v A_{[2p]}{}_{M_1\ldots M_p}=\frac{v^M}{2^{2n-1}} \mathrm{Tr}\left(\Sigma_{M_1\ldots M_{2p}M}\Xi_{2-+}+(\eta_{M_{2p} M}\Sigma_{M_1\ldots M_{2p-1}}+ \mathrm{perm.})\Xi_{2-+}\right).
\eeqa
\noi
where by {\it perm.} we mean sum on all permutations, with the sign corresponding to the signature.
Now, using once more \eqref{p-spin2} (but this time to recover $\tA_{[p+1]}$ and $\tA_{[p-1]}$) one has
\beqa
\label{muie3}
\delta_v A_{[2p]}{}_{M_1\ldots M_{2p}}=v^M \tA_{[2p+1]}{}_{M_1\ldots M_{2p}, M}+(v_{M_p} \tA_{[2p-1]}{}_{M_1\ldots M_{2p-1}}+ \mathrm{perm.}).
\eeqa
One can now use the definitions \eqref{iner2} and \eqref{ext2} to recover the inner and the exterior products
\beqa
\label{muie4}
\delta_v A_{[2p]}=i_v \tA_{[2p+1]} + \tA_{[2p-1]}\wedge v.
\eeqa
The border cases $p=0$ and $p=n$ need a particular look. When $p=0$, repeating the same procedure as above, one has $\delta_v A_{[0]}=i_v \tA_{[1]}$ (since $\delta_v A_{[0]}=\frac{v^M}{2^{2n-1}} \mathrm{Tr}\left(\delta_v\Xi_{1++}\right)$).


Let us now focus on the more elaborate situation of $p=n$. As above one has
\beqa
\label{muie1-2n}
\delta_v A_{[2n]}{}_{M_1\ldots M_{2n}}=\frac{1}{2^{2n-1}} \mathrm{Tr}\left(\Sigma_{M_1\ldots M_{2n}}\delta_v\Xi_{1++}\right).
\eeqa
\noi
Inserting as above \eqref{transfo-spin} to \eqref{muie1-2n} one gets
\beqa
\label{muie11-2n}
\delta_v A_{[2n]}{}_{M_1,\ldots ,M_{2n}}=\frac{v_M}{2^{2n-1}} \mathrm{Tr}\left(\Sigma_{M_1\ldots M_{2p}}\Sigma^M\Xi_{2+-}\right).
\eeqa
\noi
We now compute the quantity $\Sigma^M\Xi_{2+-}$. For this we make use of the decomposition \eqref{spin-p2} of  $\Xi_{2+-}$; thus one has
\beqa
\label{2n-cant}
\Sigma^M\Xi_{2+-}=\Sigma^M \left(\sum_{p=0}^{n-1} \tA_{[2p+1]}{}_{N_1\dots N_{2p+1}}\bar \Sigma^{N_1\dots N_{2p+1}}\right).
\eeqa
\noi
One makes now use again of identities of type \eqref{identities} to express $\Sigma^M \bar \Sigma^{N_1,\dots, N_{2p+1}}$; thus one has
\beqa
\label{2n-cant-2}
\Sigma^M\Xi_{2+-}=\sum_{p=0}^{n-1} \tA_{[2p+1]}{}_{N_1\dots N_{2p+1}}\left( \Sigma^{MN_1\dots N_{2p+1}}+(\eta^{MN_1}\Sigma^{N_2\dots N_{2p+1}}+{\rm perm.})\right).
\eeqa
\noi
The next step is to insert \eqref{2n-cant-2} to the original expression \eqref{muie11-2n}. One can now make use of the trace identity \eqref{puscarie} which, among all the sum over $p$ of \eqref{2n-cant-2} will keep just the last term ($p=n-1$) and consequently just $ \Sigma^{MN_1,\dots, N_{2p+1}}$. This writes explicitly as
\beqa
\label{2n-munca}
\delta_v A_{[2n]}{}_{M_1\dots M_{2n}}=({v_M}\tA_{[2p+1]}{}_{N_1\dots N_{2p+1}}\left(\delta_{M_1\dots M_{2n}}^{N_{2n-1}\dots N_1 M}-{i}\e_{M_1\dots M_2nP_{2n-1}\dots P_1 P} \eta^{MP}\eta^{N_1P_{2n-1}}\dots\eta^{N_{2n-1}P_1}\right)\nonumber\\
\eeqa
\noi
which gives further 
\beqa
\label{2n-comp}
\delta_v A_{[2n]}{}_{M_1\dots M_{2n}}=\left(v_{M_{2n}}\tA_{[2n-1]}{}_{M_1\dots M_{2n-1}}+{\rm perm.}\right) - i \e_{M_1\dots M_2nP_{2n-1}\dots P_1 P}v^P \tA_{[2n-1]}^{P_{2n-1}\dots P_1}.
\eeqa
\noi
Using now the definitions \eqref{ext2} and \eqref{hodge} of the exterior product and of the Hodge dual one has
$$\delta_v A_{[k]}=  {\tilde A}_{[k-1]} \wedge v -i {}^\star \hskip -.1truecm
\left (\tilde A_{[k]} \wedge v\right).$$

For the $p-$forms of gradation $0$, the calculus is done  as above (equations \eqref{muie1} to \eqref{muie3} with the use of the second of the identities \eqref{identities}), thus leading to 
\beqa
\label{muie5}
\delta_v \tA_{[2p+1]}=i_v \ttA_{[2p+2]} + \ttA_{[2p]}\wedge v, \mbox{ with }p=0,\dots,n-1
\eeqa

For the $p-$forms of the last sector, of gradation $1$, the calculation is simpler, leading similarly to 
$$\delta_v \ttA_{[2p]}=v A_{[2p]}=v^M\partial_M A_{[2p]},\ p=0,\dots,n.$$

Therefore, the transformation laws for the ${\mathbf \Xi}_{++}$ multiplet write

\beqa
\label{transfo-form}
&&\begin{array}{llllll}
\delta_v A_{[0]}&=& i_v {\tilde A}_{[1]}&
\delta_v {\tilde A}_{[1]}&=& i_v {\tilde {\tilde A}}_{[2]}
+ \ttA_{[0]} \wedge v \cr
&\vdots && &\vdots \cr
\delta_v A_{[2p]}&=& i_v {\tilde A}_{[2p+1]} + {\tilde A}_{[2p-1]} 
\wedge v &
\delta_v {\tilde A}_{[2p+1]}&=& i_v {\tilde {\tilde A}}_{[2p+2]} + {\tilde 
{\tilde A}}_{[2p]} \wedge v  \cr
&   \vdots && &  \vdots \cr
\delta_v A_{[2n]}^{(+)}&=&  {\tilde A}_{[2n-1]} \wedge v
-i
 \ {}^\star \hskip -.1truecm
\left
(\tilde A_{[2n-1]} \wedge v\right)
 &
\delta_v {\tilde A}_{[2n-1]}&=& i_v {\tilde {\tilde A}}_{[2n]}^{(+)}+
{\tilde {\tilde A}}_{[2n-2]} \wedge v 
 \cr
\end{array} \nonumber \\
\\
&&\ \ \delta_v {\tilde {\tilde A}}_{[0]}= v A_{[0]}, \ \ \ \ \ \ldots
 \ \ \ \ \ 
\delta_v {\tilde {\tilde A}}_{[2n]}^{(+)}= v A_{[2n]}^{(+)}. \nonumber 
\eeqa
\noi
One  notices that the term $-i\ {}^\star \hskip -.1truecm
\left
(\tilde A_{[k-1]} \wedge v\right)$ in $\delta_v A_{[k]}^{(+)}$ preserves
the  self-dual character of $A_{[k]}^{(+)}$. 
Since the decomposition \eqref{bi-spin2} imposes the existence of (anti-)self-dual $k$-forms, the (anti-)self-duality condition \eqref{self} demands that these forms are complex. From the transformation laws \eqref{transfo-form}, one sees that in this case all the $p-$form are complex. 

Similar laws are obtained for the other multiplets. When one has the transformations of anti-self-dual $k-$forms, the $-i$ in \eqref{transfo-form} becomes $i$ thus being in agreement with the complex conjugation prescription. Furthermore,
 the choice of minimal  field content done in the previous subsection (${\mathbf \Xi}_{++}^*={\mathbf \Xi}_{--}$ and  ${\mathbf \Xi}_{+-}^*={\mathbf \Xi}_{-+}$) proves to be compatible with the transformation laws.

Finally remark that these transformation laws may be geometrically interpreted in terms of inner products  \eqref{iner2}, exterior products \eqref{ext2} and action of a vector field \eqref{vector-field}.

\section{Cubic symmetry on $p-$forms}

We have seen in the previous section how, starting from the algebraic
structure \eqref{Dalgebra} in any dimension, one has the matrix representations 
which lead to the multiplets above. One then has the
 transformation laws  \eqref{transfo-form}. Everything is thus
 set to implement this at the level of invariant actions.
This allows us to obtain a new symmetry on these $p-$forms multiplets.
This purpose is achieved by coupling quadratically $p-$forms of different
multiplets. The invariant Lagrangians  obtained here are of gradation
$0$ that is, fields of gradation $-1$ are coupled to fields of
gradation $1$ and fields of gradation $0$ to themselves.

In order to obtain a real Lagrangian, one has to consider the complex conjugated multiplets of \eqref{multiplets_even2}. For the pair ${\mathbf \Xi}_{++}-{\mathbf \Xi}_{--}$ one writes
\beqa
\label{lag}
{\cal L}&=& {\cal L}({\mathbf \Xi}_{++}) + {\cal L}({\mathbf \Xi}_{--}) =
{\cal L}_{[0]} + \ldots + {\cal L}_{[k]} 
 + {\cal L}'_{[0]} \ \ + \ldots + {\cal L}'_{[k]} 
 \nonumber \\
&=&  
 d A_{[0]} d {\tilde {\tilde A}}_{[0]} +  \nonumber \\
&-& \sum_{p=0}^{n-1} \left(
\frac12  \frac{1}{(2p+2)!} d \tilde A_{[2p+1]} d \tilde A_{[2p+1]} 
+\frac12  \frac{1}{(2p)!} d^\dag \tilde A_{[2p+1]} d^\dag \tilde A_{[2p+1]}\right) 
 \nonumber \\ 
&+& \sum_{p=1}^{n-1} \left(
 \frac{1}{(2p+1)!} d  A_{[2p]} d \ttA_{[2p]} 
+  \frac{1}{(2l-1)!} d^\dag A_{[2p]} d^\dag \ttA_{[2p]}\right) \nonumber \\
&+&\frac12 \frac{1}{(2k+1)!} d A_{[k]}^{(+)} 
d {\tilde {\tilde A}}_{[k]}^{(+)} 
+\frac12
 \frac{1}{(2k-1)!} d^\dag A_{[k]}^{(+)} d^\dag {\tilde { \tilde A}}_{[k]_+} 
 \nonumber \\
&+&  d A'_{[0]} d {\tilde {\tilde A}}'_{[0]}  \nonumber \\
&-& \sum_{p=0}^{n-1} \left(
\frac12  \frac{1}{(2p+2)!} d \tilde A'_{[2p+1]} d \tilde A'_{[2p+1]} 
+\frac12  \frac{1}{(2p)!} d^\dag \tilde A'_{[2p+1]} d^\dag \tilde A'_{[2p+1]}\right) 
 \nonumber \\ 
&+& \sum_{p=1}^{n-1} \left(
 \frac{1}{(2p+1)!} d  A'_{[2p]} d \ttA'_{[2p]} 
+  \frac{1}{(2p-1)!} d^\dag A'_{[2p]} d^\dag \ttA'_{[2p]}\right) \nonumber \\
&+&\frac12 \frac{1}{(k+1)!} d A_{[k]}^{\prime (-)} 
d {\tilde {\tilde A}}_{[k]}^{\prime (-)} 
+ \frac12 \frac{1}{(k-1)!} d^\dag A_{[k]}^{\prime (-)} d^\dag {\tilde { \tilde A}}_{[k]}^{\prime (-)}.
\eeqa
\noindent
A similar Lagrangian can be written for the second pair of multiplets, ${\mathbf \Xi}_{+-}-{\mathbf \Xi}_{-+}$.
Here $\omega_{[p]} \omega'_{[p]}$ stands for
$\omega_{[p]}{}_{M_1\ldots M_p} 
$\ $\omega'_{[p]}{}^{M_1 \ldots M_P}$
, where $\omega_{[p]}$
and $\omega'_{[p]}$ are two $p-$forms.


In order to prove that the Lagrangian \eqref{lag} is invariant, we first use
\beqa
\label{fermi-D}
&& \frac{1}{ (p+1)!} d \omega_{[p]}  
d \omega'_{[p]}  +  \frac{1}{ (p-1)!} d^\dag \omega_{[p]}  
d^\dag \omega'_{[p]}  \nonumber \\
&&=  \frac{1}{ p !} \partial_{M_1} \omega_{[p]}{}_{M_2 
\ldots M_{p+1}} \partial^{M_1} \omega'_{[p]}{}^{M_2  \ldots M_{p+1}}
\nonumber \\
&&=\frac{1}{p!} \partial \omega_{[p]} \partial \omega'_{[p]}. 
\eeqa
\noi 
with $\omega_{[p]}$ and $\omega'_{[p]}$ general $p-$forms appearing in \eqref{lag}; this is actually a generalisation of the relations \eqref{ajutor}
 for forms in $4$ dimensions. 
As in the case $D=4$, the ${\mathbf \Xi}_{++}$ and ${\mathbf \Xi}_{--}$ parts of this Lagrangian do not mix under the transformation laws  \eqref{transfo-form}. Thus, we look closer at the invariance of ${\cal L}({\mathbf \Xi}_{++})$, the invariance of ${\cal L}({\mathbf \Xi}_{--})$ being analogous.


Application of the transformation laws \eqref{transfo-form} on the Lagrangian terms written with the help of \eqref{fermi-D} gives
\beqa
\label{Dinv1}
\delta_v {\cal L}_{[2p]}&=&\frac {1}{(2p)!} \left( \partial (i_v \tA_{[2p+1]} + \tA_{[2p-1]}\wedge v) \partial  \ttA _{[2p]}+\partial  A _{[2p]} \partial (v A_{[2p]})\right),\ p=1,..,n-1\nonumber\\
\delta_v {\cal L}_{[2p+1]}&=&-\frac {1}{ (2p+1)!} \left( \partial (i_v \ttA_{[2p+2]} + \ttA_{[2p]}\wedge v) \partial  \tA _{[2p+1]} \right), \ p=0,\dots,n-1.
\eeqa
\noi
Using \eqref{dual-prod}, one observes a step-by-step cancellation process: the first term of the first line cancels with the second term of the second line. Furthermore, the terms of type $\partial  A _{[2p]} \partial (v A_{[2p]})$ can be written as a total derivative, namely $\frac 12 v^M \partial_M ((\partial A_{[2p]})^2)$. 
Finally, considering all the terms above,  the Lagrangian \eqref{lag} is
invariant, that is
$$\delta_v {\cal L} =v^M \partial_m \left (\sum \frac 12 \frac{1}{(2p)!} (\partial A_{2p]})^2 \right).$$
The \3 invariance was made possible by the special form of the
transformation laws \eqref{transfo-form}, involving inner and exterior
products of the parameter $v$ of the transformations. Furthermore, the step-by-step
 compensation procedure shown above also fixed the normalisations of the different
 terms of the Lagrangian.

We now proceed to some closer investigation of  Lagrangian \eqref{lag}, a similar analysis being also possible for the case 
when one treats the pair of multiplets ${\mathbf \Xi}_{-+}-{\mathbf \Xi}_{+-}$.

As in the previous chapter, to explicitly write the Lagrangian with real fields, one has to perform a set of field redefinitions
\beqa
\label{real-D}
\begin{array}{ll}
A_1{}_{[2p]}=\frac{1}{\sqrt{2}}\left(A_{[2p]} + A'_{[2p]}\right),&
A_2{}_{[2p]}= \frac{i}{\sqrt{2}}\left(A_{[2p]} - A'_{[2p]}\right),
\cr
\tA_1{}_{[2p+1]}=\frac{1}{\sqrt{2}}
\left(\tA_{[2p+1]} + \tA'_{[2p+1]}\right),&
\tA_2{}_{[2p+1]}=\frac{i}
{\sqrt{2}}\left(\tA_{[2p+1]} - \tA'_{[2p+1]}\right),
\cr
\ttA_1{}_{[2p]}=\frac{1}{\sqrt{2}}\left(\ttA_{[2p]} + \ttA'_{[2p]}\right),&
\ttA_2{}_{[2p]}= \frac{i}{\sqrt{2}}
\left(\ttA_{[2p]} - \ttA'_{[2p]}\right).
\end{array}
\eeqa
\noindent
with $p=0,\ldots, n$. One notices now that the $k-$forms are neither self-dual nor anti-self-dual. As in the previous chapter, this is in agreement with the representation theory of the Poincar\'e algebra. Indeed, it is for the Lorentz group $SO(1,2k-1)$ that the $k-$forms are (anti-)self-duals.

Before writing the Lagrangian with the new set of fields \eqref{real-D}, another simplification can be done. One notices that
\beqa
\label{k-form}
{}^* A_{1[k]}=A_{2[k]},\ {}^* \ttA_{1[k]}=\ttA_{2[k]}.
\eeqa
\noi
This allows us once more to use relation \eqref{dd} to eliminate the $k-$forms $A_{[k]}$ and $\ttA_{[k]}$. 

Hence the  Lagrangian  writes
\beqa
\label{lag2}
{\cal L}&=&dA_{1[0]}d\ttA_{1[0]} - dA_{2[0]}d\ttA_{2[0]}\nonumber\\
&-&\sum_{p=0}^{n-1} \left( \frac 12 \frac{1}{(2p+2)!} d \tA_{1[2p+1]} d \tA_{1[2p+1]} + \frac 12 \frac{1}{(2p)!} d^\dag \tA_{1[2p+1]} d \tA_{1[2p+1]}\right)\nonumber\\
&+&\sum_{p=0}^{n-1} \left( \frac 12 \frac{1}{(2p+2)!} d \tA_{2[2p+1]} d \tA_{2[2p+1]} + \frac 12 \frac{1}{(2p)!} d^\dag \tA_{2[2p+1]} d \tA_{2[2p+1]}\right)\nonumber\\
&+&\sum_{p=1}^{n-1} \left( \frac{1}{(2p+1)!} d A_{1[2p]} d \ttA_{1[2p]} -  \frac{1}{(2p-1)!} d^\dag A_{1[2p]} d \ttA_{1[2p]}\right)\nonumber\\
&-&\sum_{p=1}^{n-1} \left( \frac{1}{(2p+1)!} d A_{2[2p]} d \ttA_{2[2p]} -  \frac{1}{(2p-1)!} d^\dag A_{2[2p]} d \ttA_{2[2p]}\right)\nonumber\\
&+&\frac{1}{(k+1)!} d A_{1[k]} d \ttA_{1[k]} + \frac{1}{(k-1)!} d^\dag A_{1[k]} d \ttA_{1[k]}.
\eeqa

As before, the terms of type 
$$\frac{1}{(2p+1)!} d A_i{}_{[2p]} d \ttA_i{}_{[2p]}
+\frac{1}{(2p-1)!} d^\dag A_i{}_{[2p]} d^\dag \ttA_i{}_{[2p]} \mbox{ with } i=1,2$$
are not diagonal so one has to perform the following redefinitions
\beqa
\label{field-diag}
&&\hA_1{}_{[2p]} =\frac{1}{\sqrt{2}}\left(A_1{}_{[2p]} + 
\ttA_1{}_{[2p]}\right),
\hhA_1{}_{[2p]}= \frac{1}{\sqrt{2}}\left(A_1{}_{[2p]} - \ttA_1{}_{[2p]}\right),
p=0,\ldots, n
 \\
&&\hA_2{}_{[2p]}= \frac{1}{\sqrt{2}}\left(A_2{}_{[2p]} + \ttA_2{}_{[2p]}\right),
\hhA_2{}_{[2p]}= \frac{1}{\sqrt{2}}\left(A_2{}_{[2p]} - \ttA_2{}_{[2p]}\right),
p=0,\ldots, n-1, \nonumber 
\eeqa
\noindent
We thus find mixtures of fields of gradations $(-1)$ and $1$. 

The Lagrangian becomes
\beqa
\label{lag3}
{\cal L}&=&\frac12 d\hA_{1[0]}d\hA_{1[0]} - \frac 12 d\hhA_{1[0]}d\hhA_{1[0]}-\frac 12 d\hA_{2[0]}d\hA_{2[0]} + \frac 12 d\hhA_{2[0]}d\hhA_{2[0]}\nonumber\\
&-&\sum_{p=0}^{n-1} \left( \frac 12 \frac{1}{(2p+2)!} d \tA_{1[2p+1]} d \tA_{1[2p+1]} + \frac 12 \frac{1}{(2p)!} d^\dag \tA_{1[2p+1]} d^\dag \tA_{1[2p+1]}\right)\nonumber\\
&+&\sum_{p=0}^{n-1} \left( \frac 12 \frac{1}{(2p+2)!} d \tA_{2[2p+1]} d \tA_{2[2p+1]} + \frac 12 \frac{1}{(2p)!} d^\dag \tA_{2[2p+1]} d^\dag\tA_{2[2p+1]}\right)\nonumber\\
&+&\sum_{p=1}^{n-1} \left( \frac12 \frac{1}{(2p+1)!} d \hA_{1[2p]} d \hA_{1[2p]} - \frac12 \frac{1}{(2p-1)!} d^\dag \hA_{1[2p]} d^\dag \hA_{1[2p]}\right)\nonumber\\
&-&\sum_{p=1}^{n-1} \left( \frac12 \frac{1}{(2p+1)!} d \hhA_{1[2p]} d \hhA_{1[2p]} - \frac12 \frac{1}{(2p-1)!} d^\dag \hhA_{1[2p]} d^\dag \hhA_{1[2p]}\right)\nonumber\\
&-&\sum_{p=1}^{n-1} \left(\frac12 \frac{1}{(2p+1)!} d \hA_{2[2p]} d \hA_{2[2p]} -  \frac12 \frac{1}{(2p-1)!} d^\dag \hA_{2[2p]} d^\dag \hA_{2[2p]}\right)\nonumber\\
&+&\sum_{p=1}^{n-1} \left( \frac12 \frac{1}{(2p+1)!} d \hhA_{2[2p]} d \hhA_{2[2p]} - \frac12  \frac{1}{(2p-1)!} d^\dag \hhA_{2[2p]} d^\dag \hhA_{2[2p]}\right)\nonumber\\
&+&\frac 12 \frac{1}{(k+1)!} d \hA_{1[k]} d \hA_{1[k]} + \frac 12 \frac{1}{(k-1)!} d^\dag \hA_{1[k]} d^\dag\hA_{1[k]}\nonumber\\
&-&\frac 12 \frac{1}{(k+1)!} d \hhA_{1[k]} d \hhA_{1[k]} - \frac 12 \frac{1}{(k-1)!} d^\dag \hhA_{1[k]} d^\dag \hhA_{1[k]}.
\eeqa

Thus, the field content is now:  
$\hA_1{}_{[2p]}, \hA_2{}_{[2p]}$,   
$\hhA_1{}_{[2p]}, \hhA_2{}_{[2p]}, \tA_1{}_{[2p+1]},\tA_2{}_{[2p+1]}, p=0,\ldots, n-1$ and $ \hA_1{}_{[k]}, \hhA_1{}_{[k]}$.

With this explicitly real and diagonalised form of the Lagrangian \eqref{lag3}, we consider the general gauge invariance

\beqa
\label{gauge}
\omega_{[p]}\rightarrow \omega_{[p]}+d\chi_{[p-1]}, \mbox{ with } p\ge 1,
\eeqa
(where $\omega_{[p]}$ is a general $p-$form appearing in \eqref{lag3} and $\chi_{[p-1]}$ is a $(p-1)-$form). The terms involving $d^\dag$ partially fix the gauge since, {\it i.e.} in order to have a gauge invariant Lagrangian, the gauge parameter $\chi_{[p-1]}$ obeys to
\beqa
\label{gf}
d^\dag d \chi_{[p-1]} = 0.
\eeqa
\noi
As already argued in subsection \ref{detoate}, the number of degrees of freedom of such a  $p-$form is 
$$ C_D^p,$$
result witch is thus different from the result of the conventional, not gauge-fixed model.

Another issue  to be discussed is the following: with the conventions used here, the boundness from below of the kinetic term of a $p-$form in $D$ dimensions requires 
$(-1)^p \frac12 \frac{1}{(p+1)!} d A_{[p]} d A_{[p]}.$ for the kinetic term. 
This can be seen by calculating the energy density $T_{00}$, as it was illustrated in subsection \ref{detoate}.
A direct inspection of the signs in \eqref{lag3} shows that one has wrong signs for the set of $p-$forms $\hhA_1{}_{[2p]}, \hA_2{}_{[2p]},$ $ \tA_2{}_{[2p+1]}, p=0,\ldots, n-1,
\hhA_1{}_{[k]}$. The solution proposed is the same as in the case for $D=4$ (see section $5.3.1$).
We use the Hodge dual to make  the substitutions
\beqa
\label{dual-field}
\hhA_1{}_{[2p]} &\to& \hhB_1{}_{[D-2p]} =   {}^\star \hhA_1{}_{[2p]},
\nonumber \\
\hA_2{}_{[2p]} &\to& \hB_2{}_{[D-2p]} =  {}^\star \hA_2{}_{[2p]}, \nonumber \\
\tA_2{}_{[2p+1]} &\to& \tB_2{}_{[D-2p-1]} = {}^\star \tA_2{}_{[2p+1]} 
 \\
\hhA_1{}_{[k]} &\to& \hhB_1{}_{[D-k]}= {}^\star \hhA_1{}_{[k]} 
\nonumber 
\eeqa
\noindent
with $p=0,\ldots, n-1$.
the number of degrees of freedom does not change.
 Since a $p-$form in this model has $C_D^p$ degrees of freedom (see above), the number of degrees of freedom does not change by this substitution.
One now uses the identities (\ref{dd}) which changes for example the kinetic term
of $\hhA_{i[p]}$ to the gauge fixing term of $\hhB_{i[D-p]}$ and {\it viceversa}. 
Hence, 
the final Lagrangian writes
\beqa
\label{lag-end}
&{\cal L}& = \frac12 d \hA_1{}_{[0]} d \hA_1{}_{[0]} +
\frac12 d \hhA_2{}_{[0]} d \hhA_2{}_{[0]}
\nonumber \\
&-&\sum_{p=0}^{n-1} \left( \frac 12 \frac{1}{(2p+2)!} d \tA_{1[2p+1]} d \tA_{1[2p+1]} + \frac 12 \frac{1}{(2p)!} d^\dag \tA_{1[2p+1]} d^\dag \tA_{1[2p+1]}\right)\nonumber\\
&+&\sum_{p=1}^{n-1} \left( \frac12 \frac{1}{(2p+1)!} d \hA_{1[2p]} d \hA_{1[2p]} - \frac12 \frac{1}{(2p-1)!} d^\dag \hA_{1[2p]} d^\dag \hA_{1[2p]}\right)\nonumber\\
&+& \frac12 \frac{1}{(k+1)!} d \hA_{1[k]}d \hA_{1[k]}+\frac12 \frac{1}{(k-1)!} d^\dag \hA_{1[k]}d^\dag \hA_{1[k]}\nonumber\\
&+& \frac 12 \frac{1}{(k+1)!} d \hB_{1[D-k]}d \hB_{1[D-k]}+\frac12 \frac{1}{(k-1)!} d^\dag \hB_{1[D-k]}d^\dag \hB_{1[D-k]}\nonumber\\
&-&\sum_{p=n}^{k-1}\left( \frac12 \frac{1}{(2p+2)!} d \tB_{2[2p+1]}d \tB_{2[2p+1]}+\frac12 \frac{1}{(2p)!} d^\dag \tB_{2[2p+1]}d^\dag \tB_{2[2p+1]}\right)\nonumber\\
&+&\sum_{p=n}^{k-1}\left( \frac12 \frac{1}{(2p+1)!} d \hhB_{1[2p]}d \hhB_{1[2p]}+\frac12 \frac{1}{(2p-1)!} d^\dag \hhB_{1[2p]}d^\dag \hhB_{1[2p-1]}\right)\nonumber\\
&+&\sum_{p=n}^{k-1}\left( \frac12 \frac{1}{(2p+1)!} d \hB_{2[2p]}d \hB_{2[2p]}+\frac12 \frac{1}{(2p-1)!} d^\dag \hB_{2[2p]}d^\dag \hB_{2[2p-1]}\right)\nonumber\\
&+&\frac12 \frac{1}{(D-1)!} d^\dag \hhB_1{}_{[D]} d^\dag \hhB_1{}_{[D]} 
+ \frac12 \frac{1}{(D-1)!} d^\dag \hB_2{}_{[D]} d^\dag \hB_2{}_{[D]}.
\eeqa

\noi
Thus, the final field content of this model is: one $1-$form, one  $3-$form, 
$\ldots$, one $(D-1)-$form in the zero-graded sector and 
two $0$-forms, two $2-$forms, $\ldots$ and two $D-$form in the
mixture of the sectors of gradation $(-1)$ and $1$.
All these $p-$forms have 
a kinetic term and a gauge fixing term; the only exceptions are 
the zero-forms,
which  have only  kinetic terms, and the $D-$forms, which
have only  gauge fixing terms. 
Furthermore notice that this mechanism demands the presence of $(D-1)-$ and $D-$forms.

\medskip

So far we have treated the massless case. Nevertheless, one can explicitly write invariant mass terms.
For the original fields of the multiplets ${\mathbf \Xi}_{++}-{\mathbf \Xi}_{--}$, 
one has
\beqa
\label{D-mass}
{\cal L}_{\mathrm{mass}}=m^2 \left(
A_{[0]}  {\tilde {\tilde A}}_{[0]} 
- \sum_{p=0}^{n-1} 
\frac12  \frac{1}{(2p+1)!} \tilde A_{[2p+1]}  \tilde A_{[2p+1]} 
+ \sum_{p=1}^{n-1} 
  \frac{1}{(2p)!}   A_{[2l]}  \ttA_{[2p]} 
+ \frac{1}{(k)!}  A_{[k]}^{(+)} {\tilde {\tilde A}}_{[2k]}^{(+)}\right. \nonumber\\ \left.
A'_{[0]}  {\tilde {\tilde A'}}_{[0]} 
- \sum_{p=0}^{n-1} 
\frac12  \frac{1}{(2p+1)!} \tilde A'_{[2p+1]}  \tilde A'_{[2p+1]} 
+ \sum_{p=1}^{n-1} 
  \frac{1}{(2p)!}   A'_{[2l]}  \ttA'_{[2p]} 
+ \frac{1}{(k)!}  A_{[k]}^{\prime(+)} {\tilde {\tilde A}}_{[2k]}^{\prime(+)}\right) . \nonumber 
\eeqa
which is also invariant. Similar invariant mass terms can be defined for the other multiplets. Obviously one can then perform the same field redefinitions as above for these terms, thus in addition having also mass terms for the redefined fields. Furthermore, the number of degrees of freedom for a massive $p-$form, is, as before, $C_D^p$, since no gauge invariance is possible.

\section{Summary of results and perspectives}

Let us make a brief recall of the main results for this chapter:

\noi
one has $4$ multiplets, ${\mathbf \Xi}_{++}$, ${\mathbf \Xi}_{--}$, ${\mathbf \Xi}_{+-}$ and ${\mathbf \Xi}_{-+}$, which form two pairs of complex conjugated multiplets - ${\mathbf \Xi}_{++}={\mathbf \Xi}_{--}^*$ and ${\mathbf \Xi}_{+-}={\mathbf \Xi}_{-+}^*$   (complex fields),

  $1.)$ for the pair ${\mathbf \Xi}_{++}-{\mathbf \Xi}_{--}$ 
  \begin{itemize}
 \item  Lagrangian: ${\cal L}({\mathbf \Xi}_{++}) + {\cal L} ({\mathbf \Xi}_{--})$,
 \item field content: $0-$graded sector - one set of $p-$forms, $p=1,\ldots, D-1$, with $p$ odd,\\
                 the mixture of $-1$ and $1$ graded sectors - two sets of $p-$ forms, $p=0,\ldots, D$ with $p$ even. 
  \end{itemize}

\noi
$2.)$ for the pair ${\mathbf \Xi}_{+-}-{\mathbf \Xi}_{-+}$ 
  \begin{itemize}
 \item  Lagrangian: ${\cal L}({\mathbf \Xi}_{+-}) + {\cal L} ({\mathbf \Xi}_{-+})$,
 \item field content: $0-$graded sector - one set of $p-$forms, $p=0,\ldots, D$, with $p$ even,\\
                 the mixture of $-1$ and $1$ graded sectors - two sets of $p-$ forms, $p=1,\ldots, D-1$ with $p$ odd. 
  \end{itemize}

  As already stated, similar results are obtained when dealing in the same way with the other cases of parities for $D$ and $[\frac D2 ]$.

\bigskip

This study in arbitrary dimensions has to be pursued in order to answer several questions. For example,
further analyses to find whether or not ghosts may be  required by the model in connexion with unitarity issues, thus leading furthermore to quantification mechanisms for our models.

Furthermore, an important question is  whether or not any interactions (between these multiplets or other multiplets, either more general bosonic multiplets or fermionic multiplets) are allowed by  \3. As we have seen in subsection $6.1.1$, $p-$forms naturally interact with extended objects, so this type of interaction may be a promising track to explore.

Connected to this issue, one may address the question of whether using a non-abelian gauge would have an impact on our model.
Moreover, for  $p-$forms with $p>1$, only a few number of consistent interactions are possible \cite{henneaux}.

Finally,  finding any other pertinent Lie algebras of order $3$ which extend the Poincar\'e symmetry may be an interesting perspective from the viewpoint of our two-sided motivation. Thus, a more detailed algebraic study seems also like a tempting future line of work.

\appendix

\chapter{Abelian Lie subalgebras of the Weyl algebra $A_1$}
\renewcommand{\theequation}{A.\arabic{equation}}   
\setcounter{equation}{0}
\label{abe}

In this appendix we give the proof of Proposition \ref{abeliana-ciudata} of subsection $3.2.1$:

\begin{proposition}
\label{abeliana-ciudata-bis}
Let $a_1=(p^2q)^3-\frac 32 (p^4q+ p^2qp^2)$ and $a_2=p^4q^2-2p^3q-2p^2$. 
Then there does not exist $a\in A_1$ such that $a_1, a_2\in \CC[a]$.
\end{proposition}
{\it Proof:} 
Firstly remark that 
$$[a_1,a_2]=0.$$
By a tedious but straightforward calculation one proves that 
\beqa
\label{aposibila}
a_1^2=a_2^3
\eeqa

We now look for an element $a\in A_1$ such that $a_2=P_2(a)$ where  
\beqa
\label{ap2}
P_2(a)=(a-\alpha_1)^{n_1} \dots (a-\alpha_k)^{n_k}
\eeqa
\noi
 is a  polynomial in $a$, with distinct complex roots $\alpha_1, \dots, \alpha_k$ (with $n_1,\dots, n_k\in\NN^*$ their multiplicity orders). Let also
\beqa
\label{aX}
X=n_1+\dots + n_k.
\eeqa

Expanding $a$ in the standard basis $p^i q^j$ of $A_1$ one has
\beqa
\label{aa}
a=\sum_{i=1}^N f_i (p) q^i.
\eeqa
\noi
where $N\in \NN$ and $f_i$ are polynomials in $p$, $f_N (p)$ being of degree $M_N$.

The equation to analyse is thus 
\beqa
\label{aecuatia}
a_2=p^4q^2-2p^3q-2p^2=(a-\alpha_1)^{n_1} \dots (a-\alpha_k)^{n_k}
\eeqa
\noi
Thus, expanding in the standard basis, the highest power of $q$ with the highest power of $p$ coming along is, in RHS, $p^{M_NX}q^{NX}$ and in the LHS, $p^4q^2$. Thus one has
\beqa
\label{asistem}
NX&=&2\nonumber\\
M_N X &=&4
\eeqa
\noi
which leads to two distinct cases:\\
$1.$ $X=2$ and thus $N=1,\ M_N=2$\\
$2.$ $X=1$ and thus $N=2,\ M_N=4$\\

\noi
$1.$ $X=2,\ N=1,\ M_N=2$. By \eqref{aX}, one distinguishes two subcases:\\
A. $k=1$ (only one distinct root of the polynomial $P_2 (a)$, with order of multiplicity $n_1=2$);\\ 
B. $k=2$ (two distinct roots of the polynomial $P_2 (a)$, with orders of multiplicity $n_1=n_2=1$).

\noi
We treat the case A, case B being similar. Equation \eqref{aecuatia} becomes
\beqa
\label{ae2}
p^4q^2-2p^3q-2p^2=(a-\alpha_1)^2=a^2 - 2\alpha_1 a + \alpha_1^2.
\eeqa
\noi
Since $N=1$ and $M_N=2$, one has $a=(K_1 p^2 + K_2 p + K_3) q + f (p)$, with $K_1, K_2, K_3\in \CC^*$, $K_1 \ne 0$ and $f (p)$ a polynomial in $p$. By similar arguments of degree counting on the two sides of equation \eqref{ae2}, one has ${\rm deg} f \le 2$. Thus one has $a=K_1 p^2q + K_2 pq + K_3q + K_4 p^2 + K_5 p + K_6$ (with $K_3, K_4, K_5\in \CC$) and \eqref{ae2} becomes
\beqa
\label{ae3}
p^4q^2-2p^3q-2p^2&=&(K_1 p^2q + K_2 pq + K_3q + K_4 p^2 + K_5 p + K_6)^2\nonumber\\&-&2\alpha_1 (K_1 p^2q + K_2 pq + K_3q + K_4 p^2 + K_5 p + K_6) + \alpha_1^2.
\eeqa
\noi
Expanding now in the standard basis $p^iq^j$ of $A_1$, one sees that the term in $p^3 q^2$ is $K_1 K_2 p^3 q^2$ in the RHS while in the LHS there is no term in $p^3 q^2$; thus $K_2=0$. Similarly, comparing the terms in $p^2 q^2$ one gets $K_3=0$, comparing the terms in $p^4q$ one gets $K_4=0$. Then, comparing the terms in $p^2q$ one gets $K_6=\alpha_1$, and then the term in $p^0 q^0$, one gets $-2\alpha_1 K_6 =0$ which means that $K_6=\alpha_1=0$.  Thus \eqref{ae3} becomes
\beqa
\label{ae4}
p^4q^2-2p^3q-2p^2=(K_1 p^2q + K_5 p )^2=K_1^2 p^4 q + 2 K_1 (K_5- K_1) p^3 q - K_5 (K_1-K_5)p^2.
\eeqa
\noi
Identifying the coefficients above, one concludes that there is no solution.

\medskip

\noi
$2.$ $X=1$. From \eqref{aX} one concludes that there is only one distinct root $\alpha_1$ of the polynomial $P_2 (a)$, with order of multiplicity $n_1=1$. Hence \eqref{aecuatia} becomes
\beqa
a_2= a-\alpha_1.
\eeqa
\noi
Hence, the only element $a\in A_1$ which allows to write $a_2$ as a polynomial in $a$ is $a=a_2+\alpha_1$. From \eqref{aposibila}, one concludes that $a_1\notin \CC[a]$. QED

\chapter{Lie groups calculations}
\renewcommand{\theequation}{B.\arabic{equation}}   
\setcounter{equation}{0}
\label{LieGroup}

In this appendix we show that 
\begin{enumerate}
\item
the Lie algebra of $R_n (i_1,\dots,i_n)$ (see {\bf E11} of section $3.3$) is isomorphic to $\rr (i_1,\dots,i_n)$ (see {\bf E8} of subsection $3.2.1$), for $i_1,\dots,i_n\NN^*$,
\item
the Lie algebra of $\tilde L_n$ (see {\bf E12} of section $3.3$) is isomorphic to $\tilde {\cal L}_n$ (see {\bf E7} of subsection $3.2.1$).
\end{enumerate} 

Before this, recall the general formulae

\beqa
\label{centrala}
[Y_a \frac{\partial}{\partial X_a}, Y_b \frac{\partial}{\partial X_b}]=Y_a \frac{\partial Y_b}{\partial X_a}\frac{\partial}{\partial X_b}- Y_b \frac{\partial Y_a}{\partial X^b} \frac{\partial}{\partial X_a}.
\eeqa

$1.$ Firstly recall from {\bf E11} the group law of $R_n (i_1,\dots,i_n)$ 
 \beqa
\label{inmultire-bis}
(a_1,\dots,a_n,v).(a'_1,\dots,a'_n,v')=(a_1+a'_1e^{-vi_1},\dots,a_n+a'_ne^{-vi_n},v+v')
\eeqa
\noi 
for any two elements of $R_n (i_1,\dots,i_n)$.

Determine a basis of left-invariant vectors, {\it i.e.} a basis of the Lie algebra of $R_n (i_1,\dots,i_n)$. Let $(x_1,\dots,x_n,w)\in R_n (i_1,\dots,i_n)$. One thus has
$$(a_1,\dots,a_n,v).(x_1,\dots,x_n,w)=(y_1,\dots,y_n,y_v)$$
where
$$ y_k=a_k+x_ke^{-vi_k},\ k=1,\dots, n \mbox{ and } y_v=v+w$$
In the coordinate system $(a_1,\dots,a_n,v)$, the left-invariant fields $\tilde X_1, \dots, \tilde X_n, \tilde h $ write
\beqa
\tilde X_k&=& \frac {\partial y_1}{\partial x_k}  \big\vert_{(x_1=0,\dots,x_n=0, w=0)}\frac{\partial}{\partial a_1} + \dots + \frac {\partial y_n}{\partial x_k}  \big\vert_{(x_1=0,\dots,x_n=0, w=0)}\frac{\partial}{\partial a_n}+ \frac {\partial y_v}{\partial x_k}  \big\vert_{(x_1=0,\dots,x_n=0, w=0)}\frac{\partial}{\partial v}, \nonumber\\
\tilde h&=& \frac {\partial y_1}{\partial w}  \big\vert_{(x_1=0,\dots,x_n=0, w=0)}\frac{\partial}{\partial a_1} + \dots + \frac {\partial y_n}{\partial w}  \big\vert_{(x_1=0,\dots,x_n=0, w=0)}\frac{\partial}{\partial a_n}+ \frac {\partial (y_v}{\partial w}  \big\vert_{(x_1=0,\dots,x_n=0, w=0)}\frac{\partial}{\partial v},\nonumber
\eeqa
with $k=1,\dots,n$. This leads to
\beqa
\tilde X_k &=& e^{-vi_k}\frac{\partial}{\partial a_k},\ k=1,\dots,n\nonumber\\
\tilde h &=& \frac{\partial}{\partial v}.
\eeqa
\noi
We now use \eqref{centrala} to calculate the Lie brackets of the above  generators. One finds the commutation relations of the Lie algebra
$\rr (i_1,\dots,i_n)$ (see {\bf E8}).

\bigskip

\noi
$2.$ Recall from now {\bf E12} the group law of $\tilde L_n$ 
\beqa
\label{grup-ln-bis}
 (a_1,\dots, a_n,t,v) . (a'_1,\dots,a'_n,t',v')= (a''_1,\dots,a''_n,t'',v'')
\eeqa
\noi where 
\beqa
\label{coef-ln-bis}
a''_k&=&a_ke^{(n-k)v'}+\sum_{j=1}^{k-1} \frac{t^{k-j}}{(k-j)!}a'_je^{-(k-j)v'}+a'_k,\nonumber\\
t''&=&t'+te^{-v'},\nonumber\\
v''&=&v+v'.\nonumber
\eeqa
\noi
We determine now a basis of left-invariant fields. Let $(x_1,\dots,x_n,t_x, w_x)\in \tilde L_n$. One thus has
$$(a_1,\dots,a_n,t,v).(x_1,\dots,x_n,t_x,w_x)=(y_1,\dots,y_n,t_y,w_y)$$
where
\beqa
\label{coef-ln-bis-2}
y_k&=&a_ke^{(n-k)w_x}+\sum_{j=1}^{k-1} \frac{t^{k-j}}{(k-j)!}x_je^{-(k-j)w_x}+x_k,\nonumber\\
t_y&=&t_x+te^{-w_x},\nonumber\\
w_y&=&v+w_x.\nonumber
\eeqa
\noi
In the coordinate system $(a_1,\dots,a_n,t,v)$, the left-invariant fields $\tilde X_1, \dots, \tilde X_n, \tilde X_0, \tilde h $ write
\beqa
\tilde X_k&=& \frac {\partial y_1}{\partial x_k}  \big\vert_{(0,\dots,0)}\frac{\partial}{\partial a_1} + \dots + \frac {\partial y_n}{\partial x_k}  \big\vert_{(0,\dots,0)}\frac{\partial}{\partial a_n}+ \frac {\partial t_y}{\partial x_k}  \big\vert_{(0,\dots,0)}\frac{\partial}{\partial t}+ \frac {\partial w_y}{\partial x_k}  \big\vert_{(0,\dots,0)}\frac{\partial}{\partial v}, \nonumber\\
\tilde X_0&=& \frac {\partial y_1}{\partial t_x}  \big\vert_{(0,\dots,0)}\frac{\partial}{\partial a_1} + \dots + \frac {\partial y_n}{\partial t_x}  \big\vert_{(0,\dots,0)}\frac{\partial}{\partial a_n}+ \frac {\partial t_y}{\partial t_x}  \big\vert_{(0,\dots,0)}\frac{\partial}{\partial t}+\frac {\partial w_y}{\partial t_x}  \big\vert_{(0,\dots,0)}\frac{\partial}{\partial v}, \nonumber\\
\tilde h&=& \frac {\partial y_1}{\partial w_x}  \big\vert_{(0,\dots,0)}\frac{\partial}{\partial a_1} + \dots + \frac {\partial y_n}{\partial w_x}  \big\vert_{(0,\dots,0)}\frac{\partial}{\partial a_n}+ \frac {\partial t_y}{\partial w_x}  \big\vert_{(0,\dots,0)}\frac{\partial}{\partial v}+ \frac {\partial w_y}{\partial w_x}  \big\vert_{(0,\dots,0)}\frac{\partial}{\partial v},\nonumber
\eeqa
with $k=1,\dots,n$. This leads to
\beqa
&&\tilde X_1 = \frac{\partial }{\partial a_1} + t\frac{\partial }{\partial a_2}+\dots + \frac{t^{n-1}}{(n-1)!} \frac{\partial} {\partial a_n} \nonumber\\
&&\tilde X_2 = \frac{\partial }{\partial a_2}+t\frac{\partial }{\partial a_3}\dots + \frac{t^{n-2}}{(n-2)!} \frac{\partial} {\partial a_n} \nonumber\\
&&\dots \nonumber \\
&&\tilde X_n=\frac{\partial }{\partial a_n}\nonumber\\
&&\tilde X_0=\frac{\partial }{\partial t}\nonumber\\
&&\tilde h = (n-1)a_1\frac{\partial }{\partial a_1} + (n-2)a_2\frac{\partial }{\partial a_2}+\dots+a_{n-1}\frac{\partial }{\partial a_{n-1}}-t \frac{\partial }{\partial t} +\frac{\partial}{\partial v}.
\eeqa
\noi
We now use \eqref{centrala} to calculate the Lie brackets of the above  generators. One finds the commutation relations of the Lie algebra
${\cal L}_n$ (see {\bf E7}).

\chapter{Useful identities on $p-$forms}
\renewcommand{\theequation}{C.\arabic{equation}}   
\setcounter{equation}{0}
\label{a-pforme}


In this appendix we give some useful identities of $p-$forms in $D$ dimensions
as well as some conventions used in this thesis.

The Levi-Civita tensors $\varepsilon_{M_1 \ldots M_{D}}$ and 
$\varepsilon^{M_1 \ldots M_{D}} = 
\varepsilon_{N_1 \ldots N_{D}} \eta^{M_1 N_1} \ldots \eta^{M_D N_D}$
are given by 

\begin{eqnarray}
\label{levi-civita}
\varepsilon_{01 \ldots (D-1)} =1, \ \ \varepsilon^{01 \ldots (D-1)} = 
(-1)^{D-1}
\end{eqnarray}

By direct component calculation one proves

\beqa
\label{star2}
{}^\star {}^\star A_{[p]}= (-1)^{(D-1)(p-1)} A_{[p]}.
\eeqa 
\noindent
and 
\beqa
\label{star3}
\frac{1}{p!} A_{[p]}A_{[p]}=\frac{(-1)^{D-1}}{(D-p-1)!} B_{[D-p]}B_{[D-p]}
\eeqa
\noi
with $B_{[D-p]}= {}^\star A_{[p]}$.

Let us now show the following relations
\beqa
\label{dd}
\frac{1}{(p+1)!} d A_{[p]}  d A_{[p]} & = &(-1)^{D-1}\frac{1}{(D-p-1)!} d^\dag  B_{[D-p]}  
d^\dag   B_{[D-p]},\nonumber\\
\frac{1}{(p-1)!} d^\dag A_{[p]}  
d^\dag A_{[p]} 
&=&(-1)^{D-1} 
\frac{1}{(D-p+ 1)!} d B_{[D-p]}  d  B_{[D-p]}, 
\eeqa 
\noi
with $B_{[D-p]}= {}^\star A_{[p]}$. We prove here the first identity, the  proof of the second being analogous. Thus, using \eqref{star2} one writes the LHS of the first equation above as
\beqa
\label{star4}
\frac{1}{(p+1)!} d A_{[p]}  d A_{[p]} & =& \frac{1}{(p+1)!} d{}^{**}A_{[p]}d{}^{**}A_{[p]}
\eeqa
\noi
Using now \eqref{star3} and  recalling that $B_{[D-p]}= {}^\star A_{[p]}$ one writes \eqref{star4} as
\beqa
\label{star5}
\frac{1}{(p+1)!} d A_{[p]}  d A_{[p]} = \frac{(-1)^{D-1}}{(D-p-1)!} {}^*d{}^{*}B_{[D-p]}{}^*d{}^{*}B_{[D-p]}
\eeqa
\noi
Inserting the definition $d^\dag=(-1)^{pD+p}{}^*d^*$ (see subsection $6.1.2$) one obtains the first equation of \eqref{dd}.




Finally, from the definition \eqref{iner2} and \eqref{ext2} of the inner and exterior product, one can check explicitly in components
\beqa
\label{dual-prod}
\frac{1}{(2p)!} (i_v A'_{[2p+1]}) A_{[2p]}= \frac{1}{(2p+1)!} (A_{[2p]}\wedge v) A'_{[2p+1]}.
\eeqa




\chapter{Identities involving (anti-)self-dual $2-$ forms}
\renewcommand{\theequation}{D.\arabic{equation}}   
\setcounter{equation}{0}
\label{a-2forme}

In this appendix we deduce some identities involving (anti-)self-dual $2-$forms in four dimensions. Obviously, this set of identities is not an exhaustive list of the properties of (anti-)self-dual $2-$forms, but rather a few such properties that we use for different calculations of chapter $5$.

Let us generically denote here by $R^{(+)}_{mn}$ (resp. $R^{(-)}_{mn}$) a generic self-dual (resp. anti-self-dual) $2-$form in four dimensions. 



Thus, let us start by explicitly calculating
$$R^{(+)\, mn}(-v_mA_n+v_nA_m+i\e_{mnpq}v^pA_q)$$
where $v_m$, $A_n$ are two generic vectors. Since $R$ is antisymmetric, we obtain
\beqa
\label{2self-1}
- 2 R^{(+)\, mn}(v_mA_n + i\e_{mnpq}R^{(+)}_{mn} v^pA_q)
\eeqa
\noi
We now  use of the definition \eqref{self}, which for the specific case of  an (anti-)self-dual $2-$form in four dimensions gives
\beqa
\label{2}
R_{mn}^{(\pm)}=\mp \frac i2 \e_{mnpq}R_{}^{(\pm)\, pq}
\eeqa
\noi
Finally, using the antisymmetry properties of the Levi-Civita tensor and inserting \eqref{2} in \eqref{2self-1}, one obtains
\beqa
 \label{dualprop1-3} 
R^{(+)\,mn}(-v_mA_n+v_nA_m+i\e_{mnpq}v^pA_q)&=&-4R^{(+)}_{mn}v_mA_n.
\eeqa
Obviously, a similar identity, namely
$$R^{(-)\,mn}(-v_mA_n+v_nA_m-i\e_{mnpq}v^pA_q)=-4R^{(+)}_{mn}v_mA_n$$
can be proven similarly for an anti-self-dual $2-$form.

\medskip

Let us now prove the following identity
\beqa
\label{dualprop3}
i \varepsilon_{m n p q}\,  \partial_r \, R^{(\pm)\, rq} &=&  \mp (\,
\partial_m R^{(\pm)}_{n p} +\partial_n R^{(\pm)}_{p m} + 
                                   \partial_p R^{(\pm)}_{m n} \,)
\eeqa
\noi
For doing this, we start with LHS and we will find the RHS. Firstly, insert \eqref{2} in the LHS of \eqref{dualprop3}; we thus obtain
\beqa
\label{dualprop3-1}
i \varepsilon_{m n p q}\,  \partial^r \, R^{(\pm)\, rq} = \mp \frac 12 \varepsilon_{m n p q} \e^{rqst} \partial_r R^{(\pm)}_{st}
\eeqa
\noi
Using the antisymmetry of the Levi-Civita tensor and the identity $\e_{mnpq}\e^{rstq}=\delta_{mnp}^{rst}$ (defined in \eqref{gros-delta}) we get
\beqa
\label{dualprop3-11}
i \varepsilon_{m n p q}\,  \partial_r \, R^{(\pm)\, rq} = \mp \frac 12 
[\delta_m^r (\delta_n^s \delta^t_p - \delta_p^s \delta^t_n)- \delta_n^r ((\delta_m^s \delta^t_p - \delta_p^s \delta^t_m)+\delta_p^r (\delta_m^s-\delta_n^t -\delta_p^s \delta_m^t)] \partial _r R^{(\pm)}_{st}.
\eeqa
\noi
Now, using the antisymmetry property of the $2-$form $R^{(\pm)}_{st}$, the result \eqref{dualprop3-1} is obtained.

\medskip

Arguing along the  same line one also proves
\beqa
\label{dualprop3-bis}
i \varepsilon_{m n p q}\,  v_r \, R^{(\pm)\, rq} &=&  \mp (\,
v_m R^{(\pm)}_{n p} +v_n R^{(\pm)}_{p m} + 
                                   v_p R^{(\pm)}_{m n} \,)
\eeqa



Now, from  \eqref{dualprop3} and  \eqref{dualprop3-bis} one gets directly 
\beqa
\label{prop}
v^r \partial_{[m} R^{(\mp)}_{n]_\pm r} -v_{[m} \partial^r R^{(\mp)}_{n]_\pm r}=0
\eeqa
\noindent
This identity is of special importance and it is used several times in the calculations of chapter $5$.

Before ending this appendix, let us also recall the well-known identity
\beqa
 \label{dualprop1} 
R^{(\pm)}_{mn}R'^{(\mp) mn}&=&0.
\eeqa

\chapter{Spinors and products of spinors in $D$ dimensions}
\renewcommand{\theequation}{E.\arabic{equation}}   
\setcounter{equation}{0}
\label{a-spinori}

In this appendix we recall some useful definitions for spinors in $D$ dimensions. Moreover, formulae for the decomposition of the product of spinors on $p-$forms are then exhibited.

In $D$ dimensions, the Clifford algebra writes
\beqa
\label{Gamma}
\{\Gamma_M, \Gamma_N \}= 2 \eta_{MN}
\eeqa
\noi where the $D-$dimensional Minkowski metric is given by
\begin{eqnarray}
\eta_{MN}= \mathrm{diag} (1, &\underbrace{-1,\ldots,-1}&)  \\
&D-1& \nonumber 
\end{eqnarray}
\noi 
This algebra has $2^{[\frac D2 ]}$-dimensional representations on the Dirac spinor space (denoted by $\cal S$).
The  $\Gamma$ matrices are thus generalisations of the Dirac matrices $\gamma_m$ in $4$ dimensions and a Dirac spinor (element of the representation space $\cal S$) is a $2^{[\frac D2]}-$dimensional object.

The $\Gamma$ matrices can be obtained from the Pauli matrices $\sigma_m$ in $4$ dimensions. 
Indeed, for $k=[\frac D2]$,
 define $2^{k-1}$-dimensional $\Sigma$ matrices by
\beqa
\label{taurile}
\Sigma_0&=&1\nonumber\\
\Sigma_{2\ell -1}&=&\sigma_3^{\otimes (\ell -1)}\otimes\sigma_1 \otimes\sigma_0^{(k-1-\ell)},\ \, \ell=1,\dots,k-1\nonumber\\
\Sigma_{2k-1}&=&\sigma_3^{\otimes (k-1)}\\
\Sigma_{2\ell}&=&\sigma_3^{\otimes (\ell -1)}\otimes\sigma_2 \otimes\sigma_0^{(k-1-\ell)},\ \, \ell=1,\dots,k\nonumber
\eeqa
\noi
where by $\sigma_3^{\otimes k}$ we have denoted the $k-$th tensorial power of the matrix $\sigma_3$, {\it etc.}. Note that one has $2k$ such $\Sigma$ matrices.
By recurrence arguments one proves that  $\{\Sigma_I, \Sigma_J\}=2 \delta_{IJ}$, $\Sigma_I$ being thus the generators of the Clifford algebra $SO(2k-1)$.
Define now $\Sigma_M=(\Sigma_0, \Sigma_I)$, $\bar \Sigma_M =(\bar \Sigma_0=\Sigma_0, \bar \Sigma_I=-\Sigma_I)$ (with $I=1,\dots,2k-1$)
(generalisation
 to arbitrary dimensions of the Pauli matrices).


Consider now even dimensions, namely
$$D=2k.$$
It is always possible to write
the Dirac matrices $\Gamma$ under the form 
\beqa
\label{Gamma-mare}
\Gamma_M=  \begin{pmatrix}0&\Sigma_M \cr \bar \Sigma_M&0 \end{pmatrix},
\eeqa
\noi

A basis of the representation space is given by the set of antisymmetric matrices
\begin{eqnarray}
\label{gamma2}
\Gamma^{(\ell)}: \Gamma_{M_1 \ldots M_\ell} =
\frac{1}{\ell !} \sum \limits_{\sigma \in S_\ell} \e(\sigma)
\Gamma_{M_{\sigma(1)} }\ldots  
\Gamma_{M_{\sigma(\ell )} }   
\end{eqnarray}   
\noindent 
where $\ell=0,\dots,D$, and we have denoted by $S_\ell$ the set of the permutations of $\ell$ elements and by $\e(\sigma)$ the signature of the permutation $\sigma$. 

Amongst these matrices of special importance are the Lorentz generators (note the difference of convention $\Gamma_{MN}=\frac12 \Gamma_{MN}^{(2)}$ )
\beqa
\label{gmn-D}
\Gamma_{MN}=\frac14 [\Gamma_{M},\Gamma_{N}],
\eeqa
\noi

Notice now that, for  $D$ even, inserting \eqref{Gamma-mare} in the expression \eqref{gamma2}, one obtains a further simplification of the $\Gamma$ matrices
\begin{eqnarray}
\label{gamma}
\Gamma_{ M_1 \ldots M_{2 \ell}  }& =& \frac{1}{\ell !}\begin{pmatrix}
\Sigma_{M_1} \bar \Sigma_{M_2} \ldots \Sigma_{M_{2 \ell-1}} 
\bar \Sigma_{M_{2 \ell}} 
+\mathrm{~perm}&0 \cr
0&  \bar  \Sigma_{M_1}  \Sigma_{M_2} \ldots \bar 
\Sigma_{M_{2 \ell-1}} \Sigma_{M_{2 \ell}}
+\mathrm{~perm}
 \end{pmatrix}
\nonumber \\
&=& \frac{1}{\ell !} \begin{pmatrix}  \Sigma_{{M_1} \ldots M_{2 \ell}}&0 \cr
0&\bar  \Sigma_{{M_1} \ldots M_{2 \ell}} \end{pmatrix} 
\nonumber \\ 
\\ \nonumber \\
 \Gamma_{ M_1 \ldots M_{2 \ell + 1 }  }& =& \begin{pmatrix} 
0& \Sigma_{M_1} \bar \Sigma_{M_2} \ldots  \bar 
\Sigma_{M_{2 \ell  } }\Sigma_{M_{2 \ell + 1 }}
+\mathrm{~perm}\cr
  \bar  \Sigma_{M_1}  \Sigma_{M_2} \ldots  \Sigma_{M_{2 \ell  }}
\bar  \Sigma_{M_{2 \ell + 1}}
+\mathrm{~perm}&0
 \end{pmatrix} \nonumber \\ 
&=&\frac{1}{\ell !}\begin{pmatrix}  0 &  \Sigma_{{M_1} \ldots M_{2 \ell +1 }} \cr
\bar  \Sigma_{{M_1} \ldots M_{2 \ell + 1}} &0 \end{pmatrix}.
\nonumber
\end{eqnarray} 

\noi 
where, as before, {\it perm.} means sum on all permutations with the sign corresponding 
to the signature. Furthermore,
 the definitions of the $\Sigma^{(\ell)}, \bar \Sigma^{(\ell)}$ matrices can be deduced 
from the equalities above.
  For instance,
 $\Sigma_{M_1 \ldots M_{2\ell}} =
\sum \limits_{\sigma \in  S_\ell} 
\Big( \prod \limits_{i=1}^\ell \Sigma_{M_{\sigma(2i-1)}}  
\bar \Sigma_{M_{\sigma(2 i)}} \Big)$
and similarly for the other matrices.

The matrices \eqref{gamma2} are also subject to the identity
\beqa
\label{prodGamma}
\Gamma^{M_1 \ldots M_\ell} \Gamma^M = 
\Gamma^{M_1 \ldots M_\ell M} + \eta^{M_\ell M} \Gamma^{M_1\ldots  M_{\ell -1}}+
\ldots + (-1)^{\ell -1} \eta^{M_1 M} \Gamma^{M_2 \ldots M_\ell}
\eeqa

Using the relations \eqref{gamma} between the $\Gamma^{(\ell)}$ and the $\Sigma^{(\ell)}$ matrices, similar identities hold
\beqa
\label{identities}
\Sigma_{M_1 \ldots M_{2k}}  \Sigma_{M_{2k+1}}&=&
\Sigma_{M_1 \ldots M_{2k} M_{2k+1}} +
 \eta_{M_{2k} M_{2k+1}}\Sigma_{M_1 \ldots M_{2k-1}} + {\rm perm.}
\nonumber \\
\Sigma_{M_1 \ldots M_{2k+1}}  \bar \Sigma_{M_{2k+2}}&=&
\Sigma_{M_1 \ldots M_{2k+1} M_{2k+2}} +
 \eta_{M_{2k+1} M_{2k+2}}\Sigma_{M_1 \ldots M_{2k}} + {\rm perm.}
\eeqa
\noindent 
for $k <n$ 
Similar relations hold for the $\bar \Sigma^{(k)}$ matrices.

Denote now by ${\cal S}^*$ the dual representation of ${\cal S}$ (on which the matrices $-\Gamma^{t}_{MN}$ act in the same way the matrices $\Gamma_{MN}$ act on $\cal S$). One can find an element
\beqa
\label{calC-D}
{\cal C}=\Gamma_0 \Gamma_2 \dots \Gamma_{2k-2}
\eeqa
 of ${\rm End}({\cal S})$ ($\cal C$ being called the charge conjugation matrix) such that 
$${\cal C} \Gamma_{M}{\cal C}^{-1}=-\Gamma_{M}^t$$
and hence 
\beqa
\label{dual-spinor-D}
{\cal C} \Gamma_{MN}{\cal C}^{-1}=-\Gamma_{MN}^t.
\eeqa
\noi
Hence, these representations ($\cal S$ and $\cal S^*$) are in fact equivalent and
$\psi^t\cal C\in {\cal S}^*$ for any Dirac spinor $\psi\in \cal S$. One can thus see the charge conjugation matrix $\cal C$ as some intertwining operator.

Furthermore, if a spinor $\psi$ transforms under a Lorentz transformation by $S(\Lambda)=e^{\frac 12 \Lambda^{mn}\Gamma_{mn}}$ (with $\Lambda^{mn}$ the parameter of the transformation) then $\Xi^t{\cal C}\in {\cal S}^*$ and transforms under a Lorentz transformation by $S(\Lambda)^{-1}$. Hence
\beqa
\label{sc-D}
\Xi^t{\cal C}\psi
\eeqa
\noi
is a spinor invariant. Hence
\beqa
\label{L-TENSOR-D}
\Xi^t{\cal C}\Gamma^{(\ell)}_{M_1\dots M_\ell}\psi
\eeqa
\noi
transforms as an antisymmetric tensor of order $\ell$.

\medskip

Now, since ${\rm End}({\cal S})\cong {\cal S}\otimes {\cal S}^*$, one can decompose any product of Dirac spinors $\psi\otimes \psi'^t {\cal C}$ on the basis \eqref{gamma2}.
Since the coefficients of this development are 
  the antisymmetric tensors of order $\ell$ (with $\ell=0,\dots,D$), one  writes this schematically as
\beqa
\label{4-mare-D}
\psi \otimes \psi'^t{\cal C}=[0]\oplus\dots\oplus [D].
\eeqa
\noi
where $[\ell]$ denotes an $\ell-$form. 
Using the Hogde equivalence $[\ell]\equiv[4-\ell]$, one can choose to write \eqref{4-mare-D} as
\beqa
\label{4-mic-D}
\psi \otimes \psi'^t{\cal C}=[0]^2\oplus\dots\oplus[k-1]^2\oplus[k].
\eeqa
\noi

\medskip

Let us now the chirality matrix $\Gamma_D$ as
\beqa
\label{gama5-D}
\Gamma_{D+1}=i^{\frac D2 +1}\Gamma_0\dots\Gamma_{D-1}.
\eeqa
\noi
From this definition and \eqref{Gamma} one proves
\beqa
\label{prop-gama5-1-D}
\{\Gamma_M,\Gamma_{D+1}\}=0,\ \Gamma_{D+1}^2=1
\eeqa
\noi
which, using \eqref{gamma2}, gives
\beqa
\label{prop-gama5-2-D}
\Gamma^{(\ell)}\Gamma_{D+1}=(-1)^{\ell}\Gamma_{D+1} \Gamma^{(\ell)}.
\eeqa
\noi
Moreover, in our representation \eqref{Gamma-mare} it follows from the definition \eqref{gama5-D}, that  $\Gamma_{D+1}$ is block-diagonal 
\beqa
\label{gama5-Weyl-D}
\Gamma_{D+1}=\begin{pmatrix} -1 & 0 \\ 0 & 1 \end{pmatrix}.
\eeqa
\noi
A further property of $\Gamma_{D+1}$ is
\beqa
\label{prop-gama5-3-D}
\Gamma_{D+1}^t=\Gamma_{D+1}.
\eeqa
\noi

\medskip

As in $4$ dimensions, the chirality matrix is used to define the irreducible $2^{\frac D2 -1}-$dimensional LH and RH Weyl spinors; a $2^{\frac D2}-$dimensional Dirac spinor decomposes as 
$$ \psi_D= \begin{pmatrix} \psi_+ \\  \psi'_- \end{pmatrix} $$
where $\psi_+$ (and resp. $\psi'_-$) is a  LH (resp. RH) Weyl spinor. 
From \eqref{gama5-Weyl-D} one sees that a Dirac spinor with only LH (resp. RH) components is an eigenstate of $\Gamma_{D+1}$ with eigenvalue $-1$ (resp. $+1$); this writes
\beqa
\label{adevaru-D}
 \Gamma_{D+1}\psi_\e = -\e \psi_{D\e}, \mbox{ with } \e=\pm.
\eeqa

\medskip

Let us now  calculate products of such two distinct Weyl spinors, as we did in subsection $5.3.3$ for $D=4$. Consider
\beqa
\label{tensor-D}
T_{M_1\dots M_\ell }= \psi'^t_{D\e_2} {\cal C} \Gamma^{(\ell)}_{M_1\dots M_\ell}\psi_{D\e_1}.
\eeqa
\noi
(Recall that by \eqref{L-TENSOR-D}, \eqref{tensor-D} above transforms like a tensor of order $\ell$). 
Using \eqref{adevaru-D} one has
\beqa
T_{M_1 \dots M_\ell }=-\e_1 \psi'^t_{D \e_2} {\cal C} \gamma^\ell_{M_1\dots M_\ell}\Gamma_{D+1}\psi_{D \e_1}.
\eeqa
\noi
Using now \eqref{prop-gama5-2-D} one gets
\beqa
T_{M_1\dots M_\ell }=-(-1)^\ell\e_1 \psi'^t_{D \e_2} {\cal C} \Gamma_{D+1}\Gamma^\ell_{M_1 \dots M_\ell}\psi_{D \e_1}.
\eeqa
\noi
which furthermore writes
\beqa
T_{M_1 \dots M_\ell}=-(-1)^{\ell+k}\e_1 \psi'^t_{D \e_2} \Gamma_{D+1}{\cal C}\Gamma^\ell_{M_1\dots M_\ell}\psi_{D \e_1}.
\eeqa
\noi
Using now \eqref{prop-gama5-3-D} one has
\beqa
T_{M_1 \dots M_\ell}=-(-1)^{\ell+k}\e_1 (\Gamma_{D+1} \psi'_{D \e_2})^t {\cal C}\Gamma^\ell_{M_1 \dots M_\ell}\psi_{D\e_1}.
\eeqa
\noi
Finally, making use again of \eqref{adevaru-D} one finds
\beqa
T_{M_1 \dots M_\ell}=(-1)^\ell\e_1 \e_2\psi'^t_{D \e_2} {\cal C}\Gamma^{(\ell)}_{M_1\dots M_\ell}\psi_{D \e_1}.
\eeqa
\noi
Hence $T_{M_1 \dots M_\ell}$ is vanishing if $(-1)^{\ell+k}\e_1 \e_2=-1$.
Consider from now on the case 
$$k=2n,$$ 
that is 
$$D=4n.$$
Hence, the decomposition \eqref{4-mic-D} reduces, at the level of Weyl spinor products at
\beqa
\label{bi-spin2}
\psi_+ \otimes  \psi_+^{\prime t}{\cal C} &=&
 [0]  \oplus [2]\oplus\dots \oplus [2n]^{(+)}\nonumber \\
{\psi}_- \otimes  {\psi}_-^{\prime t}\cal C &=&
 [0] \oplus [2]\oplus\dots \oplus [2n]^{(-)}\\
{\psi}_+ \otimes  {\psi}_-^{\prime t}\cal C &=&  
 [1] \oplus [3] \oplus [2n-1] \nonumber 
\eeqa
\noindent
where 
 $[2n]^{(\pm)}$ represents an (anti-)self-dual
$(2n)-$form. Indeed, the $(2n)$-form in $4n$ dimensions is reducible, decomposing (like the $2-$form in $4$ dimensions) as
\beqa
\label{reductibil-Lorentz-D}
[2n]=[2n]^{(+)}\oplus [2n]^{(-)}.
\eeqa
\noi
This equation is also correct from the point of view of the dimension 
The same argument of dimension counting in \eqref{bi-spin2} shows that one must have considered the (anti-)self-dual $2-$forms. 


We now apply the decomposition \eqref{bi-spin2} for the case of the ${\mathbf \Xi}_{++}=\Psi_+ \otimes \Omega_+ = \begin{pmatrix} \Xi_{1++} \\ \bar
  \Xi_{2-+} \\ \Xi_{3++} \end{pmatrix}$. Using \eqref{Gamma-mare} and \eqref{gamma2} this decomposition writes explicitly
\beqa
\label{spin-p2}
\Xi_{1++}&=&\sum \limits_{p=0}^{n-1} \frac{1}{(2p)!} A_{[2 p]} \Sigma^{(2p)}
+ \frac12\frac{1}{(D/2)!} A_{[2n]^{(+)}} \Sigma^{(D/2)} \nonumber \\
\Xi_{2-+}&=&\sum \limits_{p=0}^{n-1} \frac{1}{(2p+1)!} \tA{}_{[2p+1]} \bar 
\Sigma^{(2p+1)}\nonumber\\
\Xi_{3++}&=&\sum \limits_{p=0}^{n-1} \frac{1}{(2p)!} \ttA_{[2 p]} \Sigma^{(2p)}
+ \frac12\frac{1}{(D/2)!} \ttA_{[2n]^{(+)}} \Sigma^{(D/2)}
\eeqa
\noi

Conversely, using trace identities of the $\Sigma$ matrices, one writes
\beqa
\label{p-spin2}
A_{[2p]}&=&  \frac{1}{2^{2n-1}} \mathrm{Tr}\left(\Sigma^{(2p)} \Xi_{1++}\right),
\nonumber \\
\tA_{[2p+1]}&=&  \frac{1}{2^{2n-1}} \mathrm{Tr}\left(\bar \Sigma^{{(2p+1)}} 
{\Xi}_{2+-}\right)\nonumber\\
\ttA_{[2p]}&=&  \frac{1}{2^{2n-1}} \mathrm{Tr}\left(\Sigma^{(2p)} \Xi_{3++}\right)
\eeqa
\noindent
and similar for the $\ttA_{2p}$ forms (here we have written $ \mathrm{Tr}\left(\bar \Sigma^{{(2p+1)}} {\Xi}_{i+-}\right)=
\mathrm{Tr}\left(\bar \Sigma_{M_1 \ldots M_{2p+1}} { \Xi}_{i+-}\right)$ {\it etc.} to simplify 
notations).
To prove these formulae one reinserts the original decompositions \eqref{spin-p2} thus checking \eqref{p-spin2}.

Let us end by giving a useful trace formulae in $D=4n$ dimensions
\beqa
\label{puscarie}
\frac{1}{2^{2n-1}}{\rm Tr}(\Sigma_{M_1\dots M_{2n}}\Sigma^{MN_1\dots N_{2p+1}})=
\delta_{2n}^{2p+2}\left(\delta_{M_1\dots M_{2n}}^{N_{2n-1}\dots N_1 M}- i \e_{M_1\dots M_{2n} P_{2n-1}\dots P_1 P} \eta^{MP}\eta^{N_1P_{2n-1}}\dots\eta^{N_{2n-1} P_1}\right).\nonumber \\
\eeqa


\newpage

\cleardoublepage

\end{document}